\documentclass[12pt]{article}
\pdfoutput=1
\usepackage[nosort]{cite}

\usepackage{epsfig}
\usepackage{amsfonts}
\usepackage{amscd}
\usepackage{latexsym}
\usepackage{amsmath,amssymb,mathtools}
\usepackage{verbatim}
\usepackage{setspace}
\usepackage[dvipsnames]{xcolor}
\usepackage{fancyhdr}
\usepackage{cite}
\usepackage{xcolor}
\usepackage{tikz}
\usepackage{tikz-cd}
\usetikzlibrary{decorations.markings}
\usepackage{tensor}
\usepackage{slashed}
\usepackage{multirow}

\usetikzlibrary{calc}
\usetikzlibrary{topaths}
\usetikzlibrary{decorations}
\usetikzlibrary{decorations.pathmorphing}
\usetikzlibrary{arrows,decorations.markings,cd}
\usetikzlibrary{calc,arrows,cd,decorations.markings,snakes}
\usetikzlibrary{positioning}
\tikzset{
->-/.style args={#1rotate#2}{decoration={markings, mark=at position #1 with {\arrow[scale=1.5,rotate = #2 ]{stealth}}}, postaction={decorate}}
}
\usetikzlibrary{shapes.geometric}
\usetikzlibrary{knots}
\usepackage{tikz-3dplot}

\usepackage{cancel}

\tikzset{snake it/.style={decorate, decoration=snake}}

\usepackage{hyperref}
\hypersetup{
    colorlinks=true,
    linkcolor=Maroon,
    citecolor=MidnightBlue,
    urlcolor=MidnightBlue,
    linktoc=page
}

\usepackage{draft}
\usepackage{graphicx,color,subcaption}
\usepackage{cite}
\usepackage{mciteplus}
\usepackage{skak}
\usepackage{bbm}
\usepackage[english]{babel}
\usepackage{amsthm}
\usepackage{soul}
\usepackage{makecell}
 
\numberwithin{equation}{section}

\usepackage{ytableau}
\ytableausetup{boxsize=0.8em,aligntableaux=center}

\def\bZ{\mathbb{Z}}

\def\Hom{\mathrm{Hom}}

\def\f{\phi}
\def\({\left(}
\def\){\right)}

\def\b{\mathbf{B}}

\def\f{\phi}

\def\Rep{\text{Rep}}
\def\Mod{\text{Mod}}
\def\Fer{\text{Fer}}
\def\hookleftconf{\stackrel{\text{conf}}{\hookleftarrow}}
\def\hookrightconf{\stackrel{\text{conf}}{\hookrightarrow}}

\definecolor{PPTpurple}{RGB}{112, 48, 160}
\definecolor{PPTred}{RGB}{255, 0, 0}
\definecolor{PPTblue}{RGB}{91, 155, 213}
\definecolor{PPTgreen}{RGB}{84, 130, 53}
\definecolor{PPTorange}{RGB}{197, 90, 17}
\definecolor{PPTdarkred}{RGB}{192, 0, 0}
\definecolor{PPTdarkblue}{RGB}{47,85,151}

\def\PPTdarkred#1{{\color{PPTdarkred}{#1}}}

\usepackage{graphicx}
\newcommand{\nwhookarrow}{%
  \mathrel{\rotatebox[origin=c]{135}{$\hookrightarrow$}}%
}
\newcommand{\nehookarrow}{%
  \mathrel{\rotatebox[origin=c]{45}{$\hookrightarrow$}}%
}

\tikzset{
  sw/.style={x={(-0.5cm,-0.25cm)}, y={(1cm,0cm)}, z={(0cm,1cm)},
             line join=round, line cap=round, font=\footnotesize,
             baseline={([yshift=-0.5ex]current bounding box.center)}},
  flat/.style={line join=round, line cap=round, font=\footnotesize,
             baseline={([yshift=-0.5ex]current bounding box.center)}},
  flatD/.style={flat, baseline={(0,-0.09)}},
  topoA/.style={fill=red!15,opacity=0.5,dashed},
  topoB/.style={fill=orange!25,opacity=0.5,dashed},
  chiralA/.style={fill=blue!20,opacity=0.5,dashed,preaction={draw=white,line width=3pt}},
  chiralB/.style={fill=violet!25,opacity=0.5,dashed,preaction={draw=white,line width=3pt}},
  ifc/.style={fill=red!12,opacity=0.6,draw=PPTdarkred},
  invbulk/.style={densely dashed,gray},
  dline/.style={very thick,PPTdarkred},
  oline/.style={thick,PPTdarkblue},
  marr/.style={decoration={markings, mark=at position 0.55 with {\arrow[scale=1.3]{stealth}}}, postaction=decorate},
  pt/.style={circle,fill,inner sep=1.1pt},
  tifcA/.style={fill=green!20,opacity=0.5,dashed},
  tifcB/.style={fill=teal!25,opacity=0.5,dashed},
}
\newcommand{\swwall}[4]{\draw[#1] (0,#2,#3) -- (2,#2,#3) -- (2,#2,#4) -- (0,#2,#4) -- cycle;}
\newcommand{\swedges}[3]{\foreach \zz in {0,2}{\draw[#1] (0,#2,\zz) -- (0,#3,\zz); \draw[#1] (2,#2,\zz) -- (2,#3,\zz);}}
\def\swLL{1.5}
\def\swL{3}
\newcommand{\swLeft}[1]{\swedges{invbulk}{-\swLL}{0}\node[align=center] at (2,{0.4-\swLL},1.3) {#1};}
\newcommand{\swUN}{\swLeft{$U(N)_1$}}
\newcommand{\hatchedbdy}[2]{\draw[thick] (#1,0) -- (#1,#2); \foreach \s in {0.15,0.4,...,#2} \draw (#1,\s) -- ($(#1,\s)+(-0.2,-0.2)$);}

\newcommand{\figScatteringPuzzle}{%
\begin{tikzpicture}[flat]
  \fill[blue!15] (0,0) rectangle (2.8,3.2);
  \hatchedbdy{0}{3.1}
  \node[pt,PPTdarkblue,label={[PPTdarkblue]below right:$\psi_L$}] at (2.1,0.5) {};
  \draw[thick,PPTdarkblue,-stealth] (2.1,0.5) -- (1.2,1.35);
  \node[pt] at (1.2,1.95) {};
  \draw[thick,-stealth] (1.2,1.95) -- (2.0,2.6) node[above right] {$?$};
\end{tikzpicture}}

\newcommand{\figFoldingBdy}[1]{%
  \fill[blue!15] (0,0) rectangle (2.5,3);
  \hatchedbdy{0}{2.95}
  \node at (1.3,2.6) {$N$ Dirac};
  \node[pt,PPTpurple,label={[PPTpurple]right:#1}] at (0.35,1.4) {};
  \node at (1.25,-0.4) {$\ket{\mathcal{D}}$};}
\newcommand{\figFoldingDef}[2]{%
  \fill[blue!15] (0,0) rectangle (3.4,3);
  \draw[dline,marr] (1.7,0) -- (1.7,3);
  \node at (0.85,2.6) {$N$ Weyl}; \node at (2.55,2.6) {$N$ Weyl};
  \node[pt,PPTpurple] (p) at (#1,1.4) {}; #2
  \node at (1.7,-0.4) {$\mathcal{D}$};}
\newcommand{\figFolding}{%
\begin{tikzpicture}[flat]
  \begin{scope}\figFoldingBdy{$f_1(j)$}\end{scope}
  \node at (2.95,1.5) {$=$};
  \begin{scope}[xshift=3.6cm]\figFoldingBdy{$f_2(j)$}\end{scope}
  \node at (6.8,1.5) {$\leftrightsquigarrow$};
  \begin{scope}[xshift=7.5cm]\figFoldingDef{2.0}{\node[PPTpurple,right] at (p.east) {$f_1(j)$};}\end{scope}
  \node at (11.35,1.5) {$=$};
  \begin{scope}[xshift=11.8cm]\figFoldingDef{1.4}{\node[PPTpurple,left] at (p.west) {$f_2(j)$};}\end{scope}
\end{tikzpicture}}

\newcommand{\figSandwichLeft}[1][1.4]{%
\begin{tikzpicture}[sw,scale=#1]
  \swUN
  \swwall{chiralA}{0}{0}{2}
  \draw[dline,marr] (2,0,0.9) -- (0,0,0.9);
  \node[PPTdarkred,left] at (2,0,0.85) {$a$};
  \node[pt,PPTdarkblue,label={[PPTdarkblue]right:$\mathcal{O}$}] at (1.3,0,1.6) {};
  \node[pt,PPTgreen,label={[PPTgreen]right:$f_i(j)$}] at (1.3,0,0.55) {};
  \node[below] at (2,0,0) {$\Fer(N)$};
\end{tikzpicture}}
\newcommand{\figSandwichRight}[1][1.4]{%
\begin{tikzpicture}[sw,scale=#1]
  \swUN
  \swwall{topoA}{0}{0}{2}
  \swedges{dashed}{0}{3}
  \draw[dline,marr] (2,0,0.9) -- (0,0,0.9);
  \node[PPTdarkred,left] at (2,0,0.6) {$a$};
  \node[pt,PPTdarkblue,label={[PPTdarkblue]left:$x$}] at (1.3,0,1.35) {};
  \draw[oline,marr] (1.3,0,1.35) -- (1.3,3,1.35);
  \node[PPTdarkblue,above] at (1.3,1.5,1.35) {$\lambda$};
  \swwall{chiralA}{3}{0}{2}
  \node[pt,PPTdarkblue,label={[PPTdarkblue]right:$\widetilde{\mathcal{O}}$}] at (1.3,3,1.35) {};
  \node[pt,PPTgreen,label={[PPTgreen]right:$j$}] at (1.3,3,0.55) {};
  \node[below] at (2,0,0) {$B_{L_i}$};
  \node[below] at (2,1.5,0) {$\Rep(\widehat{\mathfrak{g}})$};
  \node[below] at (2,3,0) {$B_{\widehat{\mathfrak{g}}}$};
\end{tikzpicture}}

\newcommand{\swChiralWithD}[1]{%
  \swUN
  \swwall{chiralA}{0}{0}{2}
  \draw[dline,marr] (2,0,#1) -- (0,0,#1);
  \node[PPTdarkred,left] at (2,0,#1) {$\mathcal{D}$};
  \node[below] at (2,0,0) {$\Fer(N)$};}
\newcommand{\swSplitBackGen}[6]{%
  \swLeft{#6}
  \swwall{topoA}{0}{0}{#1}
  \swwall{topoB}{0}{#1}{2}
  \swedges{dashed}{0}{\swL}
  \draw[dline,marr] (2,0,#1) -- (0,0,#1);
  \node[PPTdarkred,left] at (2,0,#1) {#2};
  \node[left] at (2,0,1.75) {#4};
  \node[left] at (2,0,0.25) {#3};
  \node[below] at (2,{\swL/2-0.1},0) {#5};}
\newcommand{\swSplitBack}[1]{\swSplitBackGen{#1}{$\mathcal{D}$}{$B_{L_1}$}{$B_{L_2}$}{$\Rep(\widehat{\mathfrak{g}})$}{$U(N)_1$}}
\newcommand{\swSplitFrontGen}[1]{%
  \swwall{chiralA}{\swL}{0}{2}
  \node[below] at (2,\swL,0) {#1};}
\newcommand{\swSplitFront}{\swSplitFrontGen{$B_{\widehat{\mathfrak{g}}}$}}

\newcommand{\figFtwelveLeft}[1][1.4]{%
\begin{tikzpicture}[sw,scale=#1]
  \swChiralWithD{1}
  \node[pt,PPTpurple,label={[PPTpurple]right:$f_2(j)$}] at (1.3,0,1.6) {};
  \node[pt,PPTpurple,label={[PPTpurple]right:$f_1(j)$}] at (1.3,0,0.5) {};
\end{tikzpicture}}
\newcommand{\figFtwelveRight}[1][1.4]{%
\begin{tikzpicture}[sw,scale=#1]
  \swSplitBack{1}
  \swSplitFront
  \node[pt,PPTpurple,label={[PPTpurple]right:$j$}] at (1.3,3,1.5) {};
  \node[pt,PPTpurple,label={[PPTpurple]right:$j$}] at (1.3,3,0.5) {};
\end{tikzpicture}}

\newcommand{\figBeforeScatter}[1][1.4]{%
\begin{tikzpicture}[sw,scale=#1]
  \swChiralWithD{1.15}
  \node[pt,PPTdarkblue,label={[PPTdarkblue]right:$\psi$}] at (1.3,0,0.55) {};
\end{tikzpicture}}
\newcommand{\figAfterScatterGen}[3]{%
\begin{tikzpicture}[sw,scale=#1]
  \swChiralWithD{0.8}
  \draw[oline,marr] (1.3,0,0.8) -- (1.3,0,1.55);
  \node[pt,PPTdarkblue,label={[PPTdarkblue]below right:$#3$}] at (1.3,0,0.8) {};
  \node[PPTdarkblue,left] at (1.3,0,1.18) {$#2$};
  \node[pt,PPTdarkblue,label={[PPTdarkblue]right:$\widetilde{\psi}$}] at (1.3,0,1.55) {};
\end{tikzpicture}}
\newcommand{\figAfterScatter}[1][1.4]{\figAfterScatterGen{#1}{b}{w}}

\newcommand{\figDerivationA}[1][1.4]{%
\begin{tikzpicture}[sw,scale=#1]
  \swSplitBack{1.25}
  \node[pt,PPTdarkblue,label={[PPTdarkblue]left:$x$}] at (1.3,0,0.6) {};
  \draw[oline,marr] (1.3,0,0.6) -- (1.3,3,0.6);
  \node[PPTdarkblue,above] at (1.3,1.5,0.6) {$\lambda$};
  \swSplitFront
  \node[pt,PPTdarkblue,label={[PPTdarkblue]right:$\widetilde{x}$}] at (1.3,3,0.6) {};
\end{tikzpicture}}
\newcommand{\figDerivationB}[1][1.4]{%
\begin{tikzpicture}[sw,scale=#1]
  \swSplitBack{0.7}
  \draw[oline,marr] (1.3,0,0.7) -- (1.3,0,1.35);
  \node[pt,PPTdarkblue,label={[PPTdarkblue]below right:$w$}] at (1.3,0,0.7) {};
  \node[PPTdarkblue,left] at (1.3,0,1.02) {$b$};
  \node[pt,PPTdarkblue,label={[PPTdarkblue]left:$z$}] at (1.3,0,1.35) {};
  \draw[oline,marr] (1.3,0,1.35) -- (1.3,3,1.35);
  \node[PPTdarkblue,above] at (1.3,1.5,1.35) {$\lambda$};
  \swSplitFront
  \node[pt,PPTdarkblue,label={[PPTdarkblue]right:$\widetilde{x}$}] at (1.3,3,1.35) {};
\end{tikzpicture}}

\newcommand{\swGeneralBack}{%
  \swUN
  \swwall{topoA}{0}{0}{1}
  \swwall{topoB}{0}{1}{2}
  \swedges{dashed}{0}{3.5}
  \draw[ifc] (0,0,1) -- (2,0,1) -- (2,3.5,1) -- (0,3.5,1) -- cycle;
  \node[PPTdarkred] at (0.25,1.0,1) {$\mathcal{I}$};
  \draw[dline,marr] (2,0,1) -- (0,0,1);
  \node[PPTdarkred,left] at (2,0,1) {$J$};
  \node[left] at (2,0,1.75) {$B_{L_2}$};
  \node[left] at (2,0,0.25) {$B_{L_1}$};}
\newcommand{\swGeneralFront}{%
  \swwall{chiralA}{3.5}{0}{1}
  \swwall{chiralB}{3.5}{1}{2}
  \draw[dline,marr] (2,3.5,1) -- (0,3.5,1);
  \node[PPTdarkred,right] at (0,3.5,1) {$\widetilde{J}$};}
\newcommand{\swGeneralLabels}{%
  \node[anchor=south] (Rtwo) at (0,1.75,2.45) {$\Rep(\widehat{\mathfrak{g}})\boxtimes \Rep(S_2)$};
  \draw[->,gray] (Rtwo.south) to[out=-90,in=90] (1,2.0,1.55);
  \node[anchor=north] (Rone) at (2,2.0,-0.45) {$\Rep(\widehat{\mathfrak{g}})\boxtimes \Rep(S_1)$};
  \draw[->,gray] (Rone.north) to[out=90,in=-90] (1,2.0,0.45);
  \node[anchor=west] at (0,3.55,1.75) {$B_{\widehat{\mathfrak{g}}\oplus S_2}$};
  \node[anchor=west] at (0,3.55,0.25) {$B_{\widehat{\mathfrak{g}}\oplus S_1}$};}
\newcommand{\figGeneralLeft}[1][1.4]{%
\begin{tikzpicture}[sw,scale=#1]
  \swChiralWithD{1}
\end{tikzpicture}}
\newcommand{\figGeneralRight}[1][1.4]{%
\begin{tikzpicture}[sw,scale=#1]
  \swGeneralBack
  \swGeneralFront
  \swGeneralLabels
\end{tikzpicture}}

\newcommand{\figDerivationMid}[1][1.4]{%
\begin{tikzpicture}[sw,scale=#1]
  \swSplitBack{0.75}
  \node[pt,PPTdarkblue,label={[PPTdarkblue]left:$x$}] at (1.3,0,0.35) {};
  \draw[oline,marr] (1.3,0,0.35) .. controls (1.3,0.95,0.35) and (1.3,0.95,1.15) .. (1.3,0,1.15);
  \node[PPTdarkblue,right] at (1.3,0.72,0.75) {$\lambda$};
  \draw[oline,marr] (1.3,0,1.15) -- (1.3,0,1.6);
  \node[pt,PPTdarkblue,label={[PPTdarkblue]left:$y$}] at (1.3,0,1.15) {};
  \node[PPTdarkblue,right] at (1.3,0,1.38) {$b$};
  \node[pt,PPTdarkblue,label={[PPTdarkblue]left:$z$}] at (1.3,0,1.6) {};
  \draw[oline,marr] (1.3,0,1.6) -- (1.3,3,1.6);
  \node[PPTdarkblue,above] at (1.3,1.5,1.6) {$\lambda$};
  \swSplitFront
  \node[pt,PPTdarkblue,label={[PPTdarkblue]right:$\widetilde{x}$}] at (1.3,3,1.6) {};
\end{tikzpicture}}

\newcommand{\swAutoBack}{%
  \def\swL{4}%
  \swUN
  \swwall{topoA}{0}{0}{2}
  \swedges{dashed}{0}{4}
  \node[left] at (2,0,0.25) {$B_{L_1}$};
  \node[below] at (2,1.9,0) {$\Rep(\widehat{\mathfrak{g}})\boxtimes\Rep(S)$};}
\newcommand{\swAutoSurface}[1]{\draw[ifc] (0,2,#1) -- (2,2,#1) -- (2,2,2.1) -- (0,2,2.1) -- cycle;}
\newcommand{\swAutoFront}{\def\swL{4}\swSplitFrontGen{$B_{\widehat{\mathfrak{g}}\oplus S}$}}
\newcommand{\figAutoSplit}[1][1.4]{%
\begin{tikzpicture}[sw,scale=#1]
  \def\swL{4}
  \swSplitBackGen{1}{$\mathcal{D}$}{$B_{L_1}$}{$B_{L_2}$}{$\Rep(\widehat{\mathfrak{g}})\boxtimes\Rep(S)$}{$U(N)_1$}
  \swAutoFront
\end{tikzpicture}}
\newcommand{\figAutoSurface}[1][1.4]{%
\begin{tikzpicture}[sw,scale=#1]
  \swAutoBack
  \swAutoSurface{1}
  \draw[dline,marr] (2,2,1) -- (0,2,1);
  \node[PPTdarkred] at (1,2,1.6) {$\mathcal{S}_{A}$};
  \node[PPTdarkred,left] at (2,2,1) {$\sigma$};
  \swAutoFront
\end{tikzpicture}}
\newcommand{\figAutoA}[1][1.4]{%
\begin{tikzpicture}[sw,scale=#1]
  \swAutoBack
  \swAutoSurface{1}
  \draw[dline,marr] (2,2,1) -- (0,2,1);
  \node[PPTdarkred,left] at (2,2,1) {{$\sigma$}};
  \node[pt,PPTdarkblue,label={[PPTdarkblue]left:$x$}] at (1.3,0,0.45) {};
  \draw[oline,marr] (1.3,0,0.45) -- (1.3,4,0.45);
  \node[PPTdarkblue,below] at (1.3,1.0,0.45) {$\lambda$};
  \swAutoFront
  \node[pt,PPTdarkblue,label={[PPTdarkblue]right:$\widetilde{x}$}] at (1.3,4,0.45) {};
\end{tikzpicture}}
\newcommand{\figAutoB}[1][1.4]{%
\begin{tikzpicture}[sw,scale=#1]
  \swAutoBack
  \swAutoSurface{1}
  \coordinate (u) at (1.75,2,1);
  \coordinate (v) at (0.75,2,1);
  \coordinate (e) at (0,2,1);
  \draw[dline,marr] (2,2,1) -- (u);
  \draw[dline,marr] (u) -- (v);
  \draw[dline,marr] (v) -- (e);
  \node[PPTdarkred,left] at (2,2,1) {{$\sigma$}};
  \node[PPTdarkred,above left] at ($(u)!0.5!(v)$) {{$\sigma'$}};
  \node[PPTdarkred,above left] at ($(v)!0.5!(e)$) {{$\sigma$}};
  \node[pt,PPTdarkblue,label={[PPTdarkblue]left:$x$}] at (1.3,0,0.4) {};
  \draw[oline,marr] (1.3,0,0.4) -- (1.3,1.3,0.4);
  \draw[oline,marr] (1.3,1.3,0.4) .. controls (1.45,1.95,0.4) and (1.75,2,0.5) .. (u);
  \draw[oline,marr] (v) .. controls (0.75,2,0.5) and (1.1,2.05,0.4) .. (1.3,2.7,0.4);
  \draw[oline,marr] (1.3,2.7,0.4) -- (1.3,4,0.4);
  \node[PPTdarkblue,below] at (1.3,0.8,0.4) {$\lambda$};
  \node[PPTdarkblue,below] at (1.3,3.3,0.4) {$\lambda$};
  \node[pt,PPTdarkred,label={[PPTdarkred]below right:$u$}] at (u) {};
  \node[pt,PPTdarkred,label={[PPTdarkred]right:$v$}] at (v) {};
  \swAutoFront
  \node[pt,PPTdarkblue,label={[PPTdarkblue]right:$\widetilde{x}$}] at (1.3,4,0.4) {};
\end{tikzpicture}}
\newcommand{\figAutoC}[1][1.4]{%
\begin{tikzpicture}[sw,scale=#1]
  \swAutoBack
  \swAutoSurface{1}
  \coordinate (zz) at (1.3,2,1);
  \coordinate (e) at (0,2,1);
  \coordinate (ww) at (1.3,2.5,1.5);
  \draw[dline,marr] (2,2,1) -- (zz);
  \draw[dline,marr] (zz) -- (e);
  \node[PPTdarkred,left] at (2,2,1) {{$\sigma$}};
  \node[PPTdarkred,below right] at (0.2,2,1) {{$\sigma$}};
  \node[pt,PPTdarkblue,label={[PPTdarkblue]left:$x$}] at (1.3,0,1.5) {};
  \draw[oline,marr] (1.3,0,1.5) -- (1.3,1.65,1.5);
  \draw[oline,densely dotted] (1.3,1.65,1.5) -- (1.3,2,1.5);
  \draw[oline,marr] (1.3,2,1.5) -- (ww);
  \node[PPTdarkblue,above] at (1.3,2.2,1.5) {{$A(\lambda)$}};
  \node[pt,PPTdarkblue,inner sep=0.8pt] at (1.3,2,1.5) {};
  \draw[oline,marr] (zz) -- (ww);
  \node[PPTdarkblue,right] at ($(zz)!0.5!(ww)$) {$a$};
  \node[pt,PPTdarkblue,label={[PPTdarkblue]below:$z$}] at (zz) {};
  \node[pt,PPTdarkblue,label={[PPTdarkblue]above right:$w$}] at (ww) {};
  \draw[oline,marr] (ww) -- (1.3,4,1.5);
  \node[PPTdarkblue,below] at (1.3,0.9,1.5) {$\lambda$};
  \node[PPTdarkblue,above] at (1.3,3.4,1.5) {$\lambda$};
  \swAutoFront
  \node[pt,PPTdarkblue,label={[PPTdarkblue]right:$\widetilde{x}$}] at (1.3,4,1.5) {};
\end{tikzpicture}}

\newcommand{\swClub}[9]{%
  \swLeft{#1}
  \swwall{topoA}{0}{0}{1}
  \swwall{topoB}{0}{1}{2}
  \swedges{dashed}{0}{5.5}
  \draw[ifc] (0,0,1) -- (2,0,1) -- (2,2.5,1) -- (0,2.5,1) -- cycle;
  \node[PPTdarkred] at (0.35,1.25,1) {$\mathcal{I}$};
  \swwall{tifcA}{2.5}{0}{1}
  \swwall{tifcB}{2.5}{1}{2}
  \draw[dline,marr] (2,0,1) -- (0,0,1);
  \node[PPTdarkred,left] at (2,0,1) {$J$};
  \draw[dline,marr] (2,2.5,1) -- (0,2.5,1);
  \node[PPTdarkred,right] at (0,2.5,1) {$\widetilde{J}$};
  \node[left] at (2,0,1.75) {#3};
  \node[left] at (2,0,0.25) {#2};
  \node at (1.05,2.5,0.45) {#4};
  \node at (0.7,2.5,1.45) {#5};
  \node[anchor=south] (Rup) at (0,1.25,2.45) {#7};
  \draw[->,gray] (Rup.south) to[out=-90,in=90] (1,1.25,1.5);
  \node[anchor=north] (Rlow) at (2,0.9,-0.45) {#6};
  \draw[->,gray] (Rlow.north) to[out=90,in=-90] (1,1.25,0.5);
  \node[anchor=north] (Rright) at (2,4.4,-0.45) {#8};
  \draw[->,gray] (Rright.north) to[out=90,in=-90] (1,4.1,0.8);
  \swwall{chiralA}{5.5}{0}{2}
  \node[anchor=west] at (0,5.6,0.9) {#9};}
\newcommand{\figClubW}[1][1.4]{%
\begin{tikzpicture}[sw,scale=#1]
  \swClub{$U(N)_1$}{$B_{L_1}$}{$B_{L_2}$}{$\mathcal{I}_{S_1|W}$}{$\mathcal{I}_{S_2|W}$}%
    {$\Rep(\widehat{\mathfrak{g}})\boxtimes\Rep(S_1)$}{$\Rep(\widehat{\mathfrak{g}})\boxtimes\Rep(S_2)$}%
    {$\Rep(\widehat{\mathfrak{g}})\boxtimes\Rep(W)$}{$B_{\widehat{\mathfrak{g}}\oplus W}$}
\end{tikzpicture}}
\newcommand{\figClubWShrunk}[1][1.4]{%
\begin{tikzpicture}[sw,scale=#1]
  \def\swL{4}
  \swSplitBackGen{1}{$\mathcal{D}$}{$B_{\widetilde{L}_1}$}{$B_{\widetilde{L}_2}$}{$\Rep(\widehat{\mathfrak{g}})\boxtimes\Rep(W)$}{$U(N)_1$}
  \swSplitFrontGen{$B_{\widehat{\mathfrak{g}}\oplus W}$}
\end{tikzpicture}}
\newcommand{\figClubSpec}[1][1.4]{%
\begin{tikzpicture}[sw,scale=#1]
  \def\swLL{2.4}
  \swClub{$U(N)_1$\\$\boxtimes\Rep(V)$}{$B_{L_1}$}{$B_{L_2}$}{$\mathcal{I}_{\widetilde{S}_1|W}$}{$\mathcal{I}_{\widetilde{S}_2|W}$}%
    {$\Rep(\widehat{\mathfrak{g}})\boxtimes\Rep(\widetilde{S}_1)$}{$\Rep(\widehat{\mathfrak{g}})\boxtimes\Rep(\widetilde{S}_2)$}%
    {$\Rep(\widehat{\mathfrak{g}})\boxtimes\Rep(W)$}{$B_{\widehat{\mathfrak{g}}\oplus W}$}
\end{tikzpicture}}
\newcommand{\figClubSpecShrunk}[1][1.4]{%
\begin{tikzpicture}[sw,scale=#1]
  \def\swLL{2.4}\def\swL{4}
  \swSplitBackGen{1}{$\widetilde{\mathcal{D}}$}{$B_{\widetilde{L}_1}$}{$B_{\widetilde{L}_2}$}{$\Rep(\widehat{\mathfrak{g}})\boxtimes\Rep(W)$}{$U(N)_1$\\$\boxtimes\Rep(V)$}
  \swSplitFrontGen{$B_{\widehat{\mathfrak{g}}\oplus W}$}
\end{tikzpicture}}
\newcommand{\figGapping}{%
\begin{tikzpicture}[flat]
  \fill[blue!15] (0,0) rectangle (3.6,3);
  \fill[green!30,opacity=0.7] (0,0) rectangle (1.3,3);
  \hatchedbdy{0}{2.95}
  \draw[thick,densely dashed,PPTgreen] (1.3,0) -- (1.3,3);
  \node[align=center,font=\scriptsize] at (0.65,2.45) {$N$ Dirac\\$\otimes\, X_V$};
  \node at (2.5,2.6) {$N$ Dirac};
  \draw[<->] (0.05,0.35) -- (1.25,0.35) node[midway,above] {$\epsilon$};
  \node at (0,-0.4) {$|\widetilde{\mathcal{D}}\rangle$};
  \node[PPTgreen] at (1.3,-0.4) {$\ket{\mu}$};
  \node at (4.3,1.5) {$\rightsquigarrow$};
  \begin{scope}[xshift=5.2cm]
    \fill[blue!15] (0,0) rectangle (2.5,3);
    \hatchedbdy{0}{2.95}
    \node at (1.3,2.6) {$N$ Dirac};
    \node at (0,-0.4) {$\ket{\mathcal{D}_\mu}$};
  \end{scope}
\end{tikzpicture}}

\newcommand{\swAutoBackGen}[1]{%
  \def\swL{4}%
  \swUN
  \swwall{topoA}{0}{0}{2}
  \swedges{dashed}{0}{4}
  \node[left] at (2,0,0.25) {$B_{L_1}$};
  \node[below] at (2,1.9,0) {#1};}
\newcommand{\figAutoSplitGen}[3][1.4]{%
\begin{tikzpicture}[sw,scale=#1]
  \def\swL{4}
  \swSplitBackGen{1}{$\mathcal{D}$}{$B_{L_1}$}{$B_{L_2}$}{#2}{$U(N)_1$}
  \swSplitFrontGen{$B_{\widehat{\mathfrak{g}}\oplus V_{N_f,n}}$}
\end{tikzpicture}}
\newcommand{\figAutoSurfaceGen}[5][1.4]{%
\begin{tikzpicture}[sw,scale=#1]
  \swAutoBackGen{#4}
  \swAutoSurface{1}
  \draw[dline,marr] (2,2,1) -- (0,2,1);
  \node[PPTdarkred] at (1,2,1.6) {$\mathcal{S}_{\mathsf{C}_{\mathfrak{u}(1)}}$};
  \node[PPTdarkred,left] at (2,2,1) {$\sigma$};
  \def\swL{4}\swSplitFrontGen{$B_{\widehat{\mathfrak{g}}\oplus V_{N_f,n}}$}
\end{tikzpicture}}
\newcommand{\figAutoCGen}[8][1.4]{%
\begin{tikzpicture}[sw,scale=#1]
  \swAutoBackGen{#6}
  \swAutoSurface{1}
  \node[PPTdarkred] at (0.35,2,1.85) {#8};
  \coordinate (zz) at (1.3,2,1);
  \coordinate (e) at (0,2,1);
  \coordinate (ww) at (1.3,2.95,1.5);
  \draw[dline,marr] (2,2,1) -- (zz);
  \draw[dline,marr] (zz) -- (e);
  \node[PPTdarkred,left] at (2,2,1) {#5};
  \node[pt,PPTdarkblue,label={[PPTdarkblue]left:$x$}] at (1.3,0,1.5) {};
  \draw[oline,marr] (1.3,0,1.5) -- (1.3,1.65,1.5);
  \draw[oline,densely dotted] (1.3,1.65,1.5) -- (1.3,2,1.5);
  \draw[oline] (1.3,2,1.5) -- (ww);
  \node[PPTdarkblue,above] at (1.3,2.5,1.5) {#3};
  \node[pt,PPTdarkblue,inner sep=0.8pt] at (1.3,2,1.5) {};
  \draw[oline,marr] (zz) -- (ww);
  \node[PPTdarkblue,below right] at ($(zz)!0.55!(ww)$) {#4};
  \node[pt,PPTdarkblue,label={[PPTdarkblue]below:$z$}] at (zz) {};
  \node[pt,PPTdarkblue,label={[PPTdarkblue]above:$w$}] at (ww) {};
  \draw[oline,marr] (ww) -- (1.3,4,1.5);
  \node[PPTdarkblue,below] at (1.3,0.9,1.5) {#2};
  \node[PPTdarkblue,above] at (1.3,3.5,1.5) {#2};
  \def\swL{4}\swSplitFrontGen{#7}
  \node[pt,PPTdarkblue,label={[PPTdarkblue]right:$\widetilde{x}$}] at (1.3,4,1.5) {};
\end{tikzpicture}}
\newcommand{\figOut}[1]{\figAfterScatterGen{0.95}{#1}{z}}
\newcommand{\figBaryonEnd}[1][1.4]{%
\begin{tikzpicture}[sw,scale=#1]
  \swAutoBackGen{$\mathcal{T}_{N_f,n}$}
  \swAutoSurface{1}
  \node[PPTdarkred] at (0.35,2,1.85) {$\mathcal{S}_A$};
  \coordinate (q) at (1.0,2,1);
  \coordinate (p) at (1.0,0,1);
  \draw[dline,marr] (2,2,1) -- (q);
  \draw[dline,marr] (q) -- (0,2,1);
  \node[PPTdarkred,left] at (2,2,1) {$\sigma_{\mathfrak u}\boxtimes\sigma_{\mathfrak{su}}$};
  \draw[oline,marr] (q) -- (p);
  \node[PPTdarkblue,above] at (1.0,1.0,1) {$B_s$};
  \node[pt,PPTdarkblue] at (q) {};
  \node[pt,PPTdarkblue] at (p) {};
  \draw[thick,PPTgreen,marr] (2,1.3,0.2) -- (0,1.3,0.2);
  \node[PPTgreen,left] at (2,1.3,0.2) {$\eta$};
  \def\swL{4}\swSplitFrontGen{$B_{\widehat{\mathfrak{g}}\oplus V_{N_f,n}}$}
\end{tikzpicture}}
\newcommand{\figEndD}{%
\begin{tikzpicture}[flatD]
  \draw[dline,marr] (0,0) -- (1.1,0);
  \draw[dline,marr] (1.1,0) -- (2.2,0) node[right] {$\mathcal{D}$};
  \node[pt,PPTdarkblue,label={[PPTdarkblue]above:$b_s$}] at (1.1,0) {};
\end{tikzpicture}}

\newcommand{\figClubSUSU}[1][1.4]{%
\begin{tikzpicture}[sw,scale=#1]
  \def\swLL{1.9}
  \swClub{$U((N{+}1)^2)_1$}{$B_{L_1}$}{$B_{L_2}$}{$\mathcal{I}_{S_1|W}$}{$\mathcal{I}_{S_2|W}$}%
    {{$\Rep(\widehat{\mathfrak{g}})\boxtimes\Rep(S_1)$}}{{$\Rep(\widehat{\mathfrak{g}})\boxtimes\Rep(S_2)$}}%
    {$\mathcal{T}_N$}{$B_{\widehat{\mathfrak{g}}\oplus W}$}
\end{tikzpicture}}
\newcommand{\figClubSUSUShrunk}[1][1.4]{%
\begin{tikzpicture}[sw,scale=#1]
  \def\swL{4}\def\swLL{1.9}
  \swSplitBackGen{1}{$\mathcal{D}$}{$B_{\widetilde{L}_1}$}{$B_{\widetilde{L}_2}$}{$\mathcal{T}_N$}{$U((N{+}1)^2)_1$}
  \swSplitFrontGen{$B_{\widehat{\mathfrak{g}}\oplus W}$}
\end{tikzpicture}}
\newcommand{\swChiralWithDGen}[3]{%
  \swLeft{#2}
  \swwall{chiralA}{0}{0}{2}
  \draw[dline,marr] (2,0,#1) -- (0,0,#1);
  \node[PPTdarkred,left] at (2,0,#1) {$\mathcal{D}$};
  \node[below] at (2,0,0) {#3};}
\newcommand{\figBeforeScatterSUSU}[2][1.4]{%
\begin{tikzpicture}[sw,scale=#1]
  \def\swLL{1.9}
  \swChiralWithDGen{1.15}{$U((N{+}1)^2)_1$}{$\Fer((N+1)^2)$}
  \node[pt,PPTdarkblue,label={[PPTdarkblue]right:$#2$}] at (1.3,0,0.55) {};
\end{tikzpicture}}
\newcommand{\figOutSUSU}[4][1.4]{%
\begin{tikzpicture}[sw,scale=#1]
  \def\swLL{1.9}
  \swChiralWithDGen{0.8}{$U((N{+}1)^2)_1$}{$\Fer((N+1)^2)$}
  \draw[oline,marr] (1.3,0,0.8) -- (1.3,0,1.55);
  \node[pt,PPTdarkblue,label={[PPTdarkblue]below right:$#3$}] at (1.3,0,0.8) {};
  \node[PPTdarkblue,left] at (1.3,0,1.18) {$#2$};
  \node[pt,PPTdarkblue,label={[PPTdarkblue]right:$#4$}] at (1.3,0,1.55) {};
\end{tikzpicture}}

\newcommand{\figClubCQED}[1][1.4]{%
\begin{tikzpicture}[sw,scale=#1]
  \swClub{$U(8)_1$}{$B_{L_1}$}{$B_{L_2}$}{$\mathcal{I}_{S_1|W}$}{$\mathcal{I}_{S_2|W}$}%
    {$\Rep(\widehat{\mathfrak{g}})\boxtimes U(1)_{480}$}{$\Rep(\widehat{\mathfrak{g}})\boxtimes O(2)_{30}$}%
    {$\mathcal{T}_{\rm cQED}$}{$B_{\widehat{\mathfrak{g}}\oplus W}$}
\end{tikzpicture}}
\newcommand{\figClubCQEDShrunk}[1][1.4]{%
\begin{tikzpicture}[sw,scale=#1]
  \def\swL{4}
  \swSplitBackGen{1}{$\mathcal{D}$}{$B_{\widetilde{L}_1}$}{$B_{\widetilde{L}_2}$}{$\mathcal{T}_{\rm cQED}$}{$U(8)_1$}
  \swSplitFrontGen{$B_{\widehat{\mathfrak{g}}\oplus W}$}
\end{tikzpicture}}
\newcommand{\figBeforeScatterCQED}[2][1.4]{%
\begin{tikzpicture}[sw,scale=#1]
  \swChiralWithDGen{1.15}{$U(8)_1$}{$\Fer(8)$}
  \node[pt,PPTdarkblue,label={[PPTdarkblue]right:$#2$}] at (1.3,0,0.55) {};
\end{tikzpicture}}
\newcommand{\figOutCQED}[4][1.4]{%
\begin{tikzpicture}[sw,scale=#1]
  \swChiralWithDGen{0.8}{$U(8)_1$}{$\Fer(8)$}
  \draw[oline,marr] (1.3,0,0.8) -- (1.3,0,1.55);
  \node[pt,PPTdarkblue,label={[PPTdarkblue]below right:$#3$}] at (1.3,0,0.8) {};
  \node[PPTdarkblue,left] at (1.3,0,1.18) {$#2$};
  \node[pt,PPTdarkblue,label={[PPTdarkblue]right:$#4$}] at (1.3,0,1.55) {};
\end{tikzpicture}}

\newcommand{\swCQEDbare}[2]{\def\swLL{0}\swSplitBackGen{1}{$\mathcal{D}$}{#1}{#2}{}{}}
\newcommand{\swCQEDfront}{\swwall{chiralA}{\swL}{0}{2}}
\newcommand{\figEchi}[3][1.0]{%
\begin{tikzpicture}[sw,scale=#1,color=black]
  \swCQEDbare{$B_{\widetilde L_1}$}{$B_{\widetilde L_2}$}
  \draw[thick,PPTgreen] (1.3,0,#2) -- (1.3,3,#2);
  \node[pt,PPTgreen] at (1.3,0,#2) {};
  \node[PPTgreen,above] at (1.3,1.5,#2) {$#3$};
  \swCQEDfront
  \node[pt,PPTgreen,label={[PPTgreen]right:$\mathcal{O}_{#3}$}] at (1.3,3,#2) {};
\end{tikzpicture}}
\newcommand{\swCross}[2]{%
  \fill[red!15] (-0.2,0) rectangle (0,1.2);
  \fill[orange!25] (-0.2,1.2) rectangle (0,2.4);
  \fill[blue!20] (2.3,0) rectangle (2.5,2.4);
  \draw[thick,black] (0,0) -- (0,2.4);
  \draw[thick,black] (2.3,0) -- (2.3,2.4);
  \node[pt,PPTdarkred] at (0,1.2) {};
  \node[PPTdarkred,left] at (-0.2,1.2) {$\mathcal{D}$};
  \node[black,left] at (-0.2,0.25) {#1};
  \node[black,left] at (-0.2,2.15) {#2};}
\newcommand{\swEarc}[2]{%
  \draw[thick,PPTgreen] (1.5,#1) .. controls (1.0,#1) and (0.6,#2) .. (0,#2);
  \node[pt,PPTgreen,inner sep=0.9pt] at (1.5,#1) {};
  \node[pt,PPTgreen,inner sep=0.9pt] at (0,#2) {};
  \node[PPTgreen,above] at (0.55,#2) {$E_\psi$};}
\newcommand{\figEargIn}[4][0.8]{%
\begin{tikzpicture}[flat,scale=#1,color=black]
  \swCross{#3}{#4}
  \draw[oline,marr] (0,0.5) -- (2.3,0.5);
  \node[pt,PPTdarkblue,label={[PPTdarkblue]below right:$x$}] at (0,0.5) {};
  \node[pt,PPTdarkblue,label={[PPTdarkblue]right:$\psi$}] at (2.3,0.5) {};
  \node[PPTdarkblue,below] at (1.9,0.5) {$\lambda_\psi$};
  #2
\end{tikzpicture}}
\newcommand{\figEargOut}[4][0.8]{%
\begin{tikzpicture}[flat,scale=#1,color=black]
  \swCross{#3}{#4}
  \draw[line width=1.6pt,PPTdarkblue] (0,1.2) -- (0,1.75);
  \node[PPTdarkblue,right] at (0,1.4) {$b$};
  \node[PPTdarkblue,below right] at (0,1.2) {$w$};
  \draw[oline,marr] (0,1.75) -- (2.3,1.75);
  \node[pt,PPTdarkblue,label={[PPTdarkblue]left:$z$}] at (0,1.75) {};
  \node[pt,PPTdarkblue] at (2.3,1.75) {};
  \node[PPTdarkblue,below] at (1.9,1.75) {$\lambda_\psi$};
  #2
\end{tikzpicture}}

\newcommand{\figClubSM}[1][1.4]{%
\begin{tikzpicture}[sw,scale=#1]
  \swClub{$U(9)_1$}{$B_{L_1}$}{$B_{L_2}$}{$\mathcal{I}_{S_1|W}$}{$\mathcal{I}_{S_2|W}$}%
    {$\Rep(\widehat{\mathfrak{g}})\boxtimes\Rep(S_1)$}{$\Rep(\widehat{\mathfrak{g}})\boxtimes\Rep(S_2)$}%
    {$\mathcal{T}_{\rm SM}$}{$B_{\widehat{\mathfrak{g}}\oplus W}$}
\end{tikzpicture}}
\newcommand{\figClubSMShrunk}[1][1.4]{%
\begin{tikzpicture}[sw,scale=#1]
  \def\swL{4}
  \swSplitBackGen{1}{$\mathcal{D}$}{$B_{\widetilde{L}_1}$}{$B_{\widetilde{L}_2}$}{$\mathcal{T}_{\rm SM}$}{$U(9)_1$}
  \swSplitFrontGen{$B_{\widehat{\mathfrak{g}}\oplus W}$}
\end{tikzpicture}}
\newcommand{\figBeforeScatterSM}[2][1.4]{%
\begin{tikzpicture}[sw,scale=#1]
  \swChiralWithDGen{1.15}{$U(9)_1$}{$\Fer(9)$}
  \node[pt,PPTdarkblue,label={[PPTdarkblue]right:$#2$}] at (1.3,0,0.55) {};
\end{tikzpicture}}
\newcommand{\figOutSM}[4][1.4]{%
\begin{tikzpicture}[sw,scale=#1]
  \swChiralWithDGen{0.8}{$U(9)_1$}{$\Fer(9)$}
  \draw[oline,marr] (1.3,0,0.8) -- (1.3,0,1.55);
  \node[pt,PPTdarkblue,label={[PPTdarkblue]below right:$#3$}] at (1.3,0,0.8) {};
  \node[PPTdarkblue,left] at (1.3,0,1.18) {$#2$};
  \node[pt,PPTdarkblue,label={[PPTdarkblue]right:$#4$}] at (1.3,0,1.55) {};
\end{tikzpicture}}

\def\fdL{2.7}\def\fdl{2.0}\def\fdr{0.4}\def\fdH{1.3}\def\fde{0.5}
\tikzset{
  diagbdy/.style={fill=green!30,opacity=0.6},
  diagV/.style={fill=teal!35,opacity=0.55},
  lab/.style={fill=white,fill opacity=0.85,text opacity=1,inner sep=1pt},
}
\newcommand{\fdBent}{%
  \begin{scope}[transparency group,opacity=0.6]
    \fill[green!30] (2,0,\fdH) -- (2,{-\fdl+\fdr},\fdH) -- (0,{-\fdl+\fdr},\fdH) -- (0,0,\fdH) -- cycle;
    \fill[green!30] plot[domain=90:180,samples=12,variable=\t] (2,{-\fdl+\fdr+\fdr*cos(\t)},{\fdH-\fdr+\fdr*sin(\t)})
      -- plot[domain=180:90,samples=12,variable=\t] (0,{-\fdl+\fdr+\fdr*cos(\t)},{\fdH-\fdr+\fdr*sin(\t)}) -- cycle;
    \fill[green!30] (2,-\fdl,{\fdH-\fdr}) -- (2,-\fdl,0) -- (0,-\fdl,0) -- (0,-\fdl,{\fdH-\fdr}) -- cycle;
  \end{scope}
  \foreach \xx in {0,2}{%
    \draw[PPTgreen!60!black,densely dashed] (\xx,0,\fdH) -- (\xx,{-\fdl+\fdr},\fdH)
      -- plot[domain=90:180,samples=12,variable=\t] (\xx,{-\fdl+\fdr+\fdr*cos(\t)},{\fdH-\fdr+\fdr*sin(\t)})
      -- (\xx,-\fdl,0);}
  \draw[PPTgreen!60!black,densely dashed] (0,-\fdl,0) -- (2,-\fdl,0);
  \foreach \xx in {0,2}{\draw[invbulk] (\xx,-\fdl,0) -- (\xx,0,0);}}
\newcommand{\fdCorner}{%
  \draw[dline,marr] (2,0,\fdH) -- (0,0,\fdH);
  \node[PPTdarkred,right] at (0,0,\fdH) {$|\widetilde{\mathcal{D}}\rangle$};}
\newcommand{\figFoldA}[1][1.2]{%
\begin{tikzpicture}[sw,scale=#1]
  \foreach \xx in {0,2} \foreach \zz in {0,{2*\fdH}} {\draw[invbulk] (\xx,-\fdL,\zz) -- (\xx,0,\zz);}
  \foreach \xx in {0,2}{\draw[densely dotted,PPTdarkred] (\xx,-\fdL,\fdH) -- (\xx,0,\fdH);}
  \node at (1,{-\fdL-0.35},\fdH) {$\cdots$};
  \node[lab] at (2,-1.3,{1.6*\fdH}) {$\mathcal{Z}_V$};
  \swwall{chiralA}{0}{0}{2*\fdH}
  \draw[dline,marr] (2,0,\fdH) -- (0,0,\fdH);
  \node[PPTdarkred,right] at (0,0,\fdH) {$\widetilde{\mathcal{D}}$};
  \node[below] at (2,0,0) {$\Fer(N)\oplus V$};
\end{tikzpicture}}
\newcommand{\figFoldB}[1][1.2]{%
\begin{tikzpicture}[sw,scale=#1]
  \foreach \xx in {0,2}{\draw[invbulk] (\xx,-\fdL,0) -- (\xx,0,0);}
  \fill[diagbdy] (2,-\fdL,\fdH) -- (2,0,\fdH) -- (0,0,\fdH) -- (0,-\fdL,\fdH) -- cycle;
  \foreach \xx in {0,2}{\draw[PPTgreen!60!black,densely dashed] (\xx,-\fdL,\fdH) -- (\xx,0,\fdH);}
  \node at (1,{-\fdL-0.35},{0.5*\fdH}) {$\cdots$};
  \node[above] at (0,-1.4,\fdH) {$B_{\rm diag}$};
  \node[lab] at (2,-1.3,{0.75*\fdH}) {$\mathcal{Z}_V\boxtimes\overline{\mathcal{Z}_V}$};
  \swwall{chiralA}{0}{0}{\fdH}
  \fdCorner
  \node[below] at (2,0,0) {$N$ Dirac $\otimes\,V\otimes\overline{V}$};
\end{tikzpicture}}
\newcommand{\figFoldC}[1][1.2]{%
\begin{tikzpicture}[sw,scale=#1]
  \fdBent
  \node[above] at (0,-1.0,\fdH) {$B_{\rm diag}$};
  \node at (2,-0.6,{0.62*\fdH}) {$\mathcal{Z}_V\boxtimes\overline{\mathcal{Z}_V}$};
  \swwall{chiralA}{0}{0}{\fdH}
  \fdCorner
  \node[below] at (2,0,0) {$N$ Dirac $\otimes\,X_V$};
\end{tikzpicture}}
\newcommand{\figFoldD}[1][1.2]{%
\begin{tikzpicture}[sw,scale=#1]
  \fdBent
  \fill[diagV] (2,0,{\fdH-\fde}) -- (2,-\fdl,{\fdH-\fde}) -- (0,-\fdl,{\fdH-\fde}) -- (0,0,{\fdH-\fde}) -- cycle;
  \foreach \xx in {0,2}{\draw[teal!70!black,densely dashed] (\xx,-\fdl,{\fdH-\fde}) -- (\xx,0,{\fdH-\fde});}
  \node[above] at (0,-1.0,\fdH) {$B_{\rm diag}$};
  \node[lab,anchor=east] (low) at (2,{-\fdl-0.25},0.15) {$U(N)_1\boxtimes\overline{U(N)_1}$};
  \draw[->,gray] (low.east) to[out=0,in=200] (1.4,-1.2,0.35);
  \node[lab,anchor=east] (tube) at (2,{-\fdl-0.25},{\fdH+0.35}) {$\mathcal{Z}_V\boxtimes\overline{\mathcal{Z}_V}$};
  \draw[->,gray] (tube.east) to[out=0,in=160] (1.2,-1.1,{\fdH-0.5*\fde});
  \swwall{chiralA}{0}{0}{\fdH}
  \draw[thick,PPTgreen!70!black,marr] (2,0,{\fdH-\fde}) -- (0,0,{\fdH-\fde});
  \node[PPTgreen!60!black,right] at (0,0,{\fdH-\fde}) {$\mu$};
  \fdCorner
  \draw[<->] (0,0.75,{\fdH-\fde}) -- (0,0.75,\fdH) node[midway,right] {$\epsilon$};
  \node[below] at (2,0,0) {$N$ Dirac};
\end{tikzpicture}}

\begin{document}

\begin{titlepage}
\begin{flushright}
  RIKEN-iTHEMS-Report-26
  \\~
\end{flushright}

\title{Intertwining Defects and Fermion-Boundary Scattering from TQFT}

\author{Kantaro Ohmori$^1$, Masataka Watanabe$^2$, Pengcheng Wei$^3$, and Yunqin Zheng$^4$}

        \address{
        ${}^{1}$ RIKEN Center for Interdisciplinary Theoretical and Mathematical Sciences (iTHEMS), RIKEN, Wako, Saitama 351-0198, Japan\\\bigskip 
        ${}^{2}$ Faculty of Science, The University of Tokyo, Tokyo 113-0033, Japan\\\bigskip 
        ${}^{3}$ Department of Physics, Nanjing University, Nanjing, Jiangsu 210093, China\\\bigskip 
        ${}^{4}$  Kavli Institute for Theoretical Sciences and School of Quantum, \\University of Chinese Academy of Sciences, Beijing, 100190, China}

\abstract

We construct conformal boundary conditions of free Dirac fermions that preserve chiral, anomaly-free symmetries, and determine the associated fermion-boundary scattering, using the TQFT/CFT framework. We apply the construction to several previously unresolved examples: higher-charge monopoles in non-chiral QED, a family of $SU(N)\times SU(N+2)$-symmetric models, the chiral QED with charges $\{1^5,3,(-2)^4\}$, and the charge-three monopole of the Standard Model. In these examples, out-going states are created by defect operators at the end of non-invertible topological lines; when such a line is non-simple, the out-going state splits into superselection sectors, and we give a formula for the branching ratios in terms of TQFT data. We also propose a general construction that adds and then gaps neutral spectators, under a technical assumption, which is potentially applicable to the 1-5-7-8-9 model.

\end{titlepage}

\eject

\setcounter{tocdepth}{3} 

\tableofcontents

\section{Introduction}
\label{sec:intro}

Massless Dirac fermions on a half-plane are among the simplest boundary conformal field theories. Nevertheless, the question of which conformal boundary conditions are compatible with a given chiral symmetry is surprisingly subtle. 

It arises in at least two contexts. The first is the relation between 't Hooft anomalies, symmetric mass generation, and symmetric boundary conditions, reviewed in Section \ref{sec:loreintro}. The second is the scattering of fermions off magnetic monopoles in four dimensions, i.e.\ the Callan-Rubakov effect, whose lowest partial waves reduce to two-dimensional fermions on a half-line, reviewed in Section \ref{sec:introscattering}. In both contexts, the difficult cases are those in which the symmetry acts chirally, so that the in-coming and the out-going fermions transform in different representations. 

In this paper, we develop a framework, based on three-dimensional topological quantum field theory (TQFT), that constructs such boundary conditions and determines the corresponding fermion-boundary scattering. We summarize the main ideas and results in Section \ref{sec:introsummary}.

\subsection[{'t Hooft Anomalies, Gappability, and Symmetric Boundary Conditions}]{'t Hooft Anomalies, Gappability, and Symmetric Boundary Conditions}
\label{sec:loreintro}

't Hooft anomaly of a global symmetry is the obstruction to gauging that symmetry \cite{tHooft:1979rat}. A dynamical consequence of 't Hooft anomaly is that a quantum field theory with an anomalous global symmetry cannot be deformed to a trivially gapped phase while preserving the symmetry.  In recent years, this property of 't Hooft anomaly has been adopted as its alternative definition \cite{Thorngren:2019iar,Choi:2023xjw,Zhang:2023wlu,Cordova:2023bja}, in accordance with the Lieb-Schultz-Mattis-Oshikawa-Hastings theorem in condensed matter physics \cite{LIEB1961407,Hastings_2004,Yao:2020xcm,Oshikawa_2000}.

Conversely, for a quantum field theory with an anomaly-free global symmetry, it has been expected that there exists a symmetry-preserving deformation, possibly after introducing additional degrees of freedom that drives the theory to a trivially gapped phase \cite{SeibergString2019Talk,Wang:2013yta}.
Finding such a deformation for a given quantum field theory is called symmetric mass generation. While a general construction was given in \cite{SMGtalk}, achieving it with a small number of additional degrees of freedom is highly non-trivial \cite{Mouland:2025dct,Tong:2021phe,Wang:2022ucy,BoyleSmith:2022duw,Razamat:2020kyf,Zeng:2022grc}.

The 't Hooft anomaly also constrains the symmetric boundary conditions in conformal field theories. In \cite{Han:2017hdv}, it was shown, among others, that a 2d conformal field theory with an anomalous symmetry cannot have a symmetric, conformal, and indecomposable boundary condition; see also \cite{Jensen:2017eof,Numasawa:2017crf,Smith:2020nuf,Thorngren:2020yht,Hellerman:2021fla,Li:2022drc}.

Conversely, if a 2d conformal field theory has an anomaly-free global symmetry, a symmetric, indecomposable, conformal boundary condition is expected to exist. One way to see this is to turn on a symmetric gapping deformation on half-space, so that we get a boundary, and then to let the boundary flow to a conformal fixed point. One may worry about whether the boundary is indecomposable. Indeed, it was observed recently that there are a few counterexamples \cite{Wei:2025zyd,Choi:2023xjw}, but all of them involve finite symmetries. If we only wish to preserve (connected) continuous symmetries, as in most of the examples in this paper, we expect that such degeneracy is in general lifted, and that there exists a symmetric, indecomposable, conformal boundary condition.

In this work, we mainly focus on $N$ massless Dirac fermions in two dimensions\footnote{While it is in principle possible to pursue the RG-boundary approach in the previous paragraph with the general symmetric mass generation construction in \cite{SMGtalk}, we propose a different method. This is because the construction in \cite{SMGtalk} involves a non-compact boson theory, which is not easily compatible with the RCFT techniques that we would like to exploit to perform concrete computations.}.
On a half-plane, we refer to the chiral components moving toward and away from the boundary as \emph{in-coming} and \emph{out-going} fermions, respectively. The theory has a $U(N)\times U(N)$ symmetry, whose two factors act on the in-coming and on the out-going fermions. This symmetry is anomalous, but its various subgroups are anomaly free. Consider a subgroup $G$, under which the in-coming and out-going fermions carry representations $R$ and $\widetilde{R}$, respectively. Let $T(R)$ be the quadratic Dynkin index\footnote{The quadratic Dynkin index is defined as $\Tr_R(T^a T^b) = T(R) \delta^{ab}$, where $a,b$ are adjoint indices of $G$. We adopt the normalization $\Tr_{\boxed{}}(T^aT^b)=\delta^{ab}/2$ for $SU(N)$; the corresponding level for the affine Lie algebra $\mathfrak{g}_k$ is $k=2T(R)$. For $U(1)$, by $T(Q)$ we mean the coefficient of the $1/z^2$ term in the current-current OPE.} associated with the representation $R$ of $G$. Then the anomaly-free condition is simply
\begin{eqnarray}
    T(R) = T(\widetilde{R}).
\end{eqnarray}

When $G=U(1)$, under which the $i$-th in-coming and out-going fermions carry charges $Q_i$ and $\tilde{Q}_i$, respectively, we have $T(Q_i) = Q_i^2$. The anomaly-free condition is
\begin{eqnarray}
    \sum_{i=1}^N Q_i^2 = \sum_{i=1}^N \widetilde{Q}_i^2
\end{eqnarray}
Boundary conditions of $N$ Dirac fermions preserving any such anomaly-free $U(1)$ have been explicitly constructed in \cite{Smith:2019jnh,Smith:2020nuf}.

When $G=SU(2)$, suppose that the in-coming fermions transform in irreducible representations of dimensions $\widetilde{n}_i$, with $\widetilde{n}_1\leq  \widetilde{n}_2\leq  \cdots \leq  \widetilde{n}_{\widetilde{s}}$. They satisfy  $\sum_{i=1}^s n_i =\sum_{i=1}^{\widetilde{s}} \widetilde{n}_i=N$. The Dynkin index is $T(n_i) = n_i (n_i^2-1)/12$, and the anomaly-free condition is
\begin{eqnarray}
    \sum_{i=1}^s n_i(n_i^2-1) = \sum_{i=1}^{\widetilde{s}} \widetilde{n}_i(\widetilde{n}_i^2-1)
\end{eqnarray}
A boundary condition preserving such an anomaly-free $SU(2)$ symmetry is in general unknown. One of the most famous examples is the 1-5-7-8-9 model \cite{vanBeest:2023dbu,vanBeest:2023mbs}, where
\begin{eqnarray}
\begin{split}
    \text{in-coming}:&\quad (n_1,n_2,n_3) = (1,5,9)\\
    \text{out-going}:&\quad (\widetilde{n}_1,\widetilde{n}_2) = (7,8)\\
\end{split}
\end{eqnarray}
It arises from the chiral $U(1)$ gauge theory with left-handed Weyl fermions of charges $1,5,9,-7,-8$ in the presence of a unit monopole, see Section \ref{sec:chiralqed}.

More generally, for a generic anomaly-free chiral symmetry $G$, no symmetric indecomposable conformal boundary condition is known explicitly. One of the goals of this work is to propose a method to construct such boundary conditions, and to carry it out explicitly in a range of examples.

\subsection[Fermion-Boundary Scattering Puzzle and Monopoles]{Fermion-Boundary Scattering Puzzle and Monopoles}
\label{sec:introscattering}

The existence of a $G$-symmetric, indecomposable, conformal boundary condition of $N$ massless Dirac fermions implies a scattering process of a fermion off the boundary that preserves the $G$ quantum numbers. This is because such a boundary condition cannot store the symmetry charges, and all the quantum numbers of the in-coming excitation should match those of the out-going excitations.

\begin{figure}
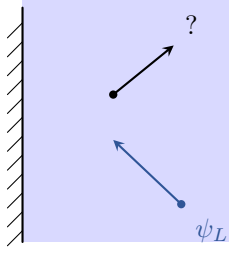

    \centering
    \figScatteringPuzzle
    \caption{Fermion-boundary scattering puzzle. }
    \label{fig:scatteringpuzzle}
\end{figure}

However, since $R \neq \widetilde{R}$, it is in general impossible to find a combination of out-going fermions that matches the representation of an in-coming fermion. This appears to conflicts with the existence of a $G$-symmetric, indecomposable, conformal boundary condition, and is known as the fermion-boundary scattering puzzle. See Figure \ref{fig:scatteringpuzzle}.

When $G$ is abelian, the scattering process has been solved in \cite{vanBeest:2023dbu,vanBeest:2023mbs,Wei:2026fsn,Arias-Tamargo:2026urw}, by using abelian bosonization. The out-going excitations are not created by local operators. Instead, they live in the twisted sectors of topological lines, which explains their fractional quantum numbers, and after unfolding, the boundary condition becomes a topological, and in general non-invertible, line defect \cite{Wei:2026fsn,Arias-Tamargo:2026urw}. However, for most cases where $G$ is a non-abelian Lie group, the scattering process has remained unknown, in accordance with the current ignorance of $G$-symmetric boundary conditions.

The problem of finding boundary conditions that preserve chiral, anomaly-free symmetries of 2d Dirac fermions arises naturally from the scattering of fermions off magnetic monopoles in 4d gauge theories, such as the Standard Model and grand unified theories. This is the Callan-Rubakov effect \cite{Callan:1982ah,Callan:1982au,Rubakov:1982fp,Callan:1982ac,Polchinski:1984uw}. 

Keeping only the lowest partial waves around the monopole and neglecting the fluctuations of the gauge field, the 4d Weyl fermions reduce to 2d chiral fermions on the half-line $r\geq 0$, and the monopole core provides a boundary condition at $r=0$. This boundary condition is expected to preserve the symmetries of the 4d theory that are free of the ABJ anomaly and do not form a 2-group with the magnetic one-form symmetry, including the rotations $SU(2)_{\rm rot}$ around the monopole \cite{vanBeest:2023dbu}, and these symmetries generically act chirally on the 2d fermions. 

The boundary conformal field theory approach to this problem was pioneered in \cite{Affleck:1993np,Maldacena:1995pq}, and the puzzle about the out-going states, sometimes called the unitarity puzzle, has been revisited from various angles in recent years \cite{Csaki:2021ozp,Kitano:2021pwt,Brennan:2021ewu,Hamada:2022eiv,Brennan:2023tae,Khoze:2024hlb,Loladze:2024ayk,Antinucci:2024izg,Loladze:2025jsq,Tachikawa:2026cxd,Hamada:2026jgj,Antinucci:2026uuh, Craig:2026rdf, Arias-Tamargo:2026urw}. In \cite{vanBeest:2023dbu,vanBeest:2023mbs}, it was argued that the out-going states are created by operators in the twisted sectors of topological lines, and the out-going states were determined when the symmetry is abelian, or when the problem can be reduced to an abelian one. 

Several examples with non-abelian chiral symmetries remained open, such as higher-charge monopoles in QED whose fermions transform as $(\square,\overline\square)$ of the flavor symmetry, and the charge-three monopole of the Standard Model \cite{vanBeest:2023dbu,vanBeest:2023mbs}. Related open problems are the $SU(N)\times SU(N+2)$-symmetric fermions \cite{Tong:2021phe,PathToQFTTalk}, and the 1-5-7-8-9 model mentioned above. Solving the 2d problem is a necessary step toward understanding monopole catalysis in these theories.

In this paper, we work with the 2d models throughout. We construct $G$-symmetric conformal boundary conditions, and solve the fermion-boundary scattering puzzle, in all the examples above except the 1-5-7-8-9 model, for which we give a construction that relies on a conjecture of \cite{Moller:2024xtt}. We leave the interpretation of our results in 4d, which requires taking into account the gauge field and the higher partial waves, for future work.

\subsection{Main Ideas and Summary of Results}
\label{sec:introsummary}

\subsubsection{Symmetric Boundaries as Intertwining Defects}

Our starting point is the unfolding trick \cite{Wei:2026fsn,Arias-Tamargo:2026urw,Oshikawa:1996ww,Bachas:2001vj,Cordova:2025zkz}. Unfolding $N$ Dirac fermions along the boundary gives $N$ Weyl fermions on the full plane, described by the holomorphic chiral algebra $\Fer(N)$, with a line defect $\mathcal{D}$ along the image of the boundary. Since $\Fer(N)$ is holomorphic, a conformal line defect is automatically topological. The $G$-currents of the in-coming and of the out-going fermions provide two embeddings
\begin{eqnarray}
    f_{1,2}:\ \widehat{\mathfrak{g}}\hookrightarrow\Fer(N)
\end{eqnarray}
of the same current algebra $\widehat{\mathfrak{g}}$, and by conformal invariance the boundary condition preserves $G$ if and only if $\mathcal{D}$ intertwines them,
\begin{eqnarray}\label{eq:introintertwin}
    f_1(j)\,\mathcal{D}=\mathcal{D}\,f_2(j),\qquad j\in\widehat{\mathfrak{g}},
\end{eqnarray}
see Section \ref{sec:2.1}. Hence, constructing a $G$-symmetric conformal boundary condition is equivalent to the kinematic problem of finding a topological line of the chiral CFT $\Fer(N)$ that intertwines two embeddings of a chiral algebra, and the fermion-boundary scattering amounts to moving a fermion across $\mathcal{D}$.

\subsubsection{Sandwich Construction}

To solve this problem, we describe chiral CFTs and their topological lines in terms of 3d TQFT \cite{Witten:1988hf,Elitzur:1989nr,Felder:1999cv,Fuchs:2002cm,Kapustin:2010if,Kong:2017hcw,Kong:2019byq}, following the symmetry-subalgebra duality \cite{Rayhaun:2023pgc,Gannon:2026ttf,Moller:2024xtt}. 

When the $f_i$ are conformal embeddings, $\Fer(N)$ is resolved into a sandwich, shown in Figure \ref{fig:sandwich}. Its bulk is the TQFT $\Rep(\widehat{\mathfrak{g}})$, one of its boundaries is chiral and supports $\widehat{\mathfrak{g}}$, and the other is a topological interface $B_{L_i}$ to the invertible spin TQFT $U(N)_1$, determined by a fermionic Lagrangian algebra $L_i$ that encodes $f_i$. The intertwining defect $\mathcal{D}$ is a topological junction between $B_{L_1}$ and $B_{L_2}$

In most of our examples, $\widehat{\mathfrak{g}}$ is not conformally embedded in $\Fer(N)$, and the coset chiral algebras $S_i$ of the two embeddings enter. They are in general different, and the problem becomes that of relating $S_1$ and $S_2$. We identify two simple cases in which this can be done explicitly.
\begin{itemize}
    \item The two embeddings are related by an automorphism of $\widehat{\mathfrak{g}}$. Then $S_1=S_2$, and $\mathcal{D}$ is obtained from a twist defect of the invertible surface that implements the automorphism in the bulk (Section \ref{sec:automorphism}).
    \item $S_1$ and $S_2$ share a common conformal subalgebra $W$. Then a club sandwich reduces the problem to the conformal case, with $\Rep(\widehat{\mathfrak{g}})$ replaced by $\Rep(\widehat{\mathfrak{g}})\boxtimes\Rep(W)$ (Section \ref{sec:commonalg}).
\end{itemize}
In general, $S_1$ and $S_2$ are (super) Witt equivalent. Assuming that they are moreover orbifold equivalent, as conjectured in \cite{Moller:2024xtt}, we propose to reduce the general case to the second one by adding $G$-neutral spectators, which are gapped at the end (Section \ref{sec:orbifoldequivalence}). 

The intertwining defects are not unique. Together with the lines on the two topological boundaries, they generate a symmetry category acting on $\Fer(N)$, which can be infinite (Section \ref{sec:symcat}).

\subsubsection{Categorical Scattering}

The sandwich also determines the scattering. An in-coming fermion corresponds to a bulk line $\lambda$ ending on $B_{L_1}$. When it is moved across $\mathcal{D}$, it becomes a superposition of out-going excitations at the ends of the topological lines $b$ contained in $F_{B_{L_2}}(\lambda)$, the image of $\lambda$ on $B_{L_2}$. The coefficients are given by generalized half-linking numbers \eqref{eq:Daction}, or, in the automorphism case, by the crossed $F$- and $R$-symbols \eqref{eq:auto45}. Compared to the abelian examples \cite{vanBeest:2023dbu,vanBeest:2023mbs,Wei:2026fsn,Arias-Tamargo:2026urw}, there are two new features:
\begin{enumerate}
    \item The lines $b$ attached to the out-going excitations are in general non-invertible. Thus, the out-going states are created neither by local operators nor by operators in twisted sectors of invertible symmetries.
    \item $F_{B_{L_2}}(\lambda)$ can be non-simple. Then the out-going state is a superposition of states in different superselection sectors, whose probabilities are fixed by TQFT data and add up to one \eqref{eq:sumto1}.
\end{enumerate}

\subsubsection{Examples}

We apply the framework to the examples summarized in Table \ref{tab:introresults}.
\begin{itemize}
    \item \textbf{Higher-charge monopoles in non-chiral QED} (Section \ref{sec:highcharge}). The two embeddings are related by charge conjugation. For $\text{QED}(\square,\square)$, we reproduce the result of \cite{vanBeest:2023dbu}. For $\text{QED}(\square,\overline\square)$, the cases $N_f,n\geq 3$ were open. The intertwining defect obtained from the twist defect is non-simple, and splits into several simple boundary conditions, each of which preserves all the symmetries, including a discrete $\bZ_n$. The in-coming fermion scatters into a superposition of two twisted sectors, attached to non-invertible lines that carry the symmetric and antisymmetric parts of $\square\otimes\square$ of $SU(N_f)$.
    \item \textbf{$SU(N)\times SU(N+2)$ model} (Section \ref{sec:susu}). The coset chiral algebras are two different rank-two lattice chiral algebras with a common conformal subalgebra $\widehat{\mathfrak{u}}(1)^2_K$, which implies that two $U(1)$ symmetries are preserved as well. Each in-coming fermion scatters with probability one into a single twisted sector of a non-invertible line.
    \item \textbf{Chiral QED with charges $\{1^5,3,(-2)^4\}$} (Section \ref{sec:chiralqed}). The in-coming fermions have an emergent $U(1)$ symmetry, which survives on the out-going side only through its charge conjugation orbifold $\widehat{\mathfrak{u}}(1)^{\bZ_2}_{480}$, the common conformal subalgebra. Each in-coming fermion scatters into a single twisted sector of a line of quantum dimension two, which is selected by the boundary condition among two candidates.
    \item \textbf{Charge-three monopole of the Standard Model} (Section \ref{sec:standardmodel}). The common conformal subalgebra is $\widehat{\mathfrak{u}}(1)_{30}\oplus\text{Vir}_{4/5}$, and the $\widehat{\mathfrak{u}}(1)_{30}$ combines with the gauge symmetries into $U(1)_{B-L}$. There are $480$ simple boundary conditions. The lepton scatters into a single non-invertible twisted sector, while the quarks scatter either into a single sector or into a superposition of two sectors, with the universal probabilities $\varphi^{-4}$ and $1-\varphi^{-4}$, $\varphi=\frac{1+\sqrt5}{2}$, which originate from a Fibonacci factor of the bulk TQFT.
    \item \textbf{1-5-7-8-9 model} (Section \ref{sec:15789}). We work out the two coset chiral algebras explicitly, and show that any zigzag chain of conformal embeddings relating them, in particular any common conformal subalgebra, must mix their rotational and abelian parts. Assuming the conjecture of \cite{Moller:2024xtt}, the general construction nevertheless gives a symmetric boundary condition, which is necessarily non-invertible.
\end{itemize}

\begin{table}[t]
    \centering
    \small
    \renewcommand{\arraystretch}{1.3}
    \begin{tabular}{|c|c|c|c|}
        \hline
        Model & Sec. & Relation between $S_1$ and $S_2$ & Out-going twisted sectors \\
        \hline
        $\text{QED}(\square,\square)$ & \ref{sec:qedss} & automorphism $\mathsf{C}_{\mathfrak{u}(1)}$ & one, invertible \\
        $\text{QED}(\square,\overline\square)$ & \ref{sec:qedsbar} & automorphism $\mathsf{C}_{\mathfrak{u}(1)}\times\mathsf{C}_{\mathfrak{su}(N_f)}$ & two, non-invertible \\
        $SU(N)\times SU(N+2)$ & \ref{sec:susu} & $W=\widehat{\mathfrak{u}}(1)^2_K$ & one, non-invertible \\
        $\{1^5,3,(-2)^4\}$ & \ref{sec:chiralqed} & $W=\widehat{\mathfrak{u}}(1)^{\bZ_2}_{480}$ & one, non-invertible \\
        SM, charge three & \ref{sec:standardmodel} & $W=\widehat{\mathfrak{u}}(1)_{30}\oplus\text{Vir}_{4/5}$ & one or two, non-invertible \\
        1-5-7-8-9 & \ref{sec:15789} & orbifold equivalence (conjectural) & not determined \\
        \hline
    \end{tabular}
    \caption{Summary of the examples. In the first two rows, $S_1=S_2$ and the two embeddings are related by an automorphism of $\widehat{\mathfrak{g}}$. In the next three rows, $S_1\neq S_2$ share a common conformal subalgebra $W$. In the last row, no common conformal subalgebra is known. The last column lists the number of twisted sectors into which an in-coming fermion scatters, for a given simple boundary condition, and whether the corresponding lines are invertible.}
    \label{tab:introresults}
\end{table}

\subsection{Outline}

The rest of the paper is organized as follows.

In Section \ref{sec:generalconstruction}, we unfold a boundary condition of Dirac fermions to a topological defect of Weyl fermions, and provide a construction for the topological defect in three simple cases, using the TQFT/CFT correspondence. We also derive the categorical scattering formula, and describe the symmetry category formed by the intertwining defects.
We  provide various examples in Sections \ref{sec:highcharge} to \ref{sec:standardmodel}, which fit into the simple cases we discussed, see Table \ref{tab:roadmap table}.
\begin{table}[t]
    \centering
    \begin{tabular}{|c|c|c|c|}
        \hline
        & Method & Features & Examples \\
        \hline
        Simple cases & Sec.\ref{sec:interdefect conformalembed} & conf. subalgebra $\widehat{\mathfrak{g}}$ & $G$: Cartan torus in $U(N)$ \\
        & Sec.\ref{sec:automorphism} & automorphic embedding & Sec.\ref{sec:highcharge} \\
        & Sec.\ref{sec:commonalg} & cosets share conf. subalgebra & Sec.\ref{sec:susu}, \ref{sec:chiralqed}, and \ref{sec:standardmodel} \\
        \hline
        General & Sec.\ref{sec:coset}, and \ref{sec:orbifoldequivalence} & & \\
        \hline
    \end{tabular}
    \caption{Summary of methods and examples. (conf.=conformal)}
    \label{tab:roadmap table}
\end{table}
In Section \ref{sec:orbifoldequivalence}, we propose how a general situation can be reduced to one of the simple cases we discussed, by adding and then gapping spectators, assuming the orbifold equivalence conjecture of \cite{Moller:2024xtt}, and apply it to the 1-5-7-8-9 model.
We summarize the results and suggest some future directions in Section \ref{sec:outlook}. 

In Appendix \ref{app:prob}, we show that the scattering probabilities add up to one. Appendix \ref{app:Anyon condensation between K matrix U(1) Chern-Simons TQFTs} reviews anyon condensation in abelian Chern-Simons theories, Appendix \ref{app:onechannel} completes the analysis of the scattering in Section \ref{sec:chiralqed}, Appendix \ref{app:More on Gapping Spectators in the Sandwich} illustrates the folding used in Section \ref{sec:orbifoldequivalence}, and Appendix \ref{app:branching data} collects the branching rules used in the main text.

\emph{Note added}: We would like to thank the authors of \cite{Arias-Tamargo:2026oct,Koundouros:2026oct} for coordinating the submission to the arXiv with us.

\section{General Construction}
\label{sec:generalconstruction}

\subsection{Symmetric Boundary Conditions as Intertwining Defects}
\label{sec:2.1}

Let us find an indecomposable conformal boundary condition of $N$ massless Dirac fermions preserving a chiral but anomaly-free global symmetry $G$. For most of our examples, $G$ is a continuous Lie group\footnote{In the cases where $G$ contains a discrete factor, for example in Section \ref{sec:highcharge}, the discrete symmetries are checked afterwards.}. We begin by setting up the conventions.

Consider $N$ massless Dirac fermions on a half-plane, with the action 
\begin{eqnarray}
    S=\frac{1}{2\pi}\int d^2z \sum_{i=1}^N \left(  \psi_i^\dagger \partial_{\bar{z}} \psi_i + \widetilde{\psi}_i^\dagger \partial_{z} \widetilde{\psi}_i \right).
\end{eqnarray}
The holomorphic and anti-holomorphic fermions transform independently under $U(N)_L\times U(N)_R$. Denoting the generators of $\mathfrak{u}(N)$ Lie algebra by $T^a$, the conserved currents are fermion bilinears, 
\begin{eqnarray}
    j^a(z)=:\psi^\dagger T^a\psi:(z), \qquad \widetilde{j}^a(\bar{z})=:\widetilde{\psi}^\dagger T^a\widetilde{\psi}:(\bar{z}).
\end{eqnarray}
These currents generate affine Lie algebra $\widehat{\mathfrak{u}}(N)_1 \oplus \overline{\widehat{\mathfrak{u}}(N)_1}$. Because of the $\widehat{\mathfrak{u}}(1)$ factor, this algebra is non-rational. It can be made strongly rational by including the fermion operators $\psi$ and $\widetilde{\psi}$ as supercurrents, and we denote the extended current (super-)algebra by $\Fer(N)\oplus \overline{\Fer(N)}$.
The non-vanishing level reflects the 't Hooft anomaly of $U(N)_L\times U(N)_R$. Consequently, $N$ Dirac fermions can not have an indecomposable conformal boundary condition preserving $U(N)_L\times U(N)_R$ symmetry.

However, a subgroup $G\subset U(N)_L \times U(N)_R$ can be anomaly free. 
Let $R$ and $\widetilde{R}$ be the representations of $G$ carried by the left- and right- moving fermions, respectively. 
They are chosen such that the $G$-anomaly of the left-movers cancels that of the right-movers. This means that $G$-currents built from the left-movers and those built from the right movers generate the same affine Lie algebra $\widehat{\mathfrak{g}}$, at the same levels. Again whenever there is a $\widehat{\mathfrak{u}}(1)$ factor, one needs to include simple (super-)currents to make the algebra rational, and to avoid cluttering we use $\widehat{\mathfrak{g}}$ for the extended current algebra.\footnote{For instance, for $\widehat{\mathfrak{u}}(1)_k$, it is generated not only by the $\mathfrak{u}(1)$ currents, but also a simple current with conformal weight $\frac{k}{2}$. The $U(1)$ anomaly is $k$ when the $\mathfrak{u}(1)$ current is normalized so that the simple current to have charge $k$. In $\Fer(k)$, the $U(1)$ acting on all the chiral fermions together with the Baryon operator and its conjugate, which are fermionic for odd $k$ and bosonic for even $k$, forms $\widehat{\mathfrak{u}}(1)_k$. In the case of $\widehat{\mathfrak{u}}(N)_1$, we denote the extended strongly rational algebra by $\Fer(N)$.}    The $\widehat{\mathfrak{g}}$-currents are embedded in $\Fer(N)$ via the maps, 
\begin{eqnarray}\label{eq:fi}
    f_i: \widehat{\mathfrak{g}} \hookrightarrow \Fer(N), \qquad i=1,2
\end{eqnarray}
where $i=1,2$ refers to the left- and right-movers, respectively, i.e. 
\begin{eqnarray}\label{eq:f12}
    f_1(j^a) = \psi^\dagger T^a_R \psi, \qquad f_2(j^a) = \widetilde{\psi}^\dagger T^a_{\widetilde{R}} \widetilde{\psi}.
\end{eqnarray}
The two embeddings are generically distinct, since $R\neq \widetilde{R}$. 

\begin{figure}
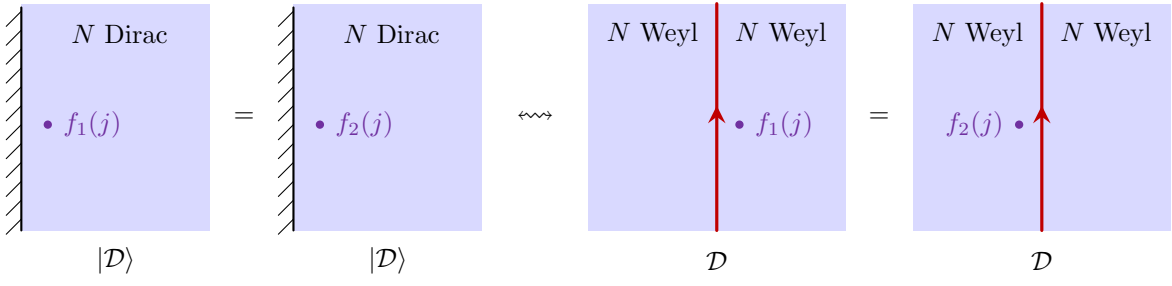

    \centering
    \figFolding
    \caption{Unfolding $N$ Dirac fermions along a conformal boundary $\ket{\mathcal{D}}$ gives $N$ Weyl fermions with a topological defect $\mathcal{D}$. }
    \label{fig:unfolding}
\end{figure}

The anomaly-free $G$-symmetry operator supported on a curve $\gamma$ is $U_1(\gamma) U_2^{\dagger}(\gamma)$, where
\begin{eqnarray}
    U_i(\gamma) = e^{i \sum_a \alpha^a \oint_{\gamma} f_i(j^a)}, \qquad g= e^{i \sum_a \alpha^a T^a} \in G
\end{eqnarray}
Since $G$ is anomaly-free, the lore reviewed in Section \ref{sec:loreintro} suggests that there exists an indecomposable, $G$-preserving conformal boundary condition $\ket{\mathcal{D}}$ along a one-dimensional locus $\gamma$, i.e.
\begin{eqnarray}\label{eq:U12D}
    U_1(\gamma) U_2^{\dagger}(\gamma) \ket{\mathcal{D}} = \ket{\mathcal{D}}.
\end{eqnarray}
Our goal is to find $\ket{\mathcal{D}}$ satisfying \eqref{eq:U12D}.\footnote{\label{footnote:compactBC}We assume the boundary state $\ket{\mathcal{D}}$ is compact. Since for most of our discussions,  the symmetry is assumed to be continuous and connected, when $\ket{\mathcal{D}}$ is decomposable into a direct sum of finite numbers of indecomposable boundary states, each component is symmetric. Therefore, we don't need to worry about indecomposability. } Taking $\alpha^a$ to be infinitesimal, this gives equality of the $G$-charges 
\begin{eqnarray}\label{eq:intf12D}
    Q_1 \ket{\mathcal{D}} = Q_2 \ket{\mathcal{D}}, \qquad Q_i= \oint_\gamma f_i(j)
\end{eqnarray}
Thanks to the conformal symmetry, this condition can be strengthened to the equality of $G$-currents. Indeed, since $Q_1-Q_2$ annihilates the boundary state, the two currents restricted to the boundary can differ at most by a total derivative along the boundary, $f_1(j^a) - f_2(j^a)= \partial_{\parallel} q^a$. As the currents have conformal dimension $1$, $q^a$ has conformal dimension $0$. On an indecomposable boundary, the only dimension-$0$ operator is the identity, whose derivative vanishes. Hence the left and right currents match on the boundary, 
\begin{eqnarray}\label{eq:f12D}
    f_1(j)|_{\gamma} \ket{\mathcal{D}}= f_2(j)|_{\gamma}  \ket{\mathcal{D}}.
\end{eqnarray}
The current-gluing condition \eqref{eq:f12D} is much stronger than the charge-matching conditions \eqref{eq:intf12D} and \eqref{eq:U12D}: it implies the conservation of infinitely many charges (all modes of the currents), whereas  \eqref{eq:intf12D} only involves zero modes. This enhancement is due to the conformal invariance.  In the rest of this work, our goal is to find boundary conditions satisfying \eqref{eq:f12D}; see the left of Figure \ref{fig:unfolding}.

Finding conformal boundary conditions seems to be a hard dynamical problem. It is, however, a kinematic, topological problem in disguise. To see this, we use the \emph{unfolding trick} \cite{Wei:2026fsn,WatanabeOhmori,Watanabetalk,Arias-Tamargo:2026urw,Cordova:2025zkz}, the inverse of the folding trick \cite{Oshikawa:1996ww,Bachas:2001vj}: unfolding the $N$ Dirac fermions along the conformal boundary $\ket{\mathcal{D}}$,  we obtain $N$ Weyl fermions on the full plane with a conformal defect $\mathcal{D}$ along $\gamma$. The theory of $N$ Weyl fermions, i.e. $\text{Fer}(N)$, is a holomorphic CFT, and any conformal line defect in a holomorphic CFT is automatically topological.\footnote{\label{footnote1} The conformal defect satisfies the gluing condition $(T(z)-\bar{T}(\bar{z}) )|_{\gamma^-} = (T(z)-\bar{T}(\bar{z})) |_{\gamma^+}$ for the stress-energy tensor, where $\gamma^\pm$ means approaching $\gamma$ from the right and left, respectively. In a holomorphic CFT, $\bar{T}(\bar{z})=0$, and the gluing condition reduces to $T(z)|_{\gamma^-} = T(z)|_{\gamma^+}$, which means that the defect $\mathcal{D}$ is topological. }

After unfolding, the current-gluing condition \eqref{eq:f12D} becomes
\begin{eqnarray}
\label{eq:f12intertwin}
    f_1(j)|_{\gamma} \mathcal{D} = \mathcal{D} f_2(j)|_{\gamma}.
\end{eqnarray}
In other words, the topological defect $\mathcal{D}$ intertwines the two realizations  \eqref{eq:f12} of the current algebra; see the right of Figure \ref{fig:unfolding}.
Our goal is to solve this problem using recent advances in generalized symmetries \cite{Gaiotto:2014kfa}, in particular the symmetry-subalgebra duality \cite{Rayhaun:2023pgc,Moller:2024xtt,Gannon:2026ttf}.

\subsection{Sandwich Construction from Conformal Embedding}
\label{sec:interdefect conformalembed}

Rather than starting from the most generic case that we can address (which we deal with in Section \ref{sec:coset} and also in Section \ref{sec:orbifoldequivalence}), we would like to motivate our construction of the intertwining defect in a very simple, restricted situation where the maps \eqref{eq:fi} are conformal embeddings
\begin{eqnarray}\label{eq:gUN1embed}
    f_i: \widehat{\mathfrak{g}} \hookrightconf \Fer(N).
\end{eqnarray}
Examples include the abelian case where $G=U(1)^N$ is the Cartan torus of $U(N)$, and the cases $n\leq 2$ and $(N_f,n)=(1,3)$ of Section \ref{sec:highcharge}.
Conformal embedding means that the Sugawara stress tensors of $\widehat{\mathfrak{g}}$ and $\Fer(N)$ coincide
\begin{eqnarray}
    T_{\widehat{\mathfrak{g}}} = T_{\Fer(N)}
\end{eqnarray}
and so $c_{\widehat{\mathfrak{g}}} = c_{\Fer(N)}=N$.

\begin{figure}[tbp]
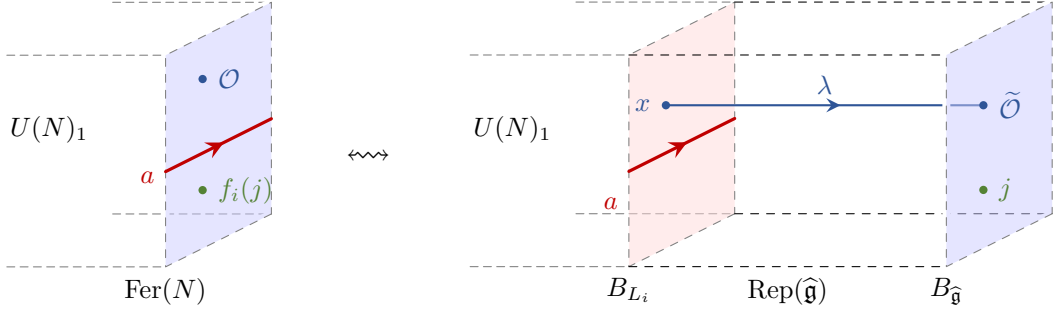

    \centering
    \figSandwichLeft\qquad $\leftrightsquigarrow$ \qquad\figSandwichRight
    \caption{Sandwich construction of chiral CFT from conformal embedding.}
    \label{fig:sandwich}
\end{figure}

We use the sandwich construction \cite{Cordova:2025zkz,Moller:2024xtt,Gannon:2026ttf,Etingof:2009yvg,Davydov:2010kfz} to find the topological defect $\mathcal{D}$ intertwining the two conformal embeddings $f_{1,2}$ now (Figure \ref{fig:sandwich}).\footnote{The description of boundary conditions and defects of rational CFTs in terms of 3d TFT goes back to \cite{Felder:1999cv,Fuchs:2002cm,Kapustin:2010if,Fuchs:2012dt}, and topological defects of rational CFTs were studied in \cite{Petkova:2000ip,Frohlich:2004ef,Frohlich:2006ch}.}
First of all, the chiral CFT $\text{Fer}(N)$ can be viewed as a boundary condition of the 3d invertible spin-TQFT $U(N)_1$. The latter contains two topological lines, the identity line and the transparent fermion line. The transparent fermion line terminates on the boundary at a fermionic operator, e.g. $\psi$. \cite{Witten:1988hf,Fuchs:2002cm,Elitzur:1989nr}. 
Then the conformal embedding \eqref{eq:gUN1embed} allows us to resolve $\Fer(N)$ into a 3d sandwich, shown on the right of Figure \ref{fig:sandwich}, which consists of the following ingredients. 
\begin{itemize}
    \item \textbf{Bulk}: The bulk is a 3d (spin) TQFT whose topological lines form the (super) modular tensor category (MTC) $\Rep(\widehat{\mathfrak{g}})$. Note that $\widehat{\mathfrak{g}}$ is extended by an appropriate simple current when $\mathfrak{g}$ contains a $\mathfrak{u}(1)$ factor.
    \item \textbf{Chiral boundary}: The boundary condition $B_{\widehat{\mathfrak{g}}}$ on the right of the sandwich is a \emph{chiral boundary} of $\Rep(\widehat{\mathfrak{g}})$. The local operators on it include  $\widehat{\mathfrak{g}}$ currents and their descendents.
    \item \textbf{Topological Interface}: The interface $B_{L_i}$ between the 3d bulk $\Rep(\widehat{\mathfrak{g}})$ and the invertible spin-TQFT $U(N)_1$ is topological. This is because, thanks to conformal embedding, the representation categories of the two chiral algebras are related by condensation of a fermionic Lagrangian algebra $L_i$ \cite{Gannon:2026ttf,Kong:2013aya}, \emph{i.e.}, $U(N)_1= \frac{\Rep(\widehat{\mathfrak{g}})}{L_i}$.
    The Lagrangian algebra $L_i$ is determined by the branching rule of the vacuum module of $\Fer(N)$ into $\widehat{\mathfrak{g}}$-modules,\footnote{Note that $\widehat{\mathfrak{u}}(N)_1$ is a super chiral algebra: its vacuum module and the module of the transparent fermion combine into a single vacuum \emph{supermodule}. In the rest of the paper, by a module we mean a supermodule whenever it does not lead to confusion. }
    \begin{eqnarray}
        \widehat{\mathfrak{u}}(N)_1|_{\widehat{\mathfrak{g}}} = \bigoplus_{\lambda \in \text{Irr}(\Rep(\widehat{\mathfrak{g}}))} M^{(i)}_\lambda \otimes V_\lambda
    \end{eqnarray}
    which gives 
    \begin{eqnarray}\label{eq:L12}
        L_i=\oplus_\lambda n^{(i)}_\lambda \lambda, \qquad n^{(i)}_\lambda = \dim(M^{(i)}_\lambda).
    \end{eqnarray}
    The lines in $L_i$ can topologically terminate on $B_{L_i}$. 
    Since the TQFT on the left side of the interface is invertible, we will refer to $B_{L_i}$ as a \emph{topological boundary}. 
\end{itemize}

Local operators of $\text{Fer}(N)$ have a simple resolution in the sandwich. A local operator, for instance a fermion $\psi$, is realized as a segment of a topological line $\lambda\in L_i$,  which ends on $B_{L_i}$ at a topological junction $x$ and on $B_{\widehat{\mathfrak{g}}}$ at a non-topological junction $\widetilde{\mathcal{O}}$. In short, 
\begin{eqnarray}
	\mathcal{O} \leftrightarrow (x,\lambda, \widetilde{\mathcal{O}})
\end{eqnarray}
See Figure \ref{fig:sandwich}. 
As a special case, a $\widehat{\mathfrak{g}}$ current of $\text{Fer}(N)$ is realized as a local current $j$ on $B_{\widehat{\mathfrak{g}}}$, attached to the trivial line. Shrinking the sandwich then realizes the embedding $f_i: \widehat{\mathfrak{g}}\to \Fer(N)$ as
\begin{eqnarray}
	f_i(j) \leftrightarrow (1,1,j).
\end{eqnarray}

The topological lines on $B_{L_i}$, such as the line $a$ in Figure \ref{fig:sandwich},  form the fusion category $\mathcal{C}_{L_i} = \Mod_{L_i}(\Rep(\widehat{\mathfrak{g}}))$ \cite{Cordova:2025zkz,Gannon:2026ttf}. After shrinking the sandwich, they become the topological lines, i.e., generalized symmetries, of $\text{Fer}(N)$.  In fact, they are \emph{all} the topological lines of $\text{Fer}(N)$ commuting with $\widehat{\mathfrak{g}}$. Some of them arise as the IR fixed point of Kondo impurities \cite{Affleck:1995ge,Affleck:1990iv}.

The above discussion is a special instance of the symmetry-subalgebra duality \cite{Moller:2024xtt,Gannon:2026ttf}, which relates a conformal embedding of chiral algebras $W\subset V$ to the symmetry category $\text{Ver}(V/W)$ of the (relative) chiral CFT with chiral algebra $V$. In our context, $V=\Fer(N), W=\widehat{\mathfrak{g}}$, and $\text{Ver}(V/W)= \mathcal{C}_{L_i}$.

While the sandwich construction is a reminiscent of the SymTFT construction\cite{Gaiotto:2020iye,Freed:2022qnc,Apruzzi:2021nmk,Ji:2019jhk}, what we do here is \emph{not} exactly identical to it. 
In the SymTFT construction, the  bulk is always the Turaev-Viro TQFT of a fusion category, in particular, the TQFT is non-chiral and admits a gapped boundary. In contrast, the bulk TQFT $\Rep(\widehat{\mathfrak{g}})$ here is in general chiral and does not admit a gapped boundary, although it admits a topological interface to an invertible \emph{spin}-TQFT. Nevertheless, folding along $B_{L_i}$ turns $B_{L_i}$ into a topological boundary of the folded TQFT $\Rep(\widehat{\mathfrak{g}})\boxtimes U(N)_{-1}$ which can then be interpreted as a SymTFT.

\subsection{Intertwining Defect from Sandwich}
\label{sec:interDefect}

\begin{figure}
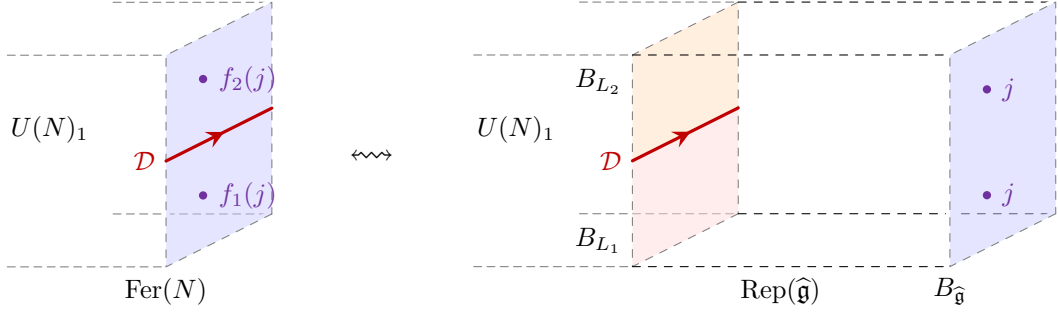

    \centering
    \figFtwelveLeft\qquad $\leftrightsquigarrow$ \qquad\figFtwelveRight
    \caption{Topological defect $\mathcal{D}$ intertwining two conformal embeddings of $\widehat{\mathfrak{g}}$ current. }
    \label{fig:f12}
\end{figure}

Having established the sandwich construction for an individual conformal embedding $f_i$, we now place two copies of $\Fer(N)$, realizing two embeddings $f_1$ and $f_2$ respectively, side by side, and  connect them by a topological interface $\mathcal{D}$, as shown on the left of Figure \ref{fig:f12}. 
Below, we use the sandwich construction to construct $\mathcal{D}$ and to show that it obeys the intertwining relation \eqref{eq:f12intertwin}.

Following Section \ref{sec:interdefect conformalembed}, we obtain two sandwiches, as shown on the right of Figure \ref{fig:f12}. The two sandwiches share the same bulk TQFT  $\Rep(\widehat{\mathfrak{g}})$ and the same chiral boundary $B_{\widehat{\mathfrak{g}}}$. However, since the embeddings $f_1$ and $f_2$ are distinct, so are the topological boundaries $B_{L_1}$ and $B_{L_2}$, which are associated with fermionic Lagrangian algebras $L_1$ and $L_2$ defined in \eqref{eq:L12}.

By \cite[Theorem 1.1]{etingof2009fusioncategorieshomotopytheory} and \cite[Proposition 4.8]{Davydov:2010kfz}, a topological line junction $\mathcal{D}$ between $B_{L_1}$ and $B_{L_2}$ always exists. Such junctions form a $(\mathcal{C}_{L_1},\mathcal{C}_{L_2})$ bimodule category, and a junction $\mathcal{D}$ is in general a duality defect for non-invertible symmetry\cite{Choi:2023vgk,Diatlyk:2023fwf}, with
\begin{eqnarray}
    \mathcal{D}^\dagger\otimes \mathcal{D} = \mathcal{A} \in \mathcal{C}_{L_1}.
\end{eqnarray}
The topological boundary $B_{L_2}$ is obtained from $B_{L_1}$ by gauging the Frobenius algebra $\mathcal{A}$ in $B_{L_1}$. The algebra $\mathcal{A}$ has a simple relation with the Lagrangian algebra $L_2$ \cite{Putrov:2024uor}: the image $F_{B_{L_1}}(L_2)$ of $L_2$ under the bulk-to-boundary functor $F_{B_{L_1}}(L_2)$ of $B_{L_1}$ contains  
\begin{eqnarray}
    \dim(\Hom(L_1, L_2)) = \sum_{\lambda} n_\lambda^{(1)} n_\lambda^{(2)}
\end{eqnarray}
copies of algebras in the Morita class of $\mathcal{A}$.  
After shrinking the sandwich, the topological junction $\mathcal{D}$ becomes a topological line defect of $\text{Fer}(N)$. 

The intertwining relation \eqref{eq:f12intertwin} follows directly from Figure \ref{fig:f12}: the current $j$ lives on the common chiral boundary and is attached to the trivial bulk line, whereas $\mathcal{D}$ lies on the opposite, topological, boundary. Hence $j$ can be moved from one side of $\mathcal{D}$ to the other without ever crossing it; after shrinking the sandwich, this turns  $f_1(j)$ into $f_2(j)$. This proves \eqref{eq:f12intertwin}. Folding back along $\mathcal{D}$ gives the desired conformal boundary condition preserving the affine symmetry $\widehat{\mathfrak{g}}$, and hence $G$.

\subsection{Categorical Scattering}

We now solve the scattering problem, i.e. we determine the out-going excitation when a fermion scatters off a $G$ preserving conformal boundary. 
In the unfolded theory, the scattering amounts to moving a chiral fermion through $\mathcal{D}$, or equivalently, acting  with $\mathcal{D}$ on the chiral fermion.  This is a purely kinematic problem, which can be solved explicitly. The result is 
\begin{eqnarray}\label{eq:Daction}
    \figBeforeScatter[1.2]
    \quad = \sum_{\substack{byzw\\b\in F_{B_{L_2}}(\lambda)}} \sqrt{\frac{d_b}{d_\lambda}} \frac{1}{d_{\mathcal{D}} \sqrt{S_{11}}} {^{B_{L_1} B_{L_2}}\widetilde{\Psi}_{\mathcal{D} \mathcal{D} (\bar{\lambda} xy)}^{(1)(\bar{b} w)} } \quad 
    \figAfterScatter[1.2]
\end{eqnarray}
Here $d_b, d_\lambda$ and $d_\mathcal{D}$ are quantum dimensions, $S_{11}$ is the vacuum entry of the modular $S$-matrix of $\Rep(\widehat{\mathfrak{g}})$, and $^{B_{L_1} B_{L_2}}\widetilde{\Psi}$ is the generalized half-linking number \cite{Choi:2024tri,Choi:2024wfm}. The sum runs over the simple lines $b$ in $F_{B_{L_2}}(\lambda)$, i.e. the lines on $B_{L_2}$ obtained by pushing the bulk line $\lambda$ onto $B_{L_2}$, and over the topological junctions $y,z,w$ that appear in the derivation below. On the right hand side, $\widetilde{\psi}$ denotes the operator in the $b$-twisted sector obtained by shrinking $(z,\lambda, \widetilde{x})$.

Let us derive \eqref{eq:Daction} by using the sandwich construction. In the sandwich, the fermion $\psi$ on the left-hand side is resolved as  $(x,\lambda, \widetilde{x})$, where $\lambda\in L_1$ is a bulk line, and $x, \widetilde{x}$ are its junctions on the two boundaries. Bending $\lambda$ towards $B_{L_2}$ and fusing it with $B_{L_2}$, we obtain 
\begin{equation}
    \figDerivationA[1.1]
    = \sum_{\substack{byz\\b\in F_{B_{L_2}}(\lambda)}} \sqrt{\frac{d_b}{d_\lambda}} \quad 
    \figDerivationMid[1.1]
\end{equation}
The half-circle formed by $\lambda$ on the right-hand side straddles $\mathcal{D}$, and can be shrunk to a junction $w$ on $\mathcal{D}$ using topological moves. This produces an additional factor, given by the generalized half-linking number \cite{Choi:2024tri,Huang:2023pyk}, and the right-hand side becomes 
\begin{eqnarray}
    \sum_{\substack{byzw\\b\in F_{B_{L_2}}(\lambda)}} \sqrt{\frac{d_b}{d_\lambda}} \frac{1}{d_{\mathcal{D}} \sqrt{S_{11}}} {^{B_{L_1} B_{L_2}}\widetilde{\Psi}_{\mathcal{D} \mathcal{D} (\bar{\lambda} xy)}^{(1)(\bar{b} w)} }  \figDerivationB[1.1]
\end{eqnarray}
Squeezing the sandwich gives the right hand side of \eqref{eq:Daction}, which completes the derivation.

We highlight some new features that are absent in the known abelian scattering examples \cite{Wei:2026fsn,Arias-Tamargo:2026urw,vanBeest:2023dbu}. 
\begin{enumerate}
    \item The out-going excitations (or, by the state-operator correspondence, the out-going state) is in general a superposition of states in different twisted sectors, labeled by topological lines $b\in F_{B_{L_2}}(\lambda)$. This is in contrast with the known scattering process, where the out-going state belongs to a single twisted sector. 
    \item The topological line $b\in F_{B_{L_2}}(\lambda)$ that labels each twisted sector is in general non-invertible. This is in contrast with the known cases, where the lines attached to the out-going excitations are invertible. 
\end{enumerate}

Equation \eqref{eq:Daction} also determines the  probability of scattering a fermion into a given twisted sector. In Appendix \ref{app:prob}, we show that the  squared norms of the coefficients sum to unity, 
\begin{eqnarray}\label{eq:sumto1}
    \sum_{\substack{bzw\\ b\in F_{B_{L_2}}(\lambda)}} \left| \sum_{y} \sqrt{\frac{d_b}{d_\lambda}} \frac{1}{d_{\mathcal{D}} \sqrt{S_{11}}} {^{B_{L_1} B_{L_2}}\widetilde{\Psi}_{\mathcal{D} \mathcal{D} (\bar{\lambda} xy)}^{(1)(\bar{b} w)} }  \right|^2 = 1.
\end{eqnarray}
Therefore, the probability of scattering a fermion into $b$-twisted sector is 
\begin{eqnarray}
    P_b = \sum_{zw} \left| \sum_{y} \sqrt{\frac{d_b}{d_\lambda}} \frac{1}{d_{\mathcal{D}} \sqrt{S_{11}}} {^{B_{L_1} B_{L_2}}\widetilde{\Psi}_{\mathcal{D} \mathcal{D} (\bar{\lambda} xy)}^{(1)(\bar{b} w)} }  \right|^2
\end{eqnarray}
and \eqref{eq:sumto1}, guarantees $\sum_b P_b=1$. Remarkably, these scattering probabilities are completely fixed by categorical data. Although the individual value of the generalized half-linking number $\widetilde{\Psi}$ is gauge-dependent on the choice of junction basis, we conjecture that the probability is not, and is a physical quantity, which would be interesting to calculate in concrete examples, and compare with numerical simulations or experiments.

\subsection{Including Coset}
\label{sec:coset}

So far we have assumed that the current algebra $\widehat{\mathfrak{g}}$ is conformally embedded in $\Fer(N)$. This is not the case in most of our examples, and then the TQFTs $\Rep(\widehat{\mathfrak{g}})$ and $U(N)_1$ are not separated by a topological interface.

To retrieve a conformal embedding, we include the  coset chiral algebra
\begin{eqnarray}\label{eq:coset}
    S_i={\Fer(N)}/{f_i(\widehat{\mathfrak{g}})}, \quad i=1,2,
\end{eqnarray}
i.e. the commutants of $f_i(\widehat{\mathfrak{g}})$ in $\Fer(N)$ for the embeddings \eqref{eq:fi}. Together with the coset chiral algebra $S_i$, $\widehat{\mathfrak{g}}$ is conformally embedded in $\Fer(N)$, 
\begin{eqnarray}\label{eq:conformalembed2}
    \widehat{\mathfrak{g}} \oplus S_i \hookrightconf \Fer(N).
\end{eqnarray}
Note that the two embeddings $f_{1}$ and $f_{2}$ in general give rise to different coset chiral algebras. Boundary conditions preserving only a subalgebra of the chiral algebra were studied in \cite{Fuchs:1999zi,Fuchs:1999xn}, and constructed from embeddings and coset chiral algebras in \cite{Maldacena:2001ky,Quella:2002ct}; cosets arising from different embeddings of the same chiral algebra appeared in \cite{Quella:2002fk}. These works start from bosonic full rational CFTs, whereas the theories of interest here are fermionic. 

\begin{figure}
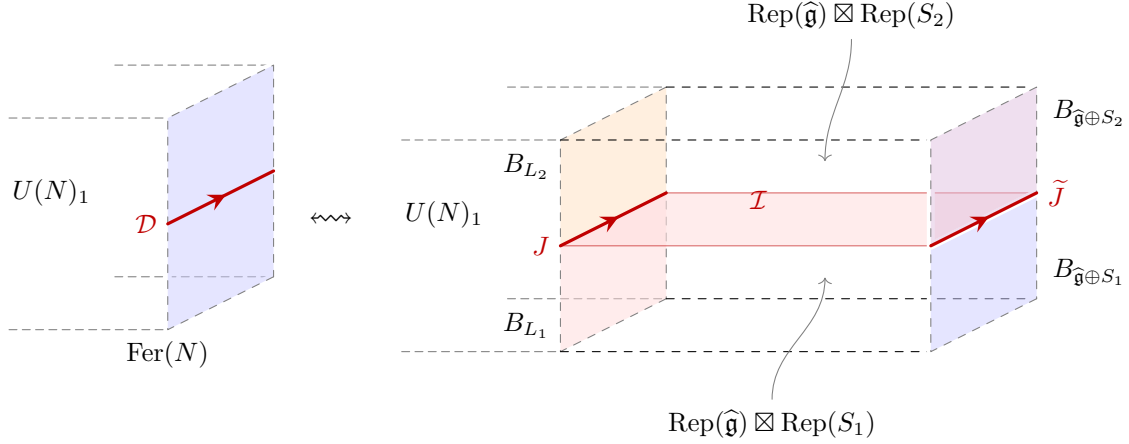

    \centering
     \figGeneralLeft\quad $\leftrightsquigarrow$ \quad\figGeneralRight\\
    \caption{Sandwich construction for two conformal embeddings $\widehat{\mathfrak{g}}\oplus S_i \subset \Fer(N)$. The intertwining defect $\mathcal{D}$ is realized as a topological interface $\mathcal{I}$ separating the two TQFTs, with two topological line junctions $J$ and $\widetilde{J}$ on two sides. }
    \label{fig:general}
\end{figure}

Using the conformal embeddings \eqref{eq:conformalembed2}, one can again construct the sandwiches following Section \ref{sec:interdefect conformalembed}. The bulk theory is now
\begin{eqnarray}
    \Rep(\widehat{\mathfrak{g}}) \boxtimes \Rep(S_i)
\end{eqnarray}
where $\Rep(S_i)$ is \emph{coset TQFT}, whose topological lines form the representation category of $S_i$ \cite{Cordova:2023jip}. The coset TQFT are in general different for $i=1,2$. The interface $B_{L_i}$ between the bulk theory and $U(N)_1$ is again topological, and is characterized by a fermionic Lagrangian algebra $L_i$. The chiral boundary $B_{\widehat{\mathfrak{g}} \oplus S_i}$ supports the chiral algebra $\widehat{\mathfrak{g}} \oplus S_i$. See Figure \ref{fig:general}.

Since the bulk theories are different, the two halves of the sandwich are separated by a topological interface $\mathcal{I}$, which is transparent to the lines of $\Rep(\widehat{\mathfrak{g}})$; The interface $\mathcal{I}$ meets $B_{L_1}$ and $B_{L_2}$ along a topological junction line $J$ on the left of the sandwich, meets the chiral boundaries $B_{\widehat{\mathfrak{g}} \oplus S_1}$ and $B_{\widehat{\mathfrak{g}} \oplus S_2}$ along a junction line $\widetilde{J}$. 

The existence of $\widetilde{J}$ is highly nontrivial, and has been studied in \cite{Moller:2024xtt} as \emph{interface equivalence}. In particular, $\widetilde{J}$ exists whenever $S_1$ and $S_2$ are orbifold equivalent \cite[Proposition 4.12]{Moller:2024xtt}, see Section \ref{sec:orbifoldequivalence}. But for the problem here, we do not have to directly work with Figure \ref{fig:general}.

To start with, in Sections \ref{sec:automorphism} and \ref{sec:commonalg}, we discuss two simple cases in which this complication is absent. In Section \ref{sec:orbifoldequivalence}, we propose that, by including spectators and using the orbifold equivalence \cite{Moller:2024xtt}, the general case can be reduced to the second simple case.

\subsection{Simple Case 1: Automorphism of Embeddings}
\label{sec:automorphism}

One simple case is when the two (non-conformal) embeddings $f_{1,2}: \widehat{\mathfrak{g}} \to \Fer(N)$ are realized by an automorphism $A$ of $\widehat{\mathfrak{g}}$, 
\begin{eqnarray}\label{eq:f12A}
    f_2=  f_1 \circ A.
\end{eqnarray}

Let $\widehat{f}_1$ be the  conformal embedding obtained from $f_1$ by including the coset chiral algebra $S\equiv S_1$, 
\begin{eqnarray}
    \widehat{f}_1: \widehat{\mathfrak{g}}\oplus S \hookrightconf \Fer(N).
\end{eqnarray}
Being conformal means that the two sides share the same stress tensor 
\begin{eqnarray}
    T_{\widehat{\mathfrak{g}}} + T_{S} = T_{\Fer(N)}.
\end{eqnarray}
Because of \eqref{eq:f12A}, we can construct another conformal embedding $\widehat{f}_2$ by composing $\widehat{f}_1$ with $A \oplus \text{id}_S$, which acts as $A$ on $\widehat{\mathfrak{g}}$ and as the identity on $S$, 
\begin{eqnarray}
    \widehat{f}_2= \widehat{f}_1\circ (A \oplus \text{id}_S): \widehat{\mathfrak{g}}\oplus S \hookrightconf \Fer(N)
\end{eqnarray}
The embedding $\widehat{f}_2$ is conformal because the automorphism $A$ leaves the Sugawara stress tensor of $\widehat{\mathfrak{g}}$ invariant, $T_{A\circ\widehat{\mathfrak{g}}}= T_{\widehat{\mathfrak{g}}}$, and $A \oplus \text{id}_S$ acts trivially on $S$. 
Therefore, the coset chiral algebras induced by $f_1$ and $f_2$ are identical, $S_1 = S_2=S$.\footnote{Note that $\widehat{f}_2$ is not necessarily obtained by composing $\widehat{f}_1$ and an automorphism of $\Fer(N)$. If this is the case, we can simply use the invertible defect implementing the automorphism on $\Fer(N)$. Interesting cases are when there is no such automorphism.}

\begin{figure}
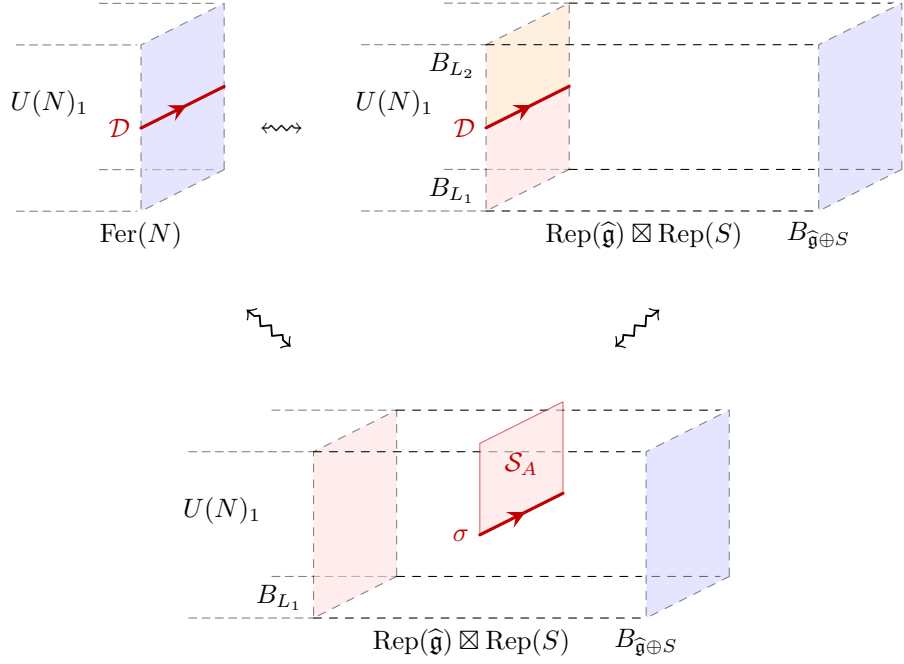

    \centering
     \figGeneralLeft[1.1]\quad $\leftrightsquigarrow$ \quad\figAutoSplit[1.1]\\[3ex]
    \rotatebox[origin=c]{-40}{\Large$\leftrightsquigarrow$}\qquad\qquad\qquad\qquad\qquad\rotatebox[origin=c]{40}{\Large$\leftrightsquigarrow$}\\[3ex]
    \figAutoSurface[1.1]
    \caption{Sandwich construction when the two embeddings are related by an automorphism, $f_2=f_1\circ A$. The two topological boundaries are related by $B_{L_2}=B_{L_1}\otimes\mathcal{S}_A$ (upper right), and the intertwining defect $\mathcal{D}$ is obtained by fusing the twist defect $\sigma$, i.e. the boundary of the truncated surface $\mathcal{S}_{A}$ (bottom), onto $B_{L_1}$. }
    \label{fig:auto}
\end{figure}

In this special case, the sandwich construction simplifies compared to Figure \ref{fig:general}: the two embeddings share the same bulk TQFT $\Rep(\widehat{\mathfrak{g}}) \boxtimes \Rep(S)$ and the same chiral boundary $B_{\widehat{\mathfrak{g}}\oplus S}$, and only the topological interfaces $B_{L_1}$ and $B_{L_2}$ between $U(N)_1$ and $\Rep(\widehat{\mathfrak{g}}) \boxtimes \Rep(S)$ differ. See the upper right of Figure \ref{fig:auto}.

The intertwining defect $\mathcal{D}$ also admits a simpler description. The automorphism $A$ of $\widehat{\mathfrak{g}}$ is realized as an invertible topological surface $\mathcal{S}_A$ in the bulk TQFT $\Rep(\widehat{\mathfrak{g}})$ (acting trivially on $\Rep(S)$), which permutes the bulk lines as $\lambda\to A(\lambda)$.  
Such surfaces can be realized by higher gauging the topological lines in $\Rep(\widehat{\mathfrak{g}})$ and are called the \emph{condensation defects} \cite{Roumpedakis:2022aik}. The two topological boundaries $B_{L_1}$ and $B_{L_2}$ are related by fusing $\mathcal{S}_{A}$ onto $B_{L_1}$, 
\begin{eqnarray}
    B_{L_2}= B_{L_1} \otimes \mathcal{S}_{A}
\end{eqnarray}
Any such condensation defects can be truncated by higher gauging only on half space, whose boundary is called the twist defect $\sigma$ of the invertible surface $\mathcal{S}_{A}$ \cite{Teo:2015xla,Kaidi:2022cpf}. In short, $\sigma$ lives on the boundary of $\mathcal{S}_{A}$. 
Consequently, the intertwining defect $\mathcal{D}$ is obtained by fusing $\sigma$ onto $B_{L_1}$,   
\begin{eqnarray}\label{eq:Dsigmaid}
    \mathcal{D} = F_{B_{L_1}}(\sigma \boxtimes \text{id}_{S}). 
\end{eqnarray}
where $\text{id}_{S}$ is the identity line in the coset TQFT, and $F_{B_{L_1}}$ is the bulk-to-boundary functor of $B_{L_1}$. 
See Figure \ref{fig:auto}. In general, the right hand side of \eqref{eq:Dsigmaid} is a direct sum of simple interfaces on $B_{L_1}$, and we need to pick a simple object to be $\mathcal{D}$. 

The categorical scattering formula \eqref{eq:Daction} also simplifies significantly. Consider again a fermion scattering off a topological line $\mathcal{D}$. The configuration is
\begin{eqnarray}\label{eq:auto1}
    \figAutoA[1.0]
\end{eqnarray}
To move the fermion across $\sigma$, we perform a sequence of topological moves. We first fuse a segment of $\lambda$ with $\sigma$, which recasts \eqref{eq:auto1} as
\begin{eqnarray}\label{eq:auto2}
    \sum_{\sigma'uv} \sqrt{\frac{d_{\sigma'}}{d_{\sigma} d_{\lambda}}}
    \figAutoB[1.0]
\end{eqnarray}
Here $\lambda=\lambda_{\widehat{\mathfrak{g}}}\boxtimes \lambda_S$ is a product of lines in $\Rep(\widehat{\mathfrak{g}})$ and $\Rep(S)$. Since the twist defect is transparent to the lines in $\Rep(S)$, all the symbols in the coefficients only involve component $\lambda_{\widehat{\mathfrak{g}}}= \lambda|_{\Rep(\widehat{\mathfrak{g}})}$. We suppress the subscript to avoid cluttering the formula.

We then use two $R$-moves to flip $\lambda$ above $\sigma$, followed by an $F$-move to trade $\sigma'$ for a topological line $a$ ending on $\sigma$. The configuration \eqref{eq:auto2} then becomes 
\begin{eqnarray}\label{eq:auto3}
    \begin{split}
    \sum_{\substack{\sigma'uv\\ u' u'' azw}} \sqrt{\frac{d_{\sigma'}}{d_{\sigma} d_{\lambda}}} (R^{\sigma'}_{\sigma\lambda})^{-1}_{uu'}  &(R^{\sigma'}_{\lambda \sigma})^{-1}_{u'u''}(F^{\sigma}_{\sigma A(\lambda) \bar{\lambda}})^{-1}_{(\sigma'u''v)(\bar{a}zw)} \\
    &\times \quad \figAutoC[1.0]
    \end{split}
\end{eqnarray}
Combining the above steps  and shrinking the sandwich, we find the categorical scattering process
\begin{equation}\label{eq:auto45}
    \figBeforeScatter[1.0] 
    = \sum_{\substack{\sigma'uv\\ u' u'' azw}} \sqrt{\frac{d_{\sigma'}}{d_{\sigma} d_{\lambda}}} (R^{\sigma'}_{\sigma\lambda})^{-1}_{uu'}  (R^{\sigma'}_{\lambda \sigma})^{-1}_{u'u''} (F^{\sigma}_{\sigma A(\lambda) \bar{\lambda}})^{-1}_{(\sigma'u''v)(\bar{a}zw)} 
    \figAfterScatterGen{1.0}{a}{z} 
\end{equation}
Here $A(\lambda)$ is the image of $\lambda$ after crossing the surface $\mathcal{S}_A$, and the $F$-move acts on the junctions $u''$ and $v$ obtained after the two $R$-moves.

We make some remarks on \eqref{eq:auto45}. 
\begin{enumerate}
    \item As for \eqref{eq:Daction}, the coefficients in \eqref{eq:auto45} are also square-normalized and admit a probability interpretation: the probability of scattering into the $a$-twisted sector is
    \begin{eqnarray}
        P_a= \sum_{zw} \left| \sum_{\sigma'uvu'u''} \sqrt{\frac{d_{\sigma'}}{d_{\sigma} d_{\lambda}}} (R^{\sigma'}_{\sigma\lambda})^{-1}_{uu'}  (R^{\sigma'}_{\lambda \sigma})^{-1}_{u'u''} (F^{\sigma}_{\sigma A(\lambda) \bar{\lambda}})^{-1}_{(\sigma'u''v)(\bar{a}zw)}  \right|^2
    \end{eqnarray}
    and satisfies $\sum_a P_a=1$. 
    \item Although the coset chiral algebra $S$ and coset TQFT $\Rep(S)$ can be very complicated, the scattering amplitudes do not depend on any topological data of the coset sector. 
    \item The coefficients only involve simpler topological data than the generalized half-linking numbers, namely the $F$- and $R$-symbols of the bulk lines and of the twist defects of the automorphism symmetry, i.e. the data of the symmetry-crossed braided extension of $\Rep(\widehat{\mathfrak{g}})$. These are more accessible to compute.   
\end{enumerate}

The scattering of fermions off higher-charge monopoles in 4d QED, discussed in Section \ref{sec:highcharge}, is of this type. This will be the first set of example we compute that go beyond the existing literature.

\subsection{Simple Case 2: Common Conformal Subalgebra}
\label{sec:commonalg}

\begin{figure}
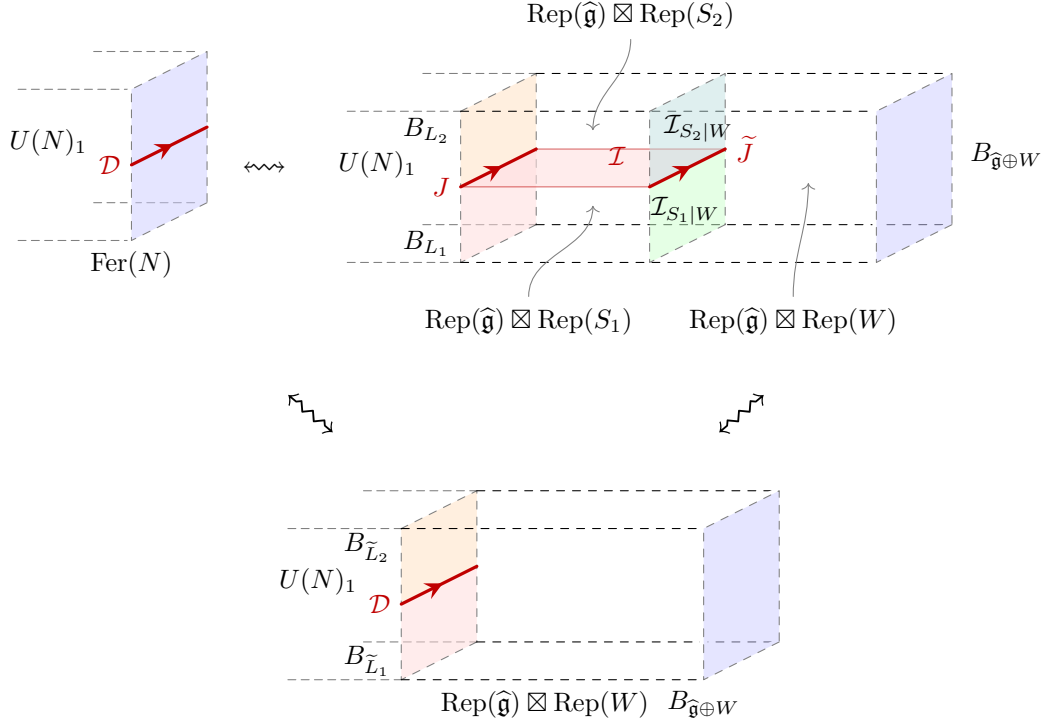

    \centering
    \figGeneralLeft[1.0]\quad $\leftrightsquigarrow$ \quad\figClubW[1.0]\\[3ex]
    \rotatebox[origin=c]{-40}{\Large$\leftrightsquigarrow$}\qquad\qquad\qquad\qquad\qquad\qquad\rotatebox[origin=c]{40}{\Large$\leftrightsquigarrow$}\\
    [3ex]
    \figClubWShrunk[1.0]
    \caption{Club sandwich (upper right) from the conformal embeddings \eqref{eq:S1S2W}. Shrinking the interfaces $\mathcal{I}_{S_i|W}$ to the right reproduces Figure \ref{fig:general}, while shrinking them to the left gives the sandwich at the bottom, in which the two topological boundaries $B_{\widetilde{L}_i}$ share the bulk TQFT $\Rep(\widehat{\mathfrak{g}})\boxtimes\Rep(W)$. }
    \label{fig:S1S2W}
\end{figure}

Another simple case is when there exists a chiral algebra $W$ that conformally embeds into both coset chiral algebras $S_1$ and $S_2$\footnote{Note that we do not demand that the two embeddings are compatible with the further embeddings \eqref{eq:conformalembed2} into $\Fer(N)$, so that $W$ could be embedded in $\Fer(N)$ in two different ways.}, 
\begin{eqnarray}\label{eq:S1S2W}
    \begin{tikzcd}[row sep=large, column sep=small, every label/.append style={sloped}]
S_1 & & S_2 \\
& W \arrow[ul, hook, "\text{conf}"] \arrow[ur, hook, "\text{conf}"].
\end{tikzcd}
\end{eqnarray}
The conformal embeddings imply that $\Rep(S_i)$ is obtained from $\Rep(W)$ by anyon condensation. Hence 
the TQFTs $\Rep(W)$ and $\Rep(S_i)$ are separated by a topological interface $\mathcal{I}_{S_i|W}, i=1,2$, which implements the condensation from the larger TQFT $\Rep(W)$ to the smaller one $\Rep(S_i)$. We therefore obtain a club sandwich \cite{Bhardwaj:2023bbf,Bhardwaj:2024qrf}, shown in the upper right of Figure \ref{fig:S1S2W}, which consists of the following ingredients.
\begin{itemize}
    \item \textbf{Right bulk}: The TQFT in the right bulk is $\Rep(\widehat{\mathfrak{g}})\boxtimes \Rep(W)$. 
    \item \textbf{Left bulks}: The TQFTs in the bottom and top left bulks are $\Rep(\widehat{\mathfrak{g}})\boxtimes \Rep(S_1)$ and $\Rep(\widehat{\mathfrak{g}})\boxtimes \Rep(S_2)$, respectively. 
    \item \textbf{Chiral boundary}: The right boundary of $\Rep(\widehat{\mathfrak{g}})\boxtimes \Rep(W)$ is a chiral boundary $B_{\widehat{\mathfrak{g}}\oplus W}$ carrying the chiral algebra $\widehat{\mathfrak{g}} \oplus W$. 
    \item \textbf{Topological interface $\mathcal{I}_{S_i|W}$}: The topological interface between $\Rep(\widehat{\mathfrak{g}})\boxtimes \Rep(W)$ and $\Rep(\widehat{\mathfrak{g}})\boxtimes \Rep(S_i)$ is $\mathcal{I}_{S_i|W}$, induced by the conformal embedding $W\hookrightconf S_i$. It is transparent to lines in $\Rep(\widehat{\mathfrak{g}})$.  
    \item \textbf{Topological interface $\mathcal{I}$}: The topological interface between $\Rep(\widehat{\mathfrak{g}})\boxtimes \Rep(S_1)$ and $\Rep(\widehat{\mathfrak{g}})\boxtimes \Rep(S_2)$ is $\mathcal{I}\equiv \text{id}_{\widehat{\mathfrak{g}}}\otimes \mathcal{I}_{S_1|S_2}$, which is again transparent to the lines in $\Rep(\widehat{\mathfrak{g}})$. It  meets the topological boundaries along the junction $J$, and the interfaces $\mathcal{I}_{S_i|W}$ along the junction $\widetilde{J}$. One can take $\mathcal{I}_{S_1|S_2}$ to be a simple summand of the composite of $\mathcal{I}_{S_1|W}$ and $\mathcal{I}_{S_2|W}$ through $\Rep(W)$, so that $\widetilde{J}$ exists by construction. The junction $J$ exists since $B_{L_1}\otimes\mathcal{I}$ and $B_{L_2}$ are topological boundaries of the same TQFT $\Rep(\widehat{\mathfrak{g}})\boxtimes\Rep(S_2)$ \cite{etingof2009fusioncategorieshomotopytheory,Davydov:2010kfz}. 
    \item \textbf{Topological boundary $B_{L_i}$}: The left boundary of $\Rep(\widehat{\mathfrak{g}})\boxtimes \Rep(S_i)$ is the topological boundary $B_{L_i}$ of Section \ref{sec:coset}. 
\end{itemize}

We can now shrink the club sandwich in two different orders. Shrinking the interfaces $\mathcal{I}_{S_1|W}$ and $\mathcal{I}_{S_2|W}$ onto the chiral boundary on the right reproduces the configuration of Figure \ref{fig:general}. In particular, this establishes the existence of the topological junction line $\widetilde{J}$ in the special case \eqref{eq:S1S2W}. Shrinking them instead onto the topological boundaries on the left gives the configuration at the bottom of Figure \ref{fig:S1S2W}, with the new topological boundaries
\begin{eqnarray}
    B_{\widetilde{L}_i} = B_{L_i} \otimes_{{\Rep(\widehat{\mathfrak{g}}) \boxtimes \Rep(S_i)}} \mathcal{I}_{S_i|W}, \quad i=1,2
\end{eqnarray}
Here $\widetilde{L}_i$ is the fermionic Lagrangian algebra of $\Rep(\widehat{\mathfrak{g}})\boxtimes \Rep(W)$ associated with the conformal embedding 
\begin{eqnarray}
    \widehat{\mathfrak{g}}\oplus W \hookrightconf \widehat{\mathfrak{g}}\oplus S_i \hookrightconf \Fer(N)
\end{eqnarray}
Now both topological boundaries share the same bulk TQFT, and the same chiral boundary, exactly as in Section \ref{sec:interDefect}. Therefore, the general formula \eqref{eq:Daction} for the categorical scattering applies, with $\Rep(\widehat{\mathfrak{g}})$ replaced by $\Rep(\widehat{\mathfrak{g}})\boxtimes \Rep(W)$ and $B_{L_i}$ by $B_{\widetilde{L}_i}$.

Examples in which the coset chiral algebras share a common conformal subalgebra turn out to be quite ubiquitous. They include the $SU(N) \times SU(N+2)$ model in Section \ref{sec:susu}, the monopole scattering problems in certain chiral QED theories in Section \ref{sec:chiralqed}, and the scattering off  the high charge monopoles in the Standard Model in Section \ref{sec:standardmodel}. More general cases where the two coset chiral algebras do not fit into the simple cases in Section \ref{sec:automorphism} and \ref{sec:commonalg} will be discussed in Section \ref{sec:orbifoldequivalence}.

\subsection{Symmetry Category of Intertwining Defects}
\label{sec:symcat}

The intertwining defect $\mathcal{D}$ constructed above is not unique, and it is naturally a member of a larger symmetry category. We describe it uniformly for the situations of Sections \ref{sec:interDefect}, \ref{sec:automorphism} and \ref{sec:commonalg}. In all of them, after shrinking the interfaces as in Figure \ref{fig:S1S2W}, the sandwich consists of a single bulk TQFT $\Rep(\widehat{\mathfrak{g}})\boxtimes\Rep(W)$, a single chiral boundary $B_{\widehat{\mathfrak{g}}\oplus W}$, and two topological boundaries $B_{L_1}$ and $B_{L_2}$ meeting along $\mathcal{D}$. Here $W$ is trivial in Section \ref{sec:interDefect}, $W=S$ in Section \ref{sec:automorphism}, and $W$ is the common conformal subalgebra in Section \ref{sec:commonalg}, where $L_i$ stands for $\widetilde{L}_i$. Correspondingly, $f_i$ extend to conformal embeddings
\begin{eqnarray}
    \widehat{f}_i: \widehat{\mathfrak{g}}\oplus W \hookrightconf \Fer(N), \qquad i=1,2,
\end{eqnarray}
with $\widehat{f}_i=f_i$ when $W$ is trivial. The general case of Section \ref{sec:orbifoldequivalence} reduces to this setting after including spectators, and the discussion below applies to the enlarged theory.

The topological junctions between $B_{L_1}$ and $B_{L_2}$ form a $(\mathcal{C}_{L_1},\mathcal{C}_{L_2})$-bimodule category $\mathcal{M}_{L_1L_2}$, which is invertible, i.e.\ it exhibits a Morita equivalence between $\mathcal{C}_{L_1}$ and $\mathcal{C}_{L_2}$ \cite{etingof2009fusioncategorieshomotopytheory,Davydov:2010kfz}. It can be described in terms of the algebra $\mathcal{A}\in\mathcal{C}_{L_1}$ of Section \ref{sec:interDefect}. Recall that $B_{L_1}$ condenses $L_1$, so that a bulk line $c$ pushed onto $B_{L_1}$ carries an action $\mu_c:L_1\otimes c\to c$, given by the junctions at which the condensed $L_1$-lines end on $c$; the lines of $\mathcal{C}_{L_1}=\Mod_{L_1}(\Rep(\widehat{\mathfrak{g}})\boxtimes\Rep(W))$ are such $L_1$-modules, with the right action obtained from $\mu_c$ by the braiding. Two lines $c,c'$ adjacent along $B_{L_1}$ fuse into $c\otimes_{L_1}c'$, the coequalizer of the two actions $c\otimes L_1\otimes c'\rightrightarrows c\otimes c'$, which identifies the condensate absorbed by either line. Recall also that $B_{L_2}$ is obtained from $B_{L_1}$ by gauging $\mathcal{A}$. Viewing $B_{L_2}$ as $B_{L_1}$ with $\mathcal{A}$ gauged, a junction between $B_{L_1}$ and $B_{L_2}$ is a line $m$ of $\mathcal{C}_{L_1}$ on which the $\mathcal{A}$-lines of the gauged region can end, i.e.\ a right $\mathcal{A}$-module in $\mathcal{C}_{L_1}$ with action $m\otimes_{L_1}\mathcal{A}\to m$. Being a morphism in $\mathcal{C}_{L_1}$, this action is compatible with the $L_1$-action, so that the condensate of $B_{L_1}$ passes freely through the junction. Similarly, a line on $B_{L_2}$ is an $\mathcal{A}$-bimodule \cite{ostrik2003module}\footnote{Alternatively and more symmetrically, since $L_1$ and $L_2$ are commutative, a module over the algebra $L_1\otimes L_2$ in $\Rep(\widehat{\mathfrak{g}})\boxtimes\Rep(W)$ is the same as an $(L_1,L_2)$-bimodule. The algebra $L_1\otimes L_2$ is in general not connected, and $\mathcal{M}_{L_1L_2}$ is the category of modules over one of its connected summands, the one containing the junction $\mathcal{A}$ underlying $\mathcal{D}$. A junction $m$ is thus an $(L_1,L_2)$-bimodule, while the lines of $\mathcal{C}_{L_i}$ are $L_i$-modules, made into bimodules by the braiding, so that the balanced products $c\otimes_{L_1}m$ and $m\otimes_{L_2}b$ make sense; the latter agrees with $m\otimes_{\mathcal{A}}b$ below.}:
\begin{eqnarray}\label{eq:MAbimod}
    \mathcal{M}_{L_1L_2}\simeq \Mod_{\mathcal{A}}(\mathcal{C}_{L_1}), \qquad \mathcal{C}_{L_2}\simeq {}_{\mathcal{A}}\text{Bimod}_{\mathcal{A}}(\mathcal{C}_{L_1}).
\end{eqnarray}
The defect $\mathcal{D}$ of the previous subsections corresponds to $m=\mathcal{A}$ itself. By the argument of Section \ref{sec:interDefect}, \emph{every} object $m\in\mathcal{M}_{L_1L_2}$ gives a topological line $\mathcal{D}_m$ of $\Fer(N)$ with
\begin{eqnarray}
    \widehat{f}_1(v)\,\mathcal{D}_m = \mathcal{D}_m\,\widehat{f}_2(v), \qquad v\in \widehat{\mathfrak{g}}\oplus W,
\end{eqnarray}
and in particular \eqref{eq:f12intertwin}. The orientation reversal $\mathcal{D}^\dagger_m$ corresponds to the dual $m^*$, a left $\mathcal{A}$-module, and these form the $(\mathcal{C}_{L_2},\mathcal{C}_{L_1})$-bimodule category $\mathcal{M}_{L_2L_1}$ of junctions from $B_{L_2}$ to $B_{L_1}$. Together with the lines on the topological boundaries, they generate the category
\begin{eqnarray}\label{eq:Xcat}
    \mathfrak{X} = \langle\, \mathcal{C}_{L_1},\ \mathcal{C}_{L_2},\ \mathcal{M}_{L_1L_2},\ \mathcal{M}_{L_2L_1}\,\rangle
\end{eqnarray}
of topological lines on a topological boundary made piecewise of $B_{L_1}$ and $B_{L_2}$. Two lines adjacent along a common region of $B_{L_1}$ (resp.\ $B_{L_2}$) fuse by $\otimes_{L_1}$ (resp.\ additionally balanced over $\mathcal{A}$). For instance, for $c\in\mathcal{C}_{L_1}$, $b\in\mathcal{C}_{L_2}$ and $m,m'\in\mathcal{M}_{L_1L_2}$,
\begin{eqnarray}
    c\otimes_{L_1} m,\ \ m\otimes_{\mathcal{A}} b\ \in\mathcal{M}_{L_1L_2}, \qquad m\otimes_{\mathcal{A}} m'^{*}\in\mathcal{C}_{L_1}, \qquad m'^{*}\otimes_{L_1} m\in\mathcal{C}_{L_2},
\end{eqnarray}
and for $m=m'=\mathcal{A}$ the fusion into $\mathcal{C}_{L_1}$ reproduces $\mathcal{D}^\dagger\otimes\mathcal{D}=\mathcal{A}$ of Section \ref{sec:interDefect}. Otherwise lines are composed freely, so that products such as $c\otimes b\otimes c'$ with $c,c'\in\mathcal{C}_{L_1}$, or $\mathcal{D}_m\otimes\mathcal{D}_{m'}$, do not reduce. Hence $\mathfrak{X}$ is in general an infinite category.

Shrinking the sandwich turns every object of $\mathfrak{X}$ into a topological line of $\Fer(N)$, and the free products above into the ordinary fusion of lines. This action of $\mathfrak{X}$ on $\Fer(N)$ is in general not faithful, and the symmetry actually realized in $\Fer(N)$ is its image, a quotient $\mathfrak{X}\twoheadrightarrow\mathfrak{X}_{\Fer(N)}$ which depends on the embeddings $\widehat{f}_{1,2}$. To characterize $\mathfrak{X}_{\Fer(N)}$, consider the subalgebra on which the two embeddings agree,
\begin{eqnarray}\label{eq:E12}
    \mathcal{E}_{12} = \{\, v\in\widehat{\mathfrak{g}}\oplus W \ |\ \widehat{f}_1(v)=\widehat{f}_2(v)\,\}.
\end{eqnarray}
Since both embeddings are conformal, $\mathcal{E}_{12}$ contains the stress tensor, and $\widehat{f}_1(\mathcal{E}_{12})\hookrightconf\Fer(N)$ is a conformal embedding; on the other hand, $\mathcal{E}_{12}$ in general contains none of the $\widehat{\mathfrak{g}}$ currents. Every line of $\mathfrak{X}_{\Fer(N)}$ commutes with $\widehat{f}_1(\mathcal{E}_{12})$: the lines of $\mathcal{C}_{L_i}$ commute with all of $\widehat{f}_i(\widehat{\mathfrak{g}}\oplus W)$, and $\mathcal{D}_m$ maps $\widehat{f}_1(v)$ to $\widehat{f}_2(v)=\widehat{f}_1(v)$ for $v\in\mathcal{E}_{12}$. Therefore, if $\mathcal{E}_{12}$ is rational, $\mathfrak{X}_{\Fer(N)}$ is contained in the fusion category of topological lines commuting with $\widehat{f}_1(\mathcal{E}_{12})$, which is finite by the symmetry-subalgebra duality \cite{Moller:2024xtt,Gannon:2026ttf}, as in Section \ref{sec:interdefect conformalembed} with $\widehat{\mathfrak{g}}$ replaced by $\mathcal{E}_{12}$. Hence $\mathfrak{X}_{\Fer(N)}$ is finite. Conversely, when $\mathcal{E}_{12}$ is not rational, we expect $\mathfrak{X}_{\Fer(N)}$ to be infinite, although we do not attempt to prove it here.

Both possibilities are realized. In the example of Section \ref{sec:highcharge}, $\widehat{f}_2=\widehat{f}_1\circ(A\oplus\text{id}_S)$ with $A=\mathsf{C}_{\mathfrak{u}(1)}$ \eqref{eq:AQEDss}, so $\mathcal{E}_{12}$ is the $A$-invariant subalgebra, in which $\widehat{\mathfrak{u}}(1)_{nN_f}$ is replaced by its $\bZ_2$ orbifold while the other factors are kept intact. Being the fixed-point subalgebra of a finite-order automorphism of a rational chiral algebra, it is rational \cite{Carnahan:2016guf}, and accordingly $\mathfrak{X}_{\Fer(nN_f)}$ is finite, consistent with $\mathcal{D}$ being a duality defect of $\bZ_{nN_f}$ or $\bZ_{nN_f/2}$ found in Section \ref{sec:qedss}. On the other hand, consider $\Fer(2)$ with the anomaly-free $U(1)$ under which the left- and right-movers carry charges $(3,4)$ and $(0,5)$ respectively \cite{Smith:2019jnh,Wei:2026fsn}. The current algebra $\widehat{\mathfrak{u}}(1)_{25}$ is completed to a conformal subalgebra $\widehat{\mathfrak{u}}(1)^2$, with $K$-matrix $\text{diag}(25,25)$ and extended by simple currents, via
\begin{eqnarray}
    f_1(j^1) = 3J_1+4J_2,\quad f_1(j^2)=-4J_1+3J_2, \qquad f_2(j^1)=5J_2,\quad f_2(j^2)=-5J_1,
\end{eqnarray}
where $J_a = :\psi_a^\dagger\psi_a:$. Moving through $\mathcal{D}$ rotates $(J_1,J_2)$ by the angle $\theta=\arctan(3/4)$. Since $\theta/\pi$ is irrational, $k\theta\notin 2\pi\bZ$ for any $k\neq 0$, so $\mathcal{D}^{\otimes k}$ ($k\geq 1$), which rotates the currents by $k\theta$, never contains the trivial line; likewise $(\mathcal{D}^{\otimes k})^\dagger\otimes\mathcal{D}^{\otimes k'}$ with $k\neq k'$. If $\mathfrak{X}_{\Fer(2)}$ were finite, two of the lines $\mathcal{D}^{\otimes k}$ would share a simple summand, and the latter product would contain the trivial line. Hence $\mathfrak{X}_{\Fer(2)}$ is infinite.

\section{Example: Higher Charge Monopole in Non-chiral QED}
\label{sec:highcharge}

\subsection{Reduction from 4d and Status of the Puzzle}

Our first example is the scattering of fermions off a Dirac monopole in 4d QED \cite{vanBeest:2023dbu}. We consider 4d QED with $N_f$ Dirac fermions of unit charge, and place a static Dirac monopole of magnetic charge $n\in \bZ$, i.e. $\int_{S^2} dA=2\pi n$ for a sphere $S^2$ surrounding the origin of the space. The worldline of the monopole is an 't Hooft line extending along time. 

After partial wave reduction on $S^2$, and neglecting the fluctuation of the gauge field, the lowest partial waves give 
\begin{eqnarray}
    N=nN_f
\end{eqnarray}
massless Dirac fermions on a 2d half-plane $r\geq 0$, with a conformal boundary condition at $r=0$, specified by the 't Hooft line. The lowest partial waves are $n$-fold degenerate, and form the $n$-dimensional irreducible representation $\boldsymbol{n}$ of the rotation group $SU(2)$, which becomes an internal symmetry in 2d. We denote the in-coming and out-going components by $\psi$ and $\widetilde{\psi}$, respectively.  See \cite{vanBeest:2023dbu} for details of the reduction. 

What symmetry does the boundary condition preserve in 2d? 
In 4d, the flavor symmetry is $SU(N_f)_L \times SU(N_f)_R$, acting on the left-handed Weyl fermions $\Psi_L$ and $\Psi_R^c$. There is also a magnetic 1-form symmetry acting on the 't Hooft line. The 1-form symmetry and the flavor symmetry form a 2-group \cite{Benini:2018reh,Cordova:2018cvg}, hence only the subgroups of the flavor symmetry that do not participate in a 2-group can be preserved by the monopole. There are two such subgroups, namely the two diagonal subgroups $SU(N_f)$ under which $(\Psi_L, \Psi_R^c)$ transform as $(\square, \square)$ or as $(\square, \overline{\square})$. This gives two different boundary conformal field theory (BCFT) problems, which are called $\text{QED}(\square,\square)$ and $\text{QED}(\square,\overline\square)$ in \cite{vanBeest:2023dbu}.\footnote{Note that if we UV-complete QED to, say $SU(2)$ gauge theory, than the Polyakov monopole in $SU(2)$ gauge theory leads to situation $\text{QED}(\square,\square)$.} The symmetry quantum numbers of the 2d fermions are 
\begin{eqnarray}\label{eq:highcharge}
    \begin{tabular}{|c|c|c|}
    \hline
        $\text{QED}(\square,\square)$ & $\psi$ & $\widetilde{\psi}$ \\
         \hline
        $U(1)$ & $1$ & $-1$\\
        $SU(N_f)$ & $\square$ & ${\square}$\\
        $SU(2)$ & $\boldsymbol{n}$ & $\boldsymbol{n}$ \\
        $\bZ_{n N_f}$ & $1$ & $1$\\
        \hline
    \end{tabular} \qquad \qquad 
    \begin{tabular}{|c|c|c|}
    \hline
        $\text{QED}(\square,\overline\square)$ & $\psi$ & $\widetilde{\psi}$ \\
         \hline
        $U(1)$ & $1$ & $-1$\\
        $SU(N_f)$ & $\square$ & $\overline{\square}$\\
        $SU(2)$ & $\boldsymbol{n}$ & $\boldsymbol{n}$ \\
        $\bZ_{n}$ & $1$ & $1$\\
        \hline
    \end{tabular}
\end{eqnarray}

In both cases, the $U(1)$ symmetry descends from the 4d gauge group, the $SU(2)$ symmetry from the spatial rotations around the monopole, and the discrete symmetry from the (non-invertible) axial symmetry in 4d \cite{Choi:2022jqy,Cordova:2022ieu}. Since $\psi$ and $\widetilde{\psi}$ carry opposite $U(1)$ charges, and the same or conjugate representations of $SU(N_f)\times SU(2)$, and the 't Hooft anomalies of $U(1)\times SU(N_f)\times SU(2)$ cancel between the in-coming and out-going fermions.

The existence of symmetric monopole suggests the existence of a conformal boundary condition preserving all the symmetries in \eqref{eq:highcharge}. Such boundary condition is also expected from the lore discussed in Section \ref{sec:loreintro}. However, due to the imbalance between the quantum number for $\psi$ and $\widetilde{\psi}$, naive linear boundary condition doesn't exist. For instance, a linear boundary condition preserving $U(1)$ must take the form $\psi = M \widetilde{\psi}^\dagger$. However, since $\psi$ and $\widetilde{\psi}^\dagger$ carry opposite discrete charges, in both cases such a boundary condition violates the discrete symmetry in \eqref{eq:highcharge} for generic $n, N_f$.

\begin{figure}
    \centering
    	\begin{tikzpicture}
			\draw[thick, ->] (0,0) -- (3,0);
			\draw[thick, ->] (0,0) -- (0,3);
			\draw[dashed] (0,0.5) -- (3,0.5);
			\draw[dashed] (0,1) -- (3,1);
			\draw[dashed] (0,1.5) -- (3,1.5);
			\draw[dashed] (0,2) -- (3,2);
			\draw[dashed] (0,2.5) -- (3,2.5);
			\draw[dashed] (0.5,0) -- (0.5, 3);
			\draw[dashed] (1,0) -- (1, 3);
			\draw[dashed] (1.5,0) -- (1.5, 3);
			\draw[dashed] (2,0) -- (2, 3);
			\draw[dashed] (2.5,0) -- (2.5, 3);
			\node[below] at (0.5, 0) {$1$};
			\node[below] at (1, 0) {$2$};
			\node[below] at (1.5, 0) {$3$};
			\node[below] at (2, 0) {$4$};
			\node[below] at (2.5, 0) {$5$};
			\node[below] at (3, 0) {$N_f$};
			\node[left] at (0, 0.5) {$1$};
			\node[left] at (0, 1) {$2$};
			\node[left] at (0, 1.5) {$3$};
			\node[left] at (0, 2) {$4$};
			\node[left] at (0, 2.5) {$5$};
			\node[left] at (0, 3) {$n$};
			\filldraw[PPTgreen] (1/2, 1/2) circle (0.1cm);
			\filldraw[PPTgreen] (2/2, 1/2) circle (0.1cm);
			\filldraw[PPTgreen] (3/2, 1/2) circle (0.1cm);
			\filldraw[PPTgreen] (4/2, 1/2) circle (0.1cm);
			\filldraw[PPTgreen] (5/2, 1/2) circle (0.1cm);
			\filldraw[PPTgreen] (1/2, 1) circle (0.1cm);
			\filldraw[PPTgreen] (2/2, 1) circle (0.1cm);
			\filldraw[PPTgreen] (3/2, 1) circle (0.1cm);
			\filldraw[PPTgreen] (4/2, 1) circle (0.1cm);
			\filldraw[PPTgreen] (5/2, 1) circle (0.1cm);
			\filldraw[PPTgreen] (1/2, 3/2) circle (0.1cm);
			\filldraw[PPTgreen] (2/2, 3/2) circle (0.1cm);
			\filldraw[PPTgreen] (3/2, 3/2) circle (0.1cm);
			\filldraw[PPTgreen] (4/2, 3/2) circle (0.1cm);
			\filldraw[PPTgreen] (5/2, 3/2) circle (0.1cm);
			\filldraw[PPTgreen] (1/2, 2) circle (0.1cm);
			\filldraw[PPTgreen] (2/2, 2) circle (0.1cm);
			\filldraw[PPTgreen] (3/2, 2) circle (0.1cm);
			\filldraw[PPTgreen] (4/2, 2) circle (0.1cm);
			\filldraw[PPTgreen] (5/2, 2) circle (0.1cm);
			\filldraw[PPTgreen] (1/2, 5/2) circle (0.1cm);
			\filldraw[PPTgreen] (2/2, 5/2) circle (0.1cm);
			\filldraw[PPTgreen] (3/2, 5/2) circle (0.1cm);
			\filldraw[PPTgreen] (4/2, 5/2) circle (0.1cm);
			\filldraw[PPTgreen] (5/2, 5/2) circle (0.1cm);
            \node[below] at (1.5, -1) {$\text{QED}(\square, {\square})$};
		\end{tikzpicture}\qquad \qquad 
    \begin{tikzpicture}
			\draw[thick, ->] (0,0) -- (3,0);
			\draw[thick, ->] (0,0) -- (0,3);
			\draw[dashed] (0,0.5) -- (3,0.5);
			\draw[dashed] (0,1) -- (3,1);
			\draw[dashed] (0,1.5) -- (3,1.5);
			\draw[dashed] (0,2) -- (3,2);
			\draw[dashed] (0,2.5) -- (3,2.5);
			\draw[dashed] (0.5,0) -- (0.5, 3);
			\draw[dashed] (1,0) -- (1, 3);
			\draw[dashed] (1.5,0) -- (1.5, 3);
			\draw[dashed] (2,0) -- (2, 3);
			\draw[dashed] (2.5,0) -- (2.5, 3);
			\node[below] at (0.5, 0) {$1$};
			\node[below] at (1, 0) {$2$};
			\node[below] at (1.5, 0) {$3$};
			\node[below] at (2, 0) {$4$};
			\node[below] at (2.5, 0) {$5$};
			\node[below] at (3, 0) {$N_f$};
			\node[left] at (0, 0.5) {$1$};
			\node[left] at (0, 1) {$2$};
			\node[left] at (0, 1.5) {$3$};
			\node[left] at (0, 2) {$4$};
			\node[left] at (0, 2.5) {$5$};
			\node[left] at (0, 3) {$n$};
			\filldraw[PPTgreen] (1/2, 1/2) circle (0.1cm);
			\filldraw[PPTgreen] (2/2, 1/2) circle (0.1cm);
			\filldraw[PPTgreen] (3/2, 1/2) circle (0.1cm);
			\filldraw[PPTgreen] (4/2, 1/2) circle (0.1cm);
			\filldraw[PPTgreen] (5/2, 1/2) circle (0.1cm);
			\filldraw[PPTgreen] (1/2, 1) circle (0.1cm);
			\filldraw[PPTgreen] (2/2, 1) circle (0.1cm);
			\filldraw[PPTgreen] (3/2, 1) circle (0.1cm);
			\filldraw[PPTgreen] (4/2, 1) circle (0.1cm);
			\filldraw[PPTgreen] (5/2, 1) circle (0.1cm);
			\filldraw[PPTgreen] (1/2, 3/2) circle (0.1cm);
			\filldraw[PPTgreen] (2/2, 3/2) circle (0.1cm);
			\node[red] at (3/2, 3/2) {$?$};
			\node[red] at  (4/2, 3/2) {$?$};
			\node[red] at (5/2, 3/2) {$?$};
			\filldraw[PPTgreen] (1/2, 2) circle (0.1cm);
			\filldraw[PPTgreen] (2/2, 2) circle (0.1cm);
			\node[red] at  (3/2, 2) {$?$};
			\node[red] at  (4/2, 2) {$?$};
			\node[red] at  (5/2, 2) {$?$};
			\filldraw[PPTgreen] (1/2, 5/2) circle (0.1cm);
			\filldraw[PPTgreen] (2/2, 5/2) circle (0.1cm);
			\node[red] at  (3/2, 5/2) {$?$};
			\node[red] at  (4/2, 5/2) {$?$};
			\node[red] at  (5/2, 5/2) {$?$};
            \node[below] at (1.5, -1) {$\text{QED}(\square, \overline{\square})$};
		\end{tikzpicture}
    \caption{Status of puzzle in fermion-monopole scattering. Green dots: cases in which a boundary condition preserving all the symmetries in \eqref{eq:highcharge}, and the out-going states, are known \cite{vanBeest:2023dbu}. Question marks: open cases, $N_f\geq 3$ and $n\geq 3$ in  $\text{QED}(\square, \overline{\square})$, which we will solve.   }
    \label{fig:status}
\end{figure}
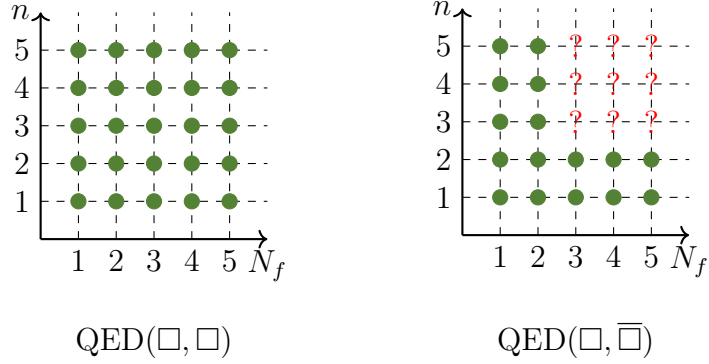

In \cite{vanBeest:2023dbu}, the out-going states were determined using abelian bosonization, which makes manifest only the Cartan torus of $U(1)\times SU(N_f)\times SU(2)$. For $\text{QED}(\square,\square)$ and all $(N_f,n)$, they found that the in-coming fermion scatters into $\widetilde{\psi}$ in the twisted sector of the invertible line $T^{-2}$, where $T$ generates a $\bZ_{nN_f}$ symmetry acting on the right-movers. For $\text{QED}(\square,\overline\square)$, the same strategy fails: no invertible topological line commuting with $SU(N_f)\times SU(2)$ can supply the quantum number needed to turn $\widetilde{\psi}$, which transforms in $\overline{\square}$, into an excitation transforming in $\square$. Instead, the authors of \cite{vanBeest:2023dbu} considered lines $T_i$ acting only on the $i$-th flavor, $\psi_i\to T_i + \widetilde{\psi}_i$, which commute only with the Cartan torus of $SU(N_f)$. Apart from $N_f\leq 2$, where $\square \simeq \overline{\square}$ and the problem reduces to $\text{QED}(\square,\square)$, and $n\leq 2$, where the linear boundary condition above is available, the problem has remained open.  See Figure \ref{fig:status} for a summary of the status.

In this section, we show that the general method of Section \ref{sec:generalconstruction} solves both problems uniformly, for all $(N_f, n)$. Both fit into the simple case of Section \ref{sec:automorphism}: the two embeddings of the current algebra are related by an automorphism (charge conjugation of $\widehat{\mathfrak{u}}(1)$ for $\text{QED}(\square, {\square})$, and $\widehat{\mathfrak{u}}(1)\oplus \widehat{\mathfrak{su}}(N_f)$ for $\text{QED}(\square, \overline{\square})$). Hence the coset chiral algebras are the same on both sides, the intertwining defect is obtained from a twist defect, which splits into several simple lines for $\text{QED}(\square, \overline{\square})$, and the scattering is governed by the crossed $F$- and $R$-symbols through \eqref{eq:auto45}. 

For $\text{QED}(\square, {\square})$, we recover the result of \cite{vanBeest:2023dbu}. For $\text{QED}(\square, \overline{\square})$, the out-going state is in general a superposition of two twisted sectors, attached to the lines  $a_{\text{S}}$ and $a_{\text{A}}$ in \eqref{eq:channel}, which carry the symmetric and antisymmetric parts of $\square\otimes \square$ and are non-invertible for $N_f, n\geq 3$. This resolves the open cases of Figure \ref{fig:status} in the 2d problem.

\subsection{$\text{QED}(\square, \square)$}
\label{sec:qedss}

We use the tool presented in Section \ref{sec:generalconstruction} to reproduce the symmetric boundary state and scattering process in \cite{vanBeest:2023dbu}.

\subsubsection{Embeddings and the Coset Chiral Algebra}

We unfold the $n N_f$ Dirac fermions on the half-plane into $nN_f$ Weyl fermions $\Fer(nN_f)$ on the full plane, with a topological line defect $\mathcal{D}$ along the image of the boundary, as in Figure \ref{fig:unfolding}. The in-coming fermions $\psi_{ia}$ live on one side of $\mathcal{D}$ and the out-going fermions $\widetilde{\psi}_{ia}$ on the other side, where $i=1, ..., N_f$ is the $SU(N_f)$ index and $a=1, ..., n$ is the $SU(2)$ index.  The currents of $G= U(1) \times SU(N_f) \times SU(2)$ are realized on the two sides by two (non-conformal) embeddings
\begin{eqnarray}
    f_{i}: \widehat{\mathfrak{g}}\hookrightarrow \Fer(nN_f).
\end{eqnarray}
From the symmetry quantum numbers of fermions \eqref{eq:highcharge}, the currents are 
\begin{equation}\label{eq:f12qed}
\begin{alignedat}{2}
    &f_1(j_{\mathfrak{u}(1)})
      = \psi^\dagger\psi,
    &&\qquad f_2(j_{\mathfrak{u}(1)})
      = -\widetilde\psi^\dagger\widetilde\psi, \\[4pt]
    &f_1(j^\alpha_{\mathfrak{su}(N_f)})
      = \psi^\dagger\,(T^\alpha_\square\otimes 1)\,\psi,
    &&\qquad f_2(j^\alpha_{\mathfrak{su}(N_f)})
      = \widetilde\psi^\dagger\,
        (T^\alpha_\square\otimes 1)\,\widetilde\psi, \\[4pt]
    &f_1(j^r_{\mathfrak{su}(2)})
      = \psi^\dagger\,(1\otimes T^r_{\boldsymbol{n}})\,\psi,
    &&\qquad f_2(j^r_{\mathfrak{su}(2)})
      = \widetilde\psi^\dagger\,
        (1\otimes T^r_{\boldsymbol{n}})\,\widetilde\psi,
\end{alignedat}
\end{equation}
where normal ordering is implied and the flavor and $SU(2)$ indices are contracted. The affine Lie algebra generated by these currents is the same for $\psi$ and $\widetilde\psi$,
\begin{eqnarray}\label{eq:currentalgebra1}
    \widehat{\mathfrak{g}} = \widehat{\mathfrak{u}}(1)_{nN_f} \oplus \widehat{\mathfrak{su}}(N_f)_{n} \oplus \widehat{\mathfrak{su}}(2)_{N_fn(n^2-1)/6}.
\end{eqnarray}

Comparing the two columns of \eqref{eq:f12qed}, we find that the in-coming and out-going modes are related by charge conjugation $\mathsf{C}_{\mathfrak{u}(1)}$ of the $\widehat{\mathfrak{u}}(1)$ currents, 
\begin{eqnarray}\label{eq:AQEDss}
    A= \mathsf{C}_{\mathfrak{u}(1)}, \qquad f_2= f_1 \circ A
\end{eqnarray}
This is precisely the situation of Section \ref{sec:automorphism}. As a consequence, the coset chiral algebras for the in-coming and out-going fermions can be chosen to be identical, $S_1=S_2=S$. 

The central charge of $\widehat{\mathfrak{g}}$ is 
\begin{eqnarray}
    c= \underbrace{1}_{\widehat{\mathfrak{u}}(1)_{nN_f}} + \underbrace{\frac{n(N_f^2-1)}{N_f+n}}_{\widehat{\mathfrak{su}}(N_f)_{n}} + \underbrace{\frac{3N_f n(n^2-1)}{N_f n (n^2-1) +12}}_{\widehat{\mathfrak{su}}(2)_{N_fn(n^2-1)/6}}
\end{eqnarray}
which is smaller than $c=nN_f$ of $\widehat{\mathfrak{u}}(nN_f)_1$ unless $n=1,2$ or $(N_f,n)=(1,3)$. In the latter case, the embedding is the conformal embedding $\widehat{\mathfrak{su}}(2)_4\hookrightconf\widehat{\mathfrak{su}}(3)_1$. Hence for $n\geq 3$ and $(N_f,n)\neq(1,3)$, $\widehat{\mathfrak{g}}$ is not conformally embedded in $\widehat{\mathfrak{u}}(nN_f)_1$, and we have to include a coset chiral algebra. 

\paragraph{Coset chiral algebra}
To find the coset chiral algebra, we use the conformal embedding 
\begin{eqnarray}\label{eq:commonconformal}
    \widehat{\mathfrak{u}}(1)_{nN_f} \oplus \widehat{\mathfrak{su}}(N_f)_{n} \oplus \widehat{\mathfrak{su}}(n)_{N_f} \hookrightconf \Fer(n N_f) 
\end{eqnarray}
in which the $nN_f$ fermions transform as $\square \boxtimes \square$ of $SU(N_f) \times SU(n)$. The first two factors of \eqref{eq:commonconformal} coincide with those of \eqref{eq:currentalgebra1}, while $\widehat{\mathfrak{su}}(2)_{N_f n(n^2-1)/6}$ is embedded in $\widehat{\mathfrak{su}}(n)_{N_f}$ via the principal embedding $\mathfrak{su}(2)\subset\mathfrak{su}(n)$, under which the fundamental representation of $\mathfrak{su}(n)$ becomes $\boldsymbol{n}$ of $\mathfrak{su}(2)$; the embedding index $n(n^2-1)/6$ accounts for the level.  Therefore, the coset chiral algebra is 
\begin{eqnarray}\label{eq:qedcoset}
    S= V_{N_f,n} \equiv \frac{\widehat{\mathfrak{su}}(n)_{N_f} }{\widehat{\mathfrak{su}}(2)_{N_f n(n^2-1)/6}}.
\end{eqnarray}
For special values of $(N_f,n)$, $V_{N_f,n}$ reduces to known chiral algebras. For instances, $V_{N_f,1}$, $V_{N_f,2}$ and $V_{1,3}$ are topological cosets \cite{Cordova:2024goh}. $V_{2,3}$ is the first member of $W_3$ minimal series, or equivalently the chiral algebra for 3-state Potts CFT \cite{Dunbar:1992gh}, which is a maverick coset \cite{Cordova:2023jip,Dunbar:1993hr}.  $V_{3,3}$ is the third member of $W_3$ minimal series \cite{Dunbar:1992gh}. $V_{1,4}$ is the Ising chiral algebra \cite{Pedrini:1999iy}, which is also a  maverick coset.

\subsubsection{Sandwich, Lagrangian Algebras, and Intertwining Defects}
\label{sec:sandwichqedss}

\begin{figure}
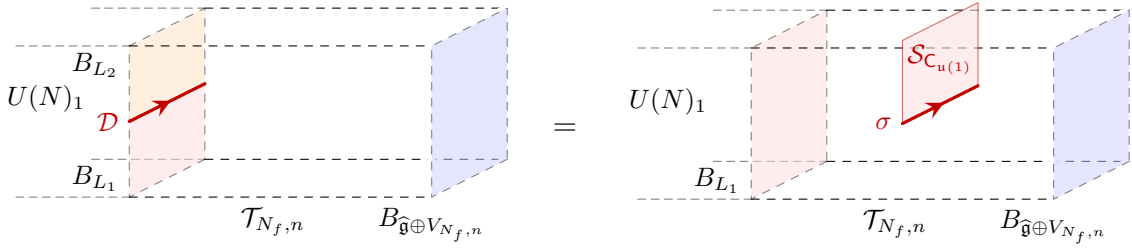

    \centering
    \figAutoSplitGen[1.0]{$\mathcal{T}_{N_f,n}$}{$B_{\widehat{\mathfrak{g}}\oplus V}$} \quad $=$ \quad \figAutoSurfaceGen[1.0]{$\mathcal{S}_{\mathsf{C}}$}{$\sigma_U$}{$\mathcal{T}_{N_f,n}$}{$B_{\widehat{\mathfrak{g}}\oplus V}$}
    \caption{Sandwich construction for $\text{QED}(\square,\square)$. The bulk TQFT is \eqref{eq:bulktqft}, and the two topological boundaries are related by $B_{L_2}=B_{L_1}\otimes\mathcal{S}_{\mathsf{C}_{{\mathfrak{u}}(1)}}$, where $\mathcal{S}_{\mathsf{C}_{{\mathfrak{u}}(1)}}$ implements the charge conjugation of $U(1)_{n N_f}$. The intertwining defect $\mathcal{D}$ between them (left) is obtained from the twist defect $\sigma$ bounding the truncated surface $\mathcal{S}_{\mathsf{C}_{\mathfrak{u}(1)}}$ (right). }
    \label{fig:qedsandwich}
\end{figure}

Following Section \ref{sec:automorphism}, the sandwich consists of the bulk TQFT 
\begin{eqnarray}\label{eq:bulktqft}
    \mathcal{T}_{N_f,n} = U(1)_{n N_f} \boxtimes SU(N_f)_n \boxtimes SU(2)_{N_f n(n^2-1)/6} \boxtimes \Rep(V_{N_f,n}).
\end{eqnarray}
The chiral boundary supports the chiral algebra $\widehat{\mathfrak{g}}\oplus V_{N_f,n}$, and the two topological interfaces $B_{L_1}$ and $B_{L_2}$ to $U(N)_1$, correspond to the two embeddings. See Figure \ref{fig:qedsandwich}.

\paragraph{Lagrangian Algebras:}
The Lagrangian algebra $L_1$ is determined by the branching of the vacuum supermodule of $\Fer(nN_f)$ into $\widehat{\mathfrak{g}}\oplus V_{N_f,n}$-supermodules. This can be done in steps, using the chain of conformal embeddings
\begin{eqnarray}\label{eq:qedembeddingchain}
    \begin{split}
    & \Fer(n N_f)\\
    \hookleftconf \quad  & \widehat{\mathfrak{u}}(1)_{nN_f} \oplus \widehat{\mathfrak{su}}(n N_f)_{1} \\
    \hookleftconf \quad  & \widehat{\mathfrak{u}}(1)_{nN_f} \oplus \widehat{\mathfrak{su}}(N_f)_{n} \oplus \widehat{\mathfrak{su}}(n)_{N_f}\\
    \hookleftconf \quad &\widehat{\mathfrak{u}}(1)_{nN_f} \oplus \widehat{\mathfrak{su}}(N_f)_{n} \oplus \widehat{\mathfrak{su}}(2)_{N_fn(n^2-1)/6} \oplus V_{N_f,n}
    \end{split}
\end{eqnarray}
We explain these conformal embeddings from the top to bottom. 
\begin{enumerate}
    \item Under the canonical embedding $\widehat{\mathfrak{u}}(1)_{nN_f} \oplus \widehat{\mathfrak{su}}(nN_f)_{1}\hookrightconf\Fer(nN_f)$, the vacuum supermodule decomposes as 
    \begin{eqnarray}\label{eq:canonicalbranching}
        1 \mapsto \bigoplus_{r\in\bZ_{nN_f}}[r]\boxtimes\Lambda_r
    \end{eqnarray}
    where $[r]$ is the line of $U(1)_{nN_f}$ with spin $\frac{r^2}{2nN_f}$ (defined modulo $nN_f$, up to the transparent fermion), and $\Lambda_r$ is the $r$-th fundamental weight of $\mathfrak{su}(nN_f)$, i.e.\ the $r$-fold antisymmetric power of $\square$. The states in $[r]\boxtimes\Lambda_r$ carry fermion number $r$ mod $nN_f$.
    \item Under the level-rank embedding \cite[Theorem 4.2]{Hasegawa:1989}\cite{Naculich:1990hg,Hsin:2016blu} $\widehat{\mathfrak{su}}(N_f)_{n} \oplus \widehat{\mathfrak{su}}(n)_{N_f}\hookrightconf \widehat{\mathfrak{su}}(nN_f)_1$, we have
    \begin{eqnarray}\label{eq:levelrankembedding}
        \Lambda_r\mapsto \bigoplus_{(\mu,\nu)\in \mathcal{P}_r}\mu\boxtimes\nu
    \end{eqnarray}
    where $\mathcal{P}_r$ is the set of pairs of integrable weights
    \begin{eqnarray}\label{eq:Prqed}
        \mathcal{P}_r=\Big\{\big(\mu,\,J^{(r-|\mu|)/N_f}(\mu^{T})\big)\ \Big|\ |\mu|\equiv r \ \text{mod}\ N_f\Big\}.
    \end{eqnarray}
    Here $\mu$ runs over the integrable weights of $\widehat{\mathfrak{su}}(N_f)_n$, identified with Young diagrams with fewer than $N_f$ rows and at most $n$ columns, and $|\mu|$ is the number of boxes. $\mu^T$ is the transposed Young diagram, regarded as a weight of $\widehat{\mathfrak{su}}(n)_{N_f}$ after removing the columns of height $n$, and $J:[a_1,\dots,a_{n-1}]\mapsto[N_f-\sum_ia_i,a_1,\dots,a_{n-2}]$ is the generating simple current of $\widehat{\mathfrak{su}}(n)_{N_f}$. The pairs with $|\mu|=r$ are the pairs $(Y,Y^T)$ of Young diagrams with $r$ boxes, which describe the lowest states of fermion number $r$. The other pairs describe states at higher grades; for instance, for $N_f=n=2$, $\Lambda_0\mapsto(0,0)\oplus(1,1)$, where $(1,1)$ comes from the fermion bilinears in the adjoint of $SU(2)\times SU(2)$.
    \item Under the principal embedding $\widehat{\mathfrak{su}}(2)_{N_f n(n^2-1)/6} \oplus V_{N_f,n}\hookrightconf \widehat{\mathfrak{su}}(n)_{N_f}$, 
    \begin{eqnarray}\label{eq:principleembed}
        \nu\mapsto\bigoplus_{j} j\boxtimes\phi_{\nu,j}
    \end{eqnarray}
    where $j$ runs over the spins of $\widehat{\mathfrak{su}}(2)_{N_f n(n^2-1)/6}$ and $\phi_{\nu,j}$ are the corresponding (possibly reducible) $V_{N_f,n}$-modules.
\end{enumerate}
Combining the three steps, we find 
\begin{eqnarray}\label{eq:L1qed}
    L_1= \bigoplus_{r\in \bZ_{nN_f}} \bigoplus_{(\mu,\nu)\in \mathcal{P}_r} \bigoplus_{j} [r] \boxtimes \mu \boxtimes j \boxtimes \phi_{\nu,j}
\end{eqnarray}
In particular, the fermion $\psi_{ia}$ has $r=1$, and $(\mu,\nu)=(\square, \square)$, and its lowest component has spin $j=\frac{n-1}{2}$. Thus, in the sandwich $\psi_{ia}$ is resolved as $(x,\lambda_{\psi}, \widetilde{x})$ with 
\begin{eqnarray}\label{eq:lampsi}
    \lambda_\psi = [1]\boxtimes\square\boxtimes \tfrac{n-1}{2}\boxtimes\phi_{\square,\frac{n-1}{2}}.
\end{eqnarray}

The Lagrangian algebra for the second embedding is the image of $L_1$ under $\mathsf{C}_{{\mathfrak{u}}(1)}$, which acts on $\mathcal{T}_{N_f,n}$ by $[r]\mapsto[-r]$ and trivially on the other factors,
\begin{eqnarray}\label{eq:L2qed}
    L_2 = \mathsf{C}_{{\mathfrak{u}}(1)}(L_1)= \bigoplus_{r\in\bZ_{nN_f}}\ \bigoplus_{(\mu,\nu)\in\mathcal{P}_r}\ \bigoplus_{j}\ [-r]\boxtimes\mu\boxtimes j\boxtimes \phi_{\nu,j}.
\end{eqnarray}
Correspondingly, the fermion $\widetilde{\psi}_{ia}$ on the other side of $\mathcal{D}$ is resolved as a bulk line
\begin{eqnarray}\label{eq:lampsitilde}
    {\lambda}_{\widetilde\psi} \equiv \mathsf{C}_{{\mathfrak{u}}(1)}(\lambda_\psi) = [-1]\boxtimes\square\boxtimes \tfrac{n-1}{2}\boxtimes\phi_{{\square, \frac{n-1}{2}}} \in L_2,
\end{eqnarray}
which differs from $\lambda_{\psi}$ only in the $U(1)_{nN_f}$ label.

\paragraph{Intertwining Defects from Twist Defects: }
By Section \ref{sec:automorphism}, the automorphism \eqref{eq:AQEDss} is implemented by an invertible surface $\mathcal{S}_A=\mathcal{S}_{\mathsf{C}_{\mathfrak{u}(1)}}\boxtimes \mathbf{1}$ of $\mathcal{T}_{N_f,n}$, where $\mathcal{S}_{\mathsf{C}_{\mathfrak{u}(1)}}$ is the charge conjugation surface of $U(1)_{nN_f}$, acting as $[r]\mapsto[-r]$, and $\mathbf{1}$ is the trivial surface of the remaining factors. The topological surface $\mathcal{S}_{\mathsf{C}_{\mathfrak{u}(1)}}$ is a condensation defect constructed by higher gauging on the surface a Frobenius algebra \cite{Roumpedakis:2022aik}.

The charge conjugation twist defects of $U(1)_{nN_f}$ are well known \cite{Barkeshli:2012pr,Barkeshli:2014cna,Teo:2015xla,Kaidi:2022cpf}. They are in one-to-one correspondence with the $\mathsf{C}_{{\mathfrak{u}}(1)}$-invariant lines $[r]=[-r]$:
\begin{itemize}
    \item $nN_f$ odd: there is a unique twist defect $\sigma$, with quantum dimension $d_{\sigma}=\sqrt{nN_f}$ and fusion rules $\sigma\otimes[r]=\sigma$, $\overline{\sigma}\otimes\sigma=\bigoplus_{r\in\bZ_{nN_f}}[r]$.
    \item $nN_f$ even: there are two twist defects $\sigma_\pm$, with $d_{\sigma_\pm}=\sqrt{nN_f/2}$ and fusion rules $\sigma_\pm\otimes[r]=\sigma_{\pm(-1)^r}$, $\overline{\sigma}_\pm\otimes\sigma_\pm=\bigoplus_{r\in \bZ_{nN_f/2}}[2r]$.
\end{itemize}
In summary, the charge conjugation twist defect is a \emph{duality defect} of $\bZ_{nN_f}$ for odd $nN_f$ and $\bZ_{nN_f/2}$ for even $nN_f$. This agrees with the result by abelian bosonization \cite{Smith:2019jnh,Smith:2020nuf,Wei:2026fsn}.

The topological boundaries are related by $B_{L_2}=B_{L_1}\otimes \mathcal{S}_{\mathsf{C}_{\mathfrak{u}(1)}}$, and the intertwining defect is the fusion of the twist defect $\sigma$ or $\sigma_{\pm}$ onto $B_{L_1}$; see Figure \ref{fig:qedsandwich}. For simplicity, we will focus on $\sigma_+$ when $nN_f$ is even, and simply denote it as $\sigma$. The discussion for $\sigma_-$ is in parallel.  After pushing onto $B_{L_1}$, the intertwining defect is 
\begin{eqnarray}\label{eq:Dqedss}
    \mathcal{D} = F_{B_{L_1}}(\sigma\boxtimes \mathbf{1}) .
\end{eqnarray}
where $\mathbf{1}$ is the identity line from all the bulk TQFTs except the $U(1)_{nN_f}$. 

Generically, $\mathcal{D}$ as defined by fusing a twist defect in the bulk to the boundary can be non-simple. However, we claim that in the current situation,  $\mathcal{D}$ is simple.  To see this, we compute the dimension of Hom space, 
\begin{eqnarray}\label{eq:Homid}
    \Hom_{\text{bdy}}(\mathcal{D}, \mathcal{D}) = \Hom_{\mathcal{T}_{N_f,n}}(\sigma \boxtimes \mathbf{1}, L_1\otimes (\sigma\boxtimes \mathbf{1}))
\end{eqnarray}
Using \eqref{eq:L1qed} and the fusion rules above, $L_1\otimes (\sigma \boxtimes \mathbf{1})= \bigoplus_{r,(\mu,\nu),j} (\sigma\otimes [r])\boxtimes \mu\boxtimes j \boxtimes \phi_{\nu,j}$, where $\sigma \otimes [r] = \sigma$ for $nN_f$ odd, while for $nN_f$ even $\sigma \otimes [r]$ is $\sigma$ or $\sigma_-$ for $r$ even or odd.  In the direct sum, only one component coincides with $\sigma \boxtimes \mathbf{1}$, namely the one with $r=0$ (see the discussion around \eqref{eq:baryon}), hence the Hom space in \eqref{eq:Homid} one-dimensional. This shows that $\mathcal{D}$ is a simple interface between $B_{L_1}$ and $B_{L_2}$.\footnote{For $nN_f$ odd, $U(1)_{nN_f}$ is a spin TQFT, and the $r=0$ component of $L_1$ contains, besides the trivial line, the transparent fermion $[nN_f] \boxtimes \mathbf{1}$, i.e. the product of all species of fermions $\prod_{ia}\psi_{ia}$. It is absorbed by the twist defect, $\sigma \otimes [nN_f]= \sigma$, and gives an additional fermionic (odd) endomorphism of $\mathcal{D}$. Hence, for $nN_f$ odd, $\mathcal{D}$ is a simple $q$-type line, which carries an unpaired Majorana zero mode. The quantum dimension and the fusion rules of $\mathcal{D}$ below are understood up to the transparent fermion. }

\paragraph{Comment on Discrete symmetry $\bZ_{nN_f}$:}
The reason why $\mathcal{D}$ intertwines with $U(1)\times SU(N_f)\times SU(2)$ is explained in Section \ref{sec:interDefect}. Here we check the discrete $\mathbb{Z}_{nN_f}$ factor in the last line in \eqref{eq:highcharge}. In the sandwich, the $\mathbb{Z}_{nN_f}$ line $\eta$ is the bulk line
\begin{eqnarray}\label{eq:etaline}
    \eta= [1] \boxtimes \mathbf{1}, \qquad \eta^{nN_f}=1
\end{eqnarray}
placed between $B_{L_1}$ and $\mathcal{S}_{\mathsf{C}_{\mathfrak{u}(1)}}$. Below $\sigma$, it acts on $\psi$ by the braiding phase of $[1]$ with $\lambda_{\psi}$, $e^{2\pi i/nN_f}$. Above $\sigma$, it acts on the $\widetilde{\psi}$ by the same phase, since the segment of the bulk line of $\widetilde{\psi}$ between $B_{L_1}$ and $\mathcal{S}_{\mathsf{C}_{\mathfrak{u}(1)}}$ is $\mathsf{C}_{\mathfrak{u}(1)}({\lambda}_{\widetilde\psi})= \lambda_\psi$. Since $\eta$ never meets $\mathsf{C}_{\mathfrak{u}(1)}$, it can be moved across $\mathcal{D}$ freely, i.e. $\mathcal{D}$ commutes with $\eta$.

\subsubsection{Scattering}

Applying \eqref{eq:auto3} to this example, the out-going configuration is
\begin{eqnarray}\label{eq:qedssout}
    \figAutoCGen[1.3]{$\lambda_\psi$}{$\widetilde{\lambda}_\psi$}{$[2]$}{$\sigma$}{$\mathcal{T}_{N_f,n}$}{$B_{\widehat{\mathfrak{g}}\oplus V_{N_f,n}}$}{$\mathcal{S}_{\mathsf{C}_{\mathfrak{u}(1)}}$},
\end{eqnarray}
where $a=[2]\boxtimes\mathbf{1}$ is obtained by solving the fusion relation $\mathsf{C}_{{\mathfrak{u}}(1)}(\lambda_\psi)\otimes a=\lambda_\psi$. Pushing all topological objects in the bulk to the topological boundary $B_{L_1}$, line $F_{B_{L_1}}(a)$ gets identified as symmetry defect $T^{-2}$ appeared in \cite[eq.~(3.34)]{vanBeest:2023dbu}, where $T^{-1}\equiv  F_{B_{L_2}}([1]\boxtimes\mathbf{1})$, hence
\begin{eqnarray}\label{eq:qedu1scattering}
    \psi_{ia}\to T^{-2} \cdot \widetilde{\psi}_{ia}
\end{eqnarray}

Since all the crossed $F$- and $R$-symbols of $U(1)_{nN_f}$ in \eqref{eq:auto45} are phases and $\sqrt{d_{\sigma'}/(d_{\sigma}d_{\lambda})}=1$, the coefficient is a pure phase,
\begin{eqnarray}\label{eq:qedssscat}
    \figBeforeScatter[0.95]\quad = \quad e^{i\alpha}\quad \figOut{T^{-2}}
\end{eqnarray}
The value of $\alpha$ depends on the choice of junction basis, and is not gauge invariant, and hence we will not compute it.

We make a few comments.
\begin{enumerate}
    \item The out-going channel is a fixed twisted sector: the fermion scatters with probability one into the $T^{-2}$-twisted sector. The line $T^{-2}$ is abelian; it generates $\bZ_{nN_f}$ for $nN_f$ odd, and $\bZ_{nN_f/2}$ for $nN_f$ even. For $nN_f=1,2$ it is trivial, and the out-going state is an ordinary fermion.
    \item 
    
    The sandwich construction automatically preserves the conformal dimension, as well as the non-abelian symmetries $SU(N_f)\times SU(2)$, across scattering since the junction $\widetilde{x}$ on the chiral boundary is unchanged and the twist defect is transparent to the non-abelian Chern Simons factors in $\mathcal{T}_{N_f,n}$.  This is not manifest from the method in \cite{vanBeest:2023dbu}.
    \item The coset sector $\Rep(V_{N_f,n})$, although complicated, plays no role in the scattering, in accordance with Section \ref{sec:automorphism}.
\end{enumerate}

\subsection{$\text{QED}(\square,\overline\square)$}
\label{sec:qedsbar}

We now turn to $\text{QED}(\square,\overline\square)$. If we ignore the discrete symmetry $\bZ_n$, there is a linear boundary condition $\psi_{ia} = J_{ab}\widetilde{\psi}^\dagger_{ib}$ preserving $U(1)\times SU(N_f) \times SU(2)$, where $J$ is the $SU(2)$-invariant bilinear form on $\boldsymbol{n}$.\footnote{Under $SU(2)$, the linear boundary condition becomes $U_{ab} \psi_{ib} = J_{ab}U^*_{bc}\widetilde{\psi}^\dagger_{ic}$. Requiring it to be compatible with the original boundary condition, we need $U J = J U^*$, i.e. $U J U^{\mathsf{T}} = J$, which is the condition for the $SU(2)$-invariant bilinear form. } However, since $\widetilde{\psi}^\dagger$ carries charge $-1$ under $\bZ_n$, this boundary condition explicitly breaks $\bZ_n$ for $n\geq 3$. For $N_f\geq 3$, these are the cases left unsolved in \cite{vanBeest:2023dbu}; see Figure \ref{fig:status}. We will find that the method in Section \ref{sec:qedss} straightforwardly solves this open problem, and some new features appears in the scattering process.

\subsubsection{Sandwich, Lagrangian Algebras, and Intertwining Defects}

The only difference from Section \ref{sec:qedss} is that $\widetilde{\psi}$ transforms in $\overline{\square}$ of $SU(N_f)$, so that in \eqref{eq:f12qed} the $\mathfrak{su}(N_f)$ current of the right-movers is replaced by $f_2(j^\alpha_{\mathfrak{su}(N_f)})=\widetilde\psi^\dagger\,(-T^{\alpha\,\mathsf{T}}_\square\otimes1)\,\widetilde\psi$. Hence
\begin{eqnarray}\label{eq:AQEDsbar}
    f_2=f_1\circ A, \qquad A=\mathsf{C}_{\mathfrak{u}(1)}\times\mathsf{C}_{\mathfrak{su}(N_f)},
\end{eqnarray}
where $\mathsf{C}_{\mathfrak{su}(N_f)}: T^\alpha\mapsto -T^{\alpha\,\mathsf{T}}$ is the charge conjugation of $\mathfrak{su}(N_f)$.\footnote{Given an $SU(N_f)$ group element $U= e^{i\sum_\alpha \theta^\alpha T^\alpha}$, charge conjugation maps $U\to U^*$. This implies that the $\mathfrak{su}(N_f)$ Lie algebra generator transforms as $T^\alpha \to - T^{\alpha \, \mathsf{T}}$. } The embedding of $\mathfrak{su}(2)$ is unchanged, given by the principal embedding \eqref{eq:principleembed}, so the coset chiral algebra is again $V_{N_f,n}$ in \eqref{eq:qedcoset}.

\paragraph{Sandwich and Lagrangian Algebras:}
The sandwich consists of the same bulk TQFT $\mathcal{T}_{N_f,n}$ in \eqref{eq:bulktqft}, the same chiral boundary $B_{\widehat{\mathfrak{g}}\oplus V_{N_f,n}}$, and the same Lagrangian algebra $L_1$ in \eqref{eq:L1qed}.
The second Lagrangian algebra is the image of $L_1$ under $A$, which now also conjugates the $SU(N_f)_n$ label,
\begin{eqnarray}\label{eq:L2qedbar}
    L_2 = A(L_1)= \bigoplus_{r\in\bZ_{nN_f}}\ \bigoplus_{(\mu,\nu)\in\mathcal{P}_r}\ \bigoplus_{j}\ [-r]\boxtimes\overline{\mu}\boxtimes j\boxtimes \phi_{\nu,j},
\end{eqnarray}
and the fermion $\widetilde{\psi}_{ia}$ is resolved as the bulk line
\begin{eqnarray}\label{eq:lampsitildebar}
    {\lambda}_{\widetilde\psi} \equiv A(\lambda_\psi) = [-1]\boxtimes\overline{\square}\boxtimes \tfrac{n-1}{2}\boxtimes\phi_{\square,\frac{n-1}{2}} \in L_2 .
\end{eqnarray}

\paragraph{Intertwining Defects and Simplicity:}
In $\mathcal{T}_{N_f,n}$, the automorphism \eqref{eq:AQEDsbar} is implemented by the product of charge conjugation surface operators $\mathcal{S}_A=\mathcal{S}_{\mathsf{C}_{\mathfrak{u}(1)}}\boxtimes\mathcal{S}_{\mathsf{C}_{\mathfrak{su}(N_f)}}\boxtimes\mathbf{1}$, where $\mathcal{S}_{\mathsf{C}_{\mathfrak{su}(N_f)}}$ maps $\mu\mapsto\overline{\mu}$. 
The twist defect for $\mathcal{S}_{A}$ is the product of twist defects $\sigma_{\mathfrak{u}}\boxtimes \sigma_{\mathfrak{su}}$ for $\mathcal{S}_{\mathsf{C}_{\mathfrak{u}(1)}}$ and $\mathcal{S}_{\mathsf{C}_{\mathfrak{su}(N_f)}}$ respectively. 
The intertwining defect is obtained by fusing them into $B_{L_1}$,
\begin{eqnarray}\label{eq:Dqedsbar}
    \mathcal{D} = F_{B_{L_1}}\big(\sigma_{\mathfrak{u}}\boxtimes\sigma_{\mathfrak{su}}\boxtimes \mathbf{1}\big).
\end{eqnarray}
This generalizes \eqref{eq:Dqedss}.

Unlike \eqref{eq:Dqedss}, the defect \eqref{eq:Dqedsbar} is not simple. To see this, we compute the dimension of the Hom space as in \eqref{eq:Homid},
\begin{eqnarray}\label{eq:Homidbar}
    \Hom_{\text{bdy}}(\mathcal{D}, \mathcal{D}) = \Hom_{\mathcal{T}_{N_f,n}}\Big(\sigma_{\mathfrak{u}}\boxtimes\sigma_{\mathfrak{su}}\boxtimes \mathbf{1},\, L_1\otimes (\sigma_{\mathfrak{u}}\boxtimes\sigma_{\mathfrak{su}}\boxtimes \mathbf{1})\Big),
\end{eqnarray}
where, by \eqref{eq:L1qed}, $L_1\otimes (\sigma_{\mathfrak{u}}\boxtimes\sigma_{\mathfrak{su}}\boxtimes \mathbf{1})=\bigoplus_{r,(\mu,\nu),j}(\sigma_{\mathfrak{u}}\otimes[r])\boxtimes(\sigma_{\mathfrak{su}}\otimes\mu)\boxtimes j\boxtimes\phi_{\nu,j}$. A summand can contribute to \eqref{eq:Homidbar} only if $j\boxtimes\phi_{\nu,j}$ is the trivial line of $SU(2)_{N_fn(n^2-1)/6}\boxtimes\Rep(V_{N_f,n})$. The conformal embedding $\widehat{\mathfrak{su}}(2)_{N_fn(n^2-1)/6}\oplus V_{N_f,n}\hookrightconf\widehat{\mathfrak{su}}(n)_{N_f}$ implies that the vacuum module of the former only appears in the vacuum module of the latter, so $\nu$ must be the trivial weight. In the level-rank decomposition \eqref{eq:Prqed}, this requires $J^{(r-|\mu|)/N_f}(\mu^T)$ to be trivial. The non-trivial simple currents of $\widehat{\mathfrak{su}}(n)_{N_f}$ have rows of length $N_f$, which $\mu^T$ cannot have since $\mu$ has fewer than $N_f$ rows. Hence $\mu^T$ itself must be trivial, i.e.\ it consists of $n$ rows of equal length $s$, and $(r-|\mu|)/N_f\in n\bZ$. Therefore $\mu=n\Lambda_s$, which is the simple current $J^s$ of $SU(N_f)_n$, and $r=sn$ mod $nN_f$. Therefore, the lines of $L_1$ that can contribute are
\begin{eqnarray}\label{eq:baryon}
    B_s = [sn]\boxtimes J^s\boxtimes \mathbf{1}, \qquad s\in\bZ_{N_f},
\end{eqnarray}
with spin $h_{B_s}=\frac{(sn)^2}{2nN_f}+\frac{ns(N_f-s)}{2N_f}=\frac{sn}{2}$. They describe baryons: the lowest operator in the sector $B_1$ is $\epsilon^{a_1\cdots a_n}\psi_{i_1a_1}\cdots\psi_{i_na_n}$.

Hence $\dim\Hom_{\text{bdy}}(\mathcal{D},\mathcal{D})=|H|$, where
\begin{eqnarray}\label{eq:stabilizer}
    H=\big\{s\in\bZ_{N_f}\ \big|\ \sigma_{\mathfrak{u}}\subset[sn]\otimes\sigma_{\mathfrak{u}}, \ \ \sigma_{\mathfrak{su}}\subset J^s\otimes\sigma_{\mathfrak{su}}\big\}
\end{eqnarray}
is the subgroup of baryon lines that fix the twist defect. To determine $H$, we use that for a twist defect $\sigma$ and a bulk line $x$, the crossed braidings give $\sigma\otimes x\cong \mathsf{C}(x)\otimes\sigma\cong\sigma\otimes \mathsf{C}(x)$ \cite{Barkeshli:2014cna}. For an invertible line $x$, this implies $\sigma\otimes x\otimes\mathsf{C}(x)^{-1}\cong\sigma$. For $x=[r]$ and $x=J^s$, we have $x\otimes\mathsf{C}(x)^{-1}=[2r]$ and $J^{2s}$, respectively. Hence $[2r]$ and $J^{2s}$ fix every twist defect, in agreement with the fusion rules of $U(1)_{nN_f}$ in Section \ref{sec:qedss}. It follows that:
\begin{itemize}
    \item $N_f$ odd: every $s\in\bZ_{N_f}$ is twice another element, so $J^s$ fixes $\sigma_{\mathfrak{su}}$. Moreover, $[sn]$ fixes $\sigma_{\mathfrak{u}}$: for $nN_f$ odd, every line of $U(1)_{nN_f}$ does, while for $nN_f$ even, $n$ is even and so is $sn$. Hence $H=\bZ_{N_f}$.
    \item $N_f$ even: $H$ contains the subgroup $\bZ_{N_f/2}$ generated by $[2]\boxtimes J^2 \boxtimes \mathbf{1}$, associated with even $s$. 
\end{itemize}
In particular, $|H|\geq 2$ for all $N_f\geq 3$, and the intertwining defect is non-simple. 

\begin{figure}
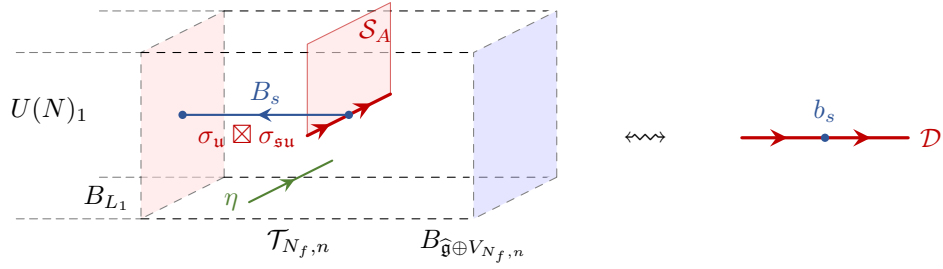

    \centering
    \figBaryonEnd[1.1]\qquad $\leftrightsquigarrow$ \qquad \figEndD
    \caption{Left: the baryon line $B_s=[sn]\boxtimes J^s\boxtimes\mathbf{1}\in L_1$, stretched between the twist defect $\sigma_{\mathfrak{u}}\boxtimes\sigma_{\mathfrak{su}}\boxtimes\mathbf{1}$ and $B_{L_1}$, gives a topological endomorphism $b_s$ of $\mathcal{D}$ after shrinking the sandwich (right). The $\bZ_{nN_f}$ generator $\eta$ \eqref{eq:etaline} braids non-trivially with $B_s$ by a phase $e^{2\pi i s/N_f}$. Only $\eta^{N_f}$ braids trivially with $B_s$ for any $s$.}
    \label{fig:baryon}
\end{figure}

The endomorphisms of $\mathcal{D}$ are realized by stretching $B_s$, $s\in H$, between the twist defect and $B_{L_1}$, as in Figure \ref{fig:baryon}. Denote by $b_s$ the resulting topological junction on $\mathcal{D}$. Since the lines $B_s$ are invertible, $b_sb_t\propto b_{s+t}$, so $\Hom_{\text{bdy}}(\mathcal{D},\mathcal{D})$ is a twisted group algebra of the cyclic group $H$. Since $H^2(H,U(1))=0$, it is isomorphic to the group algebra $\mathbb{C}[H]\simeq\mathbb{C}^{|H|}$. Therefore, the intertwining defect splits into $|H|$ distinct simple lines,
\begin{eqnarray}\label{eq:Dsplit}
    \mathcal{D}=\bigoplus_{\chi\in\widehat{H}}\mathcal{D}_\chi, \qquad \mathcal{D}_\chi = P_\chi\,\mathcal{D}, \qquad P_\chi=\frac{1}{|H|}\sum_{s\in H}\chi(s)^{-1}\,b_s,
\end{eqnarray}
where $\widehat{H}$ is the group of characters of $H$, $P_\chi$ is the projector onto the summand $\mathcal{D}_\chi$, and $d_{\mathcal{D}_\chi}=d_{\sigma_{\mathfrak{u}}}d_{\sigma_{\mathfrak{su}}}/|H|$.\footnote{For $nN_f$ odd, $U(1)_{nN_f}$ is a spin TQFT, and the transparent fermion gives in addition an odd endomorphism of $\mathcal{D}$, so that each $\mathcal{D}_\chi$ is a q-type line, carrying an unpaired Majorana zero mode. This holds for \eqref{eq:Dqedss} as well. 
} Each $\mathcal{D}_\chi$ is a summand of $\mathcal{D}$, hence commutes with $\widehat{\mathfrak{g}}\oplus V_{N_f,n}$, and folding back along $\mathcal{D}_\chi$ gives a simple conformal boundary condition preserving $U(1)\times SU(N_f)\times SU(2)$.

\paragraph{Comments on Discrete Symmetry $\bZ_n$:}
By the same argument as at the end of Section \ref{sec:sandwichqedss}, one may anticipate that  the line $\eta$ in \eqref{eq:etaline}, which generates $\bZ_{nN_f}$, commutes with the intertwining defect, and therefore $\bZ_{nN_f}$ is preserved by the conformal boundary. This is not true. 

More precisely, $\eta$ only commutes with the \emph{non-simple} intertwining defect $\mathcal{D}$ as defined in \eqref{eq:Dqedsbar}, but does not commute with individual simple components $\mathcal{D}_\chi$, since $B_s$ in \eqref{eq:baryon} braids non-trivially with $\eta$ by a phase $e^{2\pi i s/N_f}$. What braids trivially with $B_s$ is $\eta^{N_f}$, which generates a $\bZ_{n}$ symmetry. Therefore, \emph{simple} intertwining defect $\mathcal{D}_\chi$ commutes with the $\bZ_{n}$ generator $\eta^{N_f}$, consistent with the charge table \eqref{eq:highcharge}.

\subsubsection{Scattering}

We now compute the scattering by applying \eqref{eq:auto45}. The factors $SU(2)_{N_f n (n^2-1)/6}\boxtimes\Rep(V_{N_f,n})$ again contribute trivially, and the $U(1)_{nN_f}$ factor contributes the line $[2]$ and a phase, as in \eqref{eq:qedssscat}. The new ingredient is the $SU(N_f)_n$ factor. The junction $w$ between ${\lambda}_{\widetilde\psi}$, $a$ and $\lambda_\psi$ requires $\square\in\overline{\square}\otimes a$, i.e.\ the $SU(N_f)_n$ component of $a$ lies in $\square\otimes\square=\text{Sym}^2\square\oplus\wedge^2\square$. Therefore there are two out-going channels,
\begin{eqnarray}\label{eq:channel}
    a_{\rm S}= [2]\boxtimes\text{Sym}^2\square\boxtimes \mathbf{1}, \qquad a_{\rm A}= [2]\boxtimes\wedge^2\square\boxtimes \mathbf{1} .
\end{eqnarray}
In the sandwich, the out-going configuration is
\begin{eqnarray}\label{eq:qedsbarout}
    \figAutoCGen[1.25]{$\lambda_\psi$}{$\widetilde{\lambda}_\psi$}{$a$}{$\sigma_{\mathfrak{u}}\boxtimes\sigma_{\mathfrak{su}}$}{$\mathcal{T}_{N_f,n}$}{$B_{\widehat{\mathfrak{g}}\oplus V_{N_f,n}}$}{$\mathcal{S}_A$}
\end{eqnarray}
with $a=a_{\rm S}$ or $a_{\rm A}$. Shrinking the sandwich, the categorical scattering \eqref{eq:auto45} gives
\begin{eqnarray}\label{eq:qedsbarscat}
    \figBeforeScatter[0.95]\quad = \quad C_{\rm S}\quad\figOut{a_{\rm S}}\quad + \quad C_{\rm A}\quad\figOut{a_{\rm A}}
\end{eqnarray}
where, after pushing onto $B_{L_2}$, we abuse the notations $a_{\rm S}$ and $a_{\rm A}$ to denote the lines $F_{B_{L_2}}(a_{\rm S})$ and $F_{B_{L_2}}(a_{\rm A})$ of $\Fer(nN_f)$, which remain simple lines by computing the dimension of the Hom-space as we did above. The coefficients factorize into a $U(1)_{nN_f}$ part and an $SU(N_f)_n$ part,
\begin{eqnarray}\label{eq:Ca}
    C_{a} = e^{i\alpha_{\mathfrak{u}}}\sum_{\substack{\sigma'\in\sigma_{\mathfrak{su}}\otimes\square\\ uvu'u''zw}}\sqrt{\frac{d_{\sigma'}}{d_{\sigma_{\mathfrak{su}}}d_\square}} (R^{\sigma'}_{\sigma_{\mathfrak{su}}\square})^{-1}_{uu'} (R^{\sigma'}_{\square\sigma_{\mathfrak{su}}})^{-1}_{u'u''} (F_{\sigma_{\mathfrak{su}} \overline{\square} \overline{\square}}^{\sigma_{\mathfrak{su}}})^{-1}_{(\sigma' u'' v) (\bar{a} zw)}, \qquad a={\rm S},{\rm A}.
\end{eqnarray}
Here $e^{i\alpha_{\mathfrak{u}}}$ is the contribution of $U(1)_{nN_f}$ as in \eqref{eq:qedssscat}. The scattering probabilities are 
\begin{eqnarray}
    P_{\rm S}=|C_{\rm S}|^2, \qquad P_{\rm A}=|C_{\rm A}|^2.
\end{eqnarray}

The coefficients \eqref{eq:Ca} refer to the unsplit defect \eqref{eq:Dqedsbar}. The simple summands $\mathcal{D}_\chi$ are related by conjugation with the $\bZ_{nN_f}$ lines, which act on $\psi$ and on the out-going excitations by phases. Hence the scattering probabilities $P_{\rm S}$ and $P_{\rm A}$ are the same for every simple $\mathcal{D}_\chi$.

We emphasize the following features.
\begin{enumerate}
    \item The out-going state is a superposition of two twisted sectors, labeled by two different lines. The lines are generically non-invertible: their quantum dimensions are
    \begin{eqnarray}\label{eq:dimSA}
        d_{a_{\rm S}} = \frac{[N_f]_q[N_f+1]_q}{[2]_q}, \qquad d_{a_{\rm A}} = \frac{[N_f]_q[N_f-1]_q}{[2]_q},
    \end{eqnarray}
    with $[m]_q=\sin\frac{m\pi}{N_f+n}/\sin\frac{\pi}{N_f+n}$, 
    which are both larger than one for $N_f\geq 3$ and $n\geq 3$. 
    \item The non-invertible lines $a_{\rm S}$ and $a_{\rm A}$ commute with $\widehat{\mathfrak{g}}$, and in particular with $SU(N_f)\times SU(2)$; nevertheless they supply the $SU(N_f)$ quantum numbers needed to turn $\widetilde{\psi}$, which transforms in $\overline{\square}$, into an excitation transforming in $\square$. 
    This resolves the open problem of \cite{vanBeest:2023dbu} for the 2d problem: for all $(N_f,n)$, there are simple boundary conditions preserving all the symmetries in \eqref{eq:highcharge}, and the out-going states live in the twisted sectors of the non-invertible lines \eqref{eq:channel}.
    \item In the special cases, \eqref{eq:qedsbarscat} reduces to known results. For $n=1$, $\text{Sym}^2\square$ is not integrable at level one, so only $a_{\rm A}$ appears, with $P_{\rm A}=1$, and $\wedge^2\square$ is a simple current at level one, so the line is invertible. For $n=2$, $\text{Sym}^2\square$ is a simple current while $\wedge^2\square$ is not; here a simpler, linear, boundary condition is also available.
    \item The probabilities only depend on the crossed data of $SU(N_f)_n$; the coset $V_{N_f,n}$, the $SU(2)$ factor and the $U(1)$ factor drop out.
\end{enumerate}

\section{Example: $SU(N) \times SU(N+2)$ Model}
\label{sec:susu}

Our second example is a family of chiral, anomaly-free symmetries of $(N+1)^2$ Dirac fermions with non-abelian symmetry $SU(N)\times SU(N+2)$ under which, for $N\geq 2$, the in-coming and out-going fermions transform in genuinely different representations. In contrast with Section \ref{sec:highcharge}, the two embeddings of the current algebra are not related by an automorphism, and the coset chiral algebras are different. However, they share a common conformal subalgebra, so the model fits the simple case of Section \ref{sec:commonalg}.

This model was highlighted as an open problem in fermion-boundary scattering \cite{PathToQFTTalk}. Here we provide a solution.

\subsection{The Model}
\label{sec:susubackground}

We consider $(N+1)^2$ massless Dirac fermions on a half-plane, $N\geq 1$, with symmetry $G=SU(N)\times SU(N+2)$.  The in-coming components form two multiplets $\psi$ and $\chi$, and the out-going components form two multiplets $\widetilde{\psi}$ and $\widetilde{\chi}$, with symmetry quantum numbers
\begin{eqnarray}\label{eq:susucharges}
	\begin{tabular}{|c|c|c|}
		\hline
		 & $SU(N)$ & $SU(N+2)$ \\
		\hline
		$\psi$ & $\text{Sym}^2\square$ & $\mathbf{1}$ \\
		$\chi$ &  $\mathbf{1}$ & $\wedge^2\square$ \\
		\hline
		$\widetilde{\psi}$ &  $\square$ & $\overline{\square}$ \\
		$\widetilde{\chi}$ &  $\mathbf{1}$ & $\mathbf{1}$ \\
		\hline
	\end{tabular}
\end{eqnarray}
The anomalies, or equivalently the level of affine Lie algebras, are given by 
\begin{eqnarray}\label{eq:susuanomaly}
	\begin{tabular}{|c|c|c|}
		\hline
		& $SU(N)$ & $SU(N+2)$ \\
		\hline
		$\psi$ & $N+2$ & $0$ \\
		$\chi$ &  $0$ & $N$ \\
		\hline
		$\widetilde{\psi}$ &  $N+2$ & $N$ \\
		$\widetilde{\chi}$ &  $0$ & $0$ \\
		\hline
	\end{tabular}
\end{eqnarray}
Indeed, the anomalies of the in-coming fermions match/cancel those of the out-going fermions.

However, for $N\geq 2$, no linear $G$ preserving boundary condition exists, since a linear boundary condition would identify the representation of every in-coming fermion with an out-going fermion or its conjugates.

This symmetry of fermions has been studied in \cite{Tong:2021phe}, as a symmetric mass generation problem for Dirac fermions. However, their construction involves using strongly coupled dynamics, which makes the explicit form of the boundary condition, and the scattering property, obscure. This puzzle was also mentioned recently in \cite{PathToQFTTalk}.

We show that this problem falls into the case of Section \ref{sec:commonalg}, hence been solved for all $N$. The resulting intertwining defect $\mathcal{D}$ is non-invertible. Each in-coming fermion, $\psi$ or $\chi$, scatters with probability one into a single twisted sector, attached to a line of quantum dimension $1/(2\sin^2\frac{\pi}{2N+2})$, which is also non-invertible for $N\geq 2$.

\subsection{Two Embeddings}
\label{sec:susuembed}

From \eqref{eq:susuanomaly}, the currents of $G$ generate the affine Lie algebra
\begin{eqnarray}\label{eq:gsusu}
    \widehat{\mathfrak{g}} = \widehat{\mathfrak{su}}(N)_{N+2}\oplus\widehat{\mathfrak{su}}(N+2)_N,
\end{eqnarray}
and two embeddings are
\begin{equation}\label{eq:f12susu}
\begin{alignedat}{2}
    &f_1(j^\alpha_{\mathfrak{su}(N)}) = \psi^\dagger\, T^\alpha_{\text{Sym}^2\square}\,\psi,
    &&\qquad f_2(j^\alpha_{\mathfrak{su}(N)}) = \widetilde\psi^\dagger\, (T^\alpha_\square\otimes 1)\,\widetilde\psi, \\[4pt]
    &f_1(j^\beta_{\mathfrak{su}(N+2)}) = \chi^\dagger\, T^\beta_{\wedge^2\square}\,\chi,
    &&\qquad f_2(j^\beta_{\mathfrak{su}(N+2)}) = \widetilde\psi^\dagger\, (1\otimes T^\beta_{\overline\square})\,\widetilde\psi,
\end{alignedat}
\end{equation}
where $T^\beta_{\overline\square}=-T^{\beta\,\mathsf{T}}_\square$, with normal ordering implied.

The central charge of $\widehat{\mathfrak{g}}$ is
\begin{eqnarray}
    c_{\widehat{\mathfrak{g}}} = \underbrace{\frac{(N+2)(N-1)}{2}}_{\widehat{\mathfrak{su}}(N)_{N+2}} + \underbrace{\frac{N(N+3)}{2}}_{\widehat{\mathfrak{su}}(N+2)_N} = (N+1)^2-2,
\end{eqnarray}
so $\widehat{\mathfrak{g}}$ is not conformally embedded in $\Fer((N+1)^2)$, and the coset chiral algebras $S_i$ have central charge $2$. They can be worked out by a chain of canonical conformal embeddings, similar to \eqref{eq:qedembeddingchain}.

For the in-coming fermions, we have the following chain of embeddings,
\begin{eqnarray}\label{eq:susuleft}
    \begin{split}
        &\Fer((N+1)^2) \\ \hookleftconf \quad & \Fer(\tfrac{N(N+1)}{2})^{\psi} \oplus \Fer(\tfrac{(N+1)(N+2)}{2})^{\chi}\\
    \hookleftconf \quad &\widehat{\mathfrak{su}}(\tfrac{N(N+1)}{2})_1 \oplus \widehat{\mathfrak{u}}(1)_{\tfrac{N(N+1)}{2}}\oplus \widehat{\mathfrak{su}}(\tfrac{(N+1)(N+2)}{2})_1 \oplus \widehat{\mathfrak{u}}(1)_{\tfrac{(N+1)(N+2)}{2}}\\
    \hookleftconf \quad & \widehat{\mathfrak{su}}(N)_{N+2} \oplus \widehat{\mathfrak{u}}(1)_{\tfrac{N(N+1)}{2}} \oplus \widehat{\mathfrak{su}}(N+2)_N \oplus \widehat{\mathfrak{u}}(1)_{\tfrac{(N+1)(N+2)}{2}}.
    \end{split}
\end{eqnarray}
The first embedding decomposes all fermions into two multiplets associated with $\psi$ and $\chi$. The second one decomposes the respective algebras into non-abelian and abelian parts as \eqref{eq:canonicalbranching}. The last one restricts the $\square$-representation of $\mathfrak{su}(\frac{N(N+1)}{2})$ to $\text{Sym}^2 \square$-representation of $\mathfrak{su}(N)$, and the $\square$-representation of $\mathfrak{su}(\frac{(N+1)(N+2)}{2})$ to the $\wedge^2\square$-representation of $\mathfrak{su}(N+2)$. The two non-abelian factors in the last line of \eqref{eq:susuleft} form $\widehat{\mathfrak{g}}$ in \eqref{eq:gsusu}, and the remaining factors form the coset chiral algebra, 
\begin{eqnarray}\label{eq:S1susu}
    S_1= \widehat{\mathfrak{u}}(1)_{\tfrac{N(N+1)}{2}} \oplus \widehat{\mathfrak{u}}(1)_{\tfrac{(N+1)(N+2)}{2}}.
\end{eqnarray}

For the out-going fermions, we have the following chain of embeddings, 
\begin{eqnarray}\label{eq:susuright}
    \begin{split}
        &\Fer((N+1)^2)\\
        \hookleftconf \quad & \Fer(N(N+2))^{\widetilde\psi} \oplus \Fer(1)^{\widetilde\chi}\\
        \hookleftconf \quad & \widehat{\mathfrak{su}}(N(N+2))_1 \oplus \widehat{\mathfrak{u}}(1)_{N(N+2)} \oplus \widehat{\mathfrak{u}}(1)_1\\
        \hookleftconf \quad & \widehat{\mathfrak{su}}(N)_{N+2} \oplus \widehat{\mathfrak{su}}(N+2)_N \oplus \widehat{\mathfrak{u}}(1)_{N(N+2)} \oplus \widehat{\mathfrak{u}}(1)_1
    \end{split}
\end{eqnarray}
The first and second embeddings are similar to \eqref{eq:susuleft}. The last one is the level-rank embedding as in \eqref{eq:levelrankembedding}, where the $\square$-representation of ${\mathfrak{su}}(N(N+2))$ splits into the tensor product of $\square$-representation of ${\mathfrak{su}}(N)$ and $\overline{\square}$-representation of ${\mathfrak{su}}(N+2)$. The last line of \eqref{eq:susuright} gives the coset chiral algebra 
\begin{eqnarray}\label{eq:S2susu}
    S_2= \widehat{\mathfrak{u}}(1)_{N(N+2)} \oplus \widehat{\mathfrak{u}}(1)_1.
\end{eqnarray}
which is distinct from $S_1$.

\subsection{Intertwining Defect}
\label{sec:susuW}
Since the coset chiral algebras $S_1$ and $S_2$ are distinct, the system spans to a club sandwich. We first find the common conformal subalgebra, then build the club sandwich, and finally construct the intertwining defect. 
\subsubsection{Common conformal subalgebra}
Although $S_1$ and $S_2$ are not isomorphic, it turns out that $S_1$ and $S_2$ share a common conformal algebra, which fits into Section \ref{sec:commonalg}. 

We try to look for a shared chiral subalgebra $W$ of $S_1$ and $S_2$. Since $S_1$ and $S_2$ are rank-two lattice chiral algebras, we look for $W$ among the rank-two lattice chiral algebras $W=\widehat{\mathfrak{u}}(1)^2_K$ with $K$ a $2\times 2$ matrix. It was proven in \cite{Hung:2013qpa}, which we also review in Appendix \ref{app:Anyon condensation between K matrix U(1) Chern-Simons TQFTs}, that $\widehat{\mathfrak{u}}(1)^2_K$ is conformally embedded in $S_i$ if and only if\footnote{In Appendix \ref{app:Anyon condensation between K matrix U(1) Chern-Simons TQFTs}, we summarize a condition when $\widetilde{K}$ Chern Simons theory is obtained from $K$ Chern Simons theory by condensing an algebra $\mathcal{A}$, hence $\widehat{\mathfrak{u}}(1)^2_K$ is conformally embedded in $\widehat{\mathfrak{u}}(1)^2_{\widetilde{K}}$. }
\begin{eqnarray}\label{eq:KPKP}
    K=P_i^{\mathsf T}\widetilde K_iP_i, \qquad i=1,2
\end{eqnarray}
for some $P_i\in \bZ^{2\times 2}$ with $\det P_i\neq 0$, where 
\begin{eqnarray}
    \widetilde{K}_1= 
    \begin{pmatrix}
        \frac{N(N+1)}{2} & 0\\
        0 & \frac{(N+1)(N+2)}{2}
    \end{pmatrix}, \qquad  \widetilde{K}_2= 
    \begin{pmatrix}
        N(N+2) & 0 \\
        0 & 1
    \end{pmatrix}
\end{eqnarray}
are the $K$-matrices of $S_1$ and $S_2$. Equivalently, the TQFTs for $S_i$ and $W$, i.e. $\Rep(S_i)$ (i.e. $U(1)^2_{\widetilde{K}_i}$) and $\Rep(W)$ (i.e. $U(1)^2_{K}$) are related by anyon condensation if and only if \eqref{eq:KPKP} holds.

It can be checked that the $P_i$-matrices and $K$-matrix satisfying \eqref{eq:KPKP} are
\begin{eqnarray}
    K= \begin{pmatrix}
        (N+1)^2 & -(N+1) \\
        -(N+1) & (N+1)^2
    \end{pmatrix}, \qquad P_1= \begin{pmatrix} 1 & 1\\ 1 & -1\end{pmatrix}, \qquad P_2 = \begin{pmatrix} -1 & 0 \\ -1 & N+1\end{pmatrix}
\end{eqnarray}
By \eqref{eq:algebra from P}, condensing the algebra of $U(1)^2_K$, 
\begin{eqnarray}\label{eq:susuA1A2}
    \mathcal{A}_1= (0,0) \oplus (\frac{N(N+1)}{2}, \frac{N(N+1)}{2}), \qquad \mathcal{A}_2 = \bigoplus_{i=0}^N (iN(N+2),0)
\end{eqnarray}
Using the lifting functor \eqref{eq:lift fun of abelian condensation} between the TQFTs, we also get the branching functions between the affine Lie algebras
\begin{eqnarray}\label{eq:susulift1}
\begin{split}
    \widehat{\mathfrak{u}}(1)^2_{\widetilde{K}_1} &\to \widehat{\mathfrak{u}}(1)^2_K\\
    (r,s) &\mapsto \bigoplus_{m=0}^1 (r+s+m\frac{N(N+1)}{2}, r-s+m \frac{N(N+1)}{2})
\end{split}
\end{eqnarray}
where $r$ is defined modulo $N(N+1)/2$ and $s$ is defined modulo $(N+1)(N+2)/2$,
and 
\begin{eqnarray}\label{eq:susulift2}
    \begin{split}
        \widehat{\mathfrak{u}}(1)^2_{\widetilde{K}_2} &\to \widehat{\mathfrak{u}}(1)^2_K\\
    (r,0) &\mapsto \bigoplus_{j=0}^N (-r+jN(N+2),0) 
    \end{split}
\end{eqnarray}
where $r$ is modulo $N(N+2)$.

The common conformal subalgebra $\widehat{\mathfrak{u}}(1)^2_K$ implies the charge assignments of the fermions under two $U(1)$ symmetries, which we denote as $U(1)_V$ and $U(1)_F$, as 
\begin{eqnarray}\label{eq:U1charges}
    \begin{tabular}{|c|c|c|c|c|}
    \hline
         & $\psi$ & $\chi$ & $\widetilde\psi$ & $\widetilde\chi$ \\
         \hline
        $U(1)_V$ & $1$ & $1$ & $-1$ & $-1$\\
        $U(1)_F$ & $1$ & $-1$ & $0$ & $N+1$\\
        \hline
    \end{tabular}
\end{eqnarray}
Their currents, when embedded in $\widehat{\mathfrak{u}}((N+1)^2)_1$, are
\begin{equation}\label{eq:JVJF}
\begin{alignedat}{2}
    &f_1(J_V) = \psi^\dagger\psi+\chi^\dagger\chi,
    &&\qquad f_2(J_V) = -\widetilde\psi^\dagger\widetilde\psi-\widetilde\chi^\dagger\widetilde\chi, \\[4pt]
    &f_1(J_F) = \psi^\dagger\psi-\chi^\dagger\chi,
    &&\qquad f_2(J_F) = (N+1)\,\widetilde\chi^\dagger\widetilde\chi,
\end{alignedat}
\end{equation}
It is straightforward to check that these currents reproduces the desired current algebra level given by $K$.

The $U(1)_V$ symmetry acts as a vector-like symmetry. This hints that the model considered in section admits a lift to 4d non-chiral QED, where the $SU(N)$ and $SU(N+2)$ symmetries come from the flavor rotations. We will not discuss its 4d lift in this work.

\subsubsection{Sandwich from Club Sandwich}
\begin{figure}
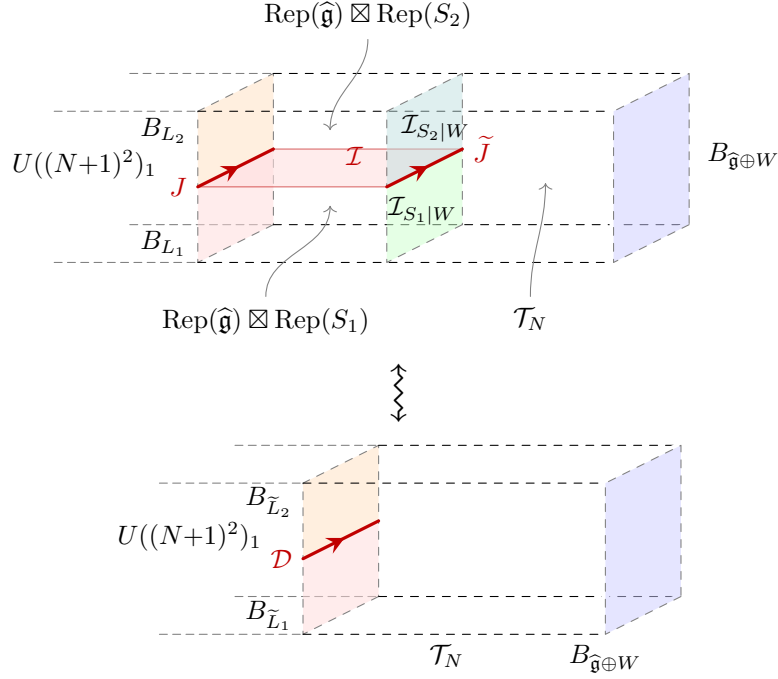

    \centering
    \figClubSUSU[1.0]\\[1ex]\rotatebox[origin=c]{-90}{\Large$\leftrightsquigarrow$}\\[1ex]
    \figClubSUSUShrunk[1.0]
    \caption{Club sandwich for the $SU(N)\times SU(N+2)$ model (top), with $\Rep(\widehat{\mathfrak{g}})=SU(N)_{N+2}\boxtimes SU(N+2)_N$. The interfaces $\mathcal{I}_{S_i|W}$ condense the algebras $\mathcal{A}_i$ in \eqref{eq:susuA1A2}. Shrinking them onto the topological interfaces $B_{L_i}$ gives the sandwich at the bottom, in which the two topological boundaries $B_{\widetilde L_i}$ share the bulk TQFT $\mathcal{T}_N$ in \eqref{eq:TN}, and are joined along the intertwining defect $\mathcal{D}$.}
    \label{fig:susuclub}
\end{figure}

Following Section \ref{sec:commonalg}, we obtain the club sandwich shown at the top of Figure \ref{fig:susuclub}. Its right bulk is the TQFT
\begin{eqnarray}\label{eq:TN}
    \mathcal{T}_N = SU(N)_{N+2}\boxtimes SU(N+2)_N\boxtimes U(1)^2_K,
\end{eqnarray}
which ends on the chiral boundary $B_{\widehat{\mathfrak{g}}\oplus W}$. The bottom and top left bulks are $SU(N)_{N+2}\boxtimes SU(N+2)_N\boxtimes U(1)_{N(N+1)/2}\boxtimes U(1)_{(N+1)(N+2)/2}$ and $SU(N)_{N+2}\boxtimes SU(N+2)_N\boxtimes U(1)_{N(N+2)}\boxtimes U(1)_1$. They are separated from $\mathcal{T}_N$ by the interfaces $\mathcal{I}_{S_i|W}$, which condense the algebras \eqref{eq:susuA1A2}, and from $U((N+1)^2)_1$ by the topological interfaces $B_{L_i}$. Shrinking $\mathcal{I}_{S_i|W}$ onto $B_{L_i}$, we obtain the sandwich at the bottom of Figure \ref{fig:susuclub}, with two topological boundaries $B_{\widetilde{L}_1}$ and $B_{\widetilde{L}_2}$ of the same bulk $\mathcal{T}_N$.

The fermionic Lagrangian algebras $\widetilde{L}_i$ follow from the branching of the vacuum supermodule of $\Fer((N+1)^2)$ along \eqref{eq:susuleft} and \eqref{eq:susuright}, followed by lifting the lines of $\Rep(S_i)$ to $U(1)^2_K$ by \eqref{eq:susulift1} and \eqref{eq:susulift2}. 
\begin{itemize}
    \item For the in-coming fermions $\psi$ and $\chi$, 
\begin{equation}\label{eq:L1susu}
    \widetilde{L}_1 = \bigoplus_{\substack{r_1\in\mathbb{Z}_{\frac{N(N+1)}{2}}\\r_2\in\mathbb{Z}_{\frac{(N+1)(N+2)}{2}} }} \bigoplus_{\substack{\mu\in\Sigma^{\rm S}_{r_1}\\\nu\in\Sigma^{\rm A}_{r_2}}} \bigoplus_{s\in\mathbb{Z}_2} \mu\boxtimes\nu\boxtimes(r_1+r_2+s\frac{N(N+1)}{2},\,r_1-r_2+s\frac{N(N+1)}{2}),
\end{equation}
where $\mu$ arises from the branching $\Lambda_{r_1}\mapsto\bigoplus_{\mu\in\Sigma^{\rm S}_{r_1}}\mu$ in $\widehat{\mathfrak{su}}(\frac{N(N+1)}{2})_1\hookleftconf \widehat{\mathfrak{su}}(N)_{N+2}$, and $\nu$ arises from the branching $\Lambda_{r_2}\mapsto\bigoplus_{\nu\in\Sigma^{\rm A}_{r_2}}\nu$ in $\widehat{\mathfrak{su}}(\frac{(N+1)(N+2)}{2})_1 \hookleftconf \widehat{\mathfrak{su}}(N+2)_{N}$. 
    \item For the out-going fermions $\widetilde\psi$ and $\widetilde\chi$, 
    \begin{eqnarray}\label{eq:L2susu}
    \widetilde{L}_2 = \bigoplus_{r\in\mathbb{Z}_{N(N+2)}}\ \bigoplus_{(\mu,\nu)\in\mathcal{P}_r}\ \bigoplus_{j\in\mathbb{Z}_{N+1}}\ \mu\boxtimes\overline{\nu}\boxtimes(-r-jN(N+2),\,0),
\end{eqnarray}
where $(\mu,\nu)$ arises from branching rule $\Lambda_r\mapsto\bigoplus_{(\mu,\nu)\in\mathcal{P}_r}\mu\boxtimes\overline{\nu}$ of the level-rank embedding $\widehat{\mathfrak{su}}(N(N+2))_1 \hookleftconf \widehat{\mathfrak{su}}(N)_{N+2}\oplus\widehat{\mathfrak{su}}(N+2)_N$.
\end{itemize}
The bulk lines attached the fermions can be read-off from the charge tables \eqref{eq:susucharges} and \eqref{eq:U1charges},
\begin{equation}\label{eq:lambdasusu}
    \begin{aligned}
    &\lambda_\psi = \text{Sym}^2\square\boxtimes\mathbf{1}\boxtimes(1,1), &&\qquad \lambda_\chi = \mathbf{1}\boxtimes\wedge^2\square\boxtimes(1,-1),\\
    &\lambda_{\widetilde\psi} = \square\boxtimes\overline{\square}\boxtimes(-1,0), &&\qquad \lambda_{\widetilde\chi} = \mathbf{1}\boxtimes\mathbf{1}\boxtimes(-1,N+1),
    \end{aligned}
\end{equation}
Indeed, $\lambda_{\psi},\lambda_{\chi}\in \widetilde{L}_1$ and ${\lambda}_{\widetilde\psi},{\lambda}_{\widetilde\chi}\in\widetilde{L}_2$, and their conformal weights are all equal to $\frac12$.\footnote{More concretely, $h_{\lambda_\psi}= \frac{(N-1)(N+2)}{2N(N+1)}+\frac{1}{N(N+1)}=\frac12, h_{\lambda_\chi} = \frac{N(N+3)}{2(N+1)(N+2)}+\frac{1}{(N+1)(N+2)}=\frac12,
    h_{\widetilde\lambda_\psi} = \frac{N-1}{4N}+\frac{N+3}{4(N+2)}+\frac{1}{2N(N+2)}=\frac12,  h_{\widetilde\lambda_\chi}=\frac12$. }

\subsubsection{Intertwining Defect}
\label{sec:susuD}

After shrinking the club sandwich, the intertwining defect $\mathcal{D}$ is a topological junction\footnote{Such a junction is not unique: the number of simple junctions is $\dim\Hom(\widetilde{L}_1,\widetilde{L}_2)$ \cite{Fuchs:2002cm}. For instance, when $N=2$,  $\dim\Hom(\widetilde{L}_1,\widetilde{L}_2)=16$. The results below hold for any simple choice.} between the two topological boundaries $B_{\widetilde L_1}$ and $B_{\widetilde L_2}$ of $\mathcal{T}_N$.
By the argument of Section \ref{sec:interDefect}, $\mathcal{D}$ commutes with $\widehat{\mathfrak{g}}\oplus W$, and hence intertwines the two embeddings \eqref{eq:f12susu}.

All such $\mathcal{D}$ is non-invertible. Indeed, an invertible line maps local operators to local operators of the same conformal weight, and since $\mathcal{D}$ intertwines \eqref{eq:f12susu}, the image of $\psi$ would be an out-going local operator of weight $\frac12$ with the same quantum numbers under $G$. The only out-going operators of weight $\frac12$ are $\widetilde\psi$, $\widetilde\chi$ and their conjugates, and none of them transforms in $\text{Sym}^2\square$ of $SU(N)$. In the sandwich, this is the statement that $\lambda_\psi, \lambda_\chi\notin\widetilde{L}_2$.

Upon folding along $\mathcal{D}$, the boundary state $\ket{\mathcal{D}}$ preserves $SU(N)\times SU(N+2)\times U(1)^2$, and is rational with respect to the chiral algebra $\widehat{\mathfrak{g}}\oplus \widehat{\mathfrak{u}}(1)^2_K$.

\subsection{Scattering}
\label{sec:sususcat}

Since $B_{\widetilde L_1}$ and $B_{\widetilde L_2}$ are boundaries of the same bulk $\mathcal{T}_N$, with the same chiral boundary, the categorical scattering formula \eqref{eq:Daction} applies, with $\Rep(\widehat{\mathfrak{g}})$ replaced by $\mathcal{T}_N$ and $B_{L_i}$ by $B_{\widetilde L_i}$. The in-coming fermion $\psi$ is resolved as $(x,\lambda_\psi,\widetilde x)$, and the out-going twisted sectors are labeled by the simple lines in $F_{B_{\widetilde{L}_2}}(\lambda_\psi)$.

$F_{B_{\widetilde L_2}}(\lambda_\psi)$ is simple for all $N$. To see this, we compute its endomorphisms on $B_{\widetilde{L}_2}$, which are counted by
\begin{eqnarray}\label{eq:EndFpsi}
    \dim\Hom_{\rm bdy}\big(F_{B_{\widetilde L_2}}(\lambda_\psi),F_{B_{\widetilde L_2}}(\lambda_\psi)\big) = \dim\Hom_{\mathcal{T}_N}\big(\lambda_\psi\otimes\overline{\lambda}_\psi,\,\widetilde L_2\big),
\end{eqnarray}
where $\lambda_\psi\otimes\overline\lambda_\psi = (\text{Sym}^2\square\otimes\overline{\text{Sym}^2\square})\boxtimes\mathbf{1}\boxtimes(0,0)$. A summand $\mu\boxtimes\overline\nu\boxtimes(-r-jN(N+2),0)$ of \eqref{eq:L2susu} contributes only if $\nu$ is trivial and $(r+jN(N+2),0)\in K\mathbb{Z}^2$. The latter requires $r+jN(N+2)\in(N+1)N(N+2)\mathbb{Z}$, i.e.\ $r=j=0$. In the level-rank decomposition, the trivial weight of $\widehat{\mathfrak{su}}(N+2)_N$ appears only in $\Lambda_{k(N+2)}$ of $\widehat{\mathfrak{su}}(N(N+2))_1$, paired with the simple current $(N+2)\Lambda_k$ of $\widehat{\mathfrak{su}}(N)_{N+2}$, $k\in\mathbb{Z}_N$; for $r=0$ only $k=0$ contributes. Hence only the trivial line of $\mathcal{T}_N$ contributes, which appears once in $\text{Sym}^2\square\otimes\overline{\text{Sym}^2\square}$, and \eqref{eq:EndFpsi} is one-dimensional. The same argument applies to $\lambda_\chi$, with the roles of the two non-abelian factors exchanged.

We denote the resulting simple lines of $\Fer((N+1)^2)$ by
\begin{eqnarray}\label{eq:bpsibchi}
    b_\psi = F_{B_{\widetilde L_2}}(\lambda_\psi), \qquad b_\chi= F_{B_{\widetilde L_2}}(\lambda_\chi).
\end{eqnarray}
Since the bulk-to-boundary functor preserves quantum dimensions, and the lines of $U(1)^2_K$ are invertible,
\begin{eqnarray}\label{eq:dbpsi}
    d_{b_\psi}=d_{\text{Sym}^2\square}=\frac{[N]_q[N+1]_q}{[2]_q}, \qquad d_{b_\chi}=d_{\wedge^2\square}=\frac{[N+2]_q[N+1]_q}{[2]_q},
\end{eqnarray}
where $d_{\text{Sym}^2\square}$ and $d_{\wedge^2\square}$ are the quantum dimensions in $SU(N)_{N+2}$ and $SU(N+2)_N$, respectively, and $[m]_q=\sin\frac{m\pi}{2N+2}/\sin\frac{\pi}{2N+2}$. Both are equal to
\begin{eqnarray}\label{eq:dbsusu}
    d_{b_\psi}=d_{b_\chi}=\frac{1}{2\sin^2\frac{\pi}{2N+2}},
\end{eqnarray}
which is larger than 1, hence non-invertible.

Therefore, $\psi$ scatters to a twisted sector operator $\mathcal{O}_\psi^{(z)}$, attached to a non-invertible line $b_\psi$, with the amplitude given by \eqref{eq:Daction}\footnote{We used $d_{\lambda_\psi}= d_{b_\psi}$. } 
\begin{eqnarray}\label{eq:sususcat}
    \figBeforeScatterSUSU[0.95]{\psi}\quad = \quad \sum_{zw}\, C^{\psi}_{zw}\quad \figOutSUSU[0.95]{b_\psi}{w}{\mathcal{O}^{(z)}_{\psi}}
\end{eqnarray}
with
\begin{eqnarray}\label{eq:Cpsisusu}
    C^\psi_{zw}= \sum_{y}\frac{1}{d_{\mathcal{D}}\sqrt{S_{11}}}\, {^{B_{\widetilde L_1}B_{\widetilde L_2}}\widetilde{\Psi}_{\mathcal{D}\mathcal{D}(\overline{\lambda}_\psi xy)}^{(1)(\overline{b}_\psi w)}},
\end{eqnarray}
where $S_{11}$ is now the vacuum entry of the $S$-matrix of $\mathcal{T}_N$, and $\mathcal{O}^{(z)}_\psi$ is obtained by shrinking $(z,\lambda_\psi,\widetilde x)$. The scattering of $\chi$ is given by the same formula, with $\lambda_\psi$ and $b_\psi$ replaced by $\lambda_\chi$ and $b_\chi$.

We highlight some features.
\begin{enumerate}
    \item The out-going excitation lives in a single twisted sector: $\psi$ scatters with probability one into the $b_\psi$-twisted sector, and $\chi$ into the $b_\chi$-twisted sector. The sum in \eqref{eq:sususcat} only runs over the topological junctions $z$ and $w$. This should be contrasted with $\text{QED}(\square,\overline\square)$ in Section \ref{sec:qedsbar}, where the out-going state is a superposition of two twisted sectors.
    \item The lines $b_\psi$ and $b_\chi$ are non-invertible, by \eqref{eq:dbsusu}. This is in contrast with the abelian examples \cite{vanBeest:2023dbu,Wei:2026fsn,Arias-Tamargo:2026urw}, where the lines attached to the out-going excitations are invertible.
    \item The out-going excitation $\mathcal{O}^{(z)}_\psi$ carries exactly the same $\widehat{\mathfrak{g}}\oplus W$ quantum numbers as $\psi$, and the same conformal weight $\frac12$, since the bulk line $\lambda_\psi$ and the chiral junction $\widetilde x$ are unchanged. As shown in Section \ref{sec:susuD}, no out-going \emph{local} operator has these quantum numbers. They are supplied by the line $b_\psi$, which commutes with $\widehat{\mathfrak{g}}\oplus W$; this is how the scattering puzzle of Section \ref{sec:susubackground} is resolved.
\end{enumerate}

\section{Example: Monopole in Chiral QED}
\label{sec:chiralqed}

Our next example is the fermion-monopole scattering in \emph{chiral} QED. In contrast with Section \ref{sec:highcharge}, the in-coming and out-going fermions now carry different $U(1)$ charges, and the flavor symmetries and the $SU(2)$ symmetry coming from spatial rotation around the monopole are also chiral. In particular, the symmetry quantum numbers of in-coming and out-going fermions are not related by automorphism, hence the symmetric boundary states are more challenging to construct.

\subsection{Enumerating Chiral QEDs}
\label{sec:cqedbackground}

We first enumerate the chiral QEDs of our interest. 
We consider a 4d $U(1)$ gauge theory with $n_q$ left-handed Weyl fermions $\Psi$ of charge $q$, where $q$ can take multiple non-zero values. As discussed in \cite{vanBeest:2023dbu}, the gauge and gravitational anomalies cancel if 
\begin{eqnarray}\label{eq:cqedanomaly}
    \sum_q n_q q =0, \qquad \sum_q n_q q^3=0
\end{eqnarray}
The theory is chiral if $n_q \neq n_{-q}$ for some $q$.

We place a Dirac monopole of unit magnetic charge at the origin.\footnote{A monopole of magnetic charge $g$ is described by replacing $q\to gq$. } After partial wave reduction and keep the lowest angular momentum modes, a Weyl fermion of charge $q$ contributes $|q|$ chiral modes on the half-plane $r\geq 0$, which form the spin $\frac{|q|-1}{2}$ representation of the rotation group $SU(2)$ and carry electric charge $q$ \cite{vanBeest:2023dbu, vanBeest:2023mbs}. They are in-coming for $q>0$ and out-going for $q<0$. The total number of in-coming and out-going chiral fermions match thanks to the first identity in \eqref{eq:cqedanomaly}, and are both 
\begin{eqnarray}
    c_{\mathcal{T}}= \sum_{q>0} n_q q
\end{eqnarray}
After reduction, we again suppress the $U(1)$ gauge field fluctuation as in Section \ref{sec:highcharge}, and treat it as a global symmetry of the 2d free fermions.

The boundary condition at $r=0$ is expected to preserve symmetries that are free of anomaly in 2d, which amounts to in 4d that do not participate in a 2-group \cite{vanBeest:2023dbu}. They are 
\begin{eqnarray}
    G=U(1)\times SU(2)\times F
\end{eqnarray}
where $U(1)\times SU(2)$ is 2-group-free thanks to \eqref{eq:cqedanomaly}. $F$ is a subgroup of flavor symmetry, such as the $SU(N_f)$ in Section \ref{sec:highcharge}. One can enumerate all possible solutions of \eqref{eq:cqedanomaly} for a given $c_{\mathcal{T}}$. Focusing on the solutions where $\min(n_q, n_{-q})=0$, i.e. the \emph{maximally chiral} case, we get Table \ref{tab:chiralqed}. In each bracket, we list the charge $q$ and the number of fermions with this charge on the exponent. For each $c_{\mathcal{T}}$, there may exist more than one solution, for instance there are two chiral QEDs with $c_{\mathcal{T}}=8$.  Other less chiral cases can be obtained by supplementing pairs of fermions with opposite charges which always satisfy \eqref{eq:cqedanomaly}.

Each model listed in Table \ref{tab:chiralqed} is interesting and (perhaps) challenging to solve, hence in this paper, we focus on two examples highlighted in red. In the remaining of this section, we focus on the first example, $\{1^5, 3, (-2)^4\}$. In Section \ref{sec:15789}, we revisit the famous 1-5-7-8-9 model, $\{1,5,9,-7,-8\}$. We leave other models for future study.

\begin{table}[t]
    \centering
    \renewcommand{\arraystretch}{1.2}
    \begin{tabular}{|c|c|}
\hline
$c_{\mathcal{T}}$ & $q^{n_q}$ \\
\hline
$8$ &
\PPTdarkred{$\{1^5,3,(-2)^4\}$},
$\{1^3,5,(-4)^2\}$ \\
\hline
$9$ &
$\{1,2^2,4,(-3)^3\}$ \\
\hline
$10$ &
$\{1^6,4,(-2)^2,(-3)^2\}$ \\
\hline
$11$ &
$\{2,3,6,-1,(-5)^2\}$ \\
\hline
$13$ &
\begin{tabular}{c}
    $\{1^4,2^2,5,(-3)^3,-4\}$, 
$\{2^4,5,(-1)^2,-3,(-4)^2\}$, \\
$\{1^2,2,3,6,(-4)^2,-5\}$, 
$\{1^6,7,-2,-5,-6\}$ 
\end{tabular}
\\
\hline
$14$ &
\begin{tabular}{c}
    $\{1^9,5,(-2)^2,(-3)^2,-4\}$, 
$\{1^8,6,-2,-3,-4,-5\}$,\\ 
$\{1,2^3,7,-3,-5,-6\}$, 
$\{3,4,7,(-1)^2,(-6)^2\}$ 
\end{tabular}
\\
\hline
$15$ &
\begin{tabular}{c}
    $\{1^{11},4,(-2)^6,-3\}$, 
$\{1^{10},5,(-3)^5\}$, \\
$\{1^7,4^2,(-3)^5\}$, 
$\{2^2,3^2,5,(-1)^3,(-4)^3\}$, \\
$\{1^3,2^3,6,(-3)^2,-4,-5\}$, 
$\{1^5,4,6,-2,-3,(-5)^2\}$, \\
$\{1^5,3,7,(-5)^3\}$, 
$\{2^4,7,(-5)^3\}$, \\
$\{2,3^2,7,-4,-5,-6\}$, 
$\{1,3^2,8,-2,-6,-7\}$, 
\PPTdarkred{$\{1,5,9,-7,-8\}$}
\end{tabular}
 \\
\hline
\end{tabular}
    \caption{Charges and multiplicities of anomaly-free 4d QED with maximal chirality }
    \label{tab:chiralqed}
\end{table}

\subsection{The $\{1^5,3,(-2)^4\}$ Model and Symmetries}

We now consider the chiral QED with ten Weyl fermions, whose charges and multiplicities are $\{1^5,3,(-2)^4\}$, i.e. five fermions of charge 1, one of charge 3, and four of charge $-2$.

Are there flavor symmetries that commute with $U(1)$ gauge group, and are preserved by the monopole? A natural candidate commuting with $U(1)$ is the $U(5)\times U(4)$ symmetry, which rotates the five and four fermions with charge 1 and $-2$, respectively. However, such symmetry participates in the 2-group with magnetic $U(1)$ 1-form symmetry.

We consider a $Spin(5)$ subgroup of $U(5)\times U(4)$, under which the five fermions of charge 1 transform as the vector $\mathbf{5}$, and the four fermions of charge $-2$ as the spinor $\mathbf{4}$. It does not participate in the 2-group, thanks to the absence of gauge-flavor-flavor anomaly \cite{Cordova:2018cvg}
\begin{eqnarray}
    \text{Tr}(q\,T^aT^b)\propto 1\cdot T(\mathbf{5})-2\cdot T(\mathbf{4})=1\cdot 1-2\cdot\tfrac12=0,
\end{eqnarray}
so the monopole can preserve it.

After reducing to 2d, the non-anomalous symmetry of the 8 massless Dirac fermions is $U(1)\times SU(2)\times Spin(5)$, which come from the $U(1)$ gauge group, the $SU(2)$ rotation around the monopole, and the $Spin(5)$ flavor symmetry  respectively. The fermions carry the following representation under this symmetry,
\begin{eqnarray}\label{eq:cqedcharges}
	\begin{tabular}{|c|c|c|c|}
		\hline
		 & $U(1)$ & $SU(2)$ & $Spin(5)$ \\
		\hline
		$\psi_a$ & $1$ & $\mathbf{1}$ & $\mathbf{5}$ \\
		$\eta_m$ &  $3$ & $\mathbf{3}$ & $\mathbf{1}$ \\
		\hline
		$\chi_{\alpha A}$ & $-2$ & $\mathbf{2}$ & $\mathbf{4}$ \\
		\hline
	\end{tabular}
\end{eqnarray}
where $a,m,\alpha, A$ are the $Spin(5)$ vector, $SU(2)$ triplet, $SU(2)$ doublet and $Spin(5)$ spinor indices.

An amusing way to view \eqref{eq:cqedcharges} is via the $Spin(8)$ triality. Under $SU(2) \times Spin(5) = Spin(3) \times Spin(5) \hookrightconf Spin(8)$, the vector and the spinor of $Spin(8)$ decomposes as $\mathbf{8}_v = (\mathbf{3},\mathbf{1}) \oplus (\mathbf{1}, \mathbf{5})$ and $\mathbf{8}_s= (\mathbf{2},\mathbf{4})$. Hence the in-coming and out-going fermions are related by triality, up to their electric charges.

\subsection{Two Embeddings}
\label{sec:cqedembed}

We unfold the eight 2d Dirac fermions into $\Fer(8)$ with a topological line $\mathcal{D}$, with $\psi, \eta$ on one side and $\chi$ on the other. 
From \eqref{eq:cqedcharges}, the currents generate the affine Lie algebra\footnote{The level for $Spin(5)$ is $2T(R)= 2 \cdot T(\mathbf{5}) = 2\cdot 1 = 2$. }
\begin{eqnarray}\label{eq:gcqed}
    \widehat{\mathfrak{g}}= \widehat{\mathfrak{u}}(1)_{32} \oplus \widehat{\mathfrak{su}}(2)_4 \oplus \widehat{\mathfrak{so}}(5)_2,
\end{eqnarray}
and the two embeddings $f_i: \widehat{\mathfrak{g}}\hookrightarrow  \widehat{\mathfrak{u}}(8)_1$ are 
\begin{equation}\label{eq:f12cqed}
\begin{alignedat}{2}
    &f_1(J) = \psi^\dagger\psi+3\,\eta^\dagger\eta,
    &&\qquad f_2(J) = -2\,\chi^\dagger\chi, \\[4pt]
    &f_1(j^r_{\mathfrak{su}(2)}) = \eta^\dagger\, T^r_{\boldsymbol{3}}\,\eta,
    &&\qquad f_2(j^r_{\mathfrak{su}(2)}) = \chi^\dagger\,(T^r_{\boldsymbol{2}}\otimes1)\,\chi, \\[4pt]
    &f_1(j^{AB}_{\mathfrak{so}(5)}) = \psi^\dagger\, M^{AB}_{\mathbf 5}\,\psi,
    &&\qquad f_2(j^{AB}_{\mathfrak{so}(5)}) = \chi^\dagger\,(1\otimes M^{AB}_{\mathbf 4})\,\chi,
\end{alignedat}
\end{equation}
where $M^{AB}_{\mathbf 5}$ and $M^{AB}_{\mathbf 4}$ are the generators of $\mathfrak{so}(5)$ in the vector and spinor representations. The normal orderings are implied.

The central charge of $\widehat{\mathfrak{g}}$ is
\begin{eqnarray}
    c_{\widehat{\mathfrak{g}}} = \underbrace{1}_{\widehat{\mathfrak{u}}(1)_{32}}+ \underbrace{\frac{4\cdot 3}{4+2}}_{\widehat{\mathfrak{su}}(2)_4 } + \underbrace{\frac{2(2\cdot 2^2+2)}{2+2\cdot 2-1}}_{\widehat{\mathfrak{so}}(5)_2}= 7
\end{eqnarray}
which is $c=1$ short of $c_{\mathcal{T}}=8$ of $\Fer(8)$, and the coset chiral algebras $S_i$ have central charge $1$. They can be worked out by a chain of conformal embeddings, similar to examples in Section \ref{sec:highcharge} and \ref{sec:susu}.

For the in-coming fermions, we have the following chain of embeddings
\begin{eqnarray}\label{eq:cqedchain1}
    \begin{split}
        &\Fer(8)\\
        \hookleftconf \quad & \Fer(5)^\psi \oplus \Fer(3)^\eta\\
        \hookleftconf \quad & \big(\widehat{\mathfrak{su}}(5)_1 \oplus \widehat{\mathfrak{u}}(1)_5\big)\oplus\big(\widehat{\mathfrak{su}}(3)_1\oplus \widehat{\mathfrak{u}}(1)_3\big)\\
        \hookleftconf \quad & \widehat{\mathfrak{so}}(5)_2 \oplus \widehat{\mathfrak{su}}(2)_4 \oplus \widehat{\mathfrak{u}}(1)_{32}\oplus\widehat{\mathfrak{u}}(1)_{480}
    \end{split}
\end{eqnarray}
The first two embeddings are standard, and have been used repeatedly in previous examples. The third embedding involves the conformal embedding $\widehat{\mathfrak{so}}(5)_2\hookrightconf\widehat{\mathfrak{su}}(5)_1$ induced by $\mathbf 5\to\mathbf 5$, and the principal embedding $\widehat{\mathfrak{su}}(2)_4\hookrightarrow\widehat{\mathfrak{su}}(3)_1$ induced by $\mathbf 3\to\boldsymbol 3$. The factor $\widehat{\mathfrak{u}}(1)_{480}$ is generated by
\begin{eqnarray}\label{eq:JCcqed}
    J_{S_1}=-9\,\psi^\dagger\psi+5\,\eta^\dagger\eta,
\end{eqnarray}
which is required to have regular OPE with $J$ in \eqref{eq:f12cqed} generating $\widehat{\mathfrak{u}}(1)_{32}$. Therefore, the $c=1$ coset chiral algebra $S_1$ is 
\begin{eqnarray}\label{eq:S1cqed}
    S_1=\widehat{\mathfrak{u}}(1)_{480}.
\end{eqnarray}
The charges of the in-coming fermions under this emergent $U(1)_{S_1}$ are $-9$ and $5$.

For the out-going fermions, we have the following chain of embeddings, 
\begin{equation}\label{eq:cqedchain2}
\begin{split}
    &\Fer(8)\\
    \hookleftconf \quad & \widehat{\mathfrak{u}}(1)_8\oplus \widehat{\mathfrak{su}}(8)_1 \\
    \hookleftconf \quad & \widehat{\mathfrak{u}}(1)_8\oplus \widehat{\mathfrak{su}}(2)_4\oplus\widehat{\mathfrak{su}}(4)_2\\
    \hookleftconf \quad & \widehat{\mathfrak{u}}(1)_{32}\oplus\widehat{\mathfrak{su}}(2)_4\oplus\widehat{\mathfrak{so}}(5)_2\oplus\widehat{\mathfrak{u}}(1)^{\bZ_2}_{30}.
\end{split}
\end{equation}
The first embedding is standard, which peels off abelian currents. The second step is the level-rank embedding. The last step uses the conformal embeddings $\widehat{\mathfrak{u}}(1)_{8} \hookleftconf \widehat{\mathfrak{u}}(1)_{32}$ and  $\widehat{\mathfrak{su}}(4)_2 \hookleftconf \widehat{\mathfrak{so}}(5)_2\oplus\widehat{\mathfrak{u}}(1)^{\bZ_2}_{30}$, where $\widehat{\mathfrak{u}}(1)^{\bZ_2}_{30}$ is the charge conjugation orbifold of $\widehat{\mathfrak{u}}(1)_{30}$.\footnote{To see the latter embedding, we use $\widehat{\mathfrak{su}}(4)_2=\widehat{\mathfrak{so}}(6)_2$ and realize it by six complex fermions $\psi_a$: the currents $i(\psi_a^\dagger\psi_b-\psi_b^\dagger\psi_a)$ generate $\widehat{\mathfrak{so}}(6)_2\hookrightconf\widehat{\mathfrak{su}}(6)_1$, and $\widehat{\mathfrak{so}}(6)_2$ is the invariant subalgebra of $\widehat{\mathfrak{su}}(6)_1$ under the charge conjugation $\psi_a\leftrightarrow\psi_a^\dagger$. Likewise $\widehat{\mathfrak{so}}(5)_2\hookrightconf\widehat{\mathfrak{su}}(5)_1$ acting on $\psi_{1},\dots,\psi_5$. Since a commutant only depends on the stress tensor of the subalgebra, the commutant of $\widehat{\mathfrak{so}}(5)_2$ in $\widehat{\mathfrak{su}}(6)_1$ is that of $\widehat{\mathfrak{su}}(5)_1$, i.e.\ $\widehat{\mathfrak{u}}(1)_{30}$ generated by $J'=\sum_{a=1}^5\psi_a^\dagger\psi_a-5\psi_6^\dagger\psi_6$ and the weight-15 operators $V_+=\psi_1^\dagger\cdots\psi_5^\dagger\,\psi_6\,\partial\psi_6\cdots\partial^4\psi_6$ and its conjugate $V_-$. The charge conjugation maps $J'\to-J'$ and $V_\pm\to V_\mp$, so the coset $\widehat{\mathfrak{so}}(6)_2/\widehat{\mathfrak{so}}(5)_2$ is the invariant part $\widehat{\mathfrak{u}}(1)_{30}^{\bZ_2}$, which contains $J'J'$ and $V_++V_-$ but not $J'$.} Therefore, the coset chiral algebra $S_2$ is 
\begin{eqnarray}
\label{eq:S2cqed}
    S_2= \widehat{\mathfrak{u}}(1)^{\bZ_2}_{30}
\end{eqnarray}
which is distinct from $S_1$.

\subsection{Intertwining Defect}

Similar to the example in Section \ref{sec:susu}, since the coset chiral algebras $S_1$ and $S_2$ are distinct, the system spans to a club sandwich. We first find the common conformal subalgebra, then build the club sandwich, and finally construct the intertwining defect. 

\subsubsection{Common Conformal Subalgebra: a $c=1$ Orbifold}
\label{sec:cqedW}

The coset TQFTs for $S_1$ and $S_2$ are\footnote{Beyond the standard integer-valued levels, the Chern Simons theories with orthogonal gauge group also has two additional discrete theta angles \cite{Cordova:2017vab}, which vanish in our cases. Hence we suppress these additional labels. }
\begin{eqnarray}
    \Rep(S_1) = U(1)_{480}, \qquad \Rep(S_2) = O(2)_{30}
\end{eqnarray}
Suppose $W$ is the common conformal subalgebra of $S_1$ and $S_2$, the coset TQFT for $W$ is a larger TQFT, which becomes $U(1)_{480}$ and $O(2)_{30}$ upon condensing appropriate condensable algebras $\mathcal{A}_1$ and $\mathcal{A}_2$ respectively. We claim that
\begin{eqnarray}\label{eq:Wcqed}
    W=\widehat{\mathfrak{u}}(1)^{\bZ_2}_{480}, \qquad \Rep(W) = O(2)_{480}.
\end{eqnarray}

To explain \eqref{eq:Wcqed}, we first recall the data for $U(1)_{2n}$ and $O(2)_{2n}$, which are collected in Appendix \ref{app:branching data}. 

$W\hookrightconf S_1$ restricts $S_1$ to its subalgebra $W$ invariant under $J_{S_1}\to-J_{S_1}$. Hence $U(1)_{480}$ is obtained from $O(2)_{480}$ by condensing $\mathcal{A}_1=\mathbf 1\oplus j$, which is implemented by the interface $\mathcal{I}_{S_1|W}$ in the club sandwich in Figure \ref{fig:cqedclub}. The branching rules from $U(1)_{480}$ to $O(2)_{480}$ is
\begin{eqnarray}\label{eq:resW1}
        [0]\mapsto\mathbf 1\oplus j,\qquad [240]\mapsto\phi^+_{240}\oplus\phi^-_{240}, \qquad [m]\mapsto
        \begin{cases}
            \phi_m,& 1\leq m \leq 239\\
            \phi_{480-m}, & 241\leq m \leq 479
        \end{cases}
\end{eqnarray}

To explain $W\hookrightconf S_2$, we first undo the $\bZ_2$ orbifold, which gives the conformal embedding $\widehat{\mathfrak{u}}(1)_{480} \hookrightconf \widehat{\mathfrak{u}}(1)_{30}$, and this embedding can be understood as a simple current extension.\footnote{$\widehat{\mathfrak{u}}(1)_{480}$ is generated by $i\partial \varphi$ and $e^{\pm i \sqrt{480} \varphi}= e^{\pm i 4\sqrt{30} \varphi}$, while $\widehat{\mathfrak{u}}(1)_{30}$ is generated by $i\partial \varphi$ and  $e^{\pm i \sqrt{30} \varphi}$. Extending the former by the simple current $e^{\pm i \sqrt{30} \varphi}$ gives the latter.}  In terms of TQFTs, the $O(2)_{30}$ is obtained from $O(2)_{480}$ by condensing $\mathcal{A}_2=\mathbf 1\oplus\phi_{120}\oplus\phi^+_{240}$, which is implemented by the interface $\mathcal{I}_{S_2|W}$ in Figure \ref{fig:cqedclub}. The branching rules from $O(2)_{30}$ to $O(2)_{480}$ is 
    \begin{eqnarray}\label{eq:resW2}
    \begin{split}
        &\mathbf 1\mapsto\mathbf 1\oplus\phi_{120}\oplus\phi^+_{240},\qquad j\mapsto j\oplus\phi_{120}\oplus\phi^-_{240},\qquad \phi_m\mapsto\bigoplus_{a\in\bZ_4}\phi_{4m+120a},\\
        &\phi^\pm_{15}\mapsto\phi_{60}\oplus\phi_{180},\qquad \sigma_i\mapsto\sigma_1,\qquad\tau_i\mapsto\tau_1 .
    \end{split}
\end{eqnarray}
Here the labels of the twisted lines are fixed as in Appendix \ref{app:branching data}. In $O(2)_{480}$, $\sigma_1$ and $\tau_1$ braid trivially with $\phi^+_{240}$, while $\sigma_2$ and $\tau_2$ braid with $\phi^+_{240}$ by $-1$. Hence $\sigma_2$ and $\tau_2$ are confined by $\mathcal{A}_2$, and they do not appear in the image of \eqref{eq:resW2}, nor in $\widetilde L_2$ in \eqref{eq:L2cqed}. The map of the twisted lines is consistent with the fusion with $\phi_m$: $\sigma_2\oplus\tau_2=\phi_1\otimes\sigma_1$ in $O(2)_{30}$, while $\phi_1\mapsto\bigoplus_{a}\phi_{4+120a}$ involves only even charges of $\widehat{\mathfrak{u}}(1)_{480}$, which do not exchange the subscripts. Therefore $\sigma_1$ and $\sigma_2$ of $O(2)_{30}$ restrict to the same twisted line of $O(2)_{480}$, and likewise for $\tau_{1,2}$.

The physical meaning of $W$ is the following. The in-coming fermions have an emergent symmetry $U(1)_{S_1}$, generated by \eqref{eq:JCcqed}, which commutes with $G$ but is not a symmetry of the out-going fermions. What survives on both sides is only the part of $S_1$ that is blind to the sign of $J_C$, such as the stress tensor $\frac{1}{960}J_C^2$ and the vertex operators $\cos(\sqrt{480}\,\phi_C)$ of weight $240$. On the out-going side, the same operators are built from the $\widehat{\mathfrak{u}}(1)^{\bZ_2}_{30}$ coset. As we will see, this is what makes the out-going lines non-invertible.

\subsubsection{Club Sandwich}

\begin{figure}
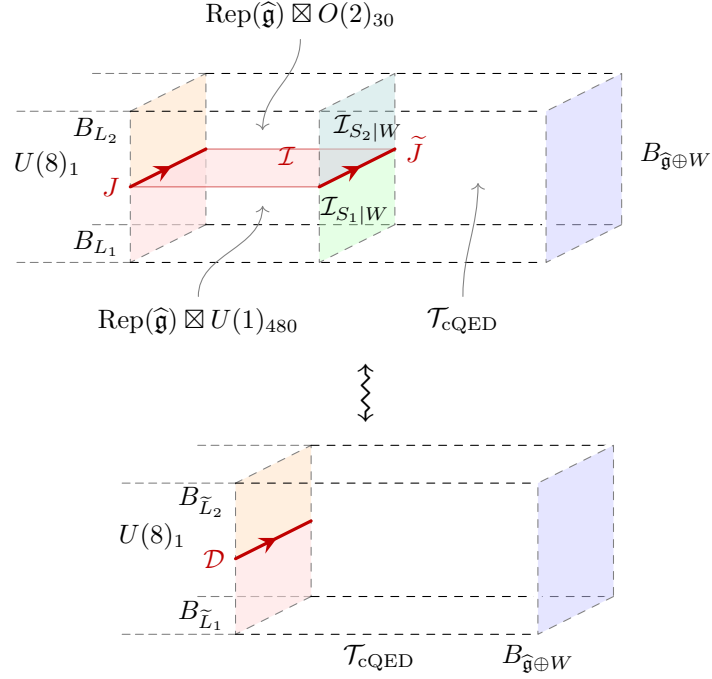

    \centering
    \figClubCQED[1.0]\\[1ex]\rotatebox[origin=c]{-90}{\Large$\leftrightsquigarrow$}\\[1ex]
    \figClubCQEDShrunk[1.0]
    \caption{Club sandwich for the $(1^5,3,(-2)^4)$ model (top), with $\Rep(\widehat{\mathfrak{g}})=SU(2)_4\boxtimes U(1)_{32}\boxtimes Spin(5)_2$, $S_1=\widehat{\mathfrak{u}}(1)_{480}$ and $S_2=\widehat{\mathfrak{u}}(1)^{\bZ_2}_{30}$. The interfaces $\mathcal{I}_{S_i|W}$ condense $\mathcal{A}_1=\mathbf 1\oplus j$ and $\mathcal{A}_2=\mathbf 1\oplus\phi_{120}\oplus\phi^+_{240}$. Shrinking them gives the sandwich at the bottom, with two topological boundaries $B_{\widetilde L_i}$ of the bulk TQFT $\mathcal{T}_{\rm cQED}$ in \eqref{eq:Tcqed}, joined along the intertwining defect $\mathcal{D}$.}
    \label{fig:cqedclub}
\end{figure}

The club sandwich is shown on the top of Figure \ref{fig:cqedclub}. Its right bulk is
\begin{eqnarray}\label{eq:Tcqed}
    \mathcal{T}_{\rm cQED}=SU(2)_4\boxtimes U(1)_{32}\boxtimes Spin(5)_2\boxtimes O(2)_{480}.
\end{eqnarray}
The data for the TQFTs in \eqref{eq:Tcqed} are collected in Appendix \ref{app:branching data}. 
Shrinking the club sandwich gives the bottom of Figure \ref{fig:cqedclub}. It remains to explain the topological boundaries $B_{\widetilde{L}_1}$ and $B_{\widetilde{L}_2}$, and the associated fermionic Lagrangian algebras follow from \eqref{eq:cqedchain1} and \eqref{eq:cqedchain2}.

For the in-coming fermions:
\begin{enumerate}
    \item The vacuum supermodules of $\widehat{\mathfrak{u}}(5)_1$ and $\widehat{\mathfrak{u}}(3)_1$ branch as $\bigoplus_{r_5\in\bZ_5}[r_5]\boxtimes\Lambda_{r_5}$ and $\bigoplus_{r_3\in\bZ_3}[r_3]\boxtimes\Lambda_{r_3}$.
    \item Under $\widehat{\mathfrak{so}}(5)_2\hookrightarrow\widehat{\mathfrak{su}}(5)_1$: $\Lambda_0\mapsto\mathbf 1\oplus\mathbf{14}$, $\Lambda_1,\Lambda_4\mapsto\mathbf 5$, $\Lambda_2,\Lambda_3\mapsto\mathbf{10}$. Under $\widehat{\mathfrak{su}}(2)_4\hookrightarrow\widehat{\mathfrak{su}}(3)_1$: $\Lambda_0\mapsto0\oplus2$, $\Lambda_1,\Lambda_2\mapsto1$.
    \item The line $[r_5]\boxtimes[r_3]$ lifts to the $32$ distinct lines $[n_5+3n_3]\boxtimes[-9n_5+5n_3]$ of $U(1)_{32}\boxtimes U(1)_{480}$, with $(n_5,n_3)\in(r_5,r_3)+5\bZ\oplus3\bZ$, which are then lifted to $\Rep(W)$ by \eqref{eq:resW1}.
\end{enumerate}
Hence
\begin{eqnarray}\label{eq:L1cqed}
    \widetilde L_1=\bigoplus_{r_5\in\bZ_5}\ \bigoplus_{r_3\in\bZ_3}\ \bigoplus_{s\in \mathcal{B}_{r_5}}\ \bigoplus_{j\in \mathcal{E}_{r_3}}\ \bigoplus_{(n_5,n_3)}\ j\boxtimes[n_5+3n_3]\boxtimes s\boxtimes\text{Res}_1(-9n_5+5n_3),
\end{eqnarray}
where $\mathcal{B}_{r_5}$ and $\mathcal{E}_{r_3}$ are the sets in step 2, the sum over $(n_5,n_3)$ runs over the $32$ classes of step 3, and $\text{Res}_1$ is \eqref{eq:resW1}.

For the out-going fermions:
\begin{enumerate}
    \item The vacuum supermodule of $\widehat{\mathfrak{u}}(8)_1$ branches as $\bigoplus_{r\in\bZ_8}[r]\boxtimes\Lambda_r$, and $[r]$ of $U(1)_8$ lifts to $[-2r]\oplus[-2r+16]$ of $U(1)_{32}$.
    \item Under the level-rank embedding, $\Lambda_r\mapsto\bigoplus_{(j,\nu)\in\mathcal{P}_r}j\boxtimes\nu$, with $\mathcal{P}_r$ provided below \eqref{eq:levelrankembedding} with $(N_f,n)=(2,4)$ (see Appendix \ref{app:branching data}).
    \item The $\widehat{\mathfrak{su}}(4)_2$ module $\nu$ branches into $\bigoplus_{(s,x)}s\boxtimes x$ with $s$ a line of $Spin(5)_2$ and $x$ a line of $\Rep(S_2)$, as listed in Appendix \ref{app:branching data}, and $x$ is lifted to $\Rep(W)$ by \eqref{eq:resW2}.
\end{enumerate}
Hence
\begin{eqnarray}\label{eq:L2cqed}
    \widetilde L_2=\bigoplus_{r\in\bZ_8}\ \bigoplus_{(j,\nu)\in\mathcal{P}_r}\ \bigoplus_{(s,x)\in\nu}\ \bigoplus_{Q\in\{-2r,-2r+16\}}\ j\boxtimes[Q]\boxtimes s\boxtimes\text{Res}_2(x).
\end{eqnarray}

The fermions correspond to the summands
\begin{equation}\label{eq:lambdacqed}
    \lambda_\psi=0\boxtimes[1]\boxtimes\mathbf 5\boxtimes\phi_9,\qquad \lambda_\eta=1\boxtimes[3]\boxtimes\mathbf 1\boxtimes\phi_5,\qquad \lambda_\chi=\tfrac12\boxtimes[-2]\boxtimes\mathbf 4\boxtimes\sigma_1,
\end{equation}
where the $W$-labels follow from the $U(1)_{S_1}$ charges $-9$ and $5$ of $\psi$ and $\eta$, while $\chi$ sits in a twisted sector of the orbifold. As a check, the conformal weights add up to $\frac12$.\footnote{$h_{\lambda_\psi}=\tfrac1{64}+\tfrac25+\tfrac{81}{960}, h_{\lambda_\eta}=\tfrac13+\tfrac9{64}+\tfrac{25}{960}, h_{\lambda_\chi}=\tfrac18+\tfrac1{16}+\tfrac14+\tfrac1{16}.$}
Their quantum dimensions are $d_{\lambda_\psi}=d_{\lambda_\eta}=4$ and $d_{\lambda_\chi}=60$.

\subsubsection{Intertwining Defect}
The number of simple topological line junctions between $B_{\widetilde L_1}$ and $B_{\widetilde L_2}$ is 
\begin{eqnarray}\label{eq:Homcqed1}
    \dim\Hom(\widetilde L_1,\widetilde L_2)=192 .
\end{eqnarray}
Each of them commutes with $\widehat{\mathfrak{g}}\oplus W$, and folding back gives a conformal boundary condition of the $8$ Dirac fermions preserving $SU(2)\times U(1)\times Spin(5)$. 
All of them are non-invertible since $\lambda_\psi\notin\widetilde L_2$. Upon folding along $\mathcal{D}$, the boundary state $\ket{\mathcal{D}}$ preserves the above symmetry and is rational with respect to $\widehat{\mathfrak{g}} \oplus W$.

\subsection{Scattering}
\label{sec:cqedscattering}

The categorical scattering formula \eqref{eq:Daction} applies with $\mathcal{T}_{\rm cQED}$ and $B_{\widetilde L_i}$. We will mainly consider the case where we scatter in $\psi$. Scattering in $\eta$ works in a similar way.

The incoming fermion $\psi$ is resolved as $(x,\lambda_\psi, \widetilde{x})$, and the out-going twisted sectors are labeled by a topological line $F_{B_{\widetilde L_2}}(\lambda_\psi)$. To see whether $F_{B_{\widetilde L_2}}(\lambda_\psi)$ is simple, we compute 
\begin{eqnarray}
\label{eq:Homcqed2}
    \Hom_{B_{\widetilde{L}_2}}\left(F_{B_{\widetilde L_2}}(\lambda_\psi), F_{B_{\widetilde L_2}}(\lambda_\psi)\right) \simeq \Hom_{\mathcal{T}_{\text{cQED}}}\left(\lambda_\psi \otimes \overline{\lambda}_{\psi}, \widetilde{L}_2\right)
\end{eqnarray}
From \eqref{eq:lambdacqed}, $\lambda_\psi\otimes \overline{\lambda}_{\psi}= 0 \boxtimes [0] \boxtimes (\mathbf{1} \oplus \mathbf{10} \oplus \mathbf{14}) \boxtimes (1\oplus j \oplus \phi_{18})$, which contains 9 simple objects. Only the trivial line $0\boxtimes [0] \boxtimes \mathbf{1} \boxtimes \mathbf{1}$ and an invertible $\bZ_2$ line
\begin{eqnarray}\label{eq:Epsi}
    E_\psi=0\boxtimes[0]\boxtimes\mathbf{14}\boxtimes j
\end{eqnarray}
belongs to $\widetilde{L}_2$. So the dimension of the Hom space \eqref{eq:Homcqed2} is 2, and $F_{B_{\widetilde L_2}}(\lambda_\psi)$ splits into two simple lines
\begin{eqnarray}
    F_{B_{\widetilde L_2}}(\lambda_\psi) = b_\psi\oplus b'_\psi.
\end{eqnarray}
The simple lines $b_\psi$ and $b'_\psi$ are realized as 
\begin{eqnarray}
    b_\psi=F_{B_{\widetilde L_2}}\big(0\boxtimes[1]\boxtimes\mathbf 1\boxtimes\phi_{15}\big), \qquad b'_\psi=F_{B_{\widetilde L_2}}\big(0\boxtimes[1]\boxtimes\mathbf 1\boxtimes\phi_{33}\big)
\end{eqnarray}
and are of quantum dimensions $d_{b_\psi} = d_{b'_{\psi}}=2$.\footnote{To determine these, we demand $b_\psi$ and $F_{B_{\widetilde L_2}}(\lambda_\psi)$ has a one-dimensional junction, and the endomorphism of $b_\psi$ is also one-dimensional.  }

A priori, \eqref{eq:Daction} gives a superposition of the $b_\psi$- and $b'_{\psi}$-twisted sectors. A closer look shows that one of the two amplitudes vanishes, and the out-going excitation is within only one twisted sector that depends on the choice of $\mathcal{D}$. We denote the attached line by $b_\psi^{(\mathcal{D})}$. The scattering process is
\begin{eqnarray}\label{eq:cqedscat}
    \figBeforeScatterCQED[0.95]{\psi}\quad = \quad \sum_{zw}C^{\psi}_{zw}\quad\figOutCQED[0.95]{b^{(\mathcal{D})}_\psi}{w}{\mathcal{O}^{(z)}_\psi},
\end{eqnarray}
We relegate the details to Appendix \ref{app:onechannel}.

We make a few comments.
\begin{enumerate}
    \item For every simple boundary condition, $\psi$ and $\eta$ scatter with probability one into a single twisted sector, labeled by a non-invertible line of quantum dimension $2$. No superposition arises--this is because the endomorphisms of $F_{B_{\widetilde L_2}}(\lambda_\psi)$ are generated by invertible common line $E_{\psi}$, which act on $\mathcal{D}$ by signs.
    \item The out-going lines carry only $U(1)$ charge and a doublet $\phi_m$ of the orbifold, and are blind to $SU(2)\times Spin(5)$. The non-abelian quantum numbers of $\psi$ and $\eta$ are carried by the excitations $\mathcal{O}_\psi$ and $\mathcal{O}_\eta$ at the end of the lines. For instance, $\lambda_\psi\subset(0\boxtimes[1]\boxtimes\mathbf 1\boxtimes\phi_{15})\otimes(0\boxtimes[0]\boxtimes\mathbf 5\boxtimes\phi_{24})$, where the second factor is the line of a neutral out-going operator in the $\mathbf 5$, such as $\chi^\dagger\Gamma^a\chi$.
\end{enumerate}

\section{Example: Higher Charge Monopole in Standard Model}
\label{sec:standardmodel}

As an application to realistic world, we study the scattering of one generation of fermions in the Standard Model, off charge-3 monopole $(n_1,n_2,m,g) =(0, 0, 1,0)$\footnote{We denote $n_1$ and $n_2$ for the $SU(3)$ magnetic fluxes, as $m$ for
the $SU(2)$ flux, and as $g$ for the $U(1)_Y$ flux, following \cite{vanBeest:2023mbs}}. This problem, and the reduced 2d version, have been studied in \cite{vanBeest:2023mbs,Khoze:2024hlb}, and was found to resist the abelian bosonization method.

We show that this problem falls into the simple case of Section \ref{sec:commonalg}, hence the intertwining defect is worked out. The quarks scatter into either single twisted sector, or superpositions of two twisted sectors, depending on different choice of $\mathcal{D}$, with probabilities being universal numbers fixed by Fibonacci category.

\subsection{Two Embeddings}
In the reduced 2d problem, the symmetry to be preserved by the boundary condition of 9 Dirac fermions is $G=SU(3)\times U(1)_W \times U(1)_Y \times SU(2)_{\text{rot}}$, where the incoming and outgoing chiral fermions are in representations
\begin{align}\label{eq:m3 sm sym rep}
	\begin{array}{|c|cccc|}
        \hline
		&  SU(3) & U(1)_W & U(1)_Y & SU(2)_\text{rot}\\
		\hline
		 q \,& \mathbf{3} & 1 & 1 & \mathbf{1} \\
		 e & \mathbf{1} & 0 & 6 &\mathbf{3}\\
		 d & \mathbf{\bar{3}} & 0 &2 &\mathbf{1} \\
		\hline
		 l^+ &\mathbf{1} &1 &-3 &\mathbf{1}\\
		l^- &\mathbf{1} &-1 &-3 &\mathbf{2}\\
		u &\mathbf{\bar{3}} &0&-4&\mathbf{2}\\
        \hline
	\end{array}
\end{align}
Here $q,e,\cdots$ are just shorthand notations. See \cite{vanBeest:2023mbs,Khoze:2024hlb} for physical discussions in standard model.

The chiral algebra generated by above symmetries is 
\begin{align}
    \widehat{\mathfrak{g}}=\widehat{\mathfrak{su}}(3)_2 \oplus \widehat{\mathfrak{su}}(2)_4 \oplus \widehat{\mathfrak{u}}(1)_K^2,\qquad K=
\begin{pmatrix}
	3&3\\
	3&123
\end{pmatrix}
\end{align}
The two embeddings are 
\begin{equation}\label{eq:f12sm}
\begin{alignedat}{2}
    &f_1(j^\alpha_{\mathfrak{su}(3)}) = q^\dagger T^\alpha_{\square}\, q+d^\dagger T^\alpha_{\overline\square}\, d,
    &&\qquad f_2(j^\alpha_{\mathfrak{su}(3)}) = u^\dagger\, (T^\alpha_{\overline\square}\otimes 1)\,u, \\[4pt]
    &f_1(J_W) = q^\dagger q,
    &&\qquad f_2(J_W) = l^{+\dagger}l^+ - l^{-\dagger}l^-, \\[4pt]
    &f_1(J_Y) = q^\dagger q+6\,e^\dagger e+2\,d^\dagger d,
    &&\qquad f_2(J_Y) = -3\,l^{+\dagger}l^+ -3\,l^{-\dagger}l^- - 4\,u^\dagger u,\\
    &f_1(j^r_{\mathfrak{su}(2)}) = e^\dagger\, T^r_{\boldsymbol{3}}\, e,
    &&\qquad f_2(j^r_{\mathfrak{su}(2)}) = l^{-\dagger}\, T^r_{\boldsymbol{2}}\, l^- + u^\dagger\, (1\otimes T^r_{\boldsymbol{2}})\,u\\[4pt]
\end{alignedat}
\end{equation}
where $T^\alpha_{\overline\square}=-T^{\alpha\,\mathsf{T}}_\square$, $T^r_{\boldsymbol{n}}$ are the generators of $SU(2)$ in the $n$-dimensional irreducible representation, and normal ordering is implied. 

Its central charge is 
\begin{eqnarray}
    c_{\widehat{\mathfrak{g}}} = \underbrace{\frac{2(2^2+2\cdot 2)}{2+2+1} }_{\widehat{\mathfrak{su}}(3)_2} + \underbrace{\frac{4(1^2+2)}{4+1+1}}_{\widehat{\mathfrak{su}}(2)_4} + \underbrace{2}_{\widehat{\mathfrak{u}}(1)_K^2} = \frac{36}{5}
\end{eqnarray}
It is $c=\frac95$ short of the total central charge 9, so $\widehat{\mathfrak{g}}$ is not conformally embedded in  $\widehat{\mathfrak{u}}(9)_1$, and the coset chiral algebras $S_i$ have central charge $\frac95$.

Now we explain how to obtain the coset, together with the way the currents are embedded. The idea is aligned with Section \ref{sec:susuembed}, by decomposing $\widehat{\mathfrak{u}}(9)_1$ and $\widehat{\mathfrak{su}}(3)_2 \oplus \widehat{\mathfrak{su}}(2)_4 \oplus \widehat{\mathfrak{u}}(1)_K^2$ by the way acting on fermion multiplets.

For the in-coming fermions, 
\begin{eqnarray}
    \begin{split}
        &\Fer(9)\\
        \hookleftconf \quad & \Fer(3)^q \oplus \Fer(3)^e \oplus \Fer(3)^d\\
        \hookleftconf \quad & \widehat{\mathfrak{u}}(1)_3^q \oplus \widehat{\mathfrak{su}}(3)_1^q \oplus \widehat{\mathfrak{u}}(1)_3^e \oplus \widehat{\mathfrak{su}}(3)_1^e \oplus \widehat{\mathfrak{u}}(1)_3^d \oplus \widehat{\mathfrak{su}}(3)_1^d\\
        \hookleftconf \quad & \widehat{\mathfrak{su}}(3)_2 \oplus \mathsf{3SP} \oplus \widehat{\mathfrak{su}}(2)_4 \oplus \widehat{\mathfrak{u}}(1)_K^2 \oplus \widehat{\mathfrak{u}}(1)_{30}
    \end{split}
\end{eqnarray}
The first conformal embedding splits the total 9 fermions into multiplets $q,e,d$. The second one peels off the abelian currents. The last one is a composition of three conformal embeddings. Firstly, $\widehat{\mathfrak{su}}(3)_1^q \oplus \widehat{\mathfrak{su}}(3)_1^d \hookleftconf \widehat{\mathfrak{su}}(3)_2 \oplus \mathsf{3SP}$, where $SU(3)$ acts diagonally on $q$ and $d$, and $\widehat{\mathfrak{su}}(3)_2$ is generated by $f_1(j^\alpha_{\mathfrak{su}(3)})$ in \eqref{eq:f12sm}, and $\mathsf{3SP}$ is the $W_3$ algebra at $c=\frac45$, i.e. the chiral algebra of the three-state Potts model \cite{Bais:1987zk}. Secondly, the principal embedding $\widehat{\mathfrak{su}}(3)_1\hookleftconf \widehat{\mathfrak{su}}(2)_4$. Lastly, the embedding of abelian algebras $\widehat{\mathfrak{u}}(1)_3^q \oplus \widehat{\mathfrak{u}}(1)_3^e \oplus \widehat{\mathfrak{u}}(1)_3^d \hookleftconf \widehat{\mathfrak{u}}(1)_K^2 \oplus \widehat{\mathfrak{u}}(1)_{30}$, where the current of $\widehat{\mathfrak{u}}(1)_{30}$ is\footnote{It is determined by demanding $f_1(J_c)$ has regular OPE with $f_1(J_W)$ and $f_1(J_Y)$. Alternatively, we can use the method in Appendix \ref{app:Anyon condensation between K matrix U(1) Chern-Simons TQFTs}, by solving $P_1^{\mathsf{T}} \begin{pmatrix}
    3 &&\\
    & 3&\\
    &&3
\end{pmatrix} P_1= \begin{pmatrix}
    K & 0\\
    0 & k
\end{pmatrix}$, which solves $k=30$, and $P_1= 
\begin{pmatrix}
    1& 1 & 0\\
    0 & 6 & 1\\
    0 & 2 & -3
\end{pmatrix}$. The three currents $f_1(J_W), f_1(J_Y), f_1(J_C)$ can be read off from the columns of $P_1$. } 
\begin{eqnarray}
    f_1(J_C)= e^\dagger e - 3 d^\dagger d
\end{eqnarray}
Therefore, the coset chiral algebra of the in-coming fermions is 
\begin{eqnarray}
    S_1= \widehat{\mathfrak{u}}(1)_{30} \oplus \mathsf{3SP}
\end{eqnarray}

For the out-going fermions, 
\begin{eqnarray}
    \begin{split}
        &\Fer(9)\\
        \hookleftconf \quad & \Fer(1)^{l^+} \oplus \Fer(2)^{l^-} \oplus \Fer(6)^u\\
        \hookleftconf \quad & \widehat{\mathfrak{u}}(1)_1^{l^+} \oplus \widehat{\mathfrak{u}}(1)_2^{l^-} \oplus \widehat{\mathfrak{su}}(2)_1^{l^-} \oplus \widehat{\mathfrak{u}}(1)_6^u \oplus \widehat{\mathfrak{su}}(6)_1^u\\
        \hookleftconf \quad & \widehat{\mathfrak{su}}(3)_2 \oplus
        \widehat{\mathfrak{su}}(2)_4 \oplus
        \text{Vir}_{4/5} \oplus \widehat{\mathfrak{u}}(1)_K^2 \oplus \widehat{\mathfrak{u}}(1)_{30}
    \end{split}
\end{eqnarray}
The first conformal embedding splits the total 9 fermions into multiplets of  $l^+, l^-,u$. The second one peels off the abelian currents. The last one is again a composition of three conformal embeddings. Firstly, $\widehat{\mathfrak{su}}(6)_1^u \hookleftconf \widehat{\mathfrak{su}}(3)_2 \oplus \widehat{\mathfrak{su}}(2)_3$ because $u$ transforms under $SU(3)\times SU(2)_{\text{rot}}$. Secondly, $\widehat{\mathfrak{su}}(2)_3 \oplus \widehat{\mathfrak{su}}(2)_1^{l^-} \hookleftconf \widehat{\mathfrak{su}}(2)_4 \oplus \text{Vir}_{4/5}$ because $SU(2)$ acts diagonally on $l^-$ and $u$. Note that this embedding is gives the coset description for the 3rd unitary minimal model, a.k.a. tetracritical Ising. Lastly, the embedding of abelian algebras $\widehat{\mathfrak{u}}(1)_1^{l^+} \oplus \widehat{\mathfrak{u}}(1)_2^{l^-}\oplus \widehat{\mathfrak{u}}(1)_6^u\hookleftconf \widehat{\mathfrak{u}}(1)_K^2 \oplus \widehat{\mathfrak{u}}(1)_{30}$, where the current of $\widehat{\mathfrak{u}}(1)_{30}$ is\footnote{The current is again determined by demanding $f_2(J_C)$ has regular OPE with $f_2(J_W)$ and $f_2(J_Y)$. Alternatively, we can again use the method in Appendix \ref{app:Anyon condensation between K matrix U(1) Chern-Simons TQFTs}, by solving $P_2^{\mathsf{T}} \begin{pmatrix}
    1 &&\\
    & 2&\\
    &&6
\end{pmatrix} P_2= \begin{pmatrix}
    K & 0\\
    0 & k
\end{pmatrix}$, which solves $k=30$, and $P_2= 
\begin{pmatrix}
    1& -3 & -4\\
    -1 & -3 & -2\\
    0 & -4 & 1
\end{pmatrix}$. The three currents $f_2(J_W), f_2(J_Y), f_2(J_C)$ can be read off from the columns of $P_2$. }
\begin{eqnarray}
    f_2(J_C)= - 4 l^{+\dagger}l^+ - 2 l^{-\dagger}l^- + u^\dagger u
\end{eqnarray}
Therefore, the coset chiral algebra of the out-going fermions is 
\begin{eqnarray}
    S_2= \widehat{\mathfrak{u}}(1)_{30}\oplus \text{Vir}_{4/5}
\end{eqnarray}

\subsection{Intertwining Defect}
\label{sec:smD}

Since the coset chiral algebras $S_1$ and $S_2$ are distinct, the system spans to a club sandwich. As in previous examples, we first find the common conformal subalgebra, then build the club sandwich, and finally construct the intertwining defect. 

\subsubsection{Common Conformal Subalgebra and $U(1)_{B-L}$}

The coset chiral algebras $S_1$ and $S_2$ share a common abelian factor $\widehat{\mathfrak{u}}(1)_{30}$. This implies that the standard model has an additional abelian symmetry $U(1)_C$, whose charges are
\begin{eqnarray}\label{eq:Csm}
    \begin{tabular}{|c|c|c|c|c|c|c|}
    \hline
         & $q$ & $e$ & $d$ & $l^+$ & $l^-$ & $u$ \\
         \hline
        $U(1)_C$ & $0$ & $1$ & $-3$ & $-4$ & $-2$ & $1$\\
        \hline
    \end{tabular}
\end{eqnarray}
Together with $U(1)_W$ and $U(1)_Y$, they generate the $U(1)_{B-L}$ symmetry whose current is 
\begin{eqnarray}
    J_{B-L}= \frac{1}{15}(3J_W + 2J_Y + 3 J_C)
\end{eqnarray}
One could include $U(1)_C$ (or usually $U(1)_{B-L}$) into the charge table \ref{eq:m3 sm sym rep}.  

To find a common conformal subalgebra, we only need to focus on the other factor, $\mathsf{3SP}$ and $\text{Vir}_{4/5}$ from $S_1$ and $S_2$, respectively. $\mathsf{3SP}$, which is the first $W_3$ algebra, is an extension of $\text{Vir}_{4/5}$ by the spin-3 current. Hence
\begin{eqnarray}\label{eq:3spvir}
    \mathsf{3SP} \hookleftconf \text{Vir}_{4/5}
\end{eqnarray}
is a conformal embedding, whose branching rules are collected in Appendix \ref{app:branching data}. This shows that, together with the abelian factor, 
\begin{eqnarray}
    W=\widehat{\mathfrak{u}}(1)_{30}\oplus \text{Vir}_{4/5}
\end{eqnarray}
is a common conformal subalgebra of $S_1$ and $S_2$, 
\begin{eqnarray}
    \begin{tikzcd}[row sep=large, column sep=small, every label/.append style={sloped}]
S_1= \widehat{\mathfrak{u}}(1)_{30}\oplus \mathsf{3SP} & & S_2= \widehat{\mathfrak{u}}(1)_{30}\oplus \text{Vir}_{4/5} \\
& W= \widehat{\mathfrak{u}}(1)_{30}\oplus\text{Vir}_{4/5} \arrow[ul, hook, "\text{conf}"] \arrow[ur, hook, "\simeq"].
\end{tikzcd}
\end{eqnarray}

\subsubsection{Club Sandwich}

\begin{figure}
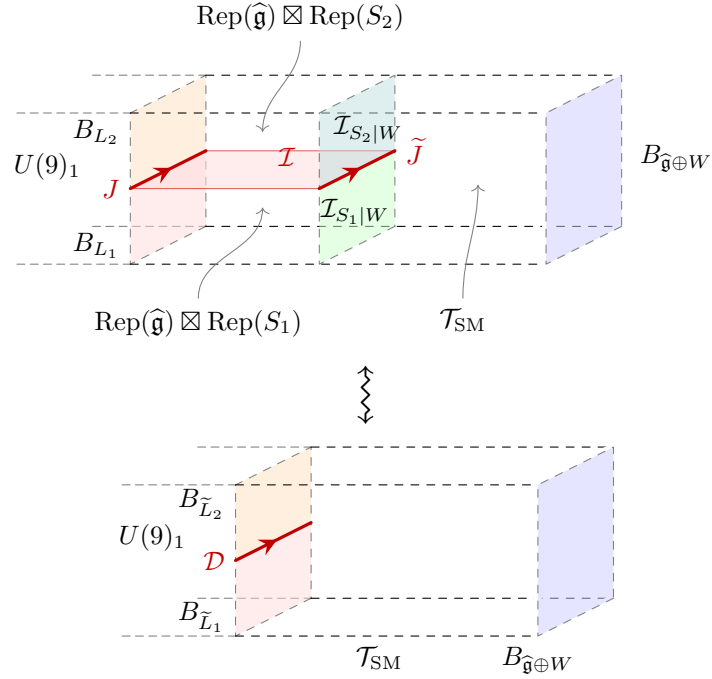

    \centering
    \figClubSM[1.0]\\[1ex]\rotatebox[origin=c]{-90}{\Large$\leftrightsquigarrow$}\\[1ex]
    \figClubSMShrunk[1.0]
    \caption{Club sandwich for the $M_3$ monopole in the Standard Model (top), with $\Rep(\widehat{\mathfrak{g}})=SU(3)_2\boxtimes SU(2)_4\boxtimes U(1)^2_K$. Since $W=S_2$, the interface $\mathcal{I}_{S_2|W}$ is trivial, while $\mathcal{I}_{S_1|W}$ condenses $\mathcal{A}_1=\phi_{1,1}\oplus\phi_{1,5}$. Shrinking the interfaces onto the topological interfaces $B_{L_i}$ gives the sandwich at the bottom, with two topological boundaries $B_{\widetilde L_i}$ of the bulk TQFT $\mathcal{T}_{\rm SM}$ in \eqref{eq:TSM}, joined along the intertwining defect $\mathcal{D}$.}
    \label{fig:smclub}
\end{figure}

Following Section \ref{sec:commonalg}, we obtain the club sandwich shown in Figure \ref{fig:smclub}. Its right bulk is the TQFT
\begin{eqnarray}\label{eq:TSM}
    \mathcal{T}_{\rm SM}= SU(3)_2\boxtimes SU(2)_4\boxtimes U(1)^2_K\boxtimes U(1)_{30}\boxtimes M(6,5).
\end{eqnarray}

The fermionic Lagrangian algebras $\widetilde L_i$ follow from the branching of the vacuum supermodule of $\widehat{\mathfrak{u}}(9)_1$. For the in-coming fermions, the steps are as follows.
\begin{enumerate}
    \item The vacuum supermodule of each $\widehat{\mathfrak{u}}(3)^x_1$ branches as \eqref{eq:canonicalbranching}, $1\mapsto\bigoplus_{r_x\in\bZ_3}[r_x]\boxtimes\Lambda_{r_x}$, $x=q,e,d$.
    \item Under $\widehat{\mathfrak{su}}(3)_2\oplus\mathsf{3SP}\hookrightarrow\widehat{\mathfrak{su}}(3)^q_1\oplus\widehat{\mathfrak{su}}(3)^d_1$, the module $\Lambda_{r_q}\boxtimes\overline{\Lambda}_{r_d}$ (the conjugation accounts for $d$ transforming in $\overline{\square}$) branches into $\bigoplus_{(\mu,\chi)\in\mathcal{B}_{r_q,r_d}}\mu\boxtimes\chi$, where $\mu$ is a line of $SU(3)_2$, $\chi$ a line of $\Rep(\mathsf{3SP})$, and each $\mathcal{B}_{r_q,r_d}$ has two elements, listed in Appendix \ref{app:branching data} (where the second entry on the left-hand side is the conjugate label $\overline{\Lambda}_{r_d}$, and the $\mathfrak{su}(2)$ weights are labeled by $2j$). The lines $\chi$ are then lifted to $M(6,5)$ by \eqref{eq:3spvir}.
    \item Under $\widehat{\mathfrak{su}}(2)_4\hookrightarrow\widehat{\mathfrak{su}}(3)^e_1$, $\Lambda_0\mapsto 0\oplus 2$ and $\Lambda_1,\Lambda_2\mapsto 1$, in terms of spins.
    \item The line $[r_q]\boxtimes[r_e]\boxtimes[r_d]$ of $U(1)_3^3$ lifts to the $20$ distinct lines
    \begin{eqnarray}\label{eq:smlift1}
        (n_q,\, n_q+6n_e+2n_d)\boxtimes[n_e-3n_d], \qquad (n_q,n_e,n_d)\in(r_q,r_e,r_d)+3\bZ^3,
    \end{eqnarray}
    of $U(1)^2_K\boxtimes U(1)_{30}$, where the two entries are the $U(1)_W\times U(1)_Y$ charges and the bracket is the $U(1)_C$ charge of $q^{n_q}e^{n_e}d^{n_d}$.
\end{enumerate}
Combining the steps,
\begin{equation}\label{eq:L1sm}
\begin{split}
    \widetilde L_1=\bigoplus_{r_q,r_e,r_d\in\bZ_3}\ \bigoplus_{(\mu,\chi)\in\mathcal{B}_{r_q,r_d}}\ \bigoplus_{j\in E_{r_e}}\ \bigoplus_{\phi\in\chi}\ \bigoplus_{(n_q,n_e,n_d)}\ &\mu\boxtimes j\boxtimes(n_q,\,n_q+6n_e+2n_d)\\
    &\boxtimes[n_e-3n_d]\boxtimes\phi,
\end{split}
\end{equation}
where $E_0=\{0,2\}$, $E_1=E_2=\{1\}$, $\phi$ runs over the Verlinde lines in $M(6,5)$, and the last sum is over the $20$ lines \eqref{eq:smlift1}. For the out-going fermions:
\begin{enumerate}
    \item The vacuum supermodules of $\widehat{\mathfrak{u}}(2)^{l^-}_1$ and $\widehat{\mathfrak{u}}(6)^u_1$ branch as $1\mapsto\bigoplus_{r_2\in\bZ_2}[r_2]\boxtimes\Lambda_{r_2}$ and $1\mapsto\bigoplus_{r_6\in\bZ_6}[r_6]\boxtimes\Lambda_{r_6}$. The free fermion $\widehat{\mathfrak{u}}(1)^{l^+}_1$ has only the trivial line, up to the transparent fermion.
    \item Under the level-rank embedding, $\Lambda_{r_6}\mapsto\bigoplus_{(j_3,\nu)\in\mathcal{P}_{r_6}}j_3\boxtimes\overline{\nu}$ (see also Appendix \ref{app:branching data}), $j_3$ is a spin of $\widehat{\mathfrak{su}}(2)_3$, and the conjugation accounts for $u$ transforming in $\overline{\square}$ of $SU(3)$.
    \item Under the diagonal embedding, $\Lambda_{r_2}\boxtimes j_3\mapsto\bigoplus_{j}j\boxtimes\phi_{2j_3+1,2j+1}$, where $j$ runs over the spins of $\widehat{\mathfrak{su}}(2)_4$ with $\frac{r_2}{2}+j_3+j\in\bZ$.
    \item The line $[r_2]\boxtimes[r_6]$ of $U(1)_2\boxtimes U(1)_6$ lifts to the $30$ distinct lines
    \begin{equation}\label{eq:smlift2}
    \begin{split}
        &(n_+-n_-,\,-3n_+-3n_--4n_u)\boxtimes[-4n_+-2n_-+n_u],\\
        &(n_+,n_-,n_u)\in(0,r_2,r_6)+\bZ\oplus 2\bZ\oplus 6\bZ,
    \end{split}
    \end{equation}
    of $U(1)^2_K\boxtimes U(1)_{30}$, given by the charges of $(l^+)^{n_+}(l^-)^{n_-}u^{n_u}$.
\end{enumerate}
Hence
\begin{equation}\label{eq:L2sm}
\begin{split}
    \widetilde L_2=\bigoplus_{r_2\in\bZ_2}\ \bigoplus_{r_6\in\bZ_6}\ \bigoplus_{(j_3,\nu)\in\mathcal{P}_{r_6}}\ \bigoplus_{j}\ \bigoplus_{(n_+,n_-,n_u)}\ &\overline\nu\boxtimes j\boxtimes(n_+-n_-,\,-3n_+-3n_--4n_u)\\
    &\boxtimes[-4n_+-2n_-+n_u]\boxtimes\phi_{2j_3+1,2j+1}.
\end{split}
\end{equation}

The in-coming and out-going fermions correspond to the summands
\begin{equation}\label{eq:lambdasm}
    \begin{aligned}
    &\lambda_q = [1,0]\boxtimes 0\boxtimes(1,1)\boxtimes[0]\boxtimes\phi_{2,3}, &&\qquad \widetilde\lambda_{l^+} = [0,0]\boxtimes 0\boxtimes(1,-3)\boxtimes[-4]\boxtimes\phi_{1,1},\\
    &\lambda_e = [0,0]\boxtimes 1\boxtimes(0,6)\boxtimes[1]\boxtimes\phi_{1,1}, &&\qquad \widetilde\lambda_{l^-} = [0,0]\boxtimes \tfrac12\boxtimes(-1,-3)\boxtimes[-2]\boxtimes\phi_{1,2},\\
    &\lambda_d = [0,1]\boxtimes 0\boxtimes(0,2)\boxtimes[-3]\boxtimes\phi_{2,3}, &&\qquad \widetilde\lambda_{u} = [0,1]\boxtimes \tfrac12\boxtimes(0,-4)\boxtimes[1]\boxtimes\phi_{2,2},
    \end{aligned}
\end{equation}
of $\widetilde L_1$ and $\widetilde L_2$, respectively. Finally,
\begin{eqnarray}\label{eq:Homsm}
    \dim\Hom(\widetilde L_1,\widetilde L_2)=480,
\end{eqnarray}
which is the number of common simple summands of $\widetilde{L}_1$ and $\widetilde{L}_2$.

\subsubsection{Intertwining Defect}

After shrinking the club sandwich, the intertwining defect $\mathcal{D}$ is a topological junction between $B_{\widetilde L_1}$ and $B_{\widetilde L_2}$, and there are $480$ simple choices by \eqref{eq:Homsm}. Any such $\mathcal{D}$ intertwines the symmetry $G=SU(3)\times U(1)_W \times U(1)_Y\times SU(2)_{\text{rot}}$ as well as $U(1)_C$ by construction from Section \ref{sec:generalconstruction}. Folding back along a simple $\mathcal{D}$ gives a conformal boundary condition of 9 Dirac fermions preserving the full symmetry $G\times U(1)_C$.

\subsection{Scattering}
\label{sec:smscat}

The scattering formula \eqref{eq:Daction} applies with $\Rep(\widehat{\mathfrak{g}})$ replaced by $\mathcal{T}_{\rm SM}$ and $B_{L_i}$ by $B_{\widetilde L_i}$. The out-going twisted sectors are labeled by the simple lines in $F_{B_{\widetilde L_2}}(\lambda)$, $\lambda=\lambda_q,\lambda_e,\lambda_d$, whose endomorphisms are counted by $\dim\Hom_{\mathcal{T}_{\rm SM}}(\lambda\otimes\overline\lambda,\widetilde L_2)$.

Let $e$ be the in-coming excitation.  Since the only summand of $\lambda_e\otimes\overline\lambda_e$ in $\widetilde L_2$ is the trivial line, 
\begin{eqnarray}
    b_e\equiv F_{B_{\widetilde L_2}}(\lambda_e), \qquad d_{b_e}=d_{\lambda_e}=2,
\end{eqnarray}
is a simple non-invertible line. So the out-going excitation after scattering in $e$ is a unique twisted sector associated with the topological line $b_e$. The scattering process is 
\begin{eqnarray}
    \figBeforeScatterSM[0.95]{e}\quad &=& \quad \sum_{zw}C^e_{zw}\quad\figOutSM[0.95]{b_e}{w}{\mathcal{O}^{(z)}_e}
\end{eqnarray}

Next let $q$ be the in-coming excitation.  Since $\lambda_q\otimes\overline\lambda_q$ contains two summands of $\widetilde L_2$, 
the trivial line and 
\begin{eqnarray}
    X=[1,1]\boxtimes 0\boxtimes(0,0)\boxtimes[0]\boxtimes\phi_{2,5}
\end{eqnarray}
So the out-going state contains two twisted sectors, associated with topological lines
\begin{eqnarray}\label{eq:bqsm}
    F_{B_{\widetilde L_2}}(\lambda_q)= b_q\oplus b_q'.
\end{eqnarray}
Note that $X$ is the product of two Fibonacci lines, $[1,1]\in SU(3)_2$, and $\phi_{3,1}\in M(6,5)$, with spin $\frac35+ \frac75=2$. This suggests that the bulk TQFT $\mathcal{T}_{\text{SM}}$ factorizes into a double-Fibonacci TQFT $\mathcal{T}_{\text{Fib}}= \text{Fib}\boxtimes \overline{\text{Fib}}$ and another TQFT $\mathcal{T}_0$.\footnote{At the level of MTC, $SU(3)_2= SU(3)_{-1}\boxtimes \overline{\text{Fib}}$, and $M(6,5)= \text{Fib}\boxtimes M_0$. Then $\mathcal{T}_{0}= SU(3)_{-1}\boxtimes SU(2)_4 \boxtimes U(1)_K^2 \boxtimes U(1)_{30} \boxtimes M_0$. } Any line factorizes, this includes $\lambda_q$ which factorizes as $\lambda_q= X \boxtimes \lambda_q'$ with $\lambda_q'= [0,2]\boxtimes0\boxtimes(1,1)\boxtimes[0]\boxtimes\phi_{1,3} \in \mathcal{T}_0$.  After pushing to $B_{\widetilde{L}_2}$, $\lambda_q'$ becomes a simple line $F_{B_{\widetilde{L}_2}}(\lambda_q')$, while $X$ splits into two simple lines $\mathbf{1}\oplus \tau$, with $\tau= F_{B_{\widetilde{L}_2}}([1,1]\boxtimes 0 \boxtimes (0,0) \boxtimes[0] \boxtimes \phi_{1,1})$. So we can identify
\begin{eqnarray}
    b_q= \mathbf{1}\boxtimes F_{B_{\widetilde{L}_2}}(\lambda_q'), \qquad b_q' =\tau \boxtimes  F_{B_{\widetilde{L}_2}}(\lambda_q')
\end{eqnarray}
Moreover, the intertwining defect $\mathcal{D}$ also factorizes as $\mathcal{D}_{\text{Fib}}\boxtimes \mathcal{D}_0$ with $\mathcal{D}_{\text{Fib}}\in \{\mathbf{1}, \tau\}$.

The scattering coefficients factorize over $\mathcal{T}_{\rm Fib}\boxtimes\mathcal{T}_0$. In $\mathcal{T}_0$, the out-going line $F_{B_{L'_2}}(\lambda'_q)$ is simple, so this factor contributes a single channel with probability one. In $\mathcal{T}_{\rm Fib}$, the in-coming quark carries the local operator at the end of $X$, i.e.\ the trivial summand of $\tau\otimes\tau$, which crosses $\mathcal{D}_{\rm Fib}$. For $\mathcal{D}_{\rm Fib}=\mathbf{1}$, it passes through unchanged. For $\mathcal{D}_{\rm Fib}=\tau$, the amplitude for it to remain untwisted is given by the action of the Fibonacci line, $\frac{ D_{\text{Fib}} S_{\tau\tau}}{d_\tau^2}=-\frac1{\varphi^2}$ which is the ratio between the Hopf link of two $\tau$-lines and two unlinked $\tau$-lines, and the remaining probability goes to the $\tau$-twisted sector by \eqref{eq:sumto1}. Altogether, the categorical scattering gives
\begin{eqnarray}\label{eq:smscat}
    \nonumber\\[1ex]
    \figBeforeScatterSM[0.95]{q}\quad &=& \quad \sum_{zw}C^q_{zw}\quad\figOutSM[0.95]{b_q}{w}{\mathcal{O}^{(z)}_q}\quad + \quad \sum_{zw}C'^q_{zw}\quad\figOutSM[0.95]{b'_q}{w}{\mathcal{O}'^{(z)}_q}
\end{eqnarray}
and similarly for $d$, with the probabilities
\begin{equation}\label{eq:Psm}
    \begin{split}
    &P(e\to b_e)=1,\\
    \mathcal{D}_{\rm Fib}=\mathbf{1}:\qquad &P(q\to b_q)=1, \qquad P(q\to b'_q)=0,\\
    \mathcal{D}_{\rm Fib}=\tau:\qquad &P(q\to b_q)=\frac{1}{\varphi^4}=\frac{7-3\sqrt5}{2}\approx 0.146,\\
    &P(q\to b'_q)=1-\frac1{\varphi^4}=\frac{3\sqrt5-5}{2}\approx0.854,
    \end{split}
\end{equation}
for any choice of $\mathcal{D}_0$. For every simple boundary condition, all the out-going lines $b_e$, $b_q$ and $b'_q$ are non-invertible. For $\mathcal{D}_{\rm Fib}=\mathbf{1}$, every in-coming fermion scatters with probability one into a single twisted sector, as in Section \ref{sec:susu}. For $\mathcal{D}_{\rm Fib}=\tau$, the quarks scatter into a superposition of two twisted sectors, as in $\text{QED}(\square,\overline\square)$ in Section \ref{sec:qedsbar}. Here, however, the superposition originates entirely from the Fibonacci factor of $\mathcal{T}_{\rm SM}$, and the probabilities \eqref{eq:Psm} are universal numbers fixed by the Fibonacci $S$-matrix.

The discussion for scattering in $d$ is parallel to $q$, so we don't repeat the exercise here.

\section{General Cases: Spectators and Orbifold Equivalence}
\label{sec:orbifoldequivalence}

We finally turn to the most general case, in which the two embeddings $f_{1,2}$ are not related by an automorphism of $\widehat{\mathfrak{g}}$, and the coset chiral algebras $S_1$ and $S_2$ do not share a common conformal subalgebra.

The key idea is to first enlarge the system by introducing \emph{spectators}, i.e. additional degrees of freedom on which $G$ acts trivially, chosen such that the enlarged problem falls into the class studied in Section \ref{sec:commonalg}. The spectators propagate with both chiralities, and at the end they are removed by turning on a gapping potential. Since $G$ does not act on the spectators, the gapping potential automatically preserves $G$, and we are left with $G$-preserving boundary condition of the original fermions. 

The underlying  mathematics that makes this construction possible is the \emph{orbifold equivalence} of chiral algebras \cite{Moller:2024xtt}.

Our construction relies on the following assumptions. (i) The coset chiral algebras $S_1$ and $S_2$, and all chiral algebras in the zigzag chain below, are positive and strongly rational. (ii) $S_1$ and $S_2$ are orbifold equivalent, as conjectured for Witt equivalent chiral algebras in \cite[Conjecture 5.6]{Moller:2024xtt}. The existence of the intertwining defect alone would already follow from the weaker \cite[Conjecture 5.7]{Moller:2024xtt} via Section \ref{sec:coset}, but the explicit chain is what makes the construction below concrete. (iii) The boundary renormalization group flow in Section 7.4 ends at a compact conformal boundary condition. The results of \cite{Moller:2024xtt} are formulated for bosonic chiral algebras, and we expect them to extend to the fermionic case.

\subsection{Witt Equivalence and Orbifold Equivalence}

Since both $\Rep(\widehat{\mathfrak{g}})\boxtimes \Rep(S_1)$ and $\Rep(\widehat{\mathfrak{g}})\boxtimes \Rep(S_2)$ admit a topological interface to the same invertible TQFT $U(N)_1$, $\Rep(S_1)$ and $\Rep(S_2)$ are (super) Witt equivalent, and are hence separated by a topological interface $\mathcal{I}_{S_1|S_2}$. This justifies the existence of the interface 
\begin{eqnarray}\label{eq:calI}
    \mathcal{I} = \text{id}_{\widehat{\mathfrak{g}}} \otimes \mathcal{I}_{S_1|S_2}
\end{eqnarray}
used in Section \ref{sec:coset}. Moreover, $S_1$ and $S_2$ have the same central charge $c_{S_1}=c_{S_2}= N-c_{\widehat{\mathfrak{g}}}$. Chiral algebras with these two properties are called \emph{Witt equivalent} \cite[Definition 3.2]{Moller:2024xtt}. 

A finer relation between chiral algebras is the \emph{orbifold equivalence} \cite[Definition 4.3]{Moller:2024xtt}. Two chiral algebras $S_1$ and $S_2$ are orbifold equivalent if there exist chiral algebras $V_2, ..., V_n$ and $W_1, ..., W_{n}$ forming a zigzag chain of conformal embeddings, 
\begin{eqnarray}\label{eq:orbifoldequiv}
    \begin{tikzcd}[row sep=large, column sep=small,every label/.append style={sloped}]
S_1 & & V_2 & \cdots & V_n & & S_2 \\
    & W_1 \arrow[ul,hook,"\text{conf}"] \arrow[ur,hook,"\text{conf}"]
    & & \cdots \arrow[ul, hook,"\text{conf}"] \arrow[ur,hook,"\text{conf}"]
    & & W_n \arrow[ul,hook,"\text{conf}"] \arrow[ur,hook,"\text{conf}"]
\end{tikzcd}
\end{eqnarray}
i.e. $W_k$ is conformally embedded in both $V_k$ and $V_{k+1}$ for $k=1, ..., n$, where we set $V_1=S_1$ and $V_{n+1}=S_2$. Each step in the chain can be viewed as a generalized orbifold, namely the gauging of a (possibly non-invertible) symmetry, hence the name. 

Orbifold equivalence implies Witt equivalence \cite[Theorem 5.1]{Moller:2024xtt}, and it has been conjectured that the converse is also true for positive, strongly rational chiral algebras \cite[Conjecture 5.6]{Moller:2024xtt}. In the following, we assume that $S_1$ and $S_2$ are orbifold equivalent.\footnote{The results of \cite{Moller:2024xtt} are formulated for bosonic and strongly rational vertex operator algebras, whereas the coset chiral algebras $S_i$ can be fermionic. We expect the statements to extend to the fermionic case. } 
Note that \eqref{eq:S1S2W} in Section \ref{sec:commonalg} is a special case  $n=1$ of \eqref{eq:orbifoldequiv}. For $n\geq 2$, $S_1$ and $S_2$ need not share any common conformal subalgebra, and the construction of Section \ref{sec:commonalg} does not directly apply.

\subsection{Spectators}

The key observation is that spectators allow us to collapse the zigzag chain \eqref{eq:orbifoldequiv} into a single wedge of the form \eqref{eq:S1S2W}.  
We introduce the intermediate chiral algebras $V_2, ..., V_n$ as spectators, and consider the enlarged chiral theory 
\begin{eqnarray}\label{eq:tensor}
    \Fer(N) \oplus V, \qquad V=\bigoplus_{k=2}^n V_k,
\end{eqnarray}
as a \emph{relative} chiral CFT, living on the chiral boundary of the 3d TQFT $U(N)_1 \boxtimes \Rep(V)$.

We embed $\widehat{\mathfrak{g}}$ into the enlarged chiral algebra such that its image in the spectator sector is zero, and denote the enlarged embedding by $\widetilde{f}_i$, 
\begin{eqnarray}\label{eq:fitilde}
    \widetilde{f}_i= f_i \oplus 0: \widehat{\mathfrak{g}} \hookrightarrow \Fer(N) \oplus V, \qquad i=1,2.
\end{eqnarray}
The coset chiral algebras of the enlarged embeddings are 
\begin{eqnarray}
    \widetilde{S}_1= S_1\oplus V_2 \oplus ... \oplus V_n, \qquad \widetilde{S}_2= V_2 \oplus ... \oplus V_n \oplus S_2
\end{eqnarray}
and $\widehat{\mathfrak{g}} \oplus \widetilde{S}_i \hookrightconf \Fer(N) \oplus V$ are conformal embeddings. Despite $S_1$ and $S_2$ may not share a common conformal subalgebra, $\widetilde{S}_1$ and $\widetilde{S}_2$ do. Indeed, the chiral algebra
\begin{eqnarray}
    W= \bigoplus_{k=1}^n W_k
\end{eqnarray}
is conformally embedded in both,
\begin{eqnarray}\label{eq:wedge}
    \begin{tikzcd}[row sep=large, column sep=small,every label/.append style={sloped}]
\widetilde{S}_1 & & \widetilde{S}_2 \\
& W \arrow[ul, hook,"\text{conf}"] \arrow[ur, hook,"\text{conf}"]
\end{tikzcd}
\end{eqnarray}
The left embedding in \eqref{eq:wedge} is the direct sum of the embeddings $\nwhookarrow$ in \eqref{eq:orbifoldequiv}, i.e. $W_1\hookrightconf S_1$ and $W_k\hookrightconf V_k$ for $k=2, ..., n$, whereas the right embedding is the direct sum of the embeddings $\nehookarrow$, i.e. $W_k\hookrightconf V_{k+1}$ for $k=1, .., n-1$ and $W_n\hookrightconf S_2$. In other words, the spectators absorb the intermediate chiral algebras of the zigzag chain, and the enlarged problem is precisely of the type studied in Section \ref{sec:commonalg}. This is a concrete realization of \cite[Proposition 4.12]{Moller:2024xtt}, which shows that orbifold equivalent chiral algebras are interface equivalent by composing the junctions along the zigzag chain \eqref{eq:orbifoldequiv}. The spectators collapse the $n$ regions of that construction into a single wedge, so that the analysis of Section \ref{sec:commonalg} can be reused.

\subsection{Club Sandwich with Spectators}

\begin{figure}
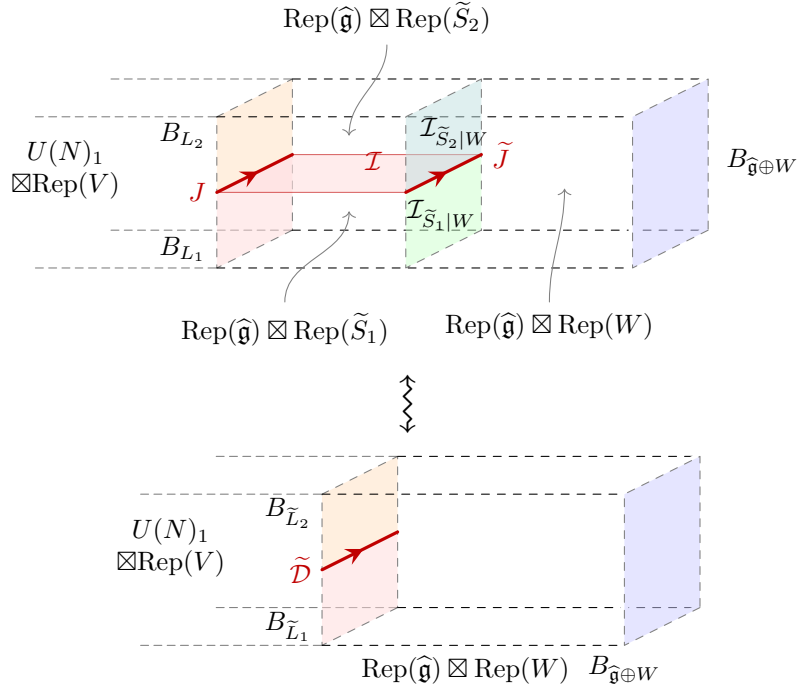

    \centering
    \figClubSpec[1.0]\\[1ex]\rotatebox[origin=c]{-90}{\Large$\leftrightsquigarrow$}\\[1ex]
    \figClubSpecShrunk[1.0]
    \caption{Club sandwich of the enlarged theory $\Fer(N)\oplus V$.  Shrinking the interfaces $\mathcal{I}_{\widetilde{S}_i|W}$ to the left, both embeddings share the same bulk $\Rep(\widehat{\mathfrak{g}})\boxtimes\Rep(W)$, giving the intertwiner $\widetilde{\mathcal{D}}$ between the two interfaces $B_{\widetilde{L}_i}$. }
    \label{fig:spectator}
\end{figure}

We can therefore apply the construction of Section \ref{sec:commonalg} to the enlarged theory \eqref{eq:tensor}, with $S_i$ replaced by $\widetilde{S}_i$. This gives the club sandwich shown at the top of Figure \ref{fig:spectator}, which consists of the following ingredients. 
\begin{itemize}
    \item \textbf{Right bulk}: The TQFT in the right bulk is $\Rep(\widehat{\mathfrak{g}})\boxtimes \Rep(W)$, with $\Rep(W) = \boxtimes_{k=1}^n \Rep(W_k)$. 
    \item \textbf{Left bulks}: The TQFTs in the bottom and top left bulks are $\Rep(\widehat{\mathfrak{g}})\boxtimes \Rep(\widetilde{S}_1)= \Rep(\widehat{\mathfrak{g}})\boxtimes \Rep(S_1) \boxtimes \Rep(V)$ and $\Rep(\widehat{\mathfrak{g}})\boxtimes \Rep(\widetilde{S}_2)=\Rep(\widehat{\mathfrak{g}})\boxtimes \Rep(V) \boxtimes \Rep(S_2)$, respectively. 
    \item \textbf{Chiral boundary}: The right boundary of $\Rep(\widehat{\mathfrak{g}})\boxtimes \Rep(W)$ is a chiral boundary $B_{\widehat{\mathfrak{g}}\oplus W}$, carrying the chiral algebra $\widehat{\mathfrak{g}}\oplus W$. 
    \item \textbf{Topological interface $\mathcal{I}$}: The topological interface between $\Rep(\widehat{\mathfrak{g}})\boxtimes \Rep(\widetilde{S}_1)$ and $\Rep(\widehat{\mathfrak{g}})\boxtimes \Rep(\widetilde{S}_2)$ is $\text{id}_{\widehat{\mathfrak{g}}}\otimes \mathcal{I}_{\widetilde{S}_1|\widetilde{S}_2}$, which exists for the same reason as \eqref{eq:calI}. By abuse of notation, we still denote it by $\mathcal{I}$. 
    \item \textbf{Topological interfaces $\mathcal{I}_{\widetilde{S}_i|W}$}: The interfaces separating the left and right bulks are induced by the conformal embeddings \eqref{eq:wedge}, by the embeddings $\nwhookarrow$ (bottom) and $\nehookarrow$ (top). 
    \item \textbf{Topological interfaces $B_{L_i}$}: Since the spectator TQFT $\Rep(V)$ is not invertible, the topological boundaries of Section \ref{sec:commonalg} are replaced by the topological interfaces between $\Rep(\widehat{\mathfrak{g}})\boxtimes \Rep(\widetilde{S}_i)$ and $U(N)_1 \boxtimes \Rep(V)$. They condense the fermionic Lagrangian algebra $L_i$ of $\Rep(\widehat{\mathfrak{g}})\boxtimes \Rep({S}_i)$, and are transparent to the lines in $\Rep(V)$. By abuse of notation, we still denote them by $B_{L_i}$. 
\end{itemize}
The junction lines $J$ and $\widetilde{J}$ exist by the same folding argument as before, now applied to topological interfaces: folding two topological interfaces between the same pair of TQFTs turns them into topological boundaries of the same folded TQFT, between which a topological junction always exists \cite{etingof2009fusioncategorieshomotopytheory,Davydov:2010kfz}.

Shrinking the interfaces $\mathcal{I}_{\widetilde{S}_i|W}$ onto the chiral boundary establishes the existence of the junction on the chiral boundary for the enlarged theory, which was assumed in Section \ref{sec:coset}. 

Shrinking them instead onto the topological interfaces $B_{L_i}$ gives the configuration at the bottom of Figure \ref{fig:spectator}, in which the interface $B_{\widetilde{L}_i}= B_{L_i} \otimes \mathcal{I}_{\widetilde{S}_i|W}$ separates the common bulk $\Rep(\widehat{\mathfrak{g}})\boxtimes \Rep(W)$ from $U(N)_1 \boxtimes \Rep(V)$, and they are joined along a topological junction line $\widetilde{\mathcal{D}}$. After further shrinking the sandwich, $\widetilde{\mathcal{D}}$ becomes a topological line of the \emph{relative} chiral CFT \eqref{eq:tensor}. By the same argument as in Section \ref{sec:interDefect}, it intertwines the two enlarged embeddings \eqref{eq:fitilde}, 
\begin{eqnarray}\label{eq:intertwinextended}
   \widetilde{f}_1(j) \widetilde{\mathcal{D}} = \widetilde{\mathcal{D}}\widetilde{f}_2(j),
\end{eqnarray}
because the current $j$ lives on the chiral boundary, and is attached to the trivial bulk line. Note that $\widetilde{\mathcal{D}}$ in general acts non-trivially on the spectators, since it also intertwines the two embeddings of $W$, which place $W_1$ in $S_1$ on one side, and in the spectator $V_2$ on the other, and $W_k$, $k\geq 2$, in $V_k$ and $V_{k+1}$, respectively.

\subsection{Gapping the Spectators}

\begin{figure}
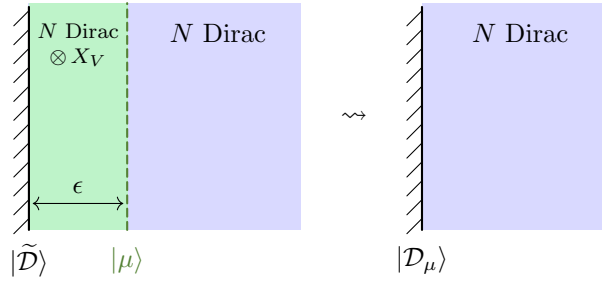

    \centering
    \figGapping
    \caption{Gapping the spectators in the folded picture. The spectators, which combine into the diagonal RCFT $X_V$, are confined to a layer of width $\epsilon$ between the boundary condition $|\widetilde{\mathcal{D}}\rangle$ and a $G$-neutral boundary condition $\ket{\mu}$ of $X_V$. The infrared fixed point of this configuration is a $G$-symmetric boundary condition $\ket{\mathcal{D}_\mu}$ of the $N$ Dirac fermions.}
    \label{fig:gapping}
\end{figure}

It remains to remove the spectators. So far, we have worked in the unfolded chiral picture. We now fold back along $\widetilde{\mathcal{D}}$, as in Section \ref{sec:2.1}. The spectators on the two sides of $\widetilde{\mathcal{D}}$ become a left-moving copy of $V$ and a right-moving copy $\overline{V}$, which combine into the diagonal, non-chiral RCFT $X_V$, since folding the bulk $\Rep(V)$ produces its canonical Lagrangian algebra $\oplus_\lambda \lambda \boxtimes \bar{\lambda}$. See Appendix \ref{app:More on Gapping Spectators in the Sandwich} for further illustration.  Therefore, $\widetilde{\mathcal{D}}$ describes a conformal boundary condition $|{\widetilde{\mathcal{D}}}\rangle$ of the stacking of $N$ Dirac fermions and $X_V$ on the half plane $x\geq 0$. By \eqref{eq:intertwinextended}, $|{\widetilde{\mathcal{D}}}\rangle$ glues the left- and right-moving $G$-currents of the fermions, and hence preserves $G$ symmetry.

We now gap out the spectators. Since $G$ acts trivially on $X_V$, any boundary condition of $X_V$ is $G$-neutral, and we can use it to confine the spectators to a thin layer near the boundary. Concretely, we impose a conformal boundary condition $\ket{\mu}$ of $X_V$ along $x=\epsilon$, so that the spectators only live on the strip $0\leq x\leq \epsilon$, while the fermions live on the whole half-plane, see Figure \ref{fig:gapping}. Physically, $\ket{\mu}$ models a $G$-neutral potential that gaps the spectators in the region $x>\epsilon$. For definiteness, we take $\ket{\mu}$ to be a Cardy state of the diagonal RCFT $X_V$. 

At distances much larger than $\epsilon$, this configuration is a boundary condition of the $N$ Dirac fermions alone. We do not take the naive limit $\epsilon\to0$, since the fusion of a conformal boundary with a non-topological interface is in general singular \cite{Bachas:2007td}; indeed, $|\widetilde{\mathcal{D}}\rangle$ preserves $W$ but not $V$, so the layer is not topological. Instead, we regard $\epsilon$ as a short-distance scale, and define $\ket{\mathcal{D}_\mu}$ as the infrared fixed point of the boundary renormalization group flow of this configuration. The flow preserves $G$, since the $G$-currents are glued at $x=0$ by $|\widetilde{\mathcal{D}}\rangle$, and neither $X_V$ nor $\ket{\mu}$ carries $G$-charge. Assuming that the flow ends at a compact conformal boundary condition, $\ket{\mathcal{D}_\mu}$ is a $G$-symmetric conformal boundary condition of the $N$ Dirac fermions. It is in general not simple, and each of its simple components is $G$-symmetric by the argument in footnote \ref{footnote:compactBC}.

Note that $\ket{\mathcal{D}_\mu}$ is not determined by $\widetilde{\mathcal{D}}$ alone, and the properties of $\widetilde{\mathcal{D}}$, such as the lines attached to the scattered excitations, need not survive the flow. On the other hand, statements that hold for every $G$-symmetric boundary condition apply to $\ket{\mathcal{D}_\mu}$.

We finally comment on the scattering. 
Before gapping, the scattering is described as in Section \ref{sec:commonalg}, with $S_i$ replaced by $\widetilde{S}_i$: across $\widetilde{\mathcal{D}}$, the $S_1$ quantum numbers of $\psi$, carried by $\lambda_{W_1}$, are passed to the spectator $V_2$, and each $\lambda_{W_k}$ to $V_{k+1}$. In the thin-strip configuration, the spectators are projected onto the ground state of the strip, so the out-going excitation contains no spectators only if every $\lambda_{W_k}$ also appears in the vacuum module of the next spectator. Heuristically, repeated reflections between $\widetilde{\mathcal{D}}$ and $\ket{\mu}$ move these labels along the zigzag chain until they reach $S_2$. We expect this to describe the boundary condition $\ket{\mathcal{D}_\mu}$ at the end of the flow. We leave it to future work to make the scattering process more explicit.

\subsection{Example: 1-5-7-8-9 Model} 
\label{sec:15789}

We finally return to the 1-5-7-8-9 model \cite{Smith:2019jnh,vanBeest:2023mbs}, i.e.\ the chiral $U(1)$ gauge theory with five left-handed Weyl fermions of charges $(1,5,9,-7,-8)$, in the presence of a unit monopole. 

After the partial wave reduction around a unit monopole, the fermions of charges $1,5,9$ give $15$ in-coming chiral fermions, and those of charges $-7,-8$ give $15$ out-going ones,
\begin{eqnarray}\label{eq:15789charges}
	\begin{tabular}{|c|c|c|}
		\hline
		 & $SU(2)_{\rm rot}$ & $U(1)$ \\
		\hline
		$\psi_1$ & $\mathbf{1}$ & $1$\\
		$\psi_5$ & $\mathbf{5}$ & $5$\\
		$\psi_9$ & $\mathbf{9}$ & $9$\\
		\hline
		$\chi_7$ & $\mathbf{7}$ & $-7$\\
		$\chi_8$ & $\mathbf{8}$ & $-8$\\
		\hline
	\end{tabular}
\end{eqnarray}
where the representations of $SU(2)_{\rm rot}$ are labeled by their dimensions.

We unfold the $15$ Dirac fermions into $\Fer(15)$ with a topological line $\mathcal{D}$. The currents of $G$ generate
\begin{eqnarray}
    \widehat{\mathfrak{g}}=\widehat{\mathfrak{su}}(2)_{140}\oplus\widehat{\mathfrak{u}}(1)_{855},\qquad c_{\widehat{\mathfrak{g}}}=\frac{210}{71}+1=\frac{281}{71},
\end{eqnarray}
The gap to $c_{\mathcal{T}}=15$ is $c_{S_i}=\frac{784}{71}$.

For the in-coming fermions, 
\begin{eqnarray}\label{eq:15789chain1}
\begin{split}
    &\Fer(15)\\
    \hookleftconf \quad & \Fer(1)^{\psi_1} \oplus \Fer(5)^{\psi_5} \oplus \Fer(9)^{\psi_9}\\
    \hookleftconf \quad & \widehat{\mathfrak{u}}(1)_1^{\psi_1} \oplus \widehat{\mathfrak{u}}(1)_5^{\psi_5} \oplus \widehat{\mathfrak{su}}(5)_1^{\psi_5} \oplus \widehat{\mathfrak{u}}(1)_9^{\psi_9} \oplus \widehat{\mathfrak{su}}(9)_1^{\psi_9} \\
    \hookleftconf \quad & \widehat{\mathfrak{su}}(2)_{140} \oplus T \oplus \widehat{\mathfrak{u}}(1)_{855} \oplus \widehat{\mathfrak{u}}(1)_{90}\oplus \widehat{\mathfrak{u}}(1)_{1710}
\end{split}
\end{eqnarray}
The first two conformal embeddings are standard. The third one is a composition of several conformal embeddings. Firstly, we use the principal embedding $\widehat{\mathfrak{su}}(5)_1^{\psi_5} \hookleftconf \widehat{\mathfrak{su}}(2)_{20}\oplus T_1$ with fundamental of $\mathfrak{su}(5)$ restricts to spin $2$ of $\mathfrak{su}(2)$, and $\widehat{\mathfrak{su}}(9)_1^{\psi_9} \hookleftconf \widehat{\mathfrak{su}}(2)_{120}\oplus T_2$ with fundamental of $\mathfrak{su}(9)$ restricts to spin $4$ of $\mathfrak{su}(2)$. Secondly, we use the diagonal embedding $\widehat{\mathfrak{su}}(2)_{20} \oplus \widehat{\mathfrak{su}}(2)_{120} \hookleftconf \widehat{\mathfrak{su}}(2)_{140} \oplus T_3$. Then $T= T_1 \oplus T_2 \oplus T_3$, 
\begin{eqnarray}
    T = \frac{\widehat{\mathfrak{su}}(5)_1}{\widehat{\mathfrak{su}}(2)_{20}} \oplus \frac{\widehat{\mathfrak{su}}(9)_1}{\widehat{\mathfrak{su}}(2)_{120}} \oplus \frac{\widehat{\mathfrak{su}}(2)_{20} \oplus \widehat{\mathfrak{su}}(2)_{120}}{\widehat{\mathfrak{su}}(2)_{140}}
\end{eqnarray}
Lastly, we also use the conformal embedding of abelian factors $\widehat{\mathfrak{u}}(1)_1^{\psi_1} \oplus \widehat{\mathfrak{u}}(1)_5^{\psi_5}\oplus \widehat{\mathfrak{u}}(1)_9^{\psi_9} \hookleftconf  \widehat{\mathfrak{u}}(1)_{855} \oplus \widehat{\mathfrak{u}}(1)_{90}\oplus \widehat{\mathfrak{u}}(1)_{1710}$. The currents in $\widehat{\mathfrak{u}}(1)_{90}\oplus \widehat{\mathfrak{u}}(1)_{1710}$ are
\begin{equation}
\begin{split}
    J_1=6\psi_1^\dagger \psi_1 + 3 \psi_5^\dagger \psi_5 - \psi_9^\dagger \psi_9, \qquad  J_2= 32\psi_1^\dagger \psi_1 -11 \psi_5^\dagger \psi_5 + 3 \psi_9^\dagger \psi_9
\end{split}
\end{equation}
So the coset chiral algebra $S_1$ is 
\begin{eqnarray}
    S_1= T \oplus \widehat{\mathfrak{u}}(1)_{90}\oplus \widehat{\mathfrak{u}}(1)_{1710}
\end{eqnarray}

For the out-going fermions, 
\begin{eqnarray}
    \begin{split}
        &\Fer(15)\\
    \hookleftconf \quad & \Fer(7)^{\chi_7} \oplus \Fer(8)^{\chi_8}\\
    \hookleftconf \quad & \widehat{\mathfrak{u}}(1)_7^{\chi_7} \oplus  \widehat{\mathfrak{su}}(7)_1^{\chi_7} \oplus \widehat{\mathfrak{u}}(1)_8^{\chi_8} \oplus \widehat{\mathfrak{su}}(8)_1^{\chi_8} \\
    \hookleftconf \quad & \widehat{\mathfrak{su}}(2)_{140} \oplus \widetilde{T} \oplus \widehat{\mathfrak{u}}(1)_{855} \oplus \widehat{\mathfrak{u}}(1)_{47880}
    \end{split}
\end{eqnarray}
The first two conformal embeddings are again standard. The third one, likewise, consists of several conformal embeddings. Firstly, we use the principal embedding $\widehat{\mathfrak{su}}(7)_1^{\chi_7} \hookleftconf \widehat{\mathfrak{su}}(2)_{56} \oplus \widetilde{T}_1$ with fundamental of $\mathfrak{su}(7)$ restricts to spin $3$ of $\mathfrak{su}(2)$, and $\widehat{\mathfrak{su}}(8)_1^{\chi_8} \hookleftconf \widehat{\mathfrak{su}}(2)_{84}\oplus \widetilde{T}_2$ with fundamental of $\mathfrak{su}(8)$ restricts to spin $\frac72$ of $\mathfrak{su}(2)$. Secondly, we use the diagonal embedding $\widehat{\mathfrak{su}}(2)_{56} \oplus \widehat{\mathfrak{su}}(2)_{84} \hookleftconf \widehat{\mathfrak{su}}(2)_{140} \oplus \widetilde{T}_3$. Then $\widetilde{T}= \widetilde{T}_1 \oplus \widetilde{T}_2 \oplus \widetilde{T}_3$,
\begin{eqnarray}
    \widetilde{T} = \frac{\widehat{\mathfrak{su}}(7)_1}{\widehat{\mathfrak{su}}(2)_{56}} \oplus \frac{\widehat{\mathfrak{su}}(8)_1}{\widehat{\mathfrak{su}}(2)_{84}} \oplus \frac{\widehat{\mathfrak{su}}(2)_{56} \oplus \widehat{\mathfrak{su}}(2)_{84}}{\widehat{\mathfrak{su}}(2)_{140}}
\end{eqnarray}
Lastly, we also use the conformal embedding of abelian factors $\widehat{\mathfrak{u}}(1)_7^{\chi_7}\oplus \widehat{\mathfrak{u}}(1)_8^{\chi_8} \hookleftconf \widehat{\mathfrak{u}}(1)_{855} \oplus \widehat{\mathfrak{u}}(1)_{47880}$. The current in $\widehat{\mathfrak{u}}(1)_{47880}$ is 
\begin{equation}
\begin{split}
    \widetilde{J} =64\chi_7^\dagger \chi_7 -49 \chi_8^\dagger \chi_8 
\end{split}
\end{equation}
So the coset chiral algebra $S_2$ is 
\begin{eqnarray}
    S_2= \widetilde{T} \oplus \widehat{\mathfrak{u}}(1)_{47880}
\end{eqnarray}

To construct the intertwining defect and solve the scattering problem, we need to find a zigzag chain connecting $S_1$ and $S_2$, as \eqref{eq:orbifoldequiv}. So far, we haven't found such a chain yet, and it would be interesting to construct it in the future.

\section{Outlook}\label{sec:outlook}

We conclude this paper with a few future directions.

\paragraph{Scattering probabilities:} 
The scattering probabilities are fixed by the data of the bulk TQFT and its topological boundaries, and can be computed analytically. For $\text{QED}(\square,\overline\square)$, they only depend on the crossed data of $SU(N_f)_n$ with respect to charge conjugation. For instance, for $N_f=n=3$, there are two twist defects in the $SU(3)_3$ theory, which we denote by $\sigma_{\frac12}$ and $\sigma_{\frac32}$. We used AI to compute the $\bZ_2$-crossed $F$- and $R$-symbols and find the probabilities
\begin{eqnarray}
    P_{\rm S}=\frac{3+\sqrt{3}}{6}\approx 0.789,\qquad P_{\rm A}= \frac{3-\sqrt{3}}{6}\approx 0.211
\end{eqnarray}
for $\sigma_{\mathfrak{su}}=\sigma_{\frac12}$, and swapped for $\sigma_{\mathfrak{su}}=\sigma_{\frac32}$. It would be nice to either find a simpler computation suitable for human understanding, or alternatively formalize the computation, and extend the result to general $(N_f,n)$. The concrete numerical data also offers a theoretical prediction that can be independently checked via numerical simulations (for instance \cite{Ueda:2026wvc,Ueda:2025ecm}) or experiments (for instance anyonic interferometry \cite{Bonderson:2007ci}).

\paragraph{Zigzag chains:}
The most immediate open problem is to find the zigzag chain for the 1-5-7-8-9 model. Finding the chain would determine the out-going states, and would provide a non-trivial test of the orbifold equivalence conjecture \cite{Moller:2024xtt}. 

More generally, it would be desirable to have a systematic method to construct zigzag chains, or common conformal subalgebras, for coset chiral algebras of this type. The other maximally chiral QED models in Table \ref{tab:chiralqed} provide natural test cases.

\paragraph{Symmetric mass generation:} It would be interesting to understand whether our boundary conditions can be reached by some particular half-space symmetric mass generation, and whether the zigzag chains, or the spectators, suggest economical sets of additional degrees of freedom for symmetric mass generation \cite{Tong:2021phe,Wang:2022ucy,BoyleSmith:2022duw,Razamat:2020kyf,Zeng:2022grc}.

\paragraph{Selection of the boundary condition:}
In examples, the intertwining defect is not unique, and they can differ in their scattering. Which of them is realized by a given monopole presumably depends on the UV completion of the monopole core. It would be interesting to determine this, and to study the boundary renormalization group flows between different boundary conditions, which are constrained by the $g$-theorem \cite{Affleck:1991tk,Friedan:2003yc}.

\paragraph{Gauge field dynamics and four dimensions:}
Throughout this work, we treated the $U(1)$ gauge field as a background, and kept only the lowest partial waves. 

More generally, lifting our results to the four-dimensional scattering requires including the higher partial waves and the gauge field couplings, and understanding the fate of the two-dimensional non-invertible lines in the presence of the 't Hooft line. In two dimensions, the out-going states are not created by local operators, and it remains to be understood how they are realized in four dimensions. 

Finally, fermion masses break the chiral symmetries explicitly, and it would be interesting to understand how the categorical scattering is modified when the fermions are massive.

\section*{Acknowledgments}
We thank Andrea Antinucci, Philip Boyle Smith, Clay Cordova, Giovanni Galati, Diego Garc\'ia-Sep\'ulveda, Maxwell L.Velasquez Cotini Hutt, Zohar Komargodski, Sakura Schafer-Nameki, Yifan Wang, Yi Zhang for valuable discussions.
K.O.~is supported by JST PRESTO Grant Number JPMJPR26A3, Japan,
JSPS KAKENHI Grant-in-Aid No.~24K00522, and RIKEN ECL program.
K.O.~also acknowledges support by the Simons Foundation Grant \#888984
(Simons Collaboration on Global Categorical Symmetries).
M.W.~is supported by a Grant-in-Aid for JSPS Fellows No.~22KJ1777 and by a Grant-in-Aid for Early-Career Scientists No.~25K17387.
Y.Z. is supported by NSFC
grant No.12505093 and the starting funds from University of Chinese Academy of Sciences
(UCAS) and from the Kavli Institute for Theoretical Sciences (KITS).

We have used large language models (Claude, ChatGPT and Gemini) for preparing figures, correcting grammar, and performing and checking calculations.

\appendix

\section{Probability Interpretation}
\label{app:prob}

In this appendix we prove \eqref{eq:sumto1}. The idea is to compute the norm of both sides of \eqref{eq:Daction} in the sandwich, by gluing each configuration to its reflection.

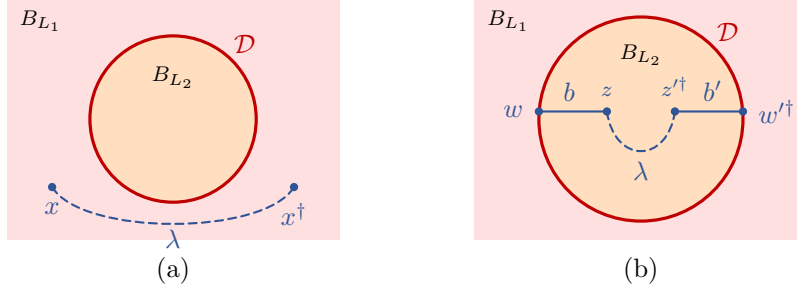
\begin{figure}
    \centering
    \begin{tikzpicture}[flat,scale=1.0,color=black]
        \fill[red!12] (-2.2,-1.6) rectangle (2.2,1.6);
        \fill[orange!25] (0,0) circle (1.1);
        \draw[dline] (0,0) circle (1.1);
        \node[PPTdarkred] at (0.95,1.0) {$\mathcal{D}$};
        \node[font=\scriptsize] at (0,0.55) {$B_{L_2}$};
        \node[font=\scriptsize] at (-1.75,1.3) {$B_{L_1}$};
        \node[pt,PPTdarkblue,label={[PPTdarkblue]below:$x$}] at (-1.6,-0.9) {};
        \node[pt,PPTdarkblue,label={[PPTdarkblue]below:$x^\dagger$}] at (1.6,-0.9) {};
        \draw[oline,densely dashed] (-1.6,-0.9) .. controls (-1.2,-1.55) and (1.2,-1.55) .. (1.6,-0.9);
        \node[PPTdarkblue,below] at (0,-1.32) {$\lambda$};
        \node at (0,-2.0) {(a)};
    \end{tikzpicture}
    \qquad\qquad
    \begin{tikzpicture}[flat,scale=1.0,color=black]
        \fill[red!12] (-2.2,-1.6) rectangle (2.2,1.6);
        \fill[orange!25] (0,0) circle (1.35);
        \draw[dline] (0,0) circle (1.35);
        \node[PPTdarkred] at (1.15,1.15) {$\mathcal{D}$};
        \node[font=\scriptsize] at (0,0.85) {$B_{L_2}$};
        \node[font=\scriptsize] at (-1.75,1.3) {$B_{L_1}$};
        \draw[oline] (-1.35,0.1) -- (-0.45,0.1);
        \draw[oline] (0.45,0.1) -- (1.35,0.1);
        \node[PPTdarkblue,above] at (-0.95,0.1) {$b$};
        \node[PPTdarkblue,above] at (0.95,0.1) {$b'$};
        \node[pt,PPTdarkblue,label={[PPTdarkblue]left:$w$}] at (-1.35,0.1) {};
        \node[pt,PPTdarkblue,label={[PPTdarkblue]right:$w'^\dagger$}] at (1.35,0.1) {};
        \node[pt,PPTdarkblue,label={[PPTdarkblue,font=\scriptsize]above:$z$}] at (-0.45,0.1) {};
        \node[pt,PPTdarkblue,label={[PPTdarkblue,font=\scriptsize]above:$z'^\dagger$}] at (0.45,0.1) {};
        \draw[oline,densely dashed] (-0.45,0.1) .. controls (-0.35,-0.6) and (0.35,-0.6) .. (0.45,0.1);
        \node[PPTdarkblue,below] at (0,-0.42) {$\lambda$};
        \node at (0,-2.0) {(b)};
    \end{tikzpicture}
    \caption{The topological boundary of the glued sandwich, a sphere drawn in the plane, divided by the loop $\mathcal{D}$ into $B_{L_1}$ (outside) and $B_{L_2}$ (inside). Dashed arcs run through the bulk, after the chiral junctions $\widetilde x$, $\widetilde x^\dagger$ have been contracted. (a) Norm of the in-coming state. (b) Inner product of two out-going states $(b,w,z)$ and $(b',w',z')$.}
    \label{fig:normsandwich}
\end{figure}

We first use the state-operator correspondence to map the local/twisted operator two sides of  \eqref{eq:Daction} into states, and it becomes 
\begin{eqnarray}\label{eq:Dactionstate}
    |\psi\rangle=\sum_{b,z,w}C_{bzw}\,|b,w,z\rangle,\qquad C_{bzw}=\sum_y\sqrt{\frac{d_b}{d_\lambda}}\,\frac{1}{d_{\mathcal{D}}\sqrt{S_{11}}}\ {^{B_{L_1}B_{L_2}}\widetilde{\Psi}_{\mathcal{D}\mathcal{D}(\bar\lambda xy)}^{(1)(\bar bw)}}.
\end{eqnarray}
where $|\psi\rangle$ and $|b,w,z\rangle$ denote the configurations on the two sides. 
The inner product of two such states is obtained by gluing one to the reflection of the other. This produces a closed sandwich whose topological boundary is a sphere, divided by the loop $\mathcal{D}$ into a $B_{L_1}$ and a $B_{L_2}$ hemisphere (Figure \ref{fig:normsandwich}). In every case below, the chiral junctions $\widetilde x$ and $\widetilde x^\dagger$ give the chiral two-point function of $\psi$, which is normalized to one, and leave a bulk arc of $\lambda$ between two points of the topological boundary.

The arc of $\lambda$ ends at $x$ and $x^\dagger$ on $B_{L_1}$ (Figure \ref{fig:normsandwich}(a)). Removing it gives $\sqrt{d_\lambda}$. Removing the empty loop $\mathcal{D}$ then gives $d_{\mathcal{D}}$:
\begin{eqnarray}\label{eq:normpsi}
    \langle\psi|\psi\rangle=d_{\mathcal{D}}\sqrt{d_\lambda}.
\end{eqnarray}

Glue $|b,w,z\rangle$ to the reflection of $|b',w',z'\rangle$ (Figure \ref{fig:normsandwich}(b)). The arc of $\lambda$ now ends at $z$ and $z'^\dagger$ on $B_{L_2}$, where it meets $b$ and $b'$.
\begin{itemize}
    \item Removing the arc gives $\delta_{bb'}\delta_{zz'}\sqrt{d_\lambda/d_b}$, and leaves a line $b$ from $w$ to $w'^\dagger$.
    \item This line is a chord of the loop $\mathcal{D}$. The resulting theta graph of $\mathcal{D}$, $\mathcal{D}^\dagger$ and $b$ gives $\delta_{ww'}\sqrt{d_{\mathcal{D}}^2d_b}$.
\end{itemize}
Hence
\begin{eqnarray}\label{eq:normout}
    \langle b',w',z'|b,w,z\rangle=\delta_{bb'}\,\delta_{ww'}\,\delta_{zz'}\,\sqrt{\frac{d_\lambda}{d_b}}\,\sqrt{d_{\mathcal{D}}^2d_b}=\delta_{bb'}\,\delta_{ww'}\,\delta_{zz'}\ d_{\mathcal{D}}\sqrt{d_\lambda}.
\end{eqnarray}

The out-going states $|b,w,z\rangle$ are mutually orthogonal, and each has the same norm \eqref{eq:normout} as the in-coming state \eqref{eq:normpsi}. Taking the norm of \eqref{eq:Dactionstate},
\begin{eqnarray}
    d_{\mathcal{D}}\sqrt{d_\lambda}=\langle\psi|\psi\rangle=\sum_{b,z,w}|C_{bzw}|^2\,d_{\mathcal{D}}\sqrt{d_\lambda},
\end{eqnarray}
so $\sum_{b,z,w}|C_{bzw}|^2=1$, which is \eqref{eq:sumto1}. Since the $|b,w,z\rangle$ are orthogonal, $|C_{bzw}|^2$ is the probability of the out-going configuration $(b,w,z)$, and $P_b=\sum_{zw}|C_{bzw}|^2$ is the probability of the $b$-twisted sector.

\section{Anyon condensation in Abelian Chern-Simons Theories}\label{app:Anyon condensation between K matrix U(1) Chern-Simons TQFTs}

In Section \ref{sec:susu}, \ref{sec:chiralqed} and $\ref{sec:standardmodel}$, we encounter the problem of the following type: given two K matrix abelian Chern-Simons TQFT, construct a larger K matrix CS theory $\mathcal{T}$ together with two condensable algebras $\mathcal{A}_1$ and $\mathcal{A}_2$ in $\mathcal{T}$, such the condensing the $\mathcal{A}_i$, $i=1,2$, in the theory $\mathcal{T}$ gives raise to the given CS theory respectively.

In this appendix, we summarize some results on the condensation between two abelian CS theories, which were discussed in \cite{Hung:2013qpa}.

Consider two rank $r$ abelian Chern-Simons (CS) TQFT described by $K,\widetilde{K}\in \mathbb{Z}^{r \times r}$,
\begin{align}
    \frac{1}{4\pi}a^TKda\quad \text{and} \quad \frac{1}{4\pi}a^T\widetilde{K}da.
\end{align}
The $\widetilde{K}$ CS theory can be obtained by condensing an algebra $\mathcal{A}$ in $K$ CS theory
\begin{align}
    \widetilde{K} \text{ CS theory} = (K \text{ CS theory})_\mathcal{A}^0
\end{align}
if there exists $P\in \bZ^{r\times r}$ with $\det P\neq 0$ such that 
\begin{align}\label{eq:P matrix}
    K=P^T\widetilde{K}P
\end{align}
and the lines to condense are
\begin{align}\label{eq:algebra from P}
    \mathcal{A}=\bigoplus _{x\in \mathbb{Z}^r/P \mathbb{Z}^r}KP^{-1}x.
\end{align}
The lifting functor (i.e. adjoint of the condensation functor) is given by
\begin{align}\label{eq:lift fun of abelian condensation}
\begin{split}
    L:\quad  U(1)^r_{\widetilde{K}} &\to U(1)^r_K\\
    n &\mapsto P^Tn\otimes \mathcal{A} = \bigoplus _{x\in \mathbb{Z}^r/P \mathbb{Z}^r} P^T(n+\widetilde{K}x)
\end{split}
\end{align}

\begin{proof}
    Recall that the lines in a $K$ CS theory are labelled by $n\in \mathbb{Z}^r/K\mathbb{Z}^r$, with fusion $n\otimes m=n+m$, braiding $B(n,m)=n^TK^{-1}m$ and topological spin $h(n)=\frac{1}{2}B(n,n)$.

    For any $\widetilde{K}$ and $P$ satisfying \eqref{eq:P matrix}, we define a map
    \begin{align}
    \begin{split}
        a:\quad \mathbb{Z}^r/P\mathbb{Z}^r&\to \mathbb{Z}^r/K\mathbb{Z}^r\\
        x&\to a(x)=KP^{-1}x.
    \end{split}
    \end{align}
    It can be checked that the definition is consistent with the quotient on both sides. For any $x\in \mathbb{Z}^r/P\mathbb{Z}^r$, the spin of $a(x)$ is in $\frac12\mathbb{Z}$,
    \begin{align}
        h(a(x))=\frac12 (KP^{-1}x)^T K^{-1}(KP^{-1}x)=\frac12 (P^{-1}x)^TK(P^{-1}x)=\frac12 x^T\widetilde{K}x\in \frac12 \mathbb{Z},
    \end{align}
    where we used \eqref{eq:P matrix} and $K^T=K$. And $a(x)$'s have trivial mutual braiding,
    \begin{align}
        B(a(x),a(y))=(KP^{-1}x)^T K^{-1}(KP^{-1}y)=(P^{-1}x)^TK(P^{-1}y)=x^T\widetilde{K}y\in \mathbb{Z}.
    \end{align}
    Hence, the image of $a$, which is \eqref{eq:algebra from P} $\mathcal{A}=\bigoplus _x a(x)$, forms a condensable algebra in $K$ CS theory, with dimension $\dim \mathcal{A}=|\det P|$.

    To identity the TQFT after condensing $\mathcal{A}$, we only need to find those lines $n\in \mathbb{Z}^r/K\mathbb{Z}^r$ which braid trivially with any $a(x)$ in $\mathcal{A}$. Note that in abelian CS TQFT, there are no fix point under $\mathcal{A}$, so no line splits. Requiring
    \begin{align}
        B(n,a(x))=n^TP^{-1}x\in \mathbb{Z},\quad \text{for any}\quad x\in \mathbb{Z}^r/P\mathbb{Z}^r,
    \end{align}
    means that $n\in P^T\mathbb{Z}$. Since $n$ is defined module $K\mathbb{Z}^r$, the lines survive under condensing $\mathcal{A}$ are $P^T\mathbb{Z}^r/K\mathbb{Z}^r$. By rewriting $\mathcal{A}$ using $a(x)=KP^{-1}x=P^T\widetilde{K}x$ as
    \begin{align}
        \mathcal{A}=P^T\widetilde{K}\mathbb{Z}^r/K\mathbb{Z}^r,
    \end{align}
    we conclude that the condensed category of lines are 
    \begin{align}
        \frac{P^T\mathbb{Z}^r/K\mathbb{Z}^r}{P^T\widetilde{K}\mathbb{Z}^r/K\mathbb{Z}^r}\simeq \mathbb{Z}^r/\widetilde{K}\mathbb{Z}^r,
    \end{align}
    where in the last isomorphism we just do a relabeling. If we collect the pre-image for any line in $\mathbb{Z}^r/\widetilde{K}\mathbb{Z}^r$, we get the lifting functor \eqref{eq:lift fun of abelian condensation}. Finally, by checking the braiding, the condensed category do is $\widetilde{K}$ CS theory.
\end{proof}

\section{More on Scattering in $\{1^5,3,(-2)^4\}$ Chiral QED}
\label{app:onechannel}

In this appendix, we provide more details on why the out-going excitation belongs to only one twisted sector among $b_\psi$ and $b_\psi'$.

In Section \ref{sec:cqedscattering}, we find that $E_\psi\in \widetilde{L}_2$, satisfying $\bZ_2$ fusion rule. We also note that $E_\psi\in \widetilde{L}_1$, since $\lambda_\psi\in \widetilde{L}_1$ and $E_\psi \in \lambda_\psi \otimes \overline{\lambda}_\psi$. Hence, $E_\psi$ is shared by both Lagrangian algebras, and can terminate topological on both $B_{\widetilde{L}_1}$ and $B_{\widetilde{L}_2}$. Bring $E_\psi$ across the topological line junction $\mathcal{D}$ produces a phase $\chi_{\mathcal{D}}(E_\psi)$, 
\begin{eqnarray}\label{eq:chiD}
    \figEchi[0.9]{0.5}{E_\psi}\quad=\quad\chi_{\mathcal D}(E_\psi)\quad\figEchi[0.9]{1.5}{E_\psi}
\end{eqnarray}

\begin{figure}
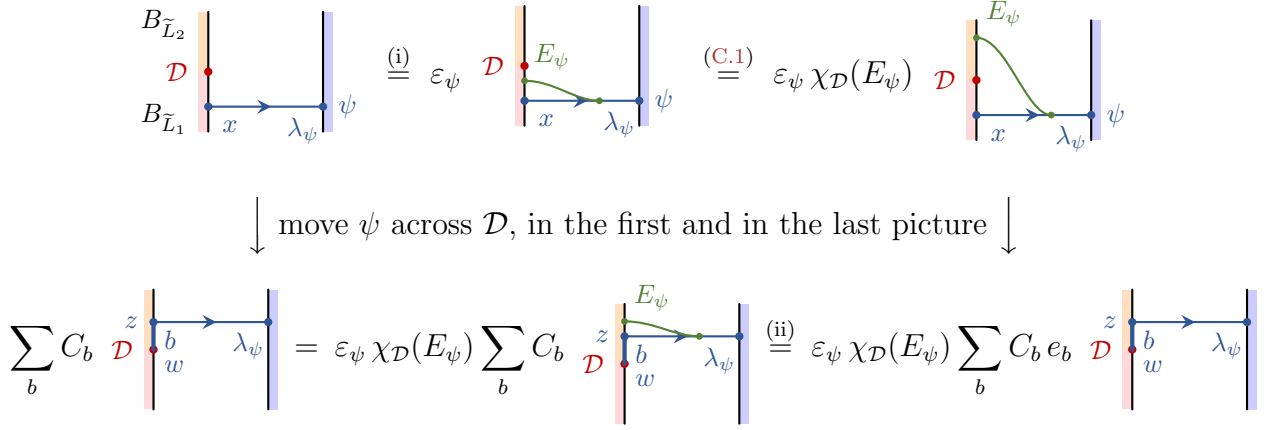

    \centering
    \[
        \figEargIn[0.66]{}{$B_{\widetilde L_1}$}{$B_{\widetilde L_2}$}
        \ \overset{\rm (i)}{=}\ \varepsilon_\psi\
        \figEargIn[0.66]{\swEarc{0.5}{0.9}}{}{}
        \ \overset{\eqref{eq:chiD}}{=}\ \varepsilon_\psi\,\chi_{\mathcal D}(E_\psi)\
        \figEargIn[0.66]{\swEarc{0.5}{2.05}}{}{}
    \]
    \[
        \Big\downarrow\ \text{move $\psi$ across $\mathcal D$, in the first and in the last picture}\ \Big\downarrow
    \]
    \[
        \sum_b C_b\ \figEargOut[0.66]{}{}{}
        \ =\ \varepsilon_\psi\,\chi_{\mathcal D}(E_\psi)\sum_b C_b\ \figEargOut[0.66]{\swEarc{1.75}{2.05}}{}{}
        \ \overset{\rm (ii)}{=}\ \varepsilon_\psi\,\chi_{\mathcal D}(E_\psi)\sum_b C_b\,e_b\ \figEargOut[0.66]{}{}{}
    \]
    \caption{Top: the end point of $\lambda_\psi$ on $B_{\widetilde L_1}$ absorbs an $E_\psi$ arc up to the number $\varepsilon_\psi$ (fact (i)), and the foot of the arc can be moved across $\mathcal D$ at the cost of $\chi_{\mathcal D}(E_\psi)$, by \eqref{eq:chiD}. Bottom: moving $\psi$ across $\mathcal D$ by \eqref{eq:Daction}, the arc is carried along. On the out-going configuration in the $b$-twisted sector it acts by $e_b$ (fact (ii)). Here $C_b$ stands for the coefficients of \eqref{eq:Daction}, summed over the junctions $z$ and $w$.}
    \label{fig:cqedEarg}
\end{figure}

We need two further facts.
\begin{enumerate}
    \item[(i)] On the in-coming side, $\lambda_\psi$ appears once in $\widetilde L_1$, so its end point $x$ on $B_{\widetilde L_1}$ is unique up to normalization. An $E_\psi$ arc between $\lambda_\psi$ and $B_{\widetilde L_1}$ produces another end point of $\lambda_\psi$, which must therefore be proportional to $x$. We write the proportionality constant as $\varepsilon_\psi\neq0$.
    \item[(ii)] On the out-going side, the same arc between $\lambda_\psi$ and $B_{\widetilde L_2}$ acts on the out-going configuration of \eqref{eq:Daction} as the endomorphism $e_\psi$ restricted to $b$, i.e.\ as the number $e_b$, with $e_{b_\psi}=+1$ and $e_{b'_\psi}=-1$.
\end{enumerate}
In pictures, with the line $b$ running from the junction $w$ on $\mathcal D$ to the junction $z$,
\begin{eqnarray}
    {\rm (i)}:&&\figEargIn[0.75]{\swEarc{0.5}{0.9}}{}{}\ =\ \varepsilon_\psi\ \figEargIn[0.75]{}{}{}, \nonumber\\[1ex]
    {\rm (ii)}:&&\figEargOut[0.75]{\swEarc{1.75}{2.05}}{}{}\ =\ e_b\ \figEargOut[0.75]{}{}{} .\label{eq:Earcs}
\end{eqnarray}
These pictures, and those of Figure \ref{fig:cqedEarg}, are cross-sections of the sandwich of Figure \ref{fig:cqedclub} (bottom) through the bulk line $\lambda_\psi$. The topological boundary is on the left, with $B_{\widetilde L_1}$ (red) below and $B_{\widetilde L_2}$ (orange) above the junction $\mathcal{D}$, which appears as a point. The chiral boundary $B_{\widehat{\mathfrak g}\oplus W}$ is on the right.

The argument is shown in Figure \ref{fig:cqedEarg}. We start from the in-coming configuration, with $\psi$ below $\mathcal{D}$. By (i), we may attach an $E_\psi$ arc to $\lambda_\psi$, at the cost of the factor $\varepsilon_\psi$. By \eqref{eq:chiD}, the foot of the arc can be moved across $\mathcal{D}$, at the cost of $\chi_{\mathcal D}(E_\psi)$. Now we move $\psi$ across $\mathcal{D}$ by \eqref{eq:Daction}, both in the first and in the last picture. The derivation of \eqref{eq:Daction} only involves the neighborhood of the end point of $\lambda_\psi$ and of $\mathcal{D}$. The arc is attached to $\lambda_\psi$ away from the boundary, and its foot is already above $\mathcal{D}$, so it is simply carried along. Finally, by (ii), the arc acts on the $b$-twisted sector by $e_b$. Comparing the two rows, and using that the out-going configurations with different $b$ are linearly independent, we find
\begin{eqnarray}\label{eq:Eselect}
    C_b\,\big(1-\varepsilon_\psi\,\chi_{\mathcal D}(E_\psi)\,e_b\big)=0,\qquad b=b_\psi,\,b'_\psi .
\end{eqnarray}
Since $e_{b_\psi}=-e_{b'_\psi}$, at most one of the two amplitudes is non-zero. Since the probabilities add up to one, exactly one is non-zero: the one with $\varepsilon_\psi\,\chi_{\mathcal D}(E_\psi)\,e_b=1$. As $e_b=\pm1$ and $\chi_{\mathcal D}(E_\psi)=\pm1$, this also shows that $\varepsilon_\psi=\pm1$, and the selected line has $e_b=\varepsilon_\psi\,\chi_{\mathcal D}(E_\psi)$. Therefore we obtain \eqref{eq:cqedscat}, where $b_\psi^{(\mathcal{D})}=b_\psi$ if $\varepsilon_\psi\chi_{\mathcal D}(E_\psi)=+1$, and $b_\psi^{(\mathcal{D})}=b'_\psi$ if $\varepsilon_\psi\chi_{\mathcal D}(E_\psi)=-1$. Simple defects $\mathcal D$ and $\mathcal D'$ with $\chi_{\mathcal D}(E_\psi)=-\chi_{\mathcal D'}(E_\psi)$ send $\psi$ into different lines.

This has a simple meaning in terms of operators. Sliding the junction of the arc in (i) along $\lambda_\psi$ to the chiral boundary brings $\mathcal{O}_{E_\psi}$ next to $\psi$. Hence $\varepsilon_\psi$ is the coefficient of $\psi$ in the operator product $\mathcal{O}_{E_\psi}\times\psi$, projected onto the sector of $\psi$, in a normalization fixed by the sandwich. Likewise, $e_b$ is the corresponding coefficient for the out-going excitation in the $b$-twisted sector, and it has opposite signs for $b_\psi$ and $b'_\psi$. Since $\mathcal{D}$ commutes with $\mathcal{O}_{E_\psi}$ up to $\chi_{\mathcal D}(E_\psi)$, it must send $\psi$ into the twisted sector in which this coefficient is $\varepsilon_\psi\chi_{\mathcal D}(E_\psi)$. The operator $\mathcal{O}_{E_\psi}$ thus acts as a $\bZ_2$-valued label that distinguishes $b_\psi$ from $b'_\psi$, and the defect fixes the value of this label.

\section{Folding Relative Chiral CFT to Absolute Non-chiral CFT}\label{app:More on Gapping Spectators in the Sandwich}
\begin{figure}
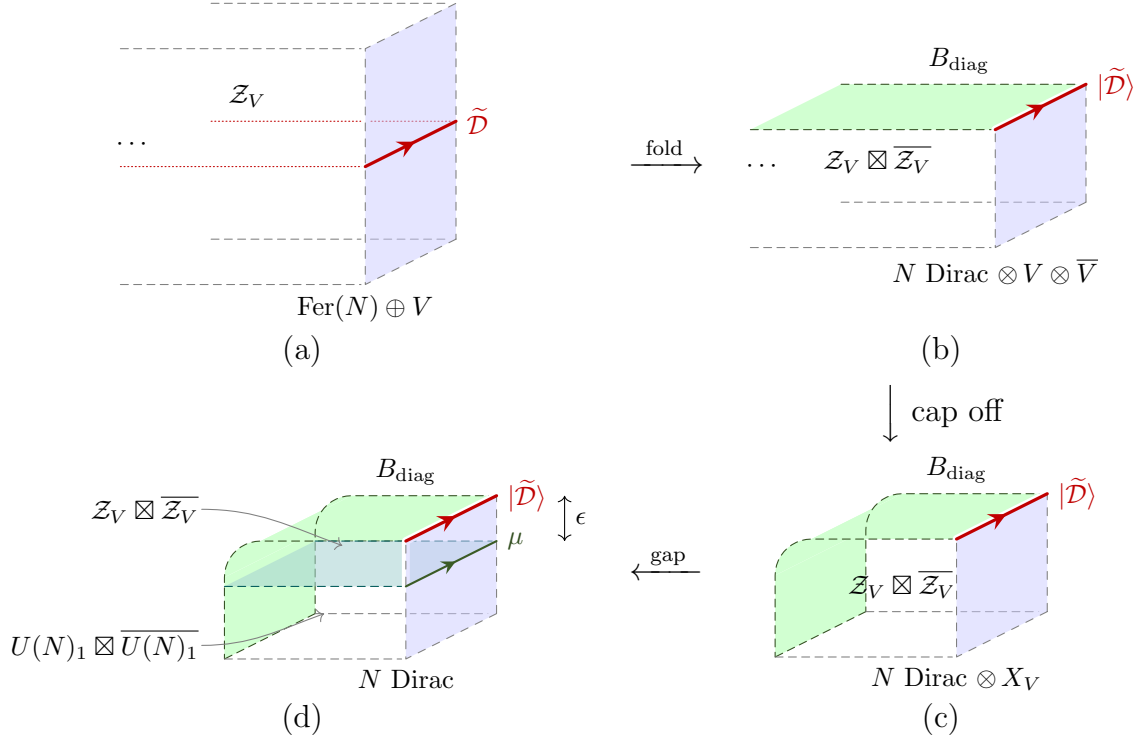

    \centering
    \begin{tabular}{ccc}
    \figFoldA & $\xrightarrow{\ \text{fold}\ }$ & \figFoldB \\[0.5ex]
    (a) & & (b) \\[1ex]
    & & $\Big\downarrow$\ \text{cap off} \\[1ex]
    \figFoldD & $\xleftarrow{\ \text{gap}\ }$ & \figFoldC \\[0.5ex]
    (d) & & (c)
    \end{tabular}
    \caption{Folding the enlarged theory along $\widetilde{\mathcal{D}}$, with $\mathcal{Z}_V\equiv U(N)_1\boxtimes\Rep(V)$. (a) After shrinking the club sandwich of Figure \ref{fig:spectator}, $\Fer(N)\oplus V$ lives on the chiral boundary of $\mathcal{Z}_V$, which fills the half-space on the left. (b) Folding along the plane through $\widetilde{\mathcal{D}}$ (dotted in (a)) gives $\mathcal{Z}_V\boxtimes\overline{\mathcal{Z}_V}$ on a quarter-space. It is bounded by the physical boundary, which carries the $N$ Dirac fermions and $V\otimes\overline{V}$, and by the folding boundary $B_{\rm diag}$, which condenses the canonical Lagrangian algebra $\bigoplus_\mu\mu\boxtimes\bar\mu$. The quarter-space still extends to infinity on the left. (c) Capping it off by a second diagonal boundary, or equivalently bending $B_{\rm diag}$ toward the left, gives a slab. Shrinking the slab gives the $N$ Dirac fermions stacked with the diagonal RCFT $X_V$ on a half-plane, with the boundary condition $|\widetilde{\mathcal{D}}\rangle$ at the corner. (d) Gapping the spectators: the factor $\Rep(V)\boxtimes\overline{\Rep(V)}$ is confined to a thin tube along $|\widetilde{\mathcal{D}}\rangle$ by its diagonal boundary (blue), which meets the physical boundary along the Verlinde line $\mu$, i.e.\ along the Cardy boundary condition $\ket{\mu}$ of $X_V$. Outside the tube, only the invertible $U(N)_1\boxtimes\overline{U(N)_1}$ remains. Shrinking the slab gives Figure \ref{fig:gapping}.}
    \label{fig:folding}
\end{figure}

In this appendix, we provide the Figure \ref{fig:folding}, to understand folding the relative chiral CFT $\Fer(N)\oplus V$ to the absolute non-chiral CFT $N$ Dirac fermions $\oplus X_V$.

\section{TQFT and Branching Data}\label{app:branching data}
In this appendix, we list some well-known TQFTs, their conformal embeddings, together with their branching rules. These data are used in the condensable algebras in the paper.

\subsection{TQFTs}

We enumerate selected data for TQFTs to setup the notation. 

\begin{enumerate}
    \item $U(1)_{2n}$: The operators are labeled by $[m]$, $m=0,1,..., 2n-1$. The spin of $[m]$ is $h= \frac{m^2}{4n}$. The quantum dimension of $[m]$ is $d_{[m]}=1$. 
    \item $O(2)_{2n}$: The operators, spins and quantum dimensions are 
\begin{eqnarray}\label{eq:orblines}
    \begin{tabular}{|c|cccccc|}
    \hline
         $O(2)_{2n}$ &  $\mathbf 1$ & $j$ & $\phi^\pm_n$ & $\phi_m$, $1\leq m\leq n-1$ & $\sigma_{1,2}$ & $\tau_{1,2}$ \\
         \hline
        $h$ & $0$ & $1$ & $\frac n4$ & $\frac{m^2}{4n}$ & $\frac1{16}$ & $\frac9{16}$ \\
        $d$ & $1$ & $1$ & $1$ & $2$ & $\sqrt n$ & $\sqrt n$\\
        \hline
    \end{tabular}
\end{eqnarray}
$O(2)_{2n}$ is obtained from $U(1)_{2n}$ by gauging the $\bZ_2$ charge conjugation, acting on $[m]$ as $[m]\to [-m \text{ mod } 2n]$. 
The line $j$ is the quantum symmetry after gauging charge conjugation. $\phi_n^\pm$ originates from $[n]$ (which is $\bZ_2$ invariant), and the superscripts are swapped under fusing with $j$, $j\otimes \phi_n^\pm= \phi_n^\mp$. The line $\phi_m$ is the $\bZ_2$-orbit of $[m]$, i.e. $\phi_m \simeq [m] \oplus [2n-m]$. $\sigma_{1,2}$ and $\tau_{1,2}$ come from the $\bZ_2$ twist defect, where the subscripts are swapped under fusing $[m]$ for odd $m$ before gauging, and $\sigma_{1,2}= j \otimes \tau_{1,2}$. We fix the labels by the full braiding with $\phi_n^+$: the full braiding of $\phi^+_n$ with $\sigma_1$ and $\tau_1$ is $e^{-\pi i n/2}$, while that with $\sigma_2$ and $\tau_2$ is $-e^{-\pi i n/2}$. Equivalently, $\phi^+_n\otimes\sigma_1$ has spin $\frac1{16}$, while $\phi^+_n\otimes\sigma_2$ has spin $\frac9{16}$. This is consistent with the fusion above, since $\phi_m$ with odd $m$ braids with $\phi_n^+$ by $-1$. The topological lines also label the modules in the $\widehat{\mathfrak{u}}(1)_{2n}^{\bZ_2}$ chiral algebra.
    \item $SU(2)_k$: the operators are labeled by the isospin $j$ of the representation, $j=0, \frac12, 1, ..., \frac{k}{2}$. The spin of $j$ is $\frac{j(j+1)}{k+2}$, and the quantum dimension is $d_j= \frac{\sin(\pi(2j+1)/(k+2))}{\sin(\pi/(k+2))}$. More generally, for $SU(N)_k$, we label the operators by their Dynkin labels $[n_1, ..., n_{N-1}]$. In the special case $N=1$, $n_1= 2j$. 

    \item $Spin(5)_2$: 
    The operators, spins and quantum dimensions are 
    \begin{eqnarray} 
    \begin{tabular}{|c|cccccc|}
    \hline
         $Spin(5)_2$ & $\mathbf 1$ & $\mathbf{14}$ & $\mathbf 5$ & $\mathbf{10}$ & $\mathbf 4$ & $\mathbf{16}$ \\
         \hline
        $h$ & $0$ & $1$ & $\frac25$ & $\frac35$ & $\frac14$ & $\frac34$ \\
        $d$ & $1$ & $1$ & $2$ & $2$ & $\sqrt5$ & $\sqrt5$\\
        \hline
    \end{tabular}
\end{eqnarray}
where $\mathbf{14}$ is the symmetric traceless tensor.
    \item $\mathrm{Rep}(\mathsf{3SP})$: The operators, spins and quantum dimensions are
    \begin{eqnarray}
    \begin{tabular}{|c|cccccc|}
    \hline
         $\mathrm{Rep}(\mathsf{3SP})$ & $\mathbf{1}$ & $\epsilon$ & $\psi$ & $\psi^\dagger$ & $\sigma$ & $\sigma^
         \dagger$ \\
         \hline
        $h$ & $0$ & $\frac25$ & $\frac23$ & $\frac23$ & $\frac{1}{15}$ & $\frac{1}{15}$ \\
        $d$ & $1$ & $\varphi$ & $1$ & $1$ & $\varphi$ & $\varphi$\\
        \hline
    \end{tabular}
    \end{eqnarray}
    where $\varphi= \frac{1+\sqrt{5}}{2}$ is the golden ratio. 

    \item $M(6,5)=\Rep(\text{Vir}_{4/5})$: The operators, spins and quantum dimensions are
    \begin{equation}
    \begin{tabular}{|c|cccccccccc|}
    \hline
         $M(6,5)=\Rep(\text{Vir}_{4/5})$ & $\phi_{1,1}$ & $\phi_{1,2}$ & $\phi_{1,3}$ & $\phi_{1,4}$ & $\phi_{1,5}$ & $\phi_{2,1}$ & $\phi_{2,2}$ & $\phi_{2,3}$ & $\phi_{2,4}$ & $\phi_{2,5}$ \\
         \hline
        $h$ & $0$ & $\frac18$ & $\frac23$ & $\frac{13}{8}$ & $3$ & $\frac25$ & $\frac{1}{40}$ & $\frac{1}{15}$ & $\frac{21}{40}$ & $\frac75$ \\
        $d$ & $1$ & $\sqrt{3}$ & $2$ & $\sqrt{3}$ & $1$ & $\varphi$ & $\sqrt{3}\varphi$ & $2\varphi$ & $\sqrt{3}\varphi$ & $\varphi$\\
        \hline
    \end{tabular}
    \end{equation}
    where $\varphi= \frac{1+\sqrt{5}}{2}$ is the golden ratio. 
\end{enumerate}

\subsection{Branching Data}

\begin{enumerate}
    \item 
$\widehat{\mathfrak{u}}(N)_1\hookleftconf \widehat{\mathfrak{u}}(1)_N\oplus \widehat{\mathfrak{su}}(N)_1$
\begin{align}
    0\to \sum_{i=0}^{N-1}(i,\omega_i)
\end{align}

    \item 
$\widehat{\mathfrak{su}}(3)_1 \hookleftconf \widehat{\mathfrak{su}}(2)_4$
\begin{align}
\begin{split}
    [0,0]&\to [0]+[2],\\
    [1,0]&\to [1],\\
    [0,1]&\to [1],
\end{split}
\end{align}

    \item
$\widehat{\mathfrak{su}}(6)_1 \hookleftconf \widehat{\mathfrak{su}}(4)_2$
\begin{align}
\begin{split}
    [0,0,0,0,0]&\to [0,0,0]+[0,2,0],\\
    [1,0,0,0,0]&\to [0,1,0],\\
    [0,1,0,0,0]&\to [1,0,1],\\
    [0,0,1,0,0]&\to [2,0,0]+[0,0,2],\\
    [0,0,0,1,0]&\to [1,0,1],\\
    [0,0,0,0,1]&\to [0,1,0].
\end{split}
\end{align}

    \item
$\widehat{\mathfrak{su}}(8)_1\hookleftconf \widehat{\mathfrak{su}}(2)_4\oplus \widehat{\mathfrak{su}}(4)_2$
\begin{align}
\begin{split}
    0&\to (0,[0,0,0])+(1,[1,0,1])+(2,[0,2,0]),\\
    1&\to (1/2,[1,0,0])+(3/2,[0,1,1]),\\
    2&\to (0,[2,0,0])+(1,[0,1,0])+(2,[0,0,2]),\\
    3&\to (1/2,[1,1,0])+(3/2,[0,0,1]),\\
    4&\to (0,[0,2,0])+(1,[1,0,1])+(2,[0,0,0]),\\
    5&\to (1/2,[0,1,1])+(3/2,[1,0,0]),\\
    6&\to (0,[0,0,2])+(1,[0,1,0])+(2,[2,0,0]),\\
    7&\to (1/2,[0,0,1])+(3/2,[1,1,0]).
\end{split}
\end{align}

\item
$\widehat{\mathfrak{su}}(4)_2\hookleftconf \widehat{\mathfrak{so}}(5)_2\oplus \widehat{\mathfrak{u}}(1)^{\bZ_2}_{30}$, with the lines of $Spin(5)_2$ and $O(2)_{30}$ labeled as in the \ref{sec:cqedW},
\begin{align}
\begin{split}
    [0,0,0]&\to (\mathbf 1,\mathbf 1)+(\mathbf{14},j)+(\mathbf 5,\phi_6)+(\mathbf{10},\phi_{12}),\\
    [0,2,0]&\to (\mathbf 1,j)+(\mathbf{14},\mathbf 1)+(\mathbf 5,\phi_6)+(\mathbf{10},\phi_{12}),\\
    [2,0,0]&\to (\mathbf 1,\phi^-_{15})+(\mathbf{14},\phi^+_{15})+(\mathbf 5,\phi_9)+(\mathbf{10},\phi_3),\\
    [0,0,2]&\to (\mathbf 1,\phi^+_{15})+(\mathbf{14},\phi^-_{15})+(\mathbf 5,\phi_9)+(\mathbf{10},\phi_3),\\
    [0,1,0]&\to (\mathbf 1,\phi_5)+(\mathbf{14},\phi_5)+(\mathbf 5,\phi_1)+(\mathbf 5,\phi_{11})+(\mathbf{10},\phi_7)+(\mathbf{10},\phi_{13}),\\
    [1,0,1]&\to (\mathbf 1,\phi_{10})+(\mathbf{14},\phi_{10})+(\mathbf 5,\phi_4)+(\mathbf 5,\phi_{14})+(\mathbf{10},\phi_2)+(\mathbf{10},\phi_8),\\
    [1,0,0]&\to (\mathbf 4,\sigma_1)+(\mathbf{16},\tau_1),\\
    [0,0,1]&\to (\mathbf 4,\sigma_2)+(\mathbf{16},\tau_2),\\
    [0,1,1]&\to (\mathbf 4,\tau_1)+(\mathbf{16},\sigma_1),\\
    [1,1,0]&\to (\mathbf 4,\tau_2)+(\mathbf{16},\sigma_2).
\end{split}
\end{align}
The only other branching consistent with the modular data is obtained by the charge conjugation $[a,b,c]\to[c,b,a]$ of $\widehat{\mathfrak{su}}(4)_2$. Since $\sigma_{1,2}$ and $\tau_{1,2}$ are mapped to $\sigma_1$ and $\tau_1$ by \eqref{eq:resW2}, this choice does not affect \eqref{eq:L2cqed}.

\item
$\widehat{\mathfrak{su}}(6)_1\hookleftconf \widehat{\mathfrak{su}}(2)_3\oplus \widehat{\mathfrak{su}}(3)_2$
\begin{align}
\begin{split}
    &0\to(0,[0,0])+(1,[1,1])\\
    &1\to(1/2,[1,0])+(3/2,[0,2])\\
    &2\to(0,[2,0])+(1,[0,1])\\
    &3\to(1/2,[1,1])+(3/2,[0,0])\\
    &4\to(0,[0,2])+(1,[1,0])\\
    &5\to(1/2,[0,1])+(3/2,[2,0])
\end{split}
\end{align}

\item
$\widehat{\mathfrak{su}}(2)_1\oplus \widehat{\mathfrak{su}}(2)_3\hookleftconf \widehat{\mathfrak{su}}(2)_4\oplus \mathrm{Vir}_{c=4/5}$
\begin{align}
\begin{split}
    &(0,0)\to(0,\phi_{1,1})+(1,\f_{1,3})+(2,\f_{1,5})\\
    &(1/2,0)\to(1/2,\f_{1,2})+(3/2,\f_{1,4})\\
    &(0,1/2)\to(1/2,\f_{2,2})+(3/2,\f_{2,4})\\
    &(1/2,1/2)\to(0,\f_{2,1})+(1,\f_{2,3})+(2,\f_{2,5})\\
    &(0,1)\to(0,\f_{2,5})+(1,\f_{2,3})+(2,\f_{2,1})\\
    &(1/2,1)\to(1/2,\f_{2,4})+(3/2,\f_{2,2})\\
    &(0,3/2)\to(1/2,\f_{1,4})+(3/2,\f_{1,2})\\
    &(1/2,3/2)\to(0,\f_{1,5})+(1,\f_{1,3})+(2,\f_{1,1})
\end{split}
\end{align}

\item
$\widehat{\mathfrak{su}}(3)_1\oplus \widehat{\mathfrak{su}}(3)_1\hookleftconf \widehat{\mathfrak{su}}(3)_2 \oplus \mathsf{3SP}$
\begin{align}
\begin{split}
    &([0,0],[0,0])
    \to ([0,0],\mathbf{1})+([1,1],\epsilon)\\
    &([1,0],[0,0])
    \to ([1,0],\sigma)+([0,2],\psi)\\
    &([0,0],[1,0])
    \to ([1,0],\sigma^\dagger)+([0,2],\psi^\dagger)\\
    &([0,1],[0,0])
    \to ([0,1],\sigma^\dagger)+([2,0],\psi^\dagger)\\
    &([0,0],[0,1])
    \to ([0,1],\sigma)+([2,0],\psi)\\
    &([1,0],[1,0])
    \to ([2,0],\mathbf{1})+([0,1],\epsilon)\\
    &([0,1],[0,1])
    \to ([0,2],\mathbf{1})+([1,0],\epsilon)\\
    &([1,0],[0,1])
    \to ([0,0],\psi^\dagger)+([1,1],\sigma^\dagger)\\
    &([0,1],[1,0])
    \to ([0,0],\psi)+([1,1],\sigma)
\end{split}
\end{align}

\item
$\mathsf{3SP} \hookleftconf \mathrm{Vir}_{c=4/5}$
\begin{align}
\begin{split}
    &\mathbf{1}\to \phi_{1,1} + \phi_{1,5}\\
    &\epsilon\to \phi_{2,1} + \phi_{2,5}\\
    &\psi\to \phi_{1,3}\\
    &\psi^\dagger\to \phi_{1,3}\\
    &\sigma\to \phi_{2,3}\\
    &\sigma^\dagger\to \phi_{2,3}
\end{split}
\end{align}
\end{enumerate}

\bibliographystyle{JHEP}
\bibliography{ref}

\providecommand{\href}[2]{#2}\begingroup\raggedright\begin{thebibliography}{100}

\bibitem{tHooft:1979rat}
G.~'t~Hooft, {\it {Naturalness, chiral symmetry, and spontaneous chiral symmetry breaking}},  {\em NATO Sci. Ser. B} {\bf 59} (1980) 135--157.

\bibitem{Thorngren:2019iar}
R.~Thorngren and Y.~Wang, {\it {Fusion category symmetry. Part I. Anomaly in-flow and gapped phases}},  {\em JHEP} {\bf 04} (2024) 132, [\href{http://arxiv.org/abs/1912.02817}{{\tt arXiv:1912.02817}}].

\bibitem{Choi:2023xjw}
Y.~Choi, B.~C. Rayhaun, Y.~Sanghavi, and S.-H. Shao, {\it {Remarks on boundaries, anomalies, and noninvertible symmetries}},  {\em Phys. Rev. D} {\bf 108} (2023), no.~12 125005, [\href{http://arxiv.org/abs/2305.09713}{{\tt arXiv:2305.09713}}].

\bibitem{Zhang:2023wlu}
C.~Zhang and C.~C{\'o}rdova, {\it {Anomalies of (1+1)-dimensional categorical symmetries}},  {\em Phys. Rev. B} {\bf 110} (2024), no.~3 035155, [\href{http://arxiv.org/abs/2304.01262}{{\tt arXiv:2304.01262}}].

\bibitem{Cordova:2023bja}
C.~Cordova, P.-S. Hsin, and C.~Zhang, {\it {Anomalies of non-invertible symmetries in (3+1)d}},  {\em SciPost Phys.} {\bf 17} (2024), no.~5 131, [\href{http://arxiv.org/abs/2308.11706}{{\tt arXiv:2308.11706}}].

\bibitem{LIEB1961407}
E.~Lieb, T.~Schultz, and D.~Mattis, {\it Two soluble models of an antiferromagnetic chain},  {\em Annals of Physics} {\bf 16} (1961), no.~3 407--466.

\bibitem{Hastings_2004}
M.~B. Hastings, {\it {Lieb-Schultz-Mattis in higher dimensions}},  {\em Phys. Rev. B} {\bf 69} (2004) 104431, [\href{http://arxiv.org/abs/cond-mat/0305505}{{\tt cond-mat/0305505}}].

\bibitem{Yao:2020xcm}
Y.~Yao and M.~Oshikawa, {\it {Twisted boundary condition and Lieb-Schultz-Mattis ingappability for discrete symmetries}},  {\em Phys. Rev. Lett.} {\bf 126} (2021), no.~21 217201, [\href{http://arxiv.org/abs/2010.09244}{{\tt arXiv:2010.09244}}].

\bibitem{Oshikawa_2000}
M.~Oshikawa, {\it {Commensurability, Excitation Gap, and Topology in Quantum Many-Particle Systems on a Periodic Lattice}},  {\em Phys. Rev. Lett.} {\bf 84} (2000), no.~7 1535.

\bibitem{SeibergString2019Talk}
N.~Seiberg, {\it {Thoughts about Quantum Field Theories}},  {\em Talk at Strings 2019}.

\bibitem{Wang:2013yta}
J.~Wang and X.-G. Wen, {\it {Nonperturbative regularization of (1+1)-dimensional anomaly-free chiral fermions and bosons: On the equivalence of anomaly matching conditions and boundary gapping rules}},  {\em Phys. Rev. B} {\bf 107} (2023), no.~1 014311, [\href{http://arxiv.org/abs/1307.7480}{{\tt arXiv:1307.7480}}].

\bibitem{SMGtalk}
P.~Boyle~Smith, {\it {SMG=MG}},  {\em Symmetric mass generation, topological phases and lattice chiral gauge theories, SCGP workshop} (2024).

\bibitem{Mouland:2025dct}
R.~Mouland, D.~Tong, and B.~Zan, {\it {Phases of 2d gauge theories and symmetric mass generation}},  {\em JHEP} {\bf 04} (2026) 154, [\href{http://arxiv.org/abs/2509.12305}{{\tt arXiv:2509.12305}}].

\bibitem{Tong:2021phe}
D.~Tong, {\it {Comments on symmetric mass generation in 2d and 4d}},  {\em JHEP} {\bf 07} (2022) 001, [\href{http://arxiv.org/abs/2104.03997}{{\tt arXiv:2104.03997}}].

\bibitem{Wang:2022ucy}
J.~Wang and Y.-Z. You, {\it {Symmetric Mass Generation}},  {\em Symmetry} {\bf 14} (2022), no.~7 1475, [\href{http://arxiv.org/abs/2204.14271}{{\tt arXiv:2204.14271}}].

\bibitem{BoyleSmith:2022duw}
P.~Boyle~Smith, {\em {Conformal Boundaries, SPTs, and the Monopole-Fermion Problem}}.
\newblock PhD thesis, Cambridge U., DAMTP, 2022.

\bibitem{Razamat:2020kyf}
S.~S. Razamat and D.~Tong, {\it {Gapped Chiral Fermions}},  {\em Phys. Rev. X} {\bf 11} (2021), no.~1 011063, [\href{http://arxiv.org/abs/2009.05037}{{\tt arXiv:2009.05037}}].

\bibitem{Zeng:2022grc}
M.~Zeng, Z.~Zhu, J.~Wang, and Y.-Z. You, {\it {Symmetric Mass Generation in the 1+1 Dimensional Chiral Fermion 3-4-5-0 Model}},  {\em Phys. Rev. Lett.} {\bf 128} (2022), no.~18 185301, [\href{http://arxiv.org/abs/2202.12355}{{\tt arXiv:2202.12355}}].

\bibitem{Han:2017hdv}
B.~Han, A.~Tiwari, C.-T. Hsieh, and S.~Ryu, {\it {Boundary conformal field theory and symmetry protected topological phases in $2+1$ dimensions}},  {\em Phys. Rev. B} {\bf 96} (2017), no.~12 125105, [\href{http://arxiv.org/abs/1704.01193}{{\tt arXiv:1704.01193}}].

\bibitem{Jensen:2017eof}
K.~Jensen, E.~Shaverin, and A.~Yarom, {\it {{\textquoteright}t Hooft anomalies and boundaries}},  {\em JHEP} {\bf 01} (2018) 085, [\href{http://arxiv.org/abs/1710.07299}{{\tt arXiv:1710.07299}}].

\bibitem{Numasawa:2017crf}
T.~Numasawa and S.~Yamaguchi, {\it {Mixed global anomalies and boundary conformal field theories}},  {\em JHEP} {\bf 11} (2018) 202, [\href{http://arxiv.org/abs/1712.09361}{{\tt arXiv:1712.09361}}].

\bibitem{Smith:2020nuf}
P.~B. Smith and D.~Tong, {\it {What Symmetries are Preserved by a Fermion Boundary State?}},  \href{http://arxiv.org/abs/2006.07369}{{\tt arXiv:2006.07369}}.

\bibitem{Thorngren:2020yht}
R.~Thorngren and Y.~Wang, {\it {Anomalous symmetries end at the boundary}},  {\em JHEP} {\bf 09} (2021) 017, [\href{http://arxiv.org/abs/2012.15861}{{\tt arXiv:2012.15861}}].

\bibitem{Hellerman:2021fla}
S.~Hellerman, D.~Orlando, and M.~Watanabe, {\it {Quantum Information Theory of the Gravitational Anomaly}},  \href{http://arxiv.org/abs/2101.03320}{{\tt arXiv:2101.03320}}.

\bibitem{Li:2022drc}
L.~Li, C.-T. Hsieh, Y.~Yao, and M.~Oshikawa, {\it {Boundary conditions and anomalies of conformal field theories in 1+1 dimensions}},  {\em Phys. Rev. B} {\bf 110} (2024), no.~4 045118, [\href{http://arxiv.org/abs/2205.11190}{{\tt arXiv:2205.11190}}].

\bibitem{Wei:2025zyd}
P.~Wei and Y.~Zheng, {\it {On the Absence of Symmetric Simple Conformal Boundary Conditions}},  \href{http://arxiv.org/abs/2512.23976}{{\tt arXiv:2512.23976}}.

\bibitem{Smith:2019jnh}
P.~B. Smith and D.~Tong, {\it {Boundary States for Chiral Symmetries in Two Dimensions}},  {\em JHEP} {\bf 09} (2020) 018, [\href{http://arxiv.org/abs/1912.01602}{{\tt arXiv:1912.01602}}].

\bibitem{vanBeest:2023dbu}
M.~van Beest, P.~Boyle~Smith, D.~Delmastro, Z.~Komargodski, and D.~Tong, {\it {Monopoles, scattering, and generalized symmetries}},  {\em JHEP} {\bf 03} (2025) 014, [\href{http://arxiv.org/abs/2306.07318}{{\tt arXiv:2306.07318}}].

\bibitem{vanBeest:2023mbs}
M.~van Beest, P.~Boyle~Smith, D.~Delmastro, R.~Mouland, and D.~Tong, {\it {Fermion-monopole scattering in the Standard Model}},  {\em JHEP} {\bf 08} (2024) 004, [\href{http://arxiv.org/abs/2312.17746}{{\tt arXiv:2312.17746}}].

\bibitem{Wei:2026fsn}
P.~Wei and Y.~Zheng, {\it {Non-invertible Symmetries in Weyl Fermions, and Applications to Fermion-Boundary Scattering Problem}},  \href{http://arxiv.org/abs/2605.19363}{{\tt arXiv:2605.19363}}.

\bibitem{Arias-Tamargo:2026urw}
G.~Arias-Tamargo, P.~Boyle~Smith, R.~Mouland, and M.~L. Vel{\'a}squez Cotini~Hutt, {\it {Non-Invertible Symmetries and Boundaries for Two-Dimensional Fermions}},  \href{http://arxiv.org/abs/2605.13952}{{\tt arXiv:2605.13952}}.

\bibitem{Callan:1982ah}
C.~G. Callan, Jr., {\it {Disappearing Dyons}},  {\em Phys. Rev. D} {\bf 25} (1982) 2141.

\bibitem{Callan:1982au}
C.~G. Callan, Jr., {\it {Dyon-Fermion Dynamics}},  {\em Phys. Rev. D} {\bf 26} (1982) 2058--2068.

\bibitem{Rubakov:1982fp}
V.~A. Rubakov, {\it {Adler-Bell-Jackiw Anomaly and Fermion Number Breaking in the Presence of a Magnetic Monopole}},  {\em Nucl. Phys. B} {\bf 203} (1982) 311--348.

\bibitem{Callan:1982ac}
C.~G. Callan, Jr., {\it {Monopole Catalysis of Baryon Decay}},  {\em Nucl. Phys. B} {\bf 212} (1983) 391--400.

\bibitem{Polchinski:1984uw}
J.~Polchinski, {\it {Monopole Catalysis: The Fermion Rotor System}},  {\em Nucl. Phys. B} {\bf 242} (1984) 345--363.

\bibitem{Affleck:1993np}
I.~Affleck and J.~Sagi, {\it {Monopole catalyzed baryon decay: A Boundary conformal field theory approach}},  {\em Nucl. Phys. B} {\bf 417} (1994) 374--402, [\href{http://arxiv.org/abs/hep-th/9311056}{{\tt hep-th/9311056}}].

\bibitem{Maldacena:1995pq}
J.~M. Maldacena and A.~W.~W. Ludwig, {\it {Majorana fermions, exact mapping between quantum impurity fixed points with four bulk fermion species, and solution of the `unitarity puzzle'}},  {\em Nucl. Phys. B} {\bf 506} (1997) 565--588, [\href{http://arxiv.org/abs/cond-mat/9502109}{{\tt cond-mat/9502109}}].

\bibitem{Csaki:2021ozp}
C.~Cs\'aki, Y.~Shirman, O.~Telem, and J.~Terning, {\it {Pairwise Multiparticle States and the Monopole Unitarity Puzzle}},  {\em Phys. Rev. Lett.} {\bf 129} (2022), no.~18 181601, [\href{http://arxiv.org/abs/2109.01145}{{\tt arXiv:2109.01145}}].

\bibitem{Kitano:2021pwt}
R.~Kitano and R.~Matsudo, {\it {Missing final state puzzle in the monopole-fermion scattering}},  {\em Phys. Lett. B} {\bf 832} (2022) 137271, [\href{http://arxiv.org/abs/2103.13639}{{\tt arXiv:2103.13639}}].

\bibitem{Brennan:2021ewu}
T.~D. Brennan, {\it {Callan-Rubakov effect and higher charge monopoles}},  {\em JHEP} {\bf 02} (2023) 159, [\href{http://arxiv.org/abs/2109.11207}{{\tt arXiv:2109.11207}}].

\bibitem{Hamada:2022eiv}
Y.~Hamada, T.~Kitahara, and Y.~Sato, {\it {Monopole-fermion scattering and varying Fock space}},  {\em JHEP} {\bf 11} (2022) 116, [\href{http://arxiv.org/abs/2208.01052}{{\tt arXiv:2208.01052}}].

\bibitem{Brennan:2023tae}
T.~D. Brennan, {\it {A new solution to the Callan Rubakov effect}},  {\em JHEP} {\bf 11} (2024) 170, [\href{http://arxiv.org/abs/2309.00680}{{\tt arXiv:2309.00680}}].

\bibitem{Khoze:2024hlb}
V.~V. Khoze, {\it {Monopoles and fermions in the Standard Model}},  {\em JHEP} {\bf 09} (2024) 146, [\href{http://arxiv.org/abs/2405.18689}{{\tt arXiv:2405.18689}}].

\bibitem{Loladze:2024ayk}
V.~Loladze and T.~Okui, {\it {Monopole-Fermion Scattering and the Solution to the Semiton{\textendash}Unitarity Puzzle}},  {\em Phys. Rev. Lett.} {\bf 134} (2025), no.~5 051602, [\href{http://arxiv.org/abs/2408.04577}{{\tt arXiv:2408.04577}}].

\bibitem{Antinucci:2024izg}
A.~Antinucci, C.~Copetti, G.~Galati, and G.~Rizi, {\it {Topological constraints on defect dynamics}},  {\em Phys. Rev. D} {\bf 111} (2025), no.~6 065025, [\href{http://arxiv.org/abs/2412.18652}{{\tt arXiv:2412.18652}}].

\bibitem{Loladze:2025jsq}
V.~Loladze, T.~Okui, and D.~Tong, {\it {Dynamics of the fermion-rotor system}},  {\em JHEP} {\bf 01} (2026) 052, [\href{http://arxiv.org/abs/2508.21059}{{\tt arXiv:2508.21059}}].

\bibitem{Tachikawa:2026cxd}
Y.~Tachikawa, K.~Tsuji, and M.~Watanabe, {\it {What happens to wavepackets of fermions when scattered by the Maldacena-Ludwig wall?}},  {\em JHEP} {\bf 10} (2026) 140, [\href{http://arxiv.org/abs/2603.25508}{{\tt arXiv:2603.25508}}].

\bibitem{Hamada:2026jgj}
Y.~Hamada, {\it {A Rotor-Dressed Semiton State in a Monopole--Fermion Model}},  \href{http://arxiv.org/abs/2608.18345}{{\tt arXiv:2608.18345}}.

\bibitem{Antinucci:2026uuh}
A.~Antinucci, C.~Copetti, G.~Galati, and G.~Rizi, {\it {A Twist on Scattering from Defect Anomalies}},  \href{http://arxiv.org/abs/2605.13961}{{\tt arXiv:2605.13961}}.

\bibitem{Craig:2026rdf}
N.~Craig and D.~Sehayek, {\it {Monopole Catalysis from Standard Model Anomalies}},  \href{http://arxiv.org/abs/2610.07150}{{\tt arXiv:2610.07150}}.

\bibitem{PathToQFTTalk}
D.~Tong, {\it {What I Don’t Understand About Monopole-Fermion Scattering}},  {\em Paths to Quantum Field Theory: June 22-26, 2026, SCGP workshop} (2026).

\bibitem{Moller:2024xtt}
S.~M{\"o}ller and B.~C. Rayhaun, {\it {Equivalence Relations on Vertex Operator Algebras, II: Witt Equivalence and Orbifolds}},  \href{http://arxiv.org/abs/2410.18166}{{\tt arXiv:2410.18166}}.

\bibitem{Oshikawa:1996ww}
M.~Oshikawa and I.~Affleck, {\it {Defect lines in the Ising model and boundary states on orbifolds}},  {\em Phys. Rev. Lett.} {\bf 77} (1996) 2604--2607, [\href{http://arxiv.org/abs/hep-th/9606177}{{\tt hep-th/9606177}}].

\bibitem{Bachas:2001vj}
C.~Bachas, J.~de~Boer, R.~Dijkgraaf, and H.~Ooguri, {\it {Permeable conformal walls and holography}},  {\em JHEP} {\bf 06} (2002) 027, [\href{http://arxiv.org/abs/hep-th/0111210}{{\tt hep-th/0111210}}].

\bibitem{Cordova:2025zkz}
C.~Cordova, D.~Garc{\'\i}a-Sep{\'u}lveda, and J.~A. Harvey, {\it {Generalized level-rank duality, holomorphic conformal field theory, and non-invertible anyon condensation}},  {\em JHEP} {\bf 09} (2026) 009, [\href{http://arxiv.org/abs/2512.24419}{{\tt arXiv:2512.24419}}].

\bibitem{Witten:1988hf}
E.~Witten, {\it {Quantum Field Theory and the Jones Polynomial}},  {\em Commun. Math. Phys.} {\bf 121} (1989) 351--399.

\bibitem{Elitzur:1989nr}
S.~Elitzur, G.~W. Moore, A.~Schwimmer, and N.~Seiberg, {\it {Remarks on the Canonical Quantization of the Chern-Simons-Witten Theory}},  {\em Nucl. Phys. B} {\bf 326} (1989) 108--134.

\bibitem{Felder:1999cv}
G.~Felder, J.~Frohlich, J.~Fuchs, and C.~Schweigert, {\it {Conformal boundary conditions and three-dimensional topological field theory}},  {\em Phys. Rev. Lett.} {\bf 84} (2000) 1659--1662, [\href{http://arxiv.org/abs/hep-th/9909140}{{\tt hep-th/9909140}}].

\bibitem{Fuchs:2002cm}
J.~Fuchs, I.~Runkel, and C.~Schweigert, {\it {TFT construction of RCFT correlators 1. Partition functions}},  {\em Nucl. Phys. B} {\bf 646} (2002) 353--497, [\href{http://arxiv.org/abs/hep-th/0204148}{{\tt hep-th/0204148}}].

\bibitem{Kapustin:2010if}
A.~Kapustin and N.~Saulina, {\it {Surface operators in 3d Topological Field Theory and 2d Rational Conformal Field Theory}},  \href{http://arxiv.org/abs/1012.0911}{{\tt arXiv:1012.0911}}.

\bibitem{Kong:2017hcw}
L.~Kong, X.-G. Wen, and H.~Zheng, {\it {Boundary-bulk relation in topological orders}},  {\em Nucl. Phys. B} {\bf 922} (2017) 62--76, [\href{http://arxiv.org/abs/1702.00673}{{\tt arXiv:1702.00673}}].

\bibitem{Kong:2019byq}
L.~Kong and H.~Zheng, {\it {A mathematical theory of gapless edges of 2d topological orders. Part I}},  {\em JHEP} {\bf 02} (2020) 150, [\href{http://arxiv.org/abs/1905.04924}{{\tt arXiv:1905.04924}}].

\bibitem{Rayhaun:2023pgc}
B.~C. Rayhaun, {\it {Bosonic rational conformal field theories in small genera, chiral fermionization, and symmetry/subalgebra duality}},  {\em J. Math. Phys.} {\bf 65} (2024), no.~5 052301, [\href{http://arxiv.org/abs/2303.16921}{{\tt arXiv:2303.16921}}].

\bibitem{Gannon:2026ttf}
T.~Gannon and B.~C. Rayhaun, {\it {Hypergroup Symmetry in Relative Quantum Field Theories and Chiral Algebras}},  \href{http://arxiv.org/abs/2606.05279}{{\tt arXiv:2606.05279}}.

\bibitem{Arias-Tamargo:2026oct}
G.~Arias-Tamargo, P.~Boyle~Smith, R.~Mouland, and M.~L. Velásquez Cotini~Hutt, {\it {Twisted scattering from Chern--Simons theory}},  {\em to appear} (2026).

\bibitem{Koundouros:2026oct}
G.~Koundouros and D.~Tong, {\it {Fermions Scattering Off Non-Abelian Rotors}},  {\em to appear} (2026).

\bibitem{WatanabeOhmori}
K.~Ohmori and M.~Watanabe, {\it {Unpublished}}, .

\bibitem{Watanabetalk}
M.~Watanabe, {\it {IPMU seminar: Callan-Rubakov effect and non-invertible defects.}}, .

\bibitem{Gaiotto:2014kfa}
D.~Gaiotto, A.~Kapustin, N.~Seiberg, and B.~Willett, {\it {Generalized Global Symmetries}},  {\em JHEP} {\bf 02} (2015) 172, [\href{http://arxiv.org/abs/1412.5148}{{\tt arXiv:1412.5148}}].

\bibitem{Etingof:2009yvg}
P.~Etingof, D.~Nikshych, V.~Ostrik, and w.~a. a. b.~E. Meir, {\it {Fusion categories and homotopy theory}},  \href{http://arxiv.org/abs/0909.3140}{{\tt arXiv:0909.3140}}.

\bibitem{Davydov:2010kfz}
A.~Davydov, M.~Mueger, D.~Nikshych, and V.~Ostrik, {\it {The Witt group of non-degenerate braided fusion categories}},  \href{http://arxiv.org/abs/1009.2117}{{\tt arXiv:1009.2117}}.

\bibitem{Fuchs:2012dt}
J.~Fuchs, C.~Schweigert, and A.~Valentino, {\it {Bicategories for boundary conditions and for surface defects in 3-d TFT}},  {\em Commun. Math. Phys.} {\bf 321} (2013) 543--575, [\href{http://arxiv.org/abs/1203.4568}{{\tt arXiv:1203.4568}}].

\bibitem{Petkova:2000ip}
V.~B. Petkova and J.~B. Zuber, {\it {Generalized twisted partition functions}},  {\em Phys. Lett. B} {\bf 504} (2001) 157--164, [\href{http://arxiv.org/abs/hep-th/0011021}{{\tt hep-th/0011021}}].

\bibitem{Frohlich:2004ef}
J.~Frohlich, J.~Fuchs, I.~Runkel, and C.~Schweigert, {\it {Kramers-Wannier duality from conformal defects}},  {\em Phys. Rev. Lett.} {\bf 93} (2004) 070601, [\href{http://arxiv.org/abs/cond-mat/0404051}{{\tt cond-mat/0404051}}].

\bibitem{Frohlich:2006ch}
J.~Frohlich, J.~Fuchs, I.~Runkel, and C.~Schweigert, {\it {Duality and defects in rational conformal field theory}},  {\em Nucl. Phys. B} {\bf 763} (2007) 354--430, [\href{http://arxiv.org/abs/hep-th/0607247}{{\tt hep-th/0607247}}].

\bibitem{Kong:2013aya}
L.~Kong, {\it {Anyon condensation and tensor categories}},  {\em Nucl. Phys. B} {\bf 886} (2014) 436--482, [\href{http://arxiv.org/abs/1307.8244}{{\tt arXiv:1307.8244}}].

\bibitem{Affleck:1995ge}
I.~Affleck, {\it {Conformal field theory approach to the Kondo effect}},  {\em Acta Phys. Polon. B} {\bf 26} (1995) 1869--1932, [\href{http://arxiv.org/abs/cond-mat/9512099}{{\tt cond-mat/9512099}}].

\bibitem{Affleck:1990iv}
I.~Affleck and A.~W.~W. Ludwig, {\it {Critical theory of overscreened Kondo fixed points}},  {\em Nucl. Phys. B} {\bf 360} (1991) 641--696.

\bibitem{Gaiotto:2020iye}
D.~Gaiotto and J.~Kulp, {\it {Orbifold groupoids}},  {\em JHEP} {\bf 02} (2021) 132, [\href{http://arxiv.org/abs/2008.05960}{{\tt arXiv:2008.05960}}].

\bibitem{Freed:2022qnc}
D.~S. Freed, G.~W. Moore, and C.~Teleman, {\it {Topological symmetry in quantum field theory}},  \href{http://arxiv.org/abs/2209.07471}{{\tt arXiv:2209.07471}}.

\bibitem{Apruzzi:2021nmk}
F.~Apruzzi, F.~Bonetti, I.~Garc{\'\i}a~Etxebarria, S.~S. Hosseini, and S.~Schafer-Nameki, {\it {Symmetry TFTs from String Theory}},  {\em Commun. Math. Phys.} {\bf 402} (2023), no.~1 895--949, [\href{http://arxiv.org/abs/2112.02092}{{\tt arXiv:2112.02092}}].

\bibitem{Ji:2019jhk}
W.~Ji and X.-G. Wen, {\it {Categorical symmetry and noninvertible anomaly in symmetry-breaking and topological phase transitions}},  {\em Phys. Rev. Res.} {\bf 2} (2020), no.~3 033417, [\href{http://arxiv.org/abs/1912.13492}{{\tt arXiv:1912.13492}}].

\bibitem{etingof2009fusioncategorieshomotopytheory}
P.~Etingof, D.~Nikshych, V.~Ostrik, and w.~a. a. b.~E. Meir, {\it {Fusion categories and homotopy theory}},  \href{http://arxiv.org/abs/0909.3140}{{\tt arXiv:0909.3140}}.

\bibitem{Choi:2023vgk}
Y.~Choi, D.-C. Lu, and Z.~Sun, {\it {Self-duality under gauging a non-invertible symmetry}},  {\em JHEP} {\bf 01} (2024) 142, [\href{http://arxiv.org/abs/2310.19867}{{\tt arXiv:2310.19867}}].

\bibitem{Diatlyk:2023fwf}
O.~Diatlyk, C.~Luo, Y.~Wang, and Q.~Weller, {\it {Gauging non-invertible symmetries: topological interfaces and generalized orbifold groupoid in 2d QFT}},  {\em JHEP} {\bf 03} (2024) 127, [\href{http://arxiv.org/abs/2311.17044}{{\tt arXiv:2311.17044}}].

\bibitem{Putrov:2024uor}
P.~Putrov and R.~Radhakrishnan, {\it {Non-anomalous non-invertible symmetries in 1+1D from gapped boundaries of SymTFTs}},  {\em JHEP} {\bf 03} (2026) 077, [\href{http://arxiv.org/abs/2405.04619}{{\tt arXiv:2405.04619}}].

\bibitem{Choi:2024tri}
Y.~Choi, B.~C. Rayhaun, and Y.~Zheng, {\it {Generalized Tube Algebras, Symmetry-Resolved Partition Functions, and Twisted Boundary States}},  {\em Commun. Math. Phys.} {\bf 407} (2026), no.~4 62, [\href{http://arxiv.org/abs/2409.02159}{{\tt arXiv:2409.02159}}].

\bibitem{Choi:2024wfm}
Y.~Choi, B.~C. Rayhaun, and Y.~Zheng, {\it {Noninvertible Symmetry-Resolved Affleck-Ludwig-Cardy Formula and Entanglement Entropy from the Boundary Tube Algebra}},  {\em Phys. Rev. Lett.} {\bf 133} (2024), no.~25 251602, [\href{http://arxiv.org/abs/2409.02806}{{\tt arXiv:2409.02806}}].

\bibitem{Huang:2023pyk}
S.-J. Huang and M.~Cheng, {\it {Topological holography, quantum criticality, and boundary states}},  {\em SciPost Phys.} {\bf 18} (2025), no.~6 213, [\href{http://arxiv.org/abs/2310.16878}{{\tt arXiv:2310.16878}}].

\bibitem{Fuchs:1999zi}
J.~Fuchs and C.~Schweigert, {\it {Symmetry breaking boundaries. 1. General theory}},  {\em Nucl. Phys. B} {\bf 558} (1999) 419--483, [\href{http://arxiv.org/abs/hep-th/9902132}{{\tt hep-th/9902132}}].

\bibitem{Fuchs:1999xn}
J.~Fuchs and C.~Schweigert, {\it {Symmetry breaking boundaries. 2. More structures: Examples}},  {\em Nucl. Phys. B} {\bf 568} (2000) 543--593, [\href{http://arxiv.org/abs/hep-th/9908025}{{\tt hep-th/9908025}}].

\bibitem{Maldacena:2001ky}
J.~M. Maldacena, G.~W. Moore, and N.~Seiberg, {\it {Geometrical interpretation of D-branes in gauged WZW models}},  {\em JHEP} {\bf 07} (2001) 046, [\href{http://arxiv.org/abs/hep-th/0105038}{{\tt hep-th/0105038}}].

\bibitem{Quella:2002ct}
T.~Quella and V.~Schomerus, {\it {Symmetry breaking boundary states and defect lines}},  {\em JHEP} {\bf 06} (2002) 028, [\href{http://arxiv.org/abs/hep-th/0203161}{{\tt hep-th/0203161}}].

\bibitem{Quella:2002fk}
T.~Quella and V.~Schomerus, {\it {Asymmetric cosets}},  {\em JHEP} {\bf 02} (2003) 030, [\href{http://arxiv.org/abs/hep-th/0212119}{{\tt hep-th/0212119}}].

\bibitem{Cordova:2023jip}
C.~Cordova and D.~Garc{\'\i}a-Sep{\'u}lveda, {\it {Non-Invertible Anyon Condensation and Level-Rank Dualities}},  {\em SciPost Phys.} {\bf 21} (2026) 089, [\href{http://arxiv.org/abs/2312.16317}{{\tt arXiv:2312.16317}}].

\bibitem{Roumpedakis:2022aik}
K.~Roumpedakis, S.~Seifnashri, and S.-H. Shao, {\it {Higher Gauging and Non-invertible Condensation Defects}},  {\em Commun. Math. Phys.} {\bf 401} (2023), no.~3 3043--3107, [\href{http://arxiv.org/abs/2204.02407}{{\tt arXiv:2204.02407}}].

\bibitem{Teo:2015xla}
J.~C.~Y. Teo, T.~L. Hughes, and E.~Fradkin, {\it {Theory of Twist Liquids: Gauging an Anyonic Symmetry}},  {\em Annals Phys.} {\bf 360} (2015) 349--445, [\href{http://arxiv.org/abs/1503.06812}{{\tt arXiv:1503.06812}}].

\bibitem{Kaidi:2022cpf}
J.~Kaidi, K.~Ohmori, and Y.~Zheng, {\it {Symmetry TFTs for Non-invertible Defects}},  {\em Commun. Math. Phys.} {\bf 404} (2023), no.~2 1021--1124, [\href{http://arxiv.org/abs/2209.11062}{{\tt arXiv:2209.11062}}].

\bibitem{Bhardwaj:2023bbf}
L.~Bhardwaj, L.~E. Bottini, D.~Pajer, and S.~Schafer-Nameki, {\it {The club sandwich: Gapless phases and phase transitions with non-invertible symmetries}},  {\em SciPost Phys.} {\bf 18} (2025), no.~5 156, [\href{http://arxiv.org/abs/2312.17322}{{\tt arXiv:2312.17322}}].

\bibitem{Bhardwaj:2024qrf}
L.~Bhardwaj, D.~Pajer, S.~Schafer-Nameki, and A.~Warman, {\it {Hasse diagrams for gapless SPT and SSB phases with non-invertible symmetries}},  {\em SciPost Phys.} {\bf 19} (2025), no.~4 113, [\href{http://arxiv.org/abs/2403.00905}{{\tt arXiv:2403.00905}}].

\bibitem{ostrik2003module}
V.~Ostrik, {\it Module categories, weak hopf algebras and modular invariants},  {\em Transformation groups} {\bf 8} (2003) 177--206.

\bibitem{Carnahan:2016guf}
S.~Carnahan and M.~Miyamoto, {\it {Regularity of fixed-point vertex operator subalgebras}},  \href{http://arxiv.org/abs/1603.05645}{{\tt arXiv:1603.05645}}.

\bibitem{Benini:2018reh}
F.~Benini, C.~C{\'o}rdova, and P.-S. Hsin, {\it {On 2-Group Global Symmetries and their Anomalies}},  {\em JHEP} {\bf 03} (2019) 118, [\href{http://arxiv.org/abs/1803.09336}{{\tt arXiv:1803.09336}}].

\bibitem{Cordova:2018cvg}
C.~C{\'o}rdova, T.~T. Dumitrescu, and K.~Intriligator, {\it {Exploring 2-Group Global Symmetries}},  {\em JHEP} {\bf 02} (2019) 184, [\href{http://arxiv.org/abs/1802.04790}{{\tt arXiv:1802.04790}}].

\bibitem{Choi:2022jqy}
Y.~Choi, H.~T. Lam, and S.-H. Shao, {\it {Noninvertible Global Symmetries in the Standard Model}},  {\em Phys. Rev. Lett.} {\bf 129} (2022), no.~16 161601, [\href{http://arxiv.org/abs/2205.05086}{{\tt arXiv:2205.05086}}].

\bibitem{Cordova:2022ieu}
C.~Cordova and K.~Ohmori, {\it {Noninvertible Chiral Symmetry and Exponential Hierarchies}},  {\em Phys. Rev. X} {\bf 13} (2023), no.~1 011034, [\href{http://arxiv.org/abs/2205.06243}{{\tt arXiv:2205.06243}}].

\bibitem{Cordova:2024goh}
C.~Cordova and D.~Garc{\'\i}a-Sep{\'u}lveda, {\it {Topological cosets via anyon condensation and applications to gapped QCD$_{2}$}},  {\em JHEP} {\bf 08} (2026) 141, [\href{http://arxiv.org/abs/2412.01877}{{\tt arXiv:2412.01877}}].

\bibitem{Dunbar:1992gh}
D.~C. Dunbar and K.~G. Joshi, {\it {Characters for coset conformal field theories}},  {\em Int. J. Mod. Phys. A} {\bf 8} (1993) 4103--4122, [\href{http://arxiv.org/abs/hep-th/9210122}{{\tt hep-th/9210122}}].

\bibitem{Dunbar:1993hr}
D.~C. Dunbar and K.~G. Joshi, {\it {Maverick examples of coset conformal field theories}},  {\em Mod. Phys. Lett. A} {\bf 8} (1993) 2803--2814, [\href{http://arxiv.org/abs/hep-th/9309093}{{\tt hep-th/9309093}}].

\bibitem{Pedrini:1999iy}
B.~Pedrini, C.~Schweigert, and J.~Walcher, {\it {New maverick coset theories}},  {\em Phys. Lett. B} {\bf 466} (1999) 206--210, [\href{http://arxiv.org/abs/hep-th/9908185}{{\tt hep-th/9908185}}].

\bibitem{Hasegawa:1989}
K.~Hasegawa, {\it {Spin Module Versions of Weyl's Reciprocity Theorem for Classical Kac-Moody Lie Algebras --- An Application to Branching Rule Duality}},  {\em Publ. Res. Inst. Math. Sci.} {\bf 25} (1989), no.~5 741--828.

\bibitem{Naculich:1990hg}
S.~G. Naculich and H.~J. Schnitzer, {\it {Duality Between SU($N$)-k and SU(k)-$N$ {WZW} Models}},  {\em Nucl. Phys. B} {\bf 347} (1990) 687--742.

\bibitem{Hsin:2016blu}
P.-S. Hsin and N.~Seiberg, {\it {Level/rank Duality and Chern-Simons-Matter Theories}},  {\em JHEP} {\bf 09} (2016) 095, [\href{http://arxiv.org/abs/1607.07457}{{\tt arXiv:1607.07457}}].

\bibitem{Barkeshli:2012pr}
M.~Barkeshli, C.-M. Jian, and X.-L. Qi, {\it {Twist defects and projective non-Abelian braiding statistics}},  {\em Phys. Rev. B} {\bf 87} (2013), no.~4 045130, [\href{http://arxiv.org/abs/1208.4834}{{\tt arXiv:1208.4834}}].

\bibitem{Barkeshli:2014cna}
M.~Barkeshli, P.~Bonderson, M.~Cheng, and Z.~Wang, {\it {Symmetry Fractionalization, Defects, and Gauging of Topological Phases}},  {\em Phys. Rev. B} {\bf 100} (2019), no.~11 115147, [\href{http://arxiv.org/abs/1410.4540}{{\tt arXiv:1410.4540}}].

\bibitem{Hung:2013qpa}
L.-Y. Hung and Y.~Wan, {\it {Symmetry Enriched Phases via Pseudo Anyon Condensation}},  {\em Int. J. Mod. Phys. B} {\bf 28} (2014) 0172, [\href{http://arxiv.org/abs/1308.4673}{{\tt arXiv:1308.4673}}].

\bibitem{Cordova:2017vab}
C.~Cordova, P.-S. Hsin, and N.~Seiberg, {\it {Global Symmetries, Counterterms, and Duality in Chern-Simons Matter Theories with Orthogonal Gauge Groups}},  {\em SciPost Phys.} {\bf 4} (2018), no.~4 021, [\href{http://arxiv.org/abs/1711.10008}{{\tt arXiv:1711.10008}}].

\bibitem{Bais:1987zk}
F.~A. Bais, P.~Bouwknegt, M.~Surridge, and K.~Schoutens, {\it {Coset Construction for Extended Virasoro Algebras}},  {\em Nucl. Phys. B} {\bf 304} (1988) 371--391.

\bibitem{Bachas:2007td}
C.~Bachas and I.~Brunner, {\it {Fusion of conformal interfaces}},  {\em JHEP} {\bf 02} (2008) 085, [\href{http://arxiv.org/abs/0712.0076}{{\tt arXiv:0712.0076}}].

\bibitem{Ueda:2026wvc}
A.~Ueda, T.~Numasawa, B.~De~Vos, and M.~Watanabe, {\it {Emergent Andreev Reflection from a Lattice Duality Defect}},  {\em Phys. Rev. Lett.} {\bf 137} (2026), no.~14 146504, [\href{http://arxiv.org/abs/2606.23684}{{\tt arXiv:2606.23684}}].

\bibitem{Ueda:2025ecm}
A.~Ueda, V.~V. Linden, B.~De~Vos, L.~Lootens, J.~Haegeman, P.~Fendley, and F.~Verstraete, {\it {Perfect particle transmission through duality defects}},  {\em Nature Phys.} {\bf 22} (2026), no.~10 1684--1690, [\href{http://arxiv.org/abs/2510.26780}{{\tt arXiv:2510.26780}}].

\bibitem{Bonderson:2007ci}
P.~Bonderson, K.~Shtengel, and J.~K. Slingerland, {\it {Interferometry of non-Abelian Anyons}},  {\em Annals Phys.} {\bf 323} (2008) 2709--2755, [\href{http://arxiv.org/abs/0707.4206}{{\tt arXiv:0707.4206}}].

\bibitem{Affleck:1991tk}
I.~Affleck and A.~W.~W. Ludwig, {\it {Universal noninteger 'ground state degeneracy' in critical quantum systems}},  {\em Phys. Rev. Lett.} {\bf 67} (1991) 161--164.

\bibitem{Friedan:2003yc}
D.~Friedan and A.~Konechny, {\it {On the boundary entropy of one-dimensional quantum systems at low temperature}},  {\em Phys. Rev. Lett.} {\bf 93} (2004) 030402, [\href{http://arxiv.org/abs/hep-th/0312197}{{\tt hep-th/0312197}}].

\end{thebibliography}\endgroup

\end{document}